\documentclass[11pt] {article}
\usepackage{epsfig, graphics, rotating, color}
\def\lt{\raisebox{0.2ex}{$<$}}
\def\gt{\raisebox{0.2ex}{$>$}}

\begin{document}
\begin{titlepage}
\begin{center}
\vspace{1.0cm}

\begin{Large}
 {\bf Are Centauros  exotic signals of the QGP?}
\end{Large} 

\vspace{2.cm}
{Ewa
G\l{}adysz-Dziadu\'s}\\
 
\vspace{0.6cm}{\it
  Institute of Nuclear Physics, Cracow, Poland}\\

\vspace{2.5cm}
{\bf Abstract}
\end{center}
\vspace*{3mm}

Exotic cosmic ray events are revieved with  special emphasis on the
connection between their hadron rich composition and their strong
penetrability in matter. 
Theoretical attempts to explain the Centauro-like phenomena are
summarized. Results of accelerator experiments looking for
Centauro-related objects are described and discussed. The discussion is
completed with an overview of the current and future collider (RHIC and
LHC)
experiments which are expected  to produce and detect such exotic
objects in the laboratory. In particular the CASTOR detector, 
 a subsystem of the ALICE experiment dedicated  to
Centauro and  strange objects research in heavy ion collisions at LHC 
energies is presented in detail. Simulations of the  passage of
 Centauros and strangelets through the deep CASTOR calorimeter are
 shown.
\vspace*{0.5cm}


\vspace{0.4cm}
\end{titlepage}
\tableofcontents
\section
{Introduction}

The main purpose of this work is to indicate and discuss 
 some
{\it unconventional signatures of the quark-gluon plasma}. The idea arised
from the analysis of super-high energy cosmic ray events detected
in  emulsion chambers exposed at the mountain altitudes.
These data reveal many unexpected features which could be understood
by means of the quark-gluon plasma picture and hence they 
could be the new and unexplored so far field of the signs of that new
state
of  matter. The new accelerator experiments should not neglect
the chance of enriching the studied signals by incorporating the unusual
 phenomena observed in cosmic ray experiments to
 the investigated  ones, despite of many
uncertainties and even their mysterious aspects.

The title of this work is rather symbolic. Centauro is a creature
being in one half of a man and in the other half of a  horse, both
parts
don't fit one to the other. It is strange and mysterious, we are not sure
if and how 
could it match to the  real word. Morever, we are not sure if it exists
at all. But we tell and think about it. Centauro is  a good symbol of 
our current
knowledge about  the cosmic ray exotic events.

 In this paper  the experimental
situation  concerning the
Centauro events and  other related phenomena is reviewed,
 with intention
 to
emphasized
 a weakly known 
 relation between Centauro species and the so--called strongly
penetrating component. The work bases as well as on the quite recent
publications
and also on the very old ones, sometimes  forgotten preprints or
conference
proceedings. The volume of available information regarding to
this subject is extremely large,
 so I am sorry if 
something essential was omitted.

The first purpose of this work was to answer the following question:
were  the Centauro related phenomena  really observed or is it some kind
of photomorgana? Each reader should answer by himself. My personal
opinion is that despite of many experimental uncertainties, some 
mess in the data  and
difficulties in their interpretation, we really observe something new,
what is outside of the extrapolation of our present knowledge.
The  future accelerator experiments (RHIC, LHC) should take advantage
of using
these hints in their investigations.  

The second question  adressed  was: where and how to
look for these phenomena?
The compositness of the answer arises from the fact that many different
aspects, such as:
the naked experimental characteristics obtained from cosmic ray
experiments,
the so far negative results of
accelerator searches and also the theoretical speculations, all of them 
should be taken
into account. In connection with it, the past, present
and future 
accelerator Centauro searches are presented and  several different
models are described. Among the numerous
 attemts  to explain Centauros, 
  some models which could be tested 
in  accelerator experiments and in particular
  these based on the quark-gluon plasma idea were chosen to
this
review.

 The most attention is given  to the  scenario
of the Centauro   strange quark matter
fireball. Its production in nucleus-nucleus collisions 
and a subsequent evolution could possibly result in the
strangelet(s)
formation, in the strangeness distillation process. The passage of
strangelets through a deep emulsion chamber (calorimeter) should give
 the specific energy deposition pattern which could be
a very spectacular quark-gluon plasma signature. This signal is
independent
of the
strangelet charge (it allows to detect also strangelets with Z=0) and
promises to detect as well as long and  short-lived
strangelets, in contrary to  other experimentally used signatures,
based mainly on Z/A ratio and sensitive only to stable strangelets.

Both  experimental Centauro characteristics and  model predictions
indicate the forward rapidity region as the most favourable  place for
 production and detection of such anomalous phenomena. This is very
essential
statement in the context of the current and future experiments,
mostly concentrated on the exploration of the midrapidity region.
This point is also strongly forced in the paper, claiming the
CASTOR detector (proposed as the
ALICE experiment  subsystem) as the good tool
 for
the new
physics studies.

\subsection{Quark-gluon plasma and its signatures}

 The understanding of the equation of state of nuclear, hadronic
and partonic matter
is an interdisciplinary interest to nuclear physics, astrophysics,
cosmology and particle physics. It is also the main motivation for
studying  relativistic heavy ion collisions which primarily search
for the so--called quark-gluon
plasma (QGP).
 A quark-gluon plasma  is a state in which quarks and gluons,
the fundamental constituents of matter, are no longer confined within the
dimensions of the nucleon, but move around over a volume in which
a high enough temperature and/or density prevails. The plasma 
exhibits the so--called `` chiral symmetry '' which in normal nuclear
matter
is spontaneously broken, resulting in effective quark masses which are
much larger than the actual masses.
 A quark-hadron phase transition is believed to have occured at about
ten micro-seconds after the Big Bang when the Universe was at a
temperature of approximately 150 to 200 MeV.
The QGP may also exist in the
cores of
dense stars at  high baryon densities.
The critical energy density for the quark-gluon plasma formation is
predicted to be $\sim$1 GeV/fm$^{3}$, or seven times the energy density of
normal
nuclear matter. The compilation of 
heavy-ion studies, looking for the quark-gluon plasma state,
 can be found for example in \cite{STAR, Bialkowska}.
More detailed reviews of theoretical and experimental topics of the
quark-matter
 are presented in 
 \cite{QM96,QM97,QM99,QM01}. 

\begin{figure}[h]
\begin{center}
\mbox{}
\epsfig{file = 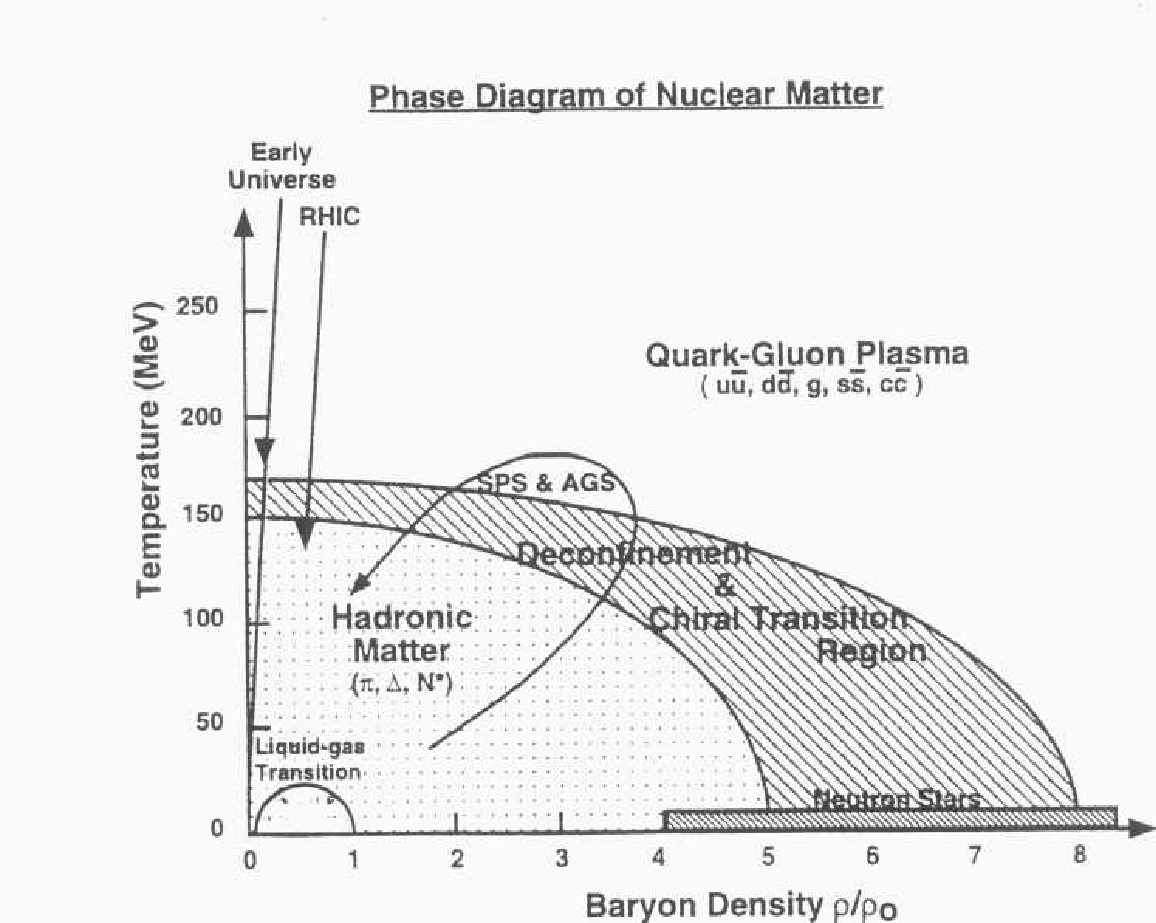,
bbllx=20,bblly=21,bburx=575,bbury=459,width=8cm}
\vspace*{4mm}
\caption{Schematic phase diagram of nuclear matter
 \cite{STAR}.}
\label{phase_diagram}
\end{center}
\end{figure}

 A schematic phase diagram of nuclear matter \cite{STAR}, showing the behaviour of
nuclear matter as a function of temperature and density is schematically
shown in Figure~\ref{phase_diagram}.

Conventional nuclear physics concerns low
temperatures and near to
normal nuclear matter density.
 This region was accessible in heavy ion studies at the
SPS accelerator facility at CERN and at the AGS accelerator  at
Brookhaven National Laboratory.
Some part of the nucleus-nucleus collisions from the CERN SPS 
accelerator as well as from the relativistic Heavy Ion
Collider (RHIC) at Brookhaven reached 
the quark-gluon plasma regime. The Large Hadron Collider
(LHC) at CERN, started at 2009, should reach temperatures 
and densities 
close to that of the early Universe.

As the QGP phase is rather complicated transient state, many interesting
or even unexpected phenomena could be observed  during its formation,
expansion, cooling,
and hadronization. There are many predictions for possible signatures of 
QGP formation and of partial restoration of chiral symmetry, grouped into
several classes. Among them the  following topics are frequently
proposed to investigate:
\begin{itemize}\vspace*{-2mm}
\item 
\begin{large} {\em Thermodynamic Variables}\\
\end{large}
 This point includes a study of  energy and entropy  densities: $\varepsilon$,
$s$ and 
a pressure $p$  as a 
function of the temperature $T$ and the baryochemical potential
$\mu_{b}$. If a phase transition to QGP occurs, a rapid rise in the 
effective number of degrees of freedom (expressed by
$\varepsilon/T^{4}$ or $s/T^{3}$) should be
observed over a small range
of $T$. The thermodynamical quantities: $T,s$ and $\varepsilon$ can be identified with
the average transverse momentum $\langle p_{T} \rangle$, the hadron
rapidity density $dN/dy$, and the transverse energy density
$dE_{T}/dy$, respectively.\vspace*{-2mm}
\item 
\begin{large}
{\em Fluctuations}\\
\end{large}
Thermodynamical instabilities during the phase transition could lead to
large fluctuations.
Critical fluctuations, appearing  in energy density, entropy
density, multiplicity, particle ratios
 etc., should  be
looked for in individual events, as a function of $p_{T}$, rapidity, 
azimuthal angles etc. 
\vspace*{-2mm} \item \begin{large}
{\em Electromagnetic signals}\\
\end{large}
Electromagnetic signals 
 can probe the interior of
the quark-gluon plasma during its hottest and earliest phase.
 The main problem is, however,
how to separate the yields for electromagnetic probes, which are
rather small relative to background processes.
\item
\begin{large}
{\em Quarkonium  suppression}
\end{large}\\
J/$\psi$  made of $c$ and $\overline {c}$ quarks 
 are expected to be suppressed in a quark-gluon plasma
 if the screening (Debye) radius, which is inversely proportional to the density of colour
charges, is smaller then the size
of the $J/\psi$ ($\simeq$ 0.5 fm).
Less tightly bound excited states of the $c \overline{c}$ system, such as
$\psi ^{'}$ and $\chi_{c}$, are more easily dissociated so they
should be  suppressed
even more than the $J/\psi$. Similar predictions concern also  the
heavier
$\Upsilon,\Upsilon^{'},\Upsilon^{''}(b\overline{b}$ systems).
\vspace*{-2mm}
\item
\begin{large}
{\em  Strangeness (heavy flavour)  enhancement}
\end{large}\\
An enhancement in the production of strange particles
was one of the first
predictions for a signature of QGP formation.
The production of strange quarks is favoured in a QGP because there the
mass of the strange quark is reduced from $\sim$ 500 MeV to
$\sim$ 150 MeV (chiral symmetry restoration).
In addition, in a baryon-rich region the creation of light $u$ and $d$
quarks
is suppresed because of 
Pauli blocking.
 Especially multi-strange baryons and anti-baryons are
predicted to be 
strongly enhanced.
The more speculative aspect is connected with the possible existence 
 of exotic matter,
i.e. the droplets of strange quark-matter.
Also an enhancement of charm can be expected.\vspace*{-2mm}
\item 
\begin{large}
{\em  Disoriented Chiral Condensate}
\end{large}\\
 DCC is a coherent
excitation of the pion field corresponding to a local
misalignment of the chiral order parameter.
 According to this speculative hypothesis the pieces of strong
interacting vacuum
with a rotated chiral order parameter might be produced in high energy
particle collisions.
 Such domains would decay into
neutral and charged pions, favouring pion ratios $N_{\pi^{0}}/N_{\pi}$
substantially different from 1/3. The formation of DCC should lead to
large fluctuations in the ratio of neutral to charged pion yields at low
$p_{T}$, and consequently to fluctuations in $\langle p_{T} \rangle$
of charged particles.
\vspace*{-2mm}
 \item
\begin{large}
{\em  High
$p_{T}$ probes of QCD}
\end{large}\\
The colour structure of QCD matter can be probed by its effects on the
propagation of a fast parton.
The characteristics of the fragmentation products of hard scattering
can provide information on the matter through which the hard scattered
parton propagates.
 \end{itemize}
\subsection{Highlights from accelerator experiments}

The experimental programmes in relativistic heavy ions using the BNL-AGS
and
CERN-SPS started in 1986. At BNL ion beams of silicon and gold,
accelerated to momenta of 14 and 11 GeV/c per nucleon respectively, have
been utilised in approximately 10 fixed-target experiments. 
 About 15 heavy ion experiments at CERN  studied heavy ion
interactions, utilizing beams of 
oxygen at 60 and 200 GeV/c per nucleon, sulphur at 200 GeV/c per nucleon
and Pb at 160 GeV/c per nucleon. 
The experiments were all optimised for measuring different signals which
might indicate if and how a quark-gluon plasma was formed. Some of them
optimised
their detectors for one rare signal, while others developed multipurpose
detectors which were sensitive to multiple signals.
Many interesting effects, concerning mainly the equilibration of hadronic
matter, chiral symmetry restoration and deconfinement, which generally
support the QGP hypothesis, have been observed in the AGS and SPS
experiments. The long awaited first results from the RHIC are now also
available indicating on very interesting physics.
\begin{itemize}
\item  First of all, it seems that 
the conditions required for 
the QGP formation have been reached. For instance,
a large degree of  stopping, resulting in the transfer of a large amount 
of energy from the relative motion into other degrees of freedom, was
observed in central collisions of the
heaviest systems at the AGS and SPS (e.g. NA49, NA44, E866, E877
experiments). At SPS, this creates the high
energy densities
($\varepsilon \sim 1.5-2.0$ GeV/fm$^{3}$) 
beyond those predicted for production of a quark-gluon
plasma. The early, very dense state (fireball) has an energy density of
$\sim$ 3-4 GeV/fm$^{3}$ what corresponds a temperature of about 240
MeV. The first estimates from RHIC \cite{RHIC_results}, obtained by using
the Bjorken formula
 with the thermalization time $\tau$ = 1 fm/c, 
give an energy density of 4.6 
GeV/fm$^{3}$, which is 60$\%$ larger than measured at the CERN-SPS.
In addition, it is suspected that the density can be  substantially higher
due to the shorter thermalization time in the higher parton density
environment and some estimates give an energy density even in the range
23.0 - 50  GeV/fm$^{3}$.
\vspace*{-2mm}
\item The data do not contradict the picture in which the created
quark-gluon plasma cools down and becomes more dilute. At an energy
density
of $\sim$ 1 GeV/fm$^{3}$ (and a temperature
of $\sim$ 170-180 MeV) the quarks and gluons condensate into hadrons, and
the
final abundancies of the different types of particles 
 are fixed. At an energy density of around 50 MeV/fm$^{3}$ (and a
temperature of
 $\sim$ 100-200 MeV) the hadrons stop interacting completely and the
fireball
freezes out.
Thermal and chemical equilibrium models are generally able to
reproduce the particle abundances and particle spectra measured in central
collisions, once the strong influence of the collective nuclear flow is
taken into account. As a result, the parameters which best fit the
particle ratios and particle spectra were found: the temperature at
chemical equilibrium $T$ = 160 - 175 (120 -140) MeV, the baryochemical
potential $\mu_{b}$ = 200 - 270 (540), and the strangeness saturation
factor $\gamma_{s}$ = 0.6 - 1.0 for SPS (AGS) energies. The preliminary
results of analysis of particle ratios \cite{RHIC_results}, from all four
RHIC experiments,
show reasonable agreement with a baryon chemical potential $\mu_{b}
\simeq$
 45-55 MeV and a chemical freeze-out temperature $T \simeq$ 160-180 MeV,
which is not significantly different from the temperature measured
at the CERN-SPS. The value of baryon chemical potential estimated from
RHIC
experiments is much lower than that obtained from the CERN SPS, showing 
a closer but not yet complete approach to a baryon free regime at
RHIC. The first measurements of the elliptic flow
 at RHIC  indicate on excellent agreement with hydrodynamical
model up to $p_{T} \simeq$ 1.5 GeV (STAR),
what may indicate a large degree of thermalization in the early stages,
validating the assumption of hydrodynamical expansion.
 Thus, there is
evidence for chemical and thermal equilibrium from the accelerator
experiments results, although the issue of strangeness saturation (i.e.
if $\gamma_{s}$ = 1), which is required for deconfinement is still to be
settled (NA44 and NA49 experiments find only partial chemical
equilibrium for strange particles).\vspace*{-2mm} 
\item An imporant aspect of ``chemical equilibrium'' is the observed
enhancement of strange particles. One can ask how many strange quarks and
antiquarks are formed relative to the newly produced up and down quarks
and antiquarks. For proton-proton and electron-positron collisions, the
fraction of extra strange quarks made is $\sim$ 0.2. But for the
nucleus-nucleus
the fraction is twice as high,
 i.e. it is seen ``global
strangeness enhancement by a factor of two'' (NA49).
 Multi-strange hadrons are
enhanced more strongly (WA97, NA49, NA50) up to a factor 15 for the
$\Omega$ hyperon and its antiparticle (WA97).
\vspace*{-2mm}
 \item An
increase in the pion to baryon ratio from approximately 1 to     
about 7, when going from the AGS to the SPS energies indicates  the rise
in entropy density.\vspace*{-2mm}
\item Measurements of charmonium production exhibit a suppression of the
yield of both the $J/\psi$ and the $\psi^{'}$ relative to Drell-Yan pair
production for central collisions of Pb+Pb at the SPS (NA50, NA38
Collaborations). The yield of the
$\psi^{'}$ relative to Drell-Yan is also observed to be suppressed
for central collisions of the lighter S+Pb system, whereas the $J/\psi$
is not. 
These results
follow the expected pattern that the Debye screening will initially affect
the $\psi^{'}$ before the $J/\psi$ and suggest that the deconfinement
regime has been reached at the SPS.
\vspace*{-2mm}
\item So far, no apparent signal from direct photons radiated by quarks in 
QGP is found. 
 For S+Au
collisions, WA80 and NA45 established that not more then 5\% of the
observed
photons are emitted directly. For Pb+Pb collisions WA98 has reported
indications for a significant direct photon contribution. Preliminary data
from NA45 are consistent with this finding, but so far not statistically
significant. NA50 has seen an excess by about a factor 2 in the dimuon
spectrum in the mass region  between $\phi$ and J/$\psi$ vector mesons.
On the other hand the data (NA45) show that in S+Au and Pb+Au collisions
the expected peak from $\rho$ vector meson
 is completely smeared out.
Simultaneously, electron pairs (virtual photons) measurements
exhibit
an enhancement at low to intermediate pair masses (in the mass region
between the mass of two pions and 1.5 GeV/c, i.e. around the $\rho$ meson)
relative to pairs
expected from hadronic decays for central collisions of the heavier
systems (S+Au and Pb+Au) at the SPS (CERES/NA45, HELIOS-3 Collaboration).
 This
effect is 
investigated by using different theorethical approaches and ideas
 such as
 $\pi\pi$ annihilation,
a decreased $\rho$ mass due to
partial chiral symmetry restoration and other effects. The $\rho$ is
of particular interest as it has a short lifetime compared to the
interaction times and it decays very quickly in the presence of the medium
(whether QGP or hadronic). Thus it should exhibit signs of possible chiral
restoration if there is a reduction of the $\rho$-meson mass. 
The clear excess in the mass spectrum of $e^{+}e^{-}$ bears the
questions: is it thermal radiation from a hot source or rather does it
indicate in-medium effects for vector meson resonances?
In the latter case, it could be a signal of the onset of chiral
symmetry
restoration as matter becomes denser.
\item To the list of interesting  phenomena, the suppression of
 transverse momentum spectra at $p_{T} \gt 3$ GeV/c, announced
 preliminary by the PHENIX  experiment \cite{PHENIX} at RHIC should be
added. It could be evidence for an  energy loss of gluon
mini-jets, the so--called ``jet--quenching'' mechanism
in high density  matter with dense colored sources \cite{RHIC_quen}.
This is in contrast to data at SPS energies, where WA98 found no
evidence for quenching in Pb + Pb collisions but a factor of two
 Cronin enhancement.
\end{itemize}
Summarizing,
the results of relativistic heavy ion experiments     
 are quite intriguing.
The data from any one experiment are not enough to give the full picture
but the combined results from all experiments rather agree and fit.
The attempts to explain all of them simultaneously, using established
particle interactions have failed, but on the other hand many of the
observations are consistent with the predicted signatures of a quark-gluon
plasma. At a special seminar on 10 February 2000, spokespersons from
the experiments on CERN's Heavy Ion programme stated that the collected
data
from seven CERN experiments (NA44, NA45, NA49, NA50, NA52, WA97, NA57 and
WA98) give compelling evidence that a new state of matter has been 
created although this evidence
is so far ``indirect'' \cite{cern_int}. The first measurements from the
RHIC were results on global observables such as: charged particle
multiplicity, transverse energy and elliptic flow. They all show
nontrivial collective behaviour. The multiplicity characteristics are
tried to understand by means of models including the mechanisms of parton
saturation \cite{Kharzeev} or string percolation \cite{percolation}.
\subsection{Experimental hints from unusual cosmic ray events}
 Cosmic-ray induced reactions can be
satisfactorily
described in terms of standard ideas below the energy of about 1000 TeV.
Above this threshold the global behaviour of cosmic-ray families
is hardly explained by smoothly extrapolating the hadron multiple
production characteristics which were learnt through CERN collider
energies. The families of hadrons and photons, mainly observed by
mountain cosmic ray laboratories,  exhibit
many surprising
features such as:  rapid
attenuation in the atmosphere,
 multi-core
structure with unexpectedly high alignement level, large lateral
spread, extremely high energy concentration in the forward 
angular region (halo) etc.
The observed flux value of cosmic ray families is smaller
than expected. According to ref. \cite{Tamada_Lodz} experimental
intensity of 
 families 
at the
Pamir is $I(\sum E_{\gamma} \geq 100$ TeV$) \simeq$ 0.35 m$^{-2}$y$^{-1}$
sr$^{-1}$. The  simulated data, based on different simulation codes,
give 3 $\sim$ 4 times higher intensity than observations.
 In addition, the spectra of high energy showers in
families are in younger stage of the shower development, what indicates
that  majority of observed families originates from main
interaction slightly above the observed levels.
 At energies $10^{15}-10^{17}$ eV appears 
the wide spectrum of
exotic events hardly explained by the known mechanisms.

 The global characteristics of  gamma-hadron 
families are affected by the primary cosmic ray chemical composition and
 show a high sensitivity to a hadronic interaction model used in the
interpretation. Thus, in principle, the problem seems to reduce to a
question of whether the data signal a change in the composition of cosmic
ray primaries at these energies, or whether they signal a change in the
hadronic interactions.
 Some unusual features of cosmic-ray events
could be, in part, understood by the use of a Fe-dominant composition
in the primary flux. However, the recent results from Chacaltaya and
Tien-Shan (``Hadron'' Experiment) instalations operating simultaneously 
the emulsion
chamber and the air shower array also
 indicate that discrepancy
between the predicted (based on conventional models)
and observed  family flux cannot be reconciled by the heavy-dominant
composition of the primary cosmic rays \cite{Ohsawa,Aguirre}. Moreover,  
experimental
results concerning  the composition of primary cosmic ray spectrum
seem to exclude this explanation. In fact,  
 the primary flux composition in the energy region
above $10^{14}$ eV can only be obtained by  using  indirect
methods of ground-based detectors looking at  showers initiated by
the interaction of primary nuclei in the upper atmosphere. Hence, there
are
substantial uncertainties on the primary cosmic ray composition
in this energy region. As  a heavy dominant composition does not have
 convincing observational support
\cite{Soudan_MACRO_Jacee},
 it seems that
{\it the introduction of some new mechanism, or  a global change
of the characteristics of  hadronic interactions at around $10^{16}$
eV, especially
in the most forward angular region
is necessary} to explain the experimental data.

 In this work   
  various anomalies  such as: exotic events called
 Centauros, Mini-Centauros, Chirons etc. were reviewed.
These events have many common features and they are
very frequently
 connected with the so--called long-flying (penetrating) component. There
are strong
indications that they are connected one to the other and 
 they  have a common origin.

 In fact, the strongly penetrating
component has two faces. At first, it has been observed in the apparatus
in the form of strongly  penetrating single  cascades, clusters of showers 
or the 
so called ``halo''. This phenomenon manifests itself  by the
characteristic
energy pattern revealed in shower development in the deep chambers
(calorimeters) indicating 
 a slow attenuation and many maxima structure (see section 2). 
The second aspect is connected with the anomalously strong penetrability
of some
objects during the passage through the atmosphere.
In principle, both items can be connected one to the other and could be 
 different faces of the same phenomenon.
  
The mentioned here  anomalies are not rare occurence but they manifest
themselves at the
5\% level or above. 
There are many theoretical attempts to explain the observed anomalies,
among them the quark-gluon plasma scenario is the very promising one.

\begin{figure}
\begin{minipage}{7.3cm}
\begin{center}
\mbox{}
\epsfig{file = 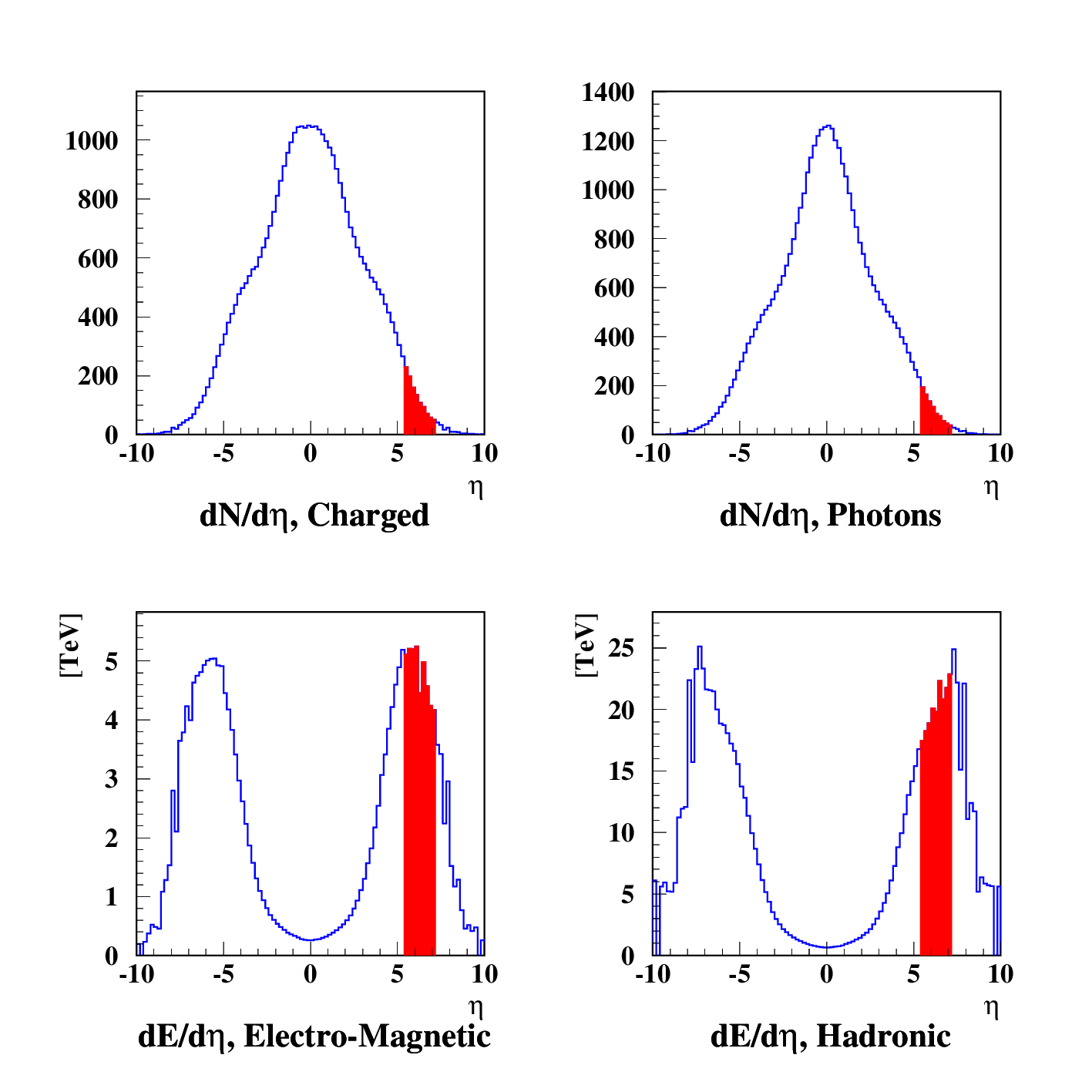,
bbllx=0,bblly=0,bburx=680,bbury=500,width=7.5cm}
\caption [Average pseudorapidity distribution of particle
multiplicities
and
energy flow. 50 central Pb+Pb collisions
at energy $\sqrt{s}$ = 5.5 A~TeV were generated by HIJING
\cite{CASTOR_Delphi}.]
{Average pseudorapidity distribution of particle
multiplicities
and
energy flow. 50 central Pb+Pb collisions
at energy $\sqrt{s}$ = 5.5 A~TeV were generated by HIJING. CASTOR
geometrical acceptance is indicated \cite{CASTOR_Delphi}.}
\label{multiplicity}
\end{center}
\end{minipage}
\vspace*{-8mm}
\hspace*{1mm}
\begin{minipage}{7.3cm}
\begin{center}
\mbox{}
\epsfig{file = 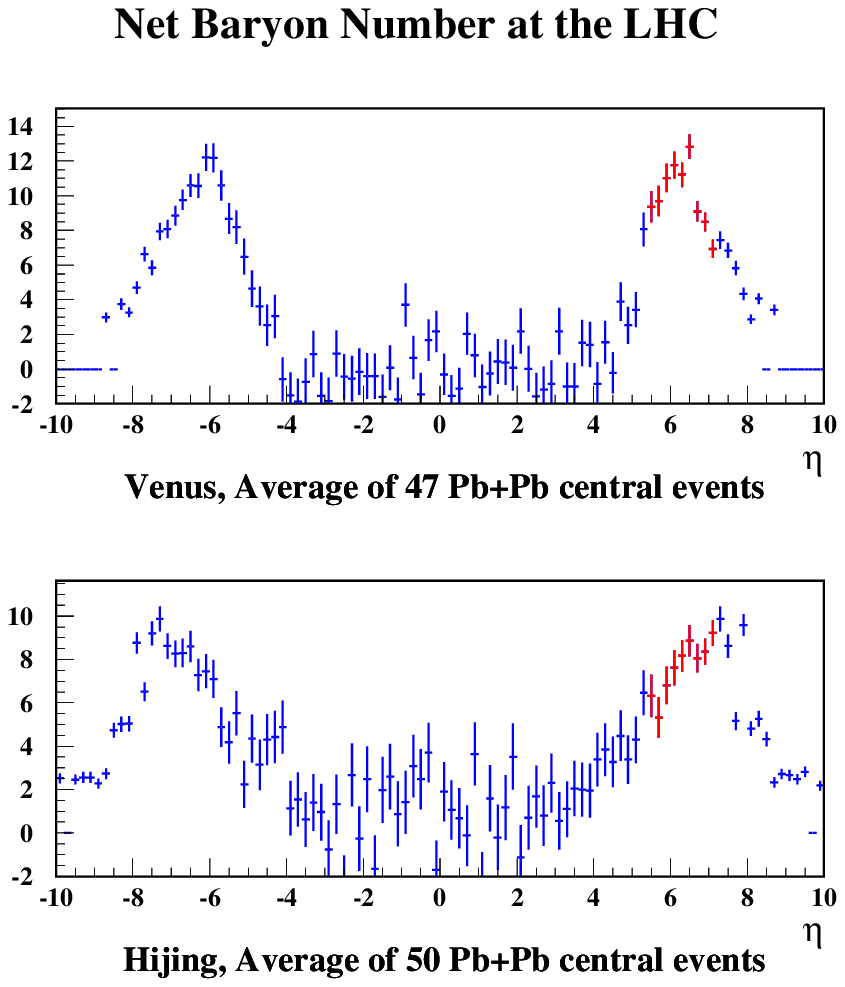,
bbllx=0,bblly=170,bburx=611,bbury=600,width=7.5cm}
\caption{Baryon number distribution for
 central Pb+Pb collisions
at energy $\sqrt{s}$ = 5.5 A~TeV  generated by HIJING and VENUS.}
\label{baryon_dist}
\end{center}
\end{minipage}
\vspace*{2mm}
\end{figure}

 Since the wide spectrum of exotic events seen in
cosmic ray experiments is observed at  {\it forward rapidities}, thus this
region seems to be a potential place for the new physics.
 Unfortunately, 
the physics in the very forward rapidity region in ultra-relativistic
nucleus-nucleus collisions has not been rigorously addressed by theory so far.
 The main
reason is the difficulty of the calculations in an enviroment of finite 
baryochemical potential. There are, however, some phenomenological and
QCD-inspired attempts to predict new phenomena or to explain unusual
phenomena already seen. It is expected that  this region will contain only 
 a small fraction of the totally produced particles and at the same time
the vast majority of the available total energy with  the baryon
density
 reaching here the maximum.
Figures~\ref{multiplicity} 
show the pseudorapidity distributions of multiplicity and energy,
 obtained in simulations (HIJING generator) of
50 central Pb+Pb collisions 
at LHC energies $\sqrt{s}$ = 5.5 TeV) \cite{CASTOR_Delphi}. There are
shown separately distributions of electromagnetic and hadronic
component, and the acceptance of the CASTOR detector (see section 6) is
marked. Figure~\ref{baryon_dist} shows  distributions of the baryon
number predicted by VENUS and HIJING for the same reactions.

  Such environment offers the possibility for the appearance of novel,
surprising
phenomena.
In particular, strangelets, the small droplets of strange quark
matter are predicted to be formed from the Quark Gluon Plasma,
predominantly in a high baryochemical potential environment.
Heavy flavour 
  \cite{Tien-Shan,Dremin} and
 Super Heavy Particles production, if they exist, may also be seen
at very forward rapidities. 
 For instance 
the new heavy vector bosons $V^{+},V^{-},V^{0}$ with masses up to 6 TeV
are theoretically predicted \cite{heavy_part}.
Another phenomenon which is attracting a lot of theoretical investigation
is colour superconductivity at finite baryon density
\cite{superconductivity}.
At relatively low temperature,  the baryon rich environment may lead to
the formation
of Deconfined Quark Matter (DQM) in the core of neutron stars
\cite{neutron_star}.
 Also  the understanding of cosmic ray exotic phenomena  in
terms of
 hadronization of DQM in an extremaly high
baryon density environment has been proposed \cite{Panagiotou,8,9}.

All these phenomena should be searched for and studied.
 Unfortunately,
 most of the present and future nucleus-nucleus experiments are
concentrated on 
  exploration of the poor baryon region and  doing 
the ''midrapidity physics''. 
  In this work  the
importance of the CASTOR project, as the unique experimental design
to explore this interesting region at very high energies planned
to be reached at the LHC nucleus-nucleus collisions will be emphasized.

 \section
{Centauro related phenomena in Chacaltaya and Pamir chambers}

  Centauro related phenomena have been discovered and analysed
 in emulsion chamber experiments investigating  cosmic ray interactions
 at  the high  mountain laboratories  at Mt. Chacaltaya (5200
m above see level
) and Pamirs ($\sim$ 4300 
or 4900 m above sea level).

 In principle the detectors used in these experiments are similar.
They  consist of three parts:  upper and lower chambers, and the 
carbon target between them. The upper part called  the gamma block 
 detects mainly
the electromagnetic component. In the lower chamber there are registered
mainly hadrons. Both the upper and the lower chambers are sandwiches
of the lead absorber and the sensitive layers which are mostly
X-ray films, sometimes emulsion plates. Chambers have rather big areas,
of the order of several tens $m^{2}$ and are exposed for a long
time periods, of about one year.

\begin{figure}
\begin{center}
\mbox{}
\epsfig{file = 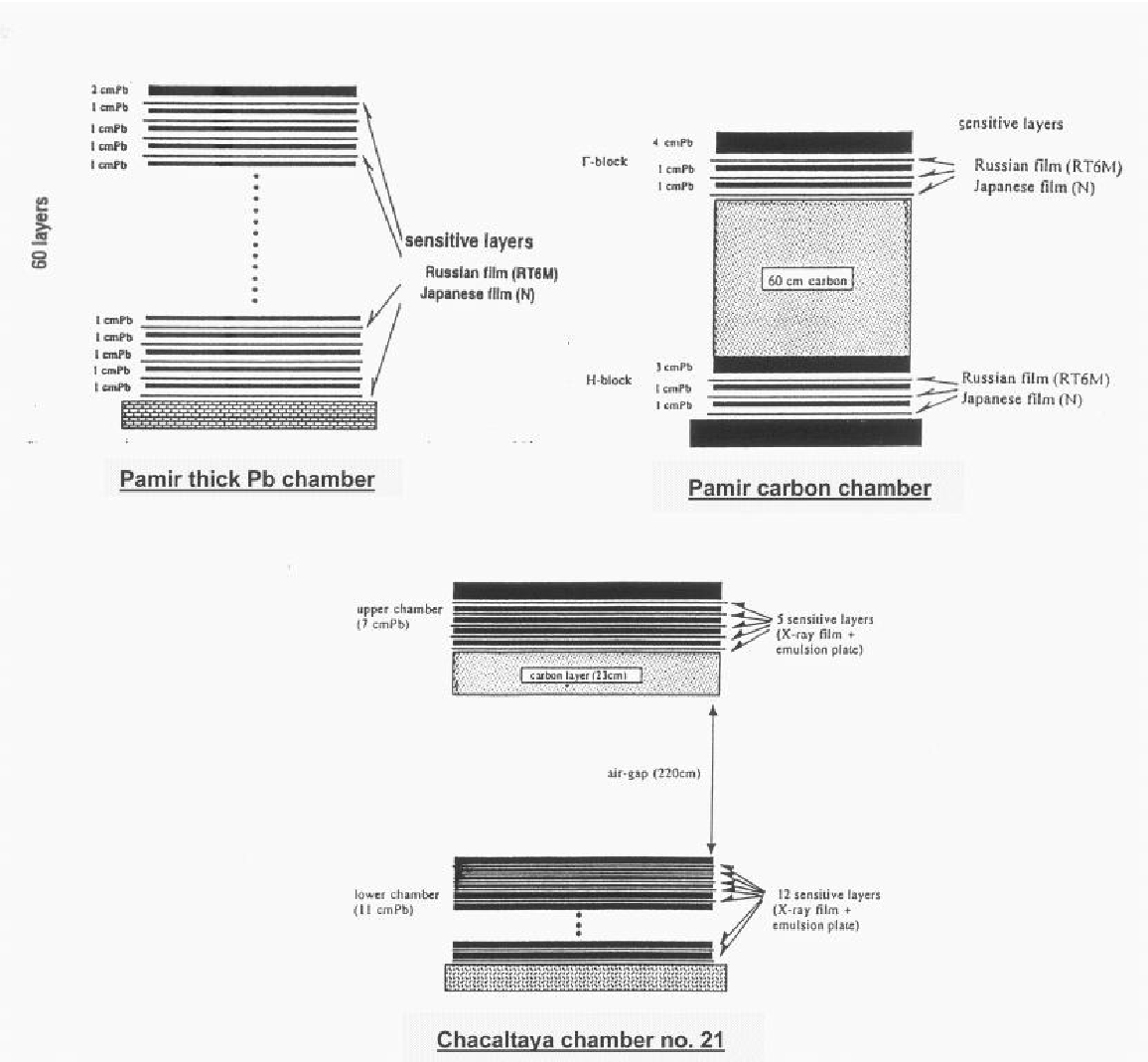, bbllx=21,bblly=20,bburx=573,bbury=532,width=15cm}
\vspace*{2mm}
\caption{Schemes of  typical emulsion chambers
\cite{Has_Tokyo}.}
\label{chambers}
\end{center}
\end{figure}

However, the direct comparison and interpretation of the experimental data
is difficult because of 
some differences in the depth and  construction of the chambers. 
 Figure~\ref{chambers} shows
the schemes of Chacaltaya two-storey chamber,
a traditional style carbon chamber employed in Pamir experiments and
also a schematic view of a homogeneous thick lead chamber (60 or 110/120
cm
thick
lead). The typical Pamir carbon chambers consist of two
(or three) parts: a  gamma-block of 5 or 6 cm Pb thick and
one (or two)
hadron(H)-blocks each consisting of carbon layer of 60 cm thick 
and 5 cm of Pb.
 The thickness of a standard Pamir-type carbon chamber
amounts to $\sim$ 1.7 $\lambda_{geo}$ \footnote{$\lambda_{geo}$ is the
geometrical collision mean
free
path of ordinary cosmic-ray baryons.}, what assures the detection
efficiency to be over $\sim$ 60--70 \%.
  Thicker carbon-type chambers (consisting of four hadron blocks)
have been also used sometimes.
The homogenous lead chambers are simply sandwiches of Pb absorber layers
(the upper layer has usually 2 cm Pb and the consecutive ones have 1 cm
Pb) and the sensitive material being mainly the X-ray films.  
The detection efficiency of thickest
Pb-chambers is $\sim$ 80--90 \%.
Typical two-storey Chacaltaya chamber consists of the upper
chamber of 6-10 cm Pb thick, the target layer 20-30 cm of carbon, 
 the air
gap (e.g. 150 cm) and the lower
chamber of 6-15  cm Pb thick.
Some chambers (as for example Chacaltaya chamber no. 19, 20, 21, 22) were
covered by
several nuclear emulsion layers over all area of both upper and lower
chamber. It enables a careful study (under the microscope), of the
shower structures. It is essential in classification of the
showers (an identification of
the photonic and hadronic cascades) and in the study of the strongly
penetrating component.

Typical detected event, called here family, is generated in
 a collision of a cosmic-ray particle (mostly  p or $\pi$)
with  air nuclei at the distance of about 500-1000 m above the
apparatus.
  The collision mean
free path of cosmic ray hadrons is 
about  1200 m and mean free path for  pair creation by
$\gamma$--quanta  is about  770 m of the
atmosphere at the top of Chacaltaya
($\lambda^{int}_{n-air} = 75$ g/cm$^{2}$,
$\lambda^{int}_{p-air} = 110$ g/cm$^{2}$, 1 c. u. = 37.7 g/cm$^{2}$).  
 Keeping in mind these numbers
we see that  particles born in a typical act of  interaction
must traverse about 0.5-1 $\lambda^{int}$ ($\approx$ 2 c.u.)
of  matter, before reaching the detector. Thus 
the families born at  large distances above  the chamber
are not  ``clean'' events but they are affected by
electromagnetic-nuclear cascade processes in the atmosphere.

Normal events contain about 30\% of $\pi^{0}$ mesons. Since each
$\pi^{0}$ decays
into $2\gamma$ there is produced rougly one $\gamma$
for each charged particle in the primary interaction. As the products
of the interaction descend towards the chamber, the fraction of their
electronic and photonic content increases through the shower formation, so
that 
 a ``usual'' family (hadrons, gammas and electrons with the common
origin)
seen in the upper detector is always several times larger
 than its continuation into the lower detector.
 In ``normal'' families, from the energy range $\Sigma E_{\gamma}$
 = 100-1000 TeV, the hadronic component constitutes $\leq 30\%$
 of the total visible energy.
Thus, a big surprise were the events with the contrary situation.
 Since the upper half of these events did not allow one to predict
their lower part, such events were named ``Centauros''.
 
At present, rather big statistics of cosmic-ray families with visible
energy
greater than 100 TeV from Chacaltaya and Pamir-joint experiments exists
 \cite{Tamada_Lodz,Has_Tokyo}.
The experimental material, suitable for investigation of  the
above mentioned
 exotic phenomena, was collected by the Brasil-Japan Chacaltaya
Collaboration, starting from the 1970's, when the Centauro I was
found. Since 1980, the joint experiment has been conducted at the Pamir,
by the Pamir and Chacaltaya groups after Nakhodka Symposium, and a part of
the  material from the standard type carbon Pamir chambers has been
reanalysed in the same manner as in
Chacaltaya experiment, with the point of view exotic phenomena. In 1991
a joint analysis, undertaken by Moscow State University and Japan
groups using the  homogenous thick lead chambers, has been also started.
 477 events with the visible energy $\Sigma E_{vis} \ge$ 100 TeV,
coming from the total exposure of $\approx 1548$ m$^{2}$yr,
is reported in
\cite{Tamada_Lodz}
where unusual
characteristics of high--energy cosmic--ray families were studied,
with CORSIKA simulation code.

Partial analysis, concerning  selected subjects, have been done
sometimes using higher statistics.
In addition some extremely interesting events have been found also
in other types of thick chambers (as well as homogenous lead chambers
and carbon type chambers consisting of several  hadron
blocks), see subsection
2.1.2. The detailed analysis of 17 super--high energy families
( $\Sigma E_{vis} \ge 700$ TeV) coming from Pamir thick lead chambers
is presented in \cite{Has_Tokyo}.

 The present experimental
 results obtained by Chacaltaya and Pamir Collaboration show that
 hadron rich families occupy more than 20\% of the whole statistics
 and indicate the existence of several types Centauro species,
  characterized by:
\vspace*{-0.2cm}
\begin{itemize}
\item {abnormal hadron dominance (both in multiplicity and in energy
   content),}\vspace*{-0.2cm}
\item {rather low total (hadron) multiplicity, in comparison with that
       expected for nucleus-nucleus collisions at that energy range,}
       \vspace*{-0.2cm}
\item {transverse momentum of produced particles  much higher than
that
      observed in ``normal'' interactions ($p_{T} \approx$ 1.7 GeV/c
      for Centauros  and 10-15 GeV/c for Chirons,  assuming the gamma
      inelasticity coefficient $K_{\gamma}\approx$
      0.2),}\vspace*{-2mm}
\item {isotropic $ \eta$ distribution.}\vspace*{-2mm}
\end{itemize}
They can be divided into several groups, such as:
\begin{enumerate}\vspace*{-2mm}
\item{Centauros of original type,}\vspace*{-2mm}
\item{Mini-Centauros,}\vspace*{-2mm}
\item{Chirons,}\vspace*{-2mm}
\item{Geminions.}\vspace*{-2mm}
\end{enumerate}

\subsection{Centauros of original type}

\subsubsection{``Classical'' Chacaltaya Centauros}

\begin{large}
{\centerline {\em History}}
\end{large}
\vspace*{1mm}

\begin{figure}
\begin{center}
\mbox{}
\epsfig{file = 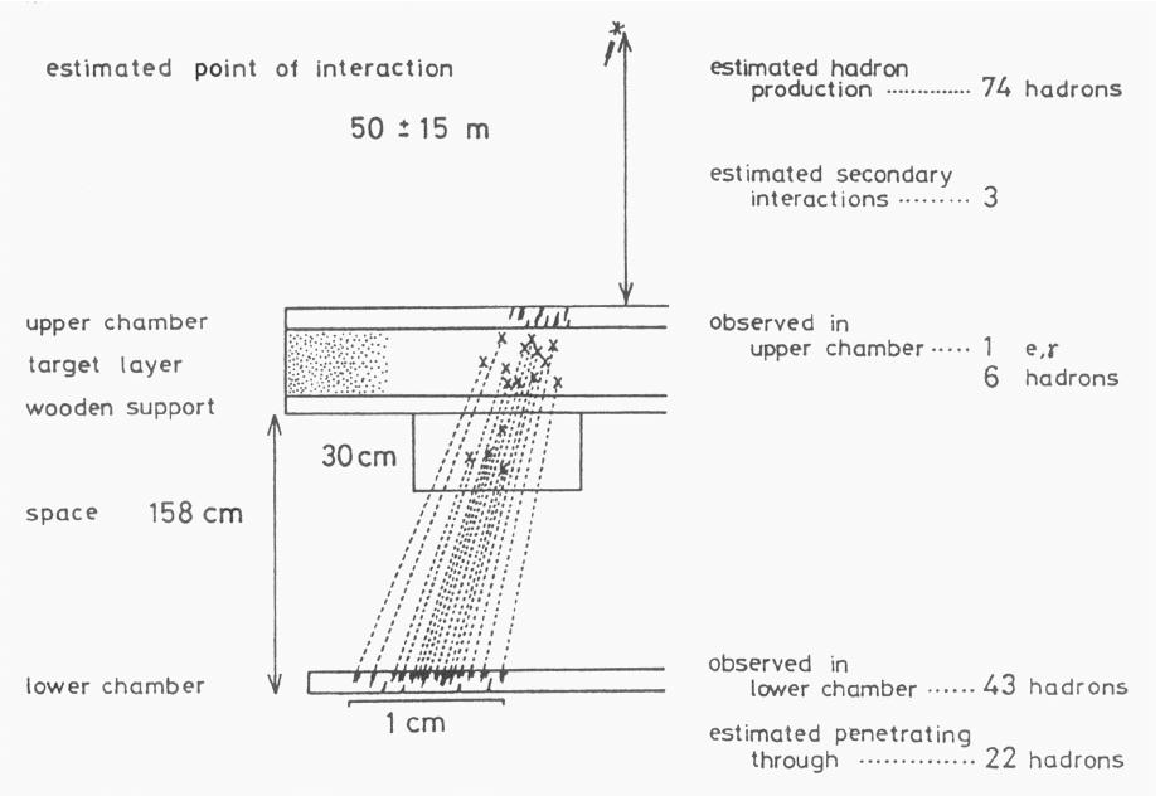,
bbllx=20,bblly=23,bburx=574,bbury=401,width=9cm}
\vspace{4mm}
\caption{Illustration of Centauro I \cite{Lattes}.}
\label{Cent1}
\end{center}
\end{figure}

 Centauro, which is mainly characterized by large imbalance between the
 hadronic  and photonic component, has been a puzzle
 for so many years. 
 The first Centauro \cite{Cent_1} was found many years ago, in 1972,
in the
 two-storey Chacaltaya chamber No. 15.
 It is the most spectacular and indisputable event of this kind. 
 The triangulation measurement (by angular divergence) of shower direction
was feasible for the
 event and it allowed to estimate the interaction height to be
 50$\pm$15 m above the chamber. It is relatively small distance and
 this is the reason that the event is clean, i.e. it didn't suffer
 from the electromagnetic and nuclear cascade in the atmospheric layer
 above the apparatus. The scheme of the event is shown in
Fig.~\ref{Cent1}. In the upper chamber there were observed 7
 cascades, identified as one atmospheric e/$\gamma$ and 6 Pb-jets. In
the
 lower part of the detector 43 observed cascades have been classified as
  7 Pb-upper-jets, 29 C-jets and 7 Pb-lower-jets.

 A ``normal'' family seen in the upper detector is always several
 times larger than its continuation in the lower detector. So, the
 Centauro I event, with the contrary situation, was a very big surprise.
 The suspicion that the event might have happened during a short
 period of assembling or removing the chamber was eliminated because
 the upper detectors were always mounted before the lower ones and the
lower
detector is always the first to be removed.

 After introducing corrections for the hadron detection efficiency
 and secondary particles generated in the  atmospheric interactions
above the chamber
 (A-jets), the  
 Centauro I can be considered as the event with the total
interaction
 energy  $\Sigma E_{\gamma}$= 330 TeV, in which
   only one electromagnetic (e/$\gamma$) particle and
74 hadrons
have been produced.
 The estimated interaction height allowed to calculate the transverse
momenta of produced secondary hadrons. The visible part of the average 
 $p_{T}$ value is $<p_{T}^{\gamma}> = 0.35\pm 0.4$ GeV/c. It results
in  
the average
transverse momentum of a produced hadron $\langle p_{T}\rangle = 1.75$
GeV/c, taking
the gamma-inelasticity factor $<K_{\gamma}>$ = 0.2 \footnote{Such value
of
 $K_{\gamma}$ factor is usually quoted for nucleons and used
by  Japan-Brasil Collaboration, basing on assumption that the hadrons
born in the Centauro interaction are nucleons.
A
little
higher value, $K_{\gamma} \simeq$ 0.4, is being preferred for pions.
In that case, preferred by 
DCC followers, $<p_{T}> \simeq 0.875\pm 
0.375$ GeV.}

 It is very exciting that in 1997, during the 
International Cosmic Ray
 Conference,  helded in Durban in South Africa, the next clean Centauro
 event has been reported \cite{Durban}.
Since the finding of the first event Centauro I a systematic search for
such surprising families
has been made in the successive exposure of the Chacaltaya chambers and a
little
 later also in the Pamir
chambers.
 The next four 
events,
 found in Chacaltaya chambers, in  several years  after the discovery
of the Centauro I,
 have been analysed and described in detail in \cite{Lattes}. These well
known events are called 
 ``classical Centauros''.

\vspace*{2mm}

\begin{large}
{\centerline {\em Multiplicities and energies}}
\end{large}

\begin{table}[h] 
\begin{center}
\caption [Characteristics of Chacaltaya Centauros.]
{Characteristics of Chacaltaya Centauros.}
\label{classical_cen}
\vskip0.3cm
\begin{scriptsize}
\begin{tabular}{lccccc}
\hline
\multicolumn{6}{c}{CHACALTAYA CENTAUROS}\\

Centauro no. & I & II & III & IV & V\\
Chamber no. & 15 &17 &17 &17 &16  \\

\hline\hline\\
\multicolumn{6}{c}{{\em Observed in the chamber}}\\
\\
$N_{\gamma}$&1&-&-&-&-\\
$N_{unid}$&-&5&26&61&34\\   
$N_{h}$&49&32&37&38&31\\
$\Sigma E_{h}^{\gamma} [TeV]$ &222&179&169&144&167\\
 $\Sigma E_{tot} [TeV]$ &231&203&270&286&285\\
$Q_{h}$&0.96&0.88&0.63&0.50&0.59\\
                                        &    &    &&&0.72$^{*}$\\
\\
\hline
\\
\multicolumn{6}{c}{{\em Estimated  at the top of chamber}}\\ 
\\

$N_{h}$& 71 & 66 & 63 & 58 & 45 \\
$\Sigma E_{h}^{\gamma} [TeV]$  & 321 & 369 & 287 &220 &242\\
\\
$N_{\gamma}$& 1&0&17&51&31\\
$\Sigma E_{\gamma} [TeV]$ &9&0&66&119&108\\
\\
\\  
\hline
\\
\multicolumn{6}{c}{{\em Estimated in Centauro interaction}}\\
\\
$N_{h}$&74&71&76&90&63\\
$\Sigma E_{h}^{\gamma} [TeV]$&330&370&350&340&350\\
$N_{\gamma}$&0&0&0&4&0\\
\\
$H [m]$&50&80&230&500&400\\
$N_{A-jets}$&3&5&13&32&18\\
\hline
\end{tabular}
\end{scriptsize}
\end{center}
{\scriptsize $^{*}$ value obtained after re-analysis of the event
\cite{Cen_V_VII}.  
It is the lower limit as  the highest energy 
hadron was excluded.}
\end{table}

 The main characteristics of the ``classical
Centauros'' are presented
 in Table~\ref{classical_cen}.
 There are shown multiplicities (above
the
energy threshold) and energies of
both
 electromagnetic and hadronic components concerning three stages
 of analysis of the events.
The observed multiplicities (upper part of the table) are the mixture
 of the particles born in the parent interaction and these
generated in secondary
  atmospheric interactions. Hadronic multiplicities are 
uncorrected for the chamber detection efficiency.  There is shown
 also a class of cascades detected in the upper chamber, labeled
 as ``unidentified'' which in principle could have  as well as
 electromagnetic as hadronic origin. 
 The fraction of the observed hadronic energy
 to the total visible energy of the event named $Q_{h} =
\Sigma E^{\gamma}_{h}/\Sigma E_{tot}$ was also calculated. This parameter
 is now widely used in the analysis of the imbalance between hadronic
and electromagnetic components of cosmic ray events.

The medium part of the table contains the above mentioned characteristics,
corrected for the hadron detection efficiency. Also 
 separation of electromagnetic and hadronic cascades from
 the ``unidentified'' group, by means of statistical procedure, was done.
These corrections can be made
by using the experimental distribution of the depths of interaction points
in the chamber (C-jets, Pb-upper-jets and Pb-lower-jets) and one can
estimate the number of 
arriving hadrons and gammas at the top of the chamber.
Knowing both the number of hadrons arriving at the top of the chamber 
and  the interaction height it is possible
to estimate the multiplicity of hadrons produced in the parent
Centauro interaction and also the number of secondary atmospheric 
nuclear interactions (A-jets) during their passage to the chamber and
in a consequence
 the number of gamma rays produced in the parent
Centauro interaction. As it is seen in the lower part of the Table~\ref{classical_cen},
the number of gammas or electrons estimated to be produced in
the Centauro interaction is practically zero, so the number
of neutral pions or other rapid-gamma-decaying hadrons must be
negligibly small.
The details of such analysis are presented for example in
  \cite{Tamada_CIV} where Centauro IV as a typical case
of the event with not too small production height has been studied.

\vspace*{2mm}

\begin{large}
{\centerline {\em Centauro production  heights}}
\end{large}
\vspace*{1mm}

Unfortunately, measurements of the heights of  interaction points, through
the
shower geometry, were possible only
for a few exotic events  \cite{Cen_VII,La Paz,Chirons_ch20}
 found in Chacaltaya chambers, among them
for Centauro
I and later for Centauro IV and Centauro VII.
The production heights for majority of  other events
were estimated comparing their respective lateral spreads with that of
Centauro I and assuming the same average $p_{T}$ for produced hadrons.
The production heights are mostly  higher than that
for Centauro I,
 so it is natural that  the events, suffering
from nuclear-electromagnetic cascade in the atmosphere, contain 
stronger admixture of atmospheric gammas and electrons. 

\vspace*{2mm}

\begin{large}
{\centerline {\em Other Centauro properties}}
\end{large}
\vspace*{-0.2cm}
\begin{itemize}\vspace*{-2mm}
\item Centauros are observed  in the  very high energy region. 
The energy threshold for their production is  about 1000
TeV. 
As the average observed energy (for 5 Chacaltaya Centauros) is 348 TeV,
the total incident energy was estimated to be  1740 TeV,
 assuming the value of inelasticity coefficient  $K_{\gamma}$
=
0.2.
 \vspace*{-2mm}
\item  Psedorapidity distributions of Centauro species 
       (Centauros, Mini - Centauros and Chirons) are consistent with
a nearly
      Gaussian distribution \cite{Tarbes},
      centered at 
     $\langle \eta_{Cent} \rangle^{exp}_{lab}$ = 9.9$\pm$0.2 for
Centauros.  
Generally,  experimental characteristics support a formation and a
subsequent
      isotropic decay of a fireball with the hadron multiplicity
       $N_{h} = 100\pm 20$ and the mass
 $M_{Cen} \simeq 180 \pm$ 60 GeV, where the
 error is mainly
 due to the uncertainty in the $p_{T}$ value.
The Centauro features  based on the  
fireball model  are described in section 4.4.2 and are  shown
in
 Tables~\ref{Cen_kin} and 11.
\vspace*{-2mm}
\item
 Centauro origin and the kinematical
region in which the phenomenon is observed are still the matter of debate.
These questions are more detaily addressed  in  section 4 concerning
 Centauro models and in section 5 regarding the accelerator searches
of the exotic phenomena. Both these
questions, however, cannot be answered on the basis of  only experimental
data.
Some additional assumptions and model speculations are necessary.
It should also be  noted that both these questions are connected
one to
the other and the answer is important for the planning of the new
collider experiments.
Table~\ref{Cen_kin}
 illustrates the problem.
\end{itemize}
\begin{table}[h]
\caption[\scriptsize{Average  characteristics of 5 Chacaltaya Centauros
assuming
nucleon-nucleon or
 nucleus-nucleus
 collisions.}]
{Average  characteristics of 5 Chacaltaya Centauros assuming
nucleon-nucleon or
 nucleus-nucleus
 collisions.}
\label{Cen_kin}
\begin{scriptsize}
\begin{center}
\begin{tabular}[l]{|ll|}\hline
 & \\
Hadron multiplicity $< N_{h} >$ & 64--90, $< 75 >$\\
$\gamma$ multiplicity & 0\\
Average total incident energy & $< E > \geq$ 1740 TeV\\
Pseudorapidity of emitted baryons & $< \eta_{lab} >$ = 9.9
$\pm$0.2\\
Width of pseudorapidity distribution & $< \Delta \eta > \simeq 1
\pm 0.2$\\
Average transverse momentum & $< p_{T} >$ = 1.75 $\pm$ 0.7 GeV/c\\
&\\
\multicolumn{2}{|c|}{\em Assuming nucleus--nucleus collision
\cite{Panagiotou,8}}\\
&\\
Total interaction energy in "60+14" c.m. & $\sqrt{s} \geq$ 6.8 TeV\\
Total interaction energy in N--N c.m. & $\sqrt{{s}_{N-N}} \geq$ 0.23 TeV\\
Incident projectile pseudorapidity  & $\eta_{lab}^{proj} \simeq$
11.\\
&  $\eta_{c.m.}^{proj} \simeq$ 4.8\\
&\\
\multicolumn{2}{|c|}{\em Assuming nucleon--nucleon collision}\\
&\\
Total interaction energy in N--N c.m. & $\sqrt{{s}_{N-N}} \geq$ 1.8 TeV\\
Incident projectile pseudorapidity  & $y_{lab}^{proj} \simeq$
15.1\\
 & $y_{c.m.}^{proj} \simeq$  7.5\\
& \\
\hline
\end{tabular}
\end{center}
\end{scriptsize}
\vspace{-2mm}
\end{table}

 In Table~\ref{Cen_kin}  are shown
energies, pseudorapidities and  maximal
projectile pseudorapidites, calculated both in the laboratory and
in the centre of  mass  system, for  two different scenarios of 
the collision:
formation  of Centauros in either nucleon-nucleon or
nucleus-nucleus collision, and assuming the projectile being the medium
primary cosmic
ray nuclei
 with a  mass $\sim$ 60 and the target with a mass $\sim$ 14, being
 a medium
atmospheric  air  nucleus.
It is clear that answering the question, if
 Centauros are produced at the
 midrapidity or
 rather in the projectile fragmentation region and where in the new
colliders should we
expect them,
 depends on a kind of a  target and a projectile,
which  are unknown objects in these experiments.
It is  seen
 that assuming nucleon-nucleon 
collision we can
expect 
Centauros 
 produced  somewhere between
the central and the fragmentation rapidity region. 
In this case, 
the threshold for  Centauro formation  is close to the Fermilab
collider energies.
 On the other hand, if Centauros are produced in nucleus-nucleus
collisions they
should be looked for in the projectile fragmentation region.
The question of the possible Centauro formation in the new collider
experiments (at RHIC
and LHC)
will be considered later.
 
\subsubsection{Other Centauro and Centauro-like events}

Besides the ``classical Chacaltaya Centauros'', a quite reasonable 
statistics of Centauro-like (or hadron rich) events has been collected
 as well  by Chacaltaya as by Pamir
Experiments.
Unfortunately, in contrast to the undisputable Centauro I, some of these
events were born at  rather large  distances from the apparatus, thus
they possess the significant fraction of electromagnetic component, which
probably
have been generated in the nuclear and electromagnetic cascade processes, 
in
the atmospheric layer above the chamber. Besides that the serious
difficulties were met during
 the  measurements and analysis of some exotic
super-familes
(with the very high visible energy, $E_{vis} \geq $ 500 TeV) which are
very
frequently
accompanied by the so called ``halo''. These are the reasons that some
events were not so detaily analysed and they are not so spectacular as the
Centauro I. Here they are named 
{\em Centauro-like events}. They are not so well known as ``classical''
Centauros although they carry a big amount of exciting experimental
information.
Here will be  listed and described examples of the most
interesting 
Centauro-like (or hadron-rich)
events.

\vspace*{0.2cm} 
1. {\em CENTAURO-NEW (C22-Sxxx-I019) \cite{Ohsawa,Durban}}.\\
      The second clean Centauro event was reported in 1997 at the
25th International Cosmic Ray Conference in Durban. The event was
found in the Chacaltaya two-storey chamber no. 22,
      consisting of the upper chamber (7 cm Pb), target layer (30 cm
      CH$_{2}$), air space (237 cm) and the lower chamber (11 cm Pb),
      i.e. of the total thickness $\sim 0.49 \lambda_{int}$.
      It is the family of 31 showers (13 with energy exceeding 1 TeV)
        of the total observed energy  57.4 TeV (or 51.2 TeV when the
energy threshold of 1 TeV is used).
No
      showers in the upper chamber
      have been found. According to ref. \cite{Durban} there is no
       trivial reason for the fact that no showers are observed
      in the upper chamber (the lower chamber was constructed only
      after the upper chamber was completed). Hence, the event
       is considered as  clean Centauro (or  Mini-Centauro
      regarding its relatively small energy), without presence
      of $\gamma$ rays. It is the second, besides the Centauro I,
     event of this type.

\vspace*{0.2cm} 
2. {\em CENTAURO VI \cite{Adelaide,Has_Tam,Navia}}.\\
      The event no. C20-107S-089I was observed in the  Chacaltaya
two-storey emulsion chamber no. 20.
      The measurements were restricted to X-ray film observation because
      the emulsion plates were not in fully satisfactory condition.
      The observed results, i.e. the domination of hadrons in the high
energy region and the exponential $E\cdot R$ distribution
     \footnote{E is an energy and R is a distance of a particle from the
     energy weighted family centre.}  can be
interpreted 
      under the hyphotesis of Centauro interaction. Assumption of the same
average transverse momentum of secondary hadrons as that measured for
Centauro I
($\langle p_{T}(\gamma)\rangle \sim$ 0.35 GeV/c) leads  to the
production height of
about  800 m.
      The estimated multiplicities and energies of photonic and hadronic
      components (for showers with visible energy greater than 4 TeV) at
the top of the chamber are: $N_{\gamma}$= 15,
      $\Sigma E_{\gamma}$ = 95.2 TeV, $N_{h}$ = 40, $\Sigma
E_{h}^{\gamma}$ = 900 TeV. The estimated number and energy of
      hadrons at the interaction point is: $N_{h}$ = 80, $\Sigma
E_{h}^{\gamma}$ =
      1500 TeV \cite{Navia}.

\vspace*{0.2cm} 
3. {\em CENTAURO VII \cite{Cen_VII,Adelaide,Has_Tam}}.\\
      The event no. C2187S75I \cite{Cen_VII,Has_Tam}
        named Centauro VII was the first Centauro with halo.
       Fortunately, the observed halo has not been developed so well and
       the energetic showers were distinguishable even on the X-ray film.
      It has been found in the Chacaltaya two-storey chamber no.
      21, consisting of the upper chamber (7 cm Pb) 
      and the lower one (11 cm Pb) 
 with the carbon target
       (23 cm pitch) and the air gap (2 m) between them.
      In this event, the detailed study of the shower core structure
      in nuclear emulsion plates was done. Also the decascading
      procedure was used to study the longitudinal and lateral
      characteristics of the event. The
family has
      an extremely  large visible energy ( $\Sigma E_{vis} \approx$ 5600
TeV + 500
TeV (halo) \cite{Cen_VII}). The interaction height, estimated by
triangulation measurements \cite{Cen_VII}, is about 2000-3000
       m above the chamber. Such value of the production height results in
      the average transverse momentum value consistent with the Centauro
type events. 
        The event is undisputable 
       the hadron-rich family with
       $Q_{h}$ value of the same order as for
``classical'' Centauros. $Q_{h}=$ 0.46 for $E_{th}=$ 2 TeV increasing to
$Q_{h}\approx
      0.8 $ when assuming the most unfavourable case that all halo energy
is 
       of electromagnetic origin and when the analysis is
       limited  to the most energetic
        cascades, with energies higher than 20 TeV (what
      seems to be justified by the large production height).
  
\begin{figure}[h]
\begin{center}
\mbox{}
\epsfig{file = 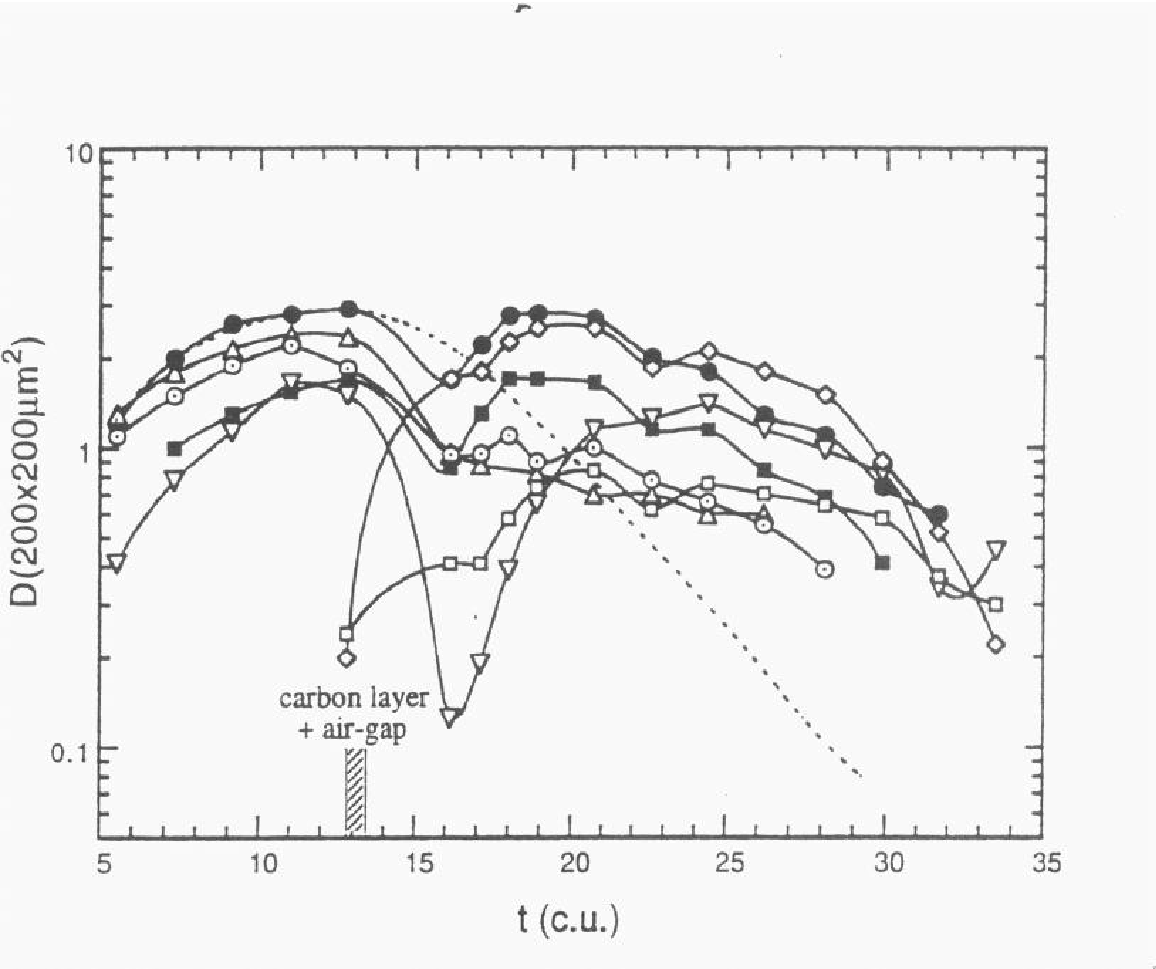,
bbllx=20,bblly=24,bburx=576,bbury=473,width=10cm}
\caption{Strongly penetrating cascades from Centauro VII
\cite{Has_Tam}.}
\label{Cen_VII_pen}
\end{center}
\end{figure}

      The family is in an old stage of development
       due to the long distance atmospheric propagation, judging from
      the existence of a large number of lower energy degraded
gamma-rays. However, the family contains, in the central region, several
 very high energy
       showers, higher than or close to 100 TeV. 
       Those are confined  within
       the  radius
       smaller than several centimeters from the centre of the family.
       All of them  started the shower development from the upper
chamber and were penetrating  deeply into the lower
       chamber, far beyond the expectation from simple electromagnetic
       cascades.  
       Their transition curves are shown in  Fig.~\ref{Cen_VII_pen}.

       This family is sometimes claimed to be the {\it Chiron-type}
interaction 
       because of  very large lateral spread of hadrons \cite{Has_Tam}
       (the spread of high energy hadrons with $E(\gamma) \ge $ 40 TeV
is as large as $\langle E(\gamma) \cdot R \rangle \simeq$ 1300
GeV$\cdot$m) 
       and the existence of the {\it strongly penetrating component},
       revealing typical {\it mini-cluster} configuration.
        The majority of very high energy penetrating cascades, given in 
        Fig. \ref{Cen_VII_pen}, show a  typical {\it mini-cluster}
structure.
 Re-analysis of the event
       \cite{Cen_V_VII} confirmed its exotic character and showed that
       characteristics of the hadronic secondaries are compatible with 
assumption of one
emitting
fire-ball.

\vspace*{0.2cm} 
4. {\em CENTAURO ``PAMIR'' \cite{Has_Tam,Cen_Pamir,Moscow}}.\\
       Centauro event, marked  in \cite{Has_Tam} as G178H178 
  or P3' C2 B178 in
\cite{Cen_Pamir,Moscow} was found in the
       carbon-type chamber no. 2 of P-3', consisting of 
       the standard gamma block (6 cm Pb) followed by the
       carbon target (60 cm C) and the hadron block (5 cm Pb), from
       USSR-Japan joint exposure at the Pamir.
 The event shows every feature of a Centauro,
       i.e. the energy is mainly released into the hadron-induced showers
and the average lateral visible-energy spread is much higher for the
hadron induced showers than for the $\gamma$-ray induced ones.
 For the hadron induced showers both
       spectra: the fractional visible energy and the lateral visible
       energy spread, can be well reproduced by exponential functions.
        The height of the interaction was estimated 
 to be H = 700 $ \pm$ 100 m,
       from the lateral visible energy spread of hadrons with
       $\Sigma E_{h}^{\gamma} \geq $4 TeV and by  using the same as for
  Centauro I
       $\langle p_{T}(\gamma) \rangle $ value.  
       Estimated numbers and energies of electromagnetic and hadronic
component at the top of the chamber are: $N_{\gamma}$ = 55, $\Sigma
E_{\gamma}$ = 372.5 TeV, $N_{h} = $ 45, $\Sigma E_{h}^{\gamma}$ =
      700 TeV. The multiplicity  and energy of  
 hadrons estimated  at the original
       interaction is  $ N_{h}$ = 77$\pm$16 and $\Sigma E_{h}^{\gamma}
=
1000$
TeV  \cite{Navia} respectively
 (for showers with the visible
energy
        greater than 2.3 TeV).

\vspace*{0.2cm} 
5. {\em ELENA \cite{Elena}}.\\
      Elena is a  superfamily with a total measured energy $\sim$ 1700
TeV,
      detected in the Pamir deep lead chamber 
      of a total thickness of 60 cm, what corresponds to $\sim 3
\lambda_{int}$.
    It has given a rather rare opportunity to observe the transition
    of particles produced in such high energy interaction through
    the deep chamber.
 The majority of
superfamilies were detected in relatively thin chambers ($\le  1.5
    \lambda_{int}$) what makes impossible the
detailed investigation
of
hadron characteristics.

 This event reveals the features of Centauro-type
     families, such as the wide energy weighted lateral  distribution
and the large fraction of
family energy transfered to
hadrons. $Q_{h} = 0.7 \pm 0.07$, when corrected for hadron detection
     efficiency,
 what locates the family well beyond the region of usual fluctuations
 on the $N_{h}$ vs. $Q_{h}$ diagram.
 Soft spectrum of gamma-rays indicates
that the family
is ``old''. The height of the initial interaction point
was
determined, assuming that $\langle p_{T}(\gamma)
\rangle$ of hadrons is 0.35 GeV/c. It gives the height  $\sim
1500-2000 $ m above
the chamber. A  similar analysis as that done for ``classical'' 
Centauros, i.e.  evaluation of the number of hadrons
generated in
the primary interaction and the number of their interactions
 in the air above the chamber, does not contradict
 the scenario of the development of a Centauro-type event occuring at
large height above the apparatus.

\begin{figure}[h]
\begin{center}
\mbox{}
\epsfig{file = 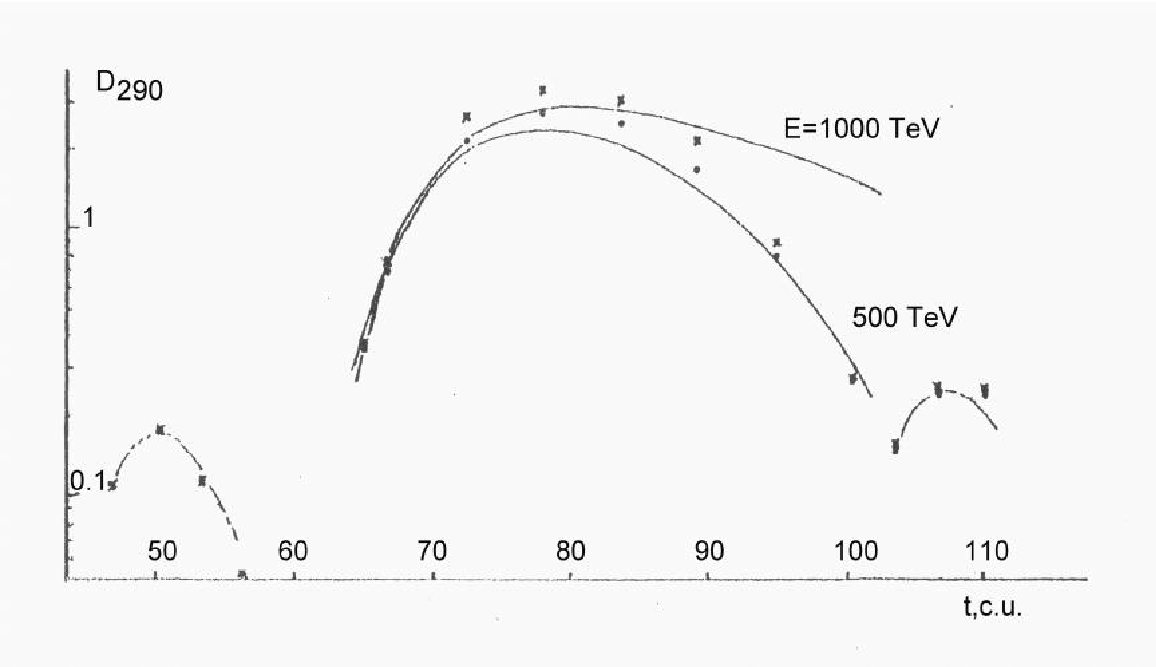,
bbllx=23,bblly=21,bburx=575,bbury=340,width=9cm}
\caption{Transition curve in terms of optical density $D_{290}$ of high
energy cascade detected in family  Elena
  \cite{Elena}. Black dots  are experimental points without methodical
   corrections, crosses are corrected experimental points.
Calculated curves for $\gamma$--quanta are also shown. }
\label{Elena_pen}
\end{center}
\end{figure}

The total number of detected hadrons is 31 and
among
     them 4 hadrons undergo in the chamber several interactions.
      Especially interesting is the ``leading'' hadron.
It was located at a distance of 7.5 mm from the family energy weighted
centre 
and  it was not accompanied by any
$\gamma
$-rays at a distance closer  than  1 mm. It started its
development rather deeply inside the chamber and escaped through
its  bottom after passing 35 cm of Pb ($\sim$ 65 c.u.).
Three separate maxima are seen in the cascade longitudinal development
 (see Fig. \ref{Elena_pen}).

Dark spots
produced on the X--ray films
by  electromagnetic showers induced by
 this hadron have large transverse dimensions, $\sim$ 3 mm,  and look
like a ``halo''. The estimated energies and the points of origin of the
showers
producing
each of the three ``humps''
have the following values:
$E_{1}$ = 9.8 TeV,
$t_{1}$ = 39.7 c.u., $E_{2}$ = 500-1000 TeV, $t_{2}$ = 57 c.u.,
$E_{3}$ = 20 TeV,
$t_{3}$ = 90 c.u.. 
The analysis presented in \cite{Elena} showed that it is
 difficult to explain such shape of a  transition curve ,
assuming the ``normal'' interaction of high energy leading hadron.
For example,  the probability of producing the observed ratio of
 released energies corresponding
to the first two humps by a  subsequent interactions
of a hadron in usual nuclear-electromagnetic cascade in  lead 
is as small as $\sim (1-4)\cdot 10^{-3}$.

\begin{figure}[h]
\begin{center}
\mbox{}
\epsfig{file = 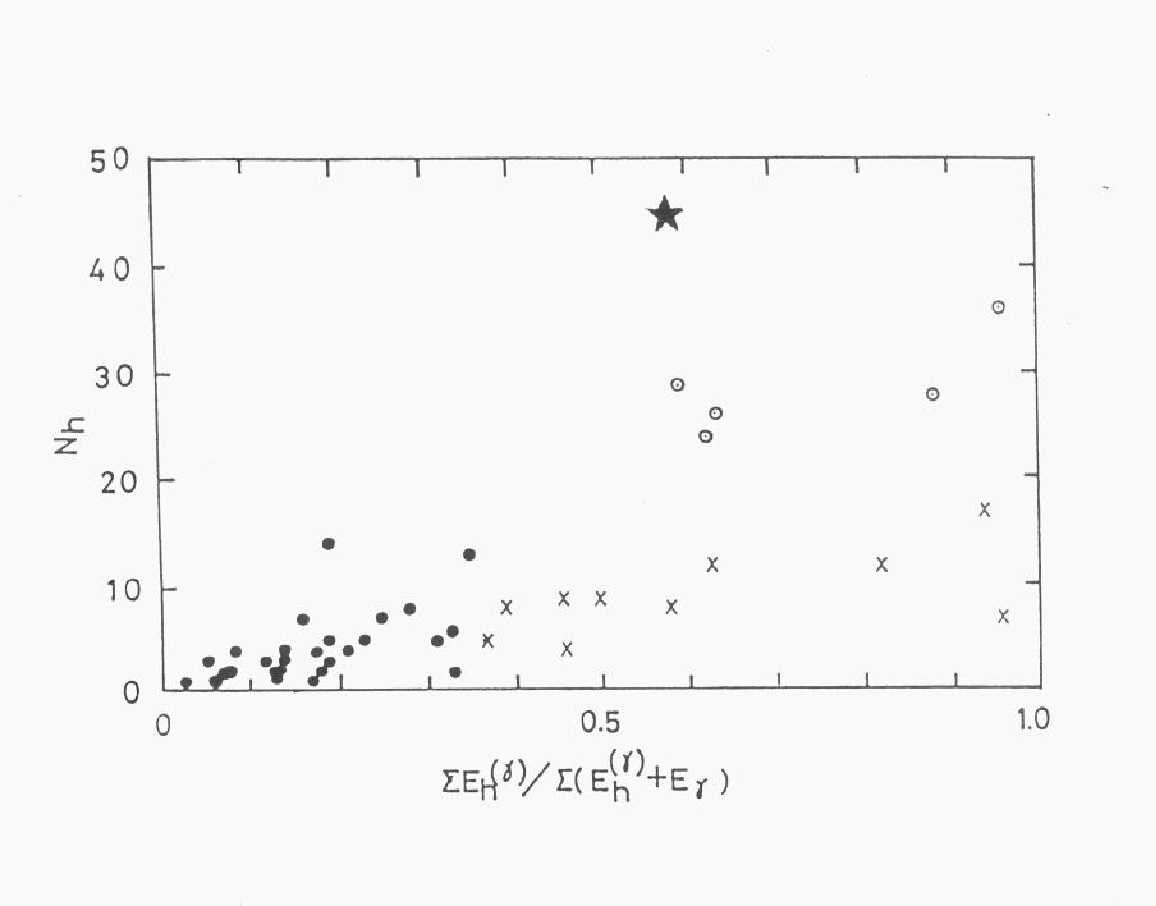,
bbllx=25,bblly=53,bburx=574,bbury=444,width=12cm}
\caption{Diagram of the number of hadrons and  hadronic energy
fraction:
Chacaltaya events with the total visible energy greater than 100 TeV
 \cite{Bayburina}:
($\circ$) Centauro, ($\times$) Mini-Centauro, ($\bullet$) others;
($\star$)  C-K
  \cite{Cen_Krakow}.}
\label{C-K_2}
\end{center}
\end{figure}

\vspace*{0.2cm}
6. {\em C-K \cite{Cen_Krakow,Cedzyna}}.\\
      The family named here C-K was found in a deep lead
chamber ``Pamir 76/77'',
      of a total thickness of 60 cm Pb.
       It has been measured and analysed by the Cracow group.
      It has been classified  as a  Centauro-like event
      accompanied by the strongly penetrating component. It was
the first hadron-rich event with so  spectacular  
evidence for penetrating cascades. Fortunately, it was found
in the homogenous type thick lead chamber which gives  possibility of  the
detailed study of the transition curve structures.
    Comparison of this family with classical Centauros  is
shown
      in 
Fig.~\ref{C-K_2}. On  the diagram
 of the number of hadrons $N_{h}$ vs. the hadron
energy fraction $Q_{h}$ for Chacaltaya events \cite{Lattes,Bayburina}
 the event C-K has been
 marked by the star.
 For the Chacaltaya events, the hadronic part
includes C-jets, Pb-jets-lower and identified Pb-jet-upper, and A-jets
 identified as a shower cluster with association of a hadron in it.
 Since the effective thickness of the Chacaltaya chambers is about 1.5
nuclear mean free paths, and the ``identified A-jets'' cover only the
A-jets with production heights less than $\sim$ 100 meters, the
``hadron energy sum'' $\Sigma E^{(\gamma)}_{h}$ will give an
under-estimate on an average. The C-K event has been registered in the
deep lead chamber, so that the loss due to penetration is negligible.
 On the other hand a relatively high threshold for hadrons in Pamir
 chambers 
 (here it was assumed to be 2 TeV) as compared with the Chacaltaya
detectors
  reduces the number of hadrons. Keeping in mind these two effects it
seems that such comparison is roughly reasonable.
The  hadron-rich nature of the event is evident.

\begin{figure}[h]
\begin{center}
\mbox{}
\epsfig{file = 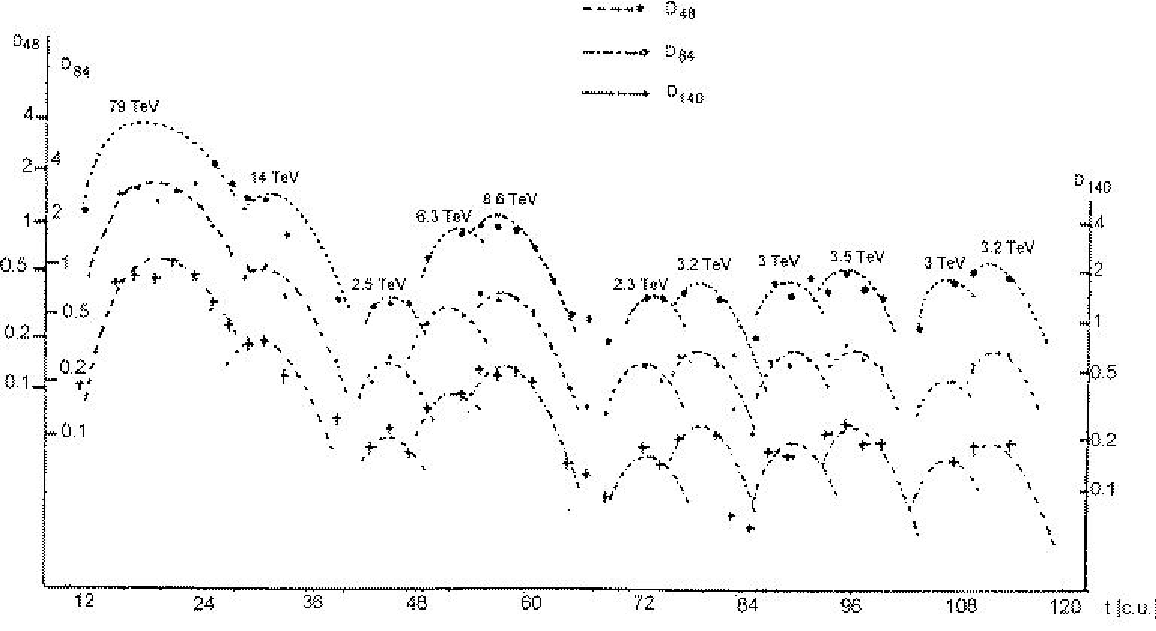,
bbllx=20,bblly=19,bburx=574,bbury=323,width=10cm}
\caption{ Transition curves in X-ray film darkness D (measured in
three diaphragms of a radius R = 48, 84 and 140 $\mu$) for cascade
no. 197.08. Energy (in TeV units) liberated into the
soft component is indicated at each hump (averaged over three estimated
values) \cite{Cen_Krakow}.}
\label{197}
\end{center}
\end{figure}

Among cascades belonging to the family, two cascades reveal
unusual features. They were observed not far from the energy weighted
centre of  the family,
in the very close distance one to the other. Both cascades 
demonstrated a multicore
structure. They started their development rather deeply inside the
chamber and after passing a  very thick layer
of  lead, escaped through the bottom of the chamber.
  The features of these cascades are summarized
in  Table~\ref{C-K_cascades}.
 Their transition curves 
 reveal surprising features.
 Both cascades have unexpectedly  long
range and  many maxima character. The longer cascade, shown in
Fig.~\ref{197}  penetrated more
than 109 cascade units and 11 maxima at its transition curve appeared.
The average distance between the neighbouring humps is very small,
equals only 10.4 $\pm$ 4 c.u.. These features are hardly explained by
simulations. According to calculations 
\cite{Iwanienko}, an average cascade
initiated by a hadron with energy $\sim$ 100 TeV can be  detectable in
such chamber only to
the
depth of
45.5 c.u. (for measurements with diaphragma of a radius of 50 $\mu$).
Morever, the many maxima cascade curves are obtained in simulations very
rarely.
But even in that case, only very limited number of  humps can be
detected (two-three) and the average distance between them is much
larger (23 $\pm$ 10 c.u.) than that observed here. 
Unexpectedly, both  unusual  cascades penetrate through
the whole chamber
practically
without
noticeable attenuation,  and a  weak growing rather than quenching of the
cascades is  observed.
From this point of view they remind the long-lived cascades
discovered by means of the Tien-Shan calorimeter \cite{Yakovlev}.

\begin{table}
\begin{center}
\caption[\scriptsize{Unusual cascades in Centauro-like event C-K.}]
{Unusual cascades in Centauro-like event C-K.}
\label{C-K_cascades}
\begin{scriptsize}
\vspace{0.3cm}
\begin{tabular}{|c|cc|cc|cc|}
\hline
& & & & & &\\
{\em Cascade} & \multicolumn{2}{c|}{{\em Starts at}} &
\multicolumn{2}{c|}{{\em Finishes
at}}&
 {\em Penetrated}& {\em Number
of}\\
{\em no.}      &{\em c.u.}& {\em layer no.}&{\em c.u.}& {\em layer
no.}&{\em c.u.}&{\em  maxima}\\
& & & & & &\\
\hline
& & & & & &\\
197.08&11.8&5&120.7& escape&108.9&11\\
748.01&48.3&23&120.7& through&72.4&5\\
      &    &    &   &  bottom  &    &\\
\hline 
\end{tabular}
\end{scriptsize}
\end{center}
\vspace*{-3mm}
\end{table}

\vspace*{0.2cm} 
7. {\em P3' C2 201 \cite{Has_Tam,Moscow,P3C2201}}.\\
       This candidate of
       Centauro type interaction of the highest energy range,
occuring at very high altitude above the apparatus, marked as
G201H201 in ref. \cite{Has_Tam},
       was detected in the Pamir joint standard  carbon type
      chamber P3'-C2.
 It has got the
      small
       blackened area (in the radius of about  2 mm  on
the X-ray film) in
the center of the family,
      being considered as the premature stage of the halo. It is
      composed of several very high energy showers strongly
      penetrating into the lower chamber. It is difficult to clearly
      resolve such a blackened area into individual shower-cores
      on X-ray film, and therefore this part was separately analysed.
      The estimated height of the family turns to be $\sim 2000$ m
      above the chamber (from the lateral spread of the high energy
      hadrons and by assuming $\langle p_{T}(\gamma) \rangle \simeq$ 0.35
GeV/c).
      Such value of the height gives the original multiplicity of hadrons
      of about 100.

\vspace*{0.2cm} 
8. {\em HALINA \cite{Has_Tam,Halina_Biel, Halina_Moscow}}.\\
       This event has been found among 7 high-energy hadron-gamma
       families of $\Sigma E_{\gamma}\geq$ 800 TeV in a systematic
       study of carbon chambers of $\simeq$ 400 m$^{2}$ year exposure
       (the Polish part). It was detected in the Pamir-79/80 carbon
chamber
C42,
       consisting of the standard gamma block (of 6 cm Pb) and two hadron
       blocks (of 5 cm Pb) interlayed the carbon target  (60 cm of
carbon), i.e.
       the effective thickness of the chamber was $\sim$ 2.3
 $\lambda{geo}$.
It is a good design for investigation of a hadronic component, as the
detection
       probability of hadrons in the chamber is larger than that of the
standard type carbon chamber and it is  close to 90\%.
It is no halo event despite of its high energy.

       This event reveals  all exotic Centauro features:\vspace*{-0.2cm}
\begin{itemize}
\item It has  hadron-rich nature: the  energy fraction of
       a hadronic
       component in the peripheral region  $Q_{h} = 0.45$. About
       30\% of its energy is carried by 106 identified hadrons 
        (with the visible energy greater than 2 TeV).\vspace*{-0.2cm} 
\item Its
       longitudinal and lateral characteristics cannot be fully
       explained by usual hadron interactions, even when the contributions
of heavy nuclei in primary
       particles are considered.\vspace*{-2mm}
\item Extraordinary wide lateral spread both in gamma-rays and in hadrons
      results in the transverse momentum of produced particles
      $p_{T}\simeq 1$ GeV/c.\vspace*{-0.2cm}
\item 
      There are 6 showers which penetrate from the gamma-block to
       both hadron blocks. \vspace*{-0.2cm}
\end{itemize}

9. {\em ANDROMEDA \cite{Yamashita}}.\\
       Andromeda was detected in the flat chamber CH-14 (11 cm Pb, 41\%
hadron
detection efficiency) of the Chacaltaya
     experiment and it is known as the first and most famous example
      of the superfamily with the huge halo of the radius of about 3.2
cm. Unfortunately, the chamber
      was not thick enough to  study the full development of the halo
transition
curve and  a hadron component of the event.
Some authors \cite{Bayburina} interpreted a  halo in Andromeda as
caused by a numerous bundle of high energy atmospheric $\gamma$ quanta.
 However, the
observed
 decrease of the intensity
      of the halo transition curve with the depth in the chamber is less
than that expected from
      a pure electron shower. Besides that the energy spectrum of
hadrons is
      harder than that of electrons/$\gamma$-rays and generally its
      characteristics are found to be inconsistent with the
hypothesis of a proton primary with pion multiple production
     under the scaling model. In ref. \cite{Yamashita} the event is
claimed
      to be  hadron rich. Estimates of the total energy of the
      photonic and 
hadronic
      component (after  corrections for detection efficiency) show that
the
      hadrons carry $\sim$ 47\% of the total energy.

\vspace*{0.2cm} 
10. {\em URSA MAIOR \cite{Has_Tam,Yamashita,Ursa,Moscow_Ursa}}.\\
      It was found in the two-storey Chacaltaya chamber CH-15 (64\% hadron
detection
      efficiency). Similarly as Andromeda it is the typical halo (with the 
multi--core
 structure) super-family.
      The hadronic component occupies the substantial part of
      the whole energy (38\% after correction for efficiency
      and above the threshold of the shower spot detection). The
      study of  correlations between hadrons and gamma rays 
      situated the family in the region of big values of hadron
number and hadron energy, and outside fluctuations expected from a scaling
type model with
     the proton primary.

\vspace*{0.2cm} 
11. {\em MINI ANDROMEDA III \cite{Has_Tam,Ursa,Moscow_Ursa}}.\\
     It was registered in Chacaltaya two-storey chamber CH-19 (59\%
      hadron detection efficiency). The wide halo (with the radius
 of about 2.2 cm) reveals the core
structure. Similarly as in the previous events the chamber is 
too thin for a study of the full longitudinal halo development
and the structures in the  halo transition curve.
But even at this limited length, the  observed attenuation of the
halo transition curve is weaker than that expected from
      a pure electromagnetic origin.
      In spite of the indications on the very high production
      altitude (wide lateral hadronic spread) this family is extremely
      rich in hadrons both in  energy fraction and in number. 
      The hadrons occupy $\sim$ 44\% of the total energy (above
the
      threshold of shower spot). 

\begin{figure}[h]
\begin{center}
\vspace*{1.cm}
\begin{turn}{90}%
\mbox{}
\epsfig{file = 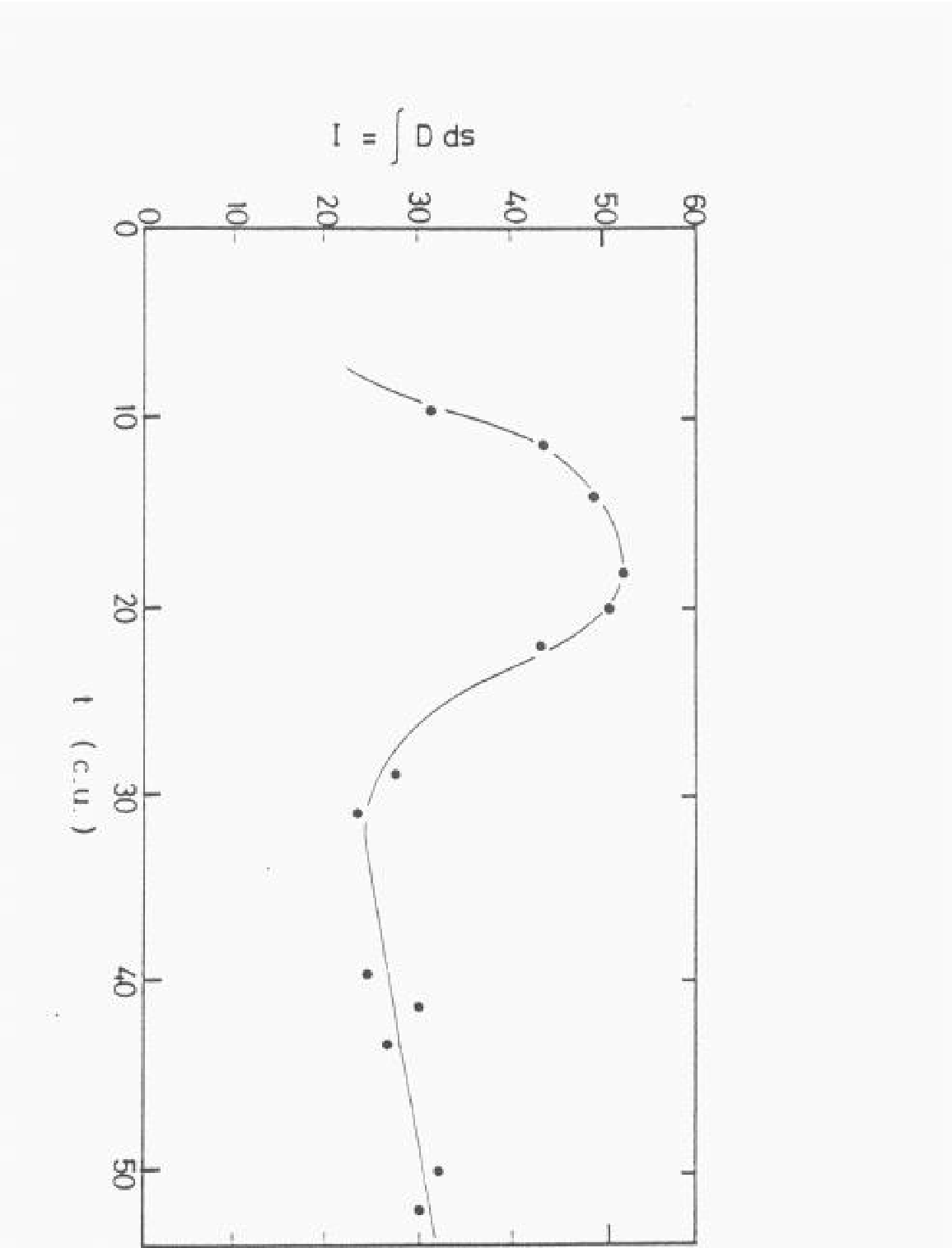,
bbllx=21,bblly=20,bburx=478,bbury=706,width=7cm}
\end{turn}
\caption{Penetrating black core in the family ``Tatyana''. The curve
presents the darkness of the core with halo on the X-ray films with the
depth in radiation units in the chamber
  \cite{Bayburina}.}
\label{Tatyana_halo}
\end{center}
\end{figure}

\vspace*{0.2cm} 
12. {\em TATYANA \cite{Bayburina,Tatyana}}.\\
     Tatyana is one of the highest energy events known in the world
statistics, extremely interesting and difficult for measurements and
analysis. The estimated visible energy of the event is about
     11 000 TeV, thus the total energy of the event was  estimated to be
     about 15 000 TeV.
     The super-family was detected in the Pamir 450-73/74 carbon type
chamber.
     In contrary to the majority of other chambers, it  is the  extremely
deep apparatus, consisting of the standard gamma
block
     and  four identical hadron blocks (each consisting  of 20 cm  
     carbon layer followed by
      5 cm of Pb) what gives in total about 5 $\lambda _{int}$ or
      55 cascade units. It makes possible the direct study of
the
      penetrating power of the family. 
      The halo, occupying the central part of the event easily
traversed  the whole chamber, revealing
      the unexpectedly strong penetrating power.
        The blackness of the core does not show
        any sign of attenuation down to the bottom of the chamber, even
       a rise of the transition curve is observed (see
Fig.~\ref{Tatyana_halo}).

 The absence of the
      attenuation of the core through the chamber indicates that the 
      secondary particles are  more penetrative than normal
hadrons.
       Outside the core halo region, of a diameter 
$\sim$  1 cm,
      there are numerous showers. There were detected  224 $\gamma$--quanta
      with $\Sigma E_{\gamma} \sim$ 3200 TeV and  66 hadrons with
      $\Sigma E^{\gamma}_{h} \sim$ 1500 TeV.
Tatyana has been classified as
       the family with high energy of the hadronic component.
 
  \vspace*{0.3cm}
 The main
features of these and other
Centauro--like/hadron--rich events are summarized in
Tables \ref{Cen_2}, \ref{Cen_3} and \ref{Cen_4}.
The hadron dominant character of families registered in Pamir-joint
chambers series P3 and P2 was spectacularly presented on the correlation
diagram ($N_{h}$ vs. $Q_{h}$) in \cite{Moscow}, also in comparison
with classical Chacaltaya Centauros and some halo families such as
Andromeda, Ursa Maior and U.M.III \cite{P3C2505_Moscow}, see also
Fig.~\ref{Centauro_statistics} in subsection 3.1.

\begin{table}
\begin{center}
\caption[\scriptsize{Centauro-like events.}]
{Centauro-like events.}
\label{Cen_2}
\begin{scriptsize}
\vskip0.3cm
\begin{tabular}{|cccccccccc|}
\hline
  & & & & & & & & &\\
 {\em Event,} & {\em Collab}., & &{\em N} &
$Energy $ &
 $Q_{h}$& $\langle ER \rangle$&$E_{halo}$&$ E_{th}$&{\em Remarks}\\
 {\em Refer.}&{\em Chamber}.& & &[{\em TeV}]& &[$GeV\cdot m$]&
[{\em TeV}]&[{\em TeV}] &  \\
 & & & & & & & & &\\
\hline\hline
 & & & & & & & & &\\
CENT. &Brasil-&$ \gamma$&0&0& & & & &\\
NEW   &Japan&h&13&51.2&1. & & & 1&\\
\cite{Ohsawa,Durban}&2-storey& & & & & & & &\\
 && & & & & & & &\\
\hline
 & & & & & & & & &\\
CENT. & Brasil-&$\gamma$&56 &361& & & & 4 & \\
VI &Japan & & 157&644 & & & & &\\
\cite{Adelaide,Has_Tam} & & & & & & && 2& \\
 &2-storey & h& 28 & 390 & & & & 4& \\
  & & & 68 & 496 & & & & 2 & \\
 & & tot& & 751 & 0.52& 735$^{1}$&& 4& \\
 & &      & & 1140& 0.44& 803$^{1}$&& 2& \\
 & & & & & & & & &\\
\hline
& & & & & & & & &\\
CENT.&Brasil-& $\gamma$&547&2978& & & & 2& Centauro \\
VII &Japan & & 265 & 2179 & & & & 4 &or Chiron \\
 \cite{Cen_VII,Adelaide, Has_Tam}& &h&129&2486& & &
&2&\\
 &2-storey & & 74 & 2328& & & & 4 & penetr.\\
 & & tot& & 5464& 0.46& & 500&2&  cascades\\
 & &    & & 4506& 0.52& 842& &4&   and\\
 & &     &  &     & 0.8& 857$^{1}$& &20&mini-\\
 & &     &  &     &              &    & &  &clusters,\\
 & &     &  &     &              &    & &  & halo\\
&&&&&&&&&\\
\hline
 & & & & & & & & & \\
CENT.&USSR-&$\gamma$&15&95& &67 & & 4& \\
PAMIR       &Japan&        &120&298& &28.6&& 1&
\\
\cite{Has_Tam,Cen_Pamir,Moscow}&&h&22&444& &244 &&4& \\
                               &standard& &37&476& &173 &&1& \\
 &carbon&tot& & 539&0.82& 495$^{1}$&&4& \\
 &        &     & &    &0.62&          &&1& \\
& & & & & & & & &\\
\hline
& & & & & & & & &\\
ELENA&Pamir&$\gamma$&78&600& &360 & &4&str.pen.\\
\cite{Elena}& &h&23&1100& &885 & &4&leading\\
 &deep&tot&&1700&0.65$\pm$0.05& & & &cascade\\          
            &carbon& h&22$^{2}$&300$^{2}$&   &475$^{2}$ &
&4&\\
            &   &  &  &   &   & & & &
\\
 & & & & & & & & &\\
\hline
 & & & & & & & & &\\
C-K&Pamir&$\gamma$& 74 &306  &  & 111 &  &$\sim$ 1 & str.pen.\\
\cite{Cen_Krakow}&&h& 55 &531
& &195 & & & cascades\\
   &deep       &          &  &382$^{3}$& &  &
&&\\
   &Pb   & tot      &  &  &0.64& & & & \\
   & & $\gamma$ &27  &198  &  &  &  & 4 & \\
   &
   & h        &22  &446& & &  &&\\
   &   &          &  &297$^{3}$ & & & & &\\
   &   &tot     &  &          &0.69 & & & &\\
&&&&&&&&&\\
\hline
\end{tabular}
\end{scriptsize}
\end{center}
{\scriptsize $^{1}$
measured by showers of E$(\gamma)\geq $ 20 TeV\\
$^{2}$ without leading cascade\\
$^{3}$ energy released only in the first peaks}
\end{table}

\begin{table}
\begin{center}
\caption[\scriptsize{Hadron-rich events.}]
{Hadron-rich events.}
\label{Cen_3}
\begin{scriptsize}
\vskip0.3cm
\begin{tabular}{|cccccccccc|}
\hline
&&&&&&&&&\\
 {\em Event,} & {\em Collab.} &  & {\em N} & {\em Energy}  &
 $Q_{h}$& $\langle ER \rangle$&$E_{halo}$&$ E_{th}$&{\em Remarks}\\
 {\em Refer.}& {\em Chamber}& & &[ {\em TeV}]& &[ $ GeV \cdot m$]&  
[{\em
 TeV}]&[ {\em TeV} ] &  \\
& & & & & & & & &\\
\hline\hline
& & & & & & & & &\\
P3'C2&USSR-&$\gamma$&132&1479& & &
&4&\\
201&Japan&h&43&1089& & & &4&penetr.casc.,\\
\cite{Has_Tam,P3C2201} &&tot&&2568&0.42&435$^{1}$&400&4&premature\\
 &standard&      & &    &              &          & & & halo        \\
 &carbon& & & & & & & &
\\
& & & & & & & & &\\
\hline
& & & & & & & & &\\
HALINA&Pamir&&&&&&&&\\   
\cite{Has_Tam,Halina_Biel,Halina_Moscow}&&$\gamma$&
171&1630& &300 & - & 4 &\\
 &deep & & 469 & 2468 & & & - & 2 & \\
& carbon& h&65&936& &583 & - &4&penetr. casc.\\
& & & 106& 1071& & & - &2& \\  
& & tot& & 2566&0.37&883$^{1}$&-&4&\\
& & & & 3540 &&&&2 &
\\
&&&&&&&&&\\
\hline
&&&&&&&&&\\
C141-&Pamir&&&&&&&&\\
G4836&&$\gamma$&157&1277& &295 & - &4& hadron
rich\\
H4784 &standard & & 346 &1807& & & - &2 & \\
\cite{Has_Tam,Halina_Biel} &carbon& h & 31 & 415 & &903 & - & 4 &\\
 & & & 50 & 482 & & & - & 2 &\\
& & tot & & 1692 & 0.25 & 1071$^{1}$&-&4& \\
& &     & & 2289 &      &           & &2& \\
&&&&&&&&&\\
\hline
&&&&&&&&&\\
ANDRO-&Chacal.&$\gamma$&627&4488& &
 &&1&\\
MEDA&&h&110
&1656& &
 & & 1& \\
\cite{Yamashita}&flat Pb & &268$^{5}$
&(4039)$^{5}$&$\sim 0.47^{5}$ & & & & \\
 &&total& & & & & $\sim$21000& & \\
&&&&&&&&&\\
\hline
&&&&&&&&&\\
URSA&Chacal.&$\gamma$&239&1074& &
 & & 2&single \\
MAIOR&& & 430&1344& & & &1&cluster \\
\cite{Has_Tam,Yamashita}&2-storey&h&38&508& &
&&2&\\
\cite{Ursa,Moscow_Ursa} && &54&532& &
 & &1&$\sim$ 1080 TeV\\
 &      & &(84)$^{5}$&(830)$^{5}$&$\sim 0.38^{5}$&&&1&\\
 & &tot& & 1582&0.32&498$^{1}$
 &$\sim$ 980&2& \\
&&&&&&&&&\\
\hline
&&&&&&&&&\\
M.A.III&Chacal.&$\gamma$&192&1701& & & &4& \\
\cite{Has_Tam,Ursa,Moscow_Ursa}&& & 441&2370& & & &2& \\
 &2-storey& & 537&2531& &
 & &1& \\
 &&h&80&1070& & & &4& \\
 &      & &112&1167&& & &2& \\
 &      & &115&1172& &
 &  &1& \\
 &      & &(195)$^{5}$&(1986)$^{5}$&$\sim 0.44^{5}$ & & & &\\
 &      &tot& &2771&0.39&842$^{1}$& &4& \\
 &      &     && 3536&0.33&866$^{}$&$\sim$5060&2& \\
&&&&&&&&&\\\hline
\end{tabular}
\end{scriptsize}
\end{center}
{\scriptsize $^{1}$ measured by showers of $E(\gamma) \geq$ 20 TeV\\
$^{5}$ after correction for detection efficiency}
\end{table}

\begin{table}[t]
\begin{center} 
\caption[\scriptsize{Hadron-rich
 events, continuation.}]
{Hadron-rich
 events, continuation.}
\label{Cen_4}
\begin{scriptsize}
\vskip0.3cm
\begin{tabular}{|cccccccccc|}
\hline
&&&&&&&&&\\
 {\em Event} & {\em Collab.} &  & $N$ & $Energy $ &
 $Q_{h}$& $\langle ER \rangle$&$E_{halo}$&$ E_{th}$&{\em Remarks}\\
 {\em Refer.}&{\em Chamber}& & &[$TeV$]& &[$GeV \cdot m$]&
[$TeV$]&[$TeV$] &  \\
&&&&&&&&&\\
\hline\hline
&&&&&&&&&\\
TAT-&Pamir&$\gamma$&224&3200& & & &$\sim$ 1& periph.\\
YANA&&h&66&1500& & & && region\\
\cite{Tatyana} &thick&tot& & $\sim$ 11000& & &$\sim$ 6000&& 
\\
 &carbon   &    & &             & & &
           & & str.penetr.\\
 &       &     & &             & &            & &&  halo\\
&&&&&&&&&\\
\hline
&&&&&&&&&\\
P2C96-&USSR-&$\gamma$&40&288& & & & 4&data for \\
125&Japan&h&20&579& & & & 4& off halo part\\
\cite{Has_Tam,Moscow,P2C96125}  & &tot&&5437&0.67&1466$^{1}$
&4570&4&(r $\geq 1.2$ cm) \\
  &standard && &   &    &          &    & &2 penetr.\\
  & carbon       && &   &    &          &    & &clusters\\
&&&&&&&&&\\
\hline
&&&&&&&&&\\
P3'C2-&USSR-&$\gamma$&41&316& & & &4& \\
168&Japan&h&15&212& & & &4& \\
\cite{Has_Tam,Moscow} &&tot& & 528& 0.40&1007$^{1}$&-&4& \\
 &standard&     & &    &     &    & & & \\
&carbon&&&&&&&&\\
&&&&&&&&&\\
\hline
&&&&&&&&&\\
P3'C5-&USSR-&$\gamma$&98&705& & & &4&data for\\
505&Japan&h&49&513& & & &4&off halo\\
\cite{Has_Tam,P3C2505_Moscow}&&tot& &10400&0.42&3235$^{1}$&9200&4&(r
$\geq$ 2.2 cm)\\
     &standard   &     & &     &               &    && &\\
     &carbon     &     & &     &               &    & &&penetr.\\
     &     &     & &     &               &    & &&halo\\
&&&&&&&&&\\
\hline
&&&&&&&&&\\
P3'C2-&USSR-&$\gamma$&30&246& & 350& &4& \\
228&Japan&h&18&236& &875 & &4& \\
\cite{Moscow} &&tot& &483&0.60&&-&4& \\
&standard&&&&&&&&\\
& carbon&&&&&&&&\\
&&&&&&&&&\\
\hline  
\end{tabular}
\end{scriptsize}
\end{center}
\hspace*{0.5cm}
{\scriptsize
$^{1}$ measured by showers of E$(\gamma) \geq $20 TeV\\
}
\vspace*{-1cm}
\end{table}

\subsection{Mini-Centauros}

Mini-Centauros  are events of the same hadron dominant
nature as Centauros,
the difference being their smaller multiplicity. There were reported by
the Brasil-Japan Collaboration
15 atmospheric  and 9 produced in the target layer Mini-Centauros
\cite{Lattes, Navia}. Among 15 atmospheric families there were two
favourable cases where the heights of the interaction vertex have
been
determined through triangulation measurement of shower positions and
directions  \cite{La Paz}. In  these events the direct measurements  of the
transverse
momenta of hadronic showers were possible, giving $\langle p_{T}(\gamma)
\rangle
\simeq$
0.35
GeV/c. In  searching for Mini-Centauros among C-jets the criterion
was imposed that any pair of showers which couple to form a particle of
neutral pion rest mass within the experimental error, i.e. $90 \le
m_{i,j}
\le 200 $ MeV, where $m_{i,j}$ is the invariant mass of the shower pair,
was not found.
 
The main features of Mini-Centauro  events are the following:
 \begin{enumerate}\vspace*{-2mm}
\item Hadron multiplicity $N_{h} \simeq$ 10-20.\vspace*{-2mm}
\item $\langle p_{T}(\gamma) \rangle$ value as large as for Centauros
($\langle p_{T}(\gamma) \rangle\simeq
      0.35 \pm 0.10$ GeV/c).\vspace*{-2mm}
\item Approximately gaussian pseudorapidity distribution.\vspace*{-2mm} 
\item Experimental characteristics in accordance with isotropic 
      decay of the fireball with the mass $M_{fb} \simeq$ 35 GeV
      and average multiplicity $\langle N_{h} \rangle = 15 \pm
3$.\vspace*{-2mm}
\item Incident energy $ <E_{0}>_{lab}\simeq$ 940 (100) TeV for atmospheric
      (carbon target) Mini-Centauros. These values correspond to
      $\sqrt{s}$ = 1.3 TeV (430 GeV) when assuming a nucleon incidence.
 \vspace*{-2mm}
\item Average pseudorapidity in cms system, assuming
      the proton incidence $\langle \eta \rangle $= 3.7 (they are
      produced in  a little more forward region then Centauros).
\vspace*{-2mm}
\end{enumerate}

The characteristics of individual Chacaltaya Mini--Centauros can be found
in \cite{Lattes}.
Similarly as Centauros they are sometimes  accompanied by
the strongly penetrating component. The detailed study of C-jets
and Pb-jets from Chacaltaya Mini-Centauros (from chambers no.
17, 18 and 19) revealed the existence of penetrative mini-clusters
among them \cite{C-jets}. This analysis gave also some suggestions
on the genetic relations among different types of exotic phenomena.

 Families of similar type were reported also by Pamir Collaboration.
 By a systematic survey in the Pamir chamber in the exposure of $\sim$
100 m$^{2}$year \cite{Bayburina,Mini-Cen_Pamir}  six
Mini-Centauro events  were found among $\sim$ 50  families
observed in the Pamir carbon-chambers C18, C19, C24 in ``Pamir 78/79''
and C34 in ``Pamir 79/80''. Comparison with simulation calculations showed
that these events are beyond fluctuations in the atmospheric nuclear
cascade process originating from normal type meson production (see
Figure 2 in ref. \cite{Mini-Cen_Pamir}). 

The examples of other than ``classical '' Chacaltaya Mini-Centauros
are shown in  Table~\ref{Mini_Cen}.

\begin{table}
\begin{center}
\caption[\scriptsize{Mini-Centauros in Pamir-joint chambers, examples.}]
{Mini-Centauros in Pamir-joint chambers, examples.}
\label{Mini_Cen}
\begin{scriptsize}
\vskip0.3cm
\begin{tabular}{|ccccccccc|}
\hline
&&&&&&&&\\
 {\em Event,} & {\em Collab.} &  & $N$ & $Energy $ &
 $Q_{h}$& $\langle ER \rangle$&$ E_{th}$&$Remarks$\\
 {\em Refer.}& $Chamber$& & &[$TeV$]& &[$GeV\cdot m$]&
[$TeV$] &  \\
&&&&&&&&\\
\hline\hline
&&&&&&&&
\\
P3'C2&USSR-&$\gamma$&3&25& &57 & 4 & \\
223&Japan&h&5&302& &172 & 4 &\\
\cite{Moscow} & &tot& & 327&0.92& &  4& \\
 &standard&&&&&&&\\
 &carbon&&&&&&&\\
&&&&&&&&\\
\hline
&&&&&&&&\\
G544H534&USSR-&$\gamma$&18&187&&&4& \\
\cite{Has_Tam}&Japan&h&11&339&&&4&\\
              &     &tot&&526&0.65&309$^{1}$&4&\\
&standard&&&&&&&\\
&carbon&&&&&&&\\
&&&&&&&&\\
\hline
\end{tabular}
\end{scriptsize}
\end{center}
\hspace*{1.5cm}
{\scriptsize $^{1}$measured by showers of $E(\gamma) \geq$ 20 TeV.}
\vspace*{-2mm}
\end{table}

\subsection{Chirons}

 Chirons are hadron-rich species characterized by   the
following
features:
\begin{enumerate}\vspace*{-2mm}
\item
The interaction energy close to that of Centauros and estimated to be
  $\sim$ 1670 TeV.\vspace*{-2mm}
\item Rather low hadron multiplicity $N_{h}\approx$ 10--20.
The characteristic feature is the existence of {\it clean hadronic
cascades}, not clustered among electromagnetic cascades from atmospheric
showering of
$\gamma$'s in the atmosphere.\vspace*{-2mm}
\item Extremely large lateral spread ( $<E(\gamma)R>$ = 10--20
GeV$\cdot$m, hence
     $\langle p_{T}(\gamma) \rangle$  being 2-3 GeV/c,
     and $\langle p_{T} \rangle \approx$ 10-15 GeV/c).\vspace*{-1mm}
\item Pseudorapidity   
distribution of high energy showers
       in accordance with  isotropic
      decay of the fireball with the mass
       $M_{Chiron} \simeq$ 180 GeV and  centered at $\eta_{cms} \simeq$
2.3 \cite{Lattes}.\vspace*{-2mm}
\item Frequent appearance of  strongly penetrating
      single cores or the so--called mini--clusters
      (i.e.  clusters
       of particles with very small lateral spread corresponding to  very
      low mutual transverse momentum $p_{T}(\gamma) \approx$ 10-20
MeV/c). They are
      observed mainly in the centre of the family.\vspace*{-2mm}
\item Existence of unusual hadronic component with the short interaction
mean
      free path, as small as $\sim$ 1/3-1/2 of  the nucleon geometrical
      collision mean free path.\vspace*{-2mm}
\end{enumerate}

The study of C-jets and Pb-jets belonging to the Chiron type
families may indicate  the existence of the chain of
genetic relations \cite{C-jets}  among such phenomena as Chiron, Centauro,
Mini-Centauro etc.

\begin{figure}
\begin{center}
\mbox{}
\epsfig{file = 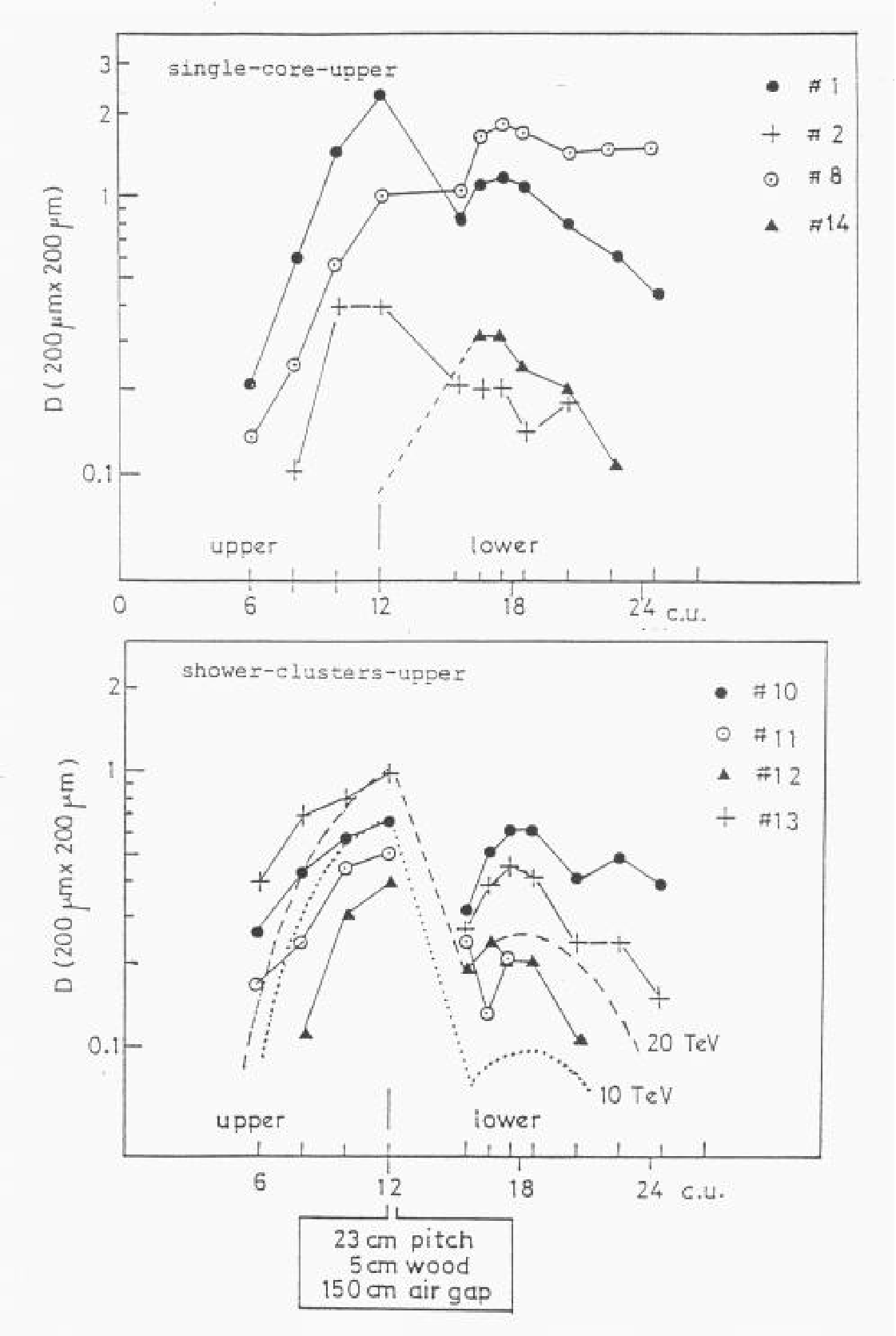,
bbllx=39,bblly=41,bburx=556,bbury=801,width=12cm}
\caption{Transition curves of high energy showers in Chiron I
  \cite{Has_Tokyo}. The curves expected for electromagnetic cascades
  are also shown by dotted and dashed lines.}
\label{Chiron_I}
\end{center}
\end{figure}

 Similarly as ``classical'' Chacaltaya Centauros 
 several clean Chiron families  have been
 reported 
 by Chacaltaya group
 \cite{Lattes}.
 The first event of this type, Chiron I
(family 198S-154I) \cite{Has_Tokyo,Bayburina},
was found in 1979 in Chacaltaya chamber no. 19.
 Similarly as Centauro~I, it is the clean family for which the altitude of
interaction vertex was measured by triangulation method to be 330 $\pm$
30 m above the chamber. This allowed to measure transverse momenta 
of
observed secondaries and resulting $\langle p_{T}(\gamma) \rangle
\sim 1.42$
GeV/c. The total visible energy of the event was estimated
to be $\sim 400$ TeV. 
The event shows extremely large transverse momenta of produced
secondaries
 and no
 $\pi^{0}$ meson emission. Besides that it  indicates the
existence of penetrating shower clusters which show similar lateral
spread with atmospheric cascades ( $\sim$ a few mm in diameter). The
transition curves of high energy
showers for both single-cores-upper and shower-clusters-upper start
development just after entering into the upper chamber, like
electromagnetic cascades, however on the other side, they show
strong penetrating power, far over than that expected from the pure
electromagnetic shower development, see
Figure~\ref{Chiron_I}.
The showers are  separated at such large
distances to each other that none of the couple of showers which
could be attributed to a $\pi^{0}$ meson decay were found. They have
been tentatively
named ``mini-clusters'' to distinguish them from pure electromagnetic
atmospheric cascades.

 Later, two other events were found for both of which
the estimation of interaction heights was possible by the triangulation
method  \cite{La Paz,Chirons_ch20}.

A systematic study of Chiron type interactions was extended to the strong
penetrative 
families with wide lateral spread detected in  other chambers.
 30 Chiron-type events found in the systematic study
of the Chacaltaya chamber no.19 are described in detail in
\cite{La Paz,
Chirons_19}.
Other events have been collected in the past several years and the
analysis of  82 Chiron-type families (selected out from 120
families
 from
Chacaltaya
chambers
no. 19, 20 and 22) is presented in \cite{Has_Tokyo,Baradzei,Arisawa}.
21 Chiron events among them were found.
Four Chiron-type events, found among $\sim$ 60  
 families with total observed energy greater than 100 TeV
from Pamir exposure ($\sim$ 120 m$^{2}$yr of carbon chambers and
$\sim$ 9 m$^{2}$yr of thick lead chambers) have been reported in
\cite{Chirons_Pamir}.

\begin{table}[h]
\begin{center}
\label{Chirons_table}   
\caption[\scriptsize{Chiron--type events, examples.}]
{Chiron--type events, examples.}
\begin{scriptsize}
\vskip0.3cm
\begin{tabular}{|cccccccccc|}
\hline
&&&&&&&&&
\\
 $Event$ &$Collab.$ &  & $N$ & $Energy $ &
 $Q_{h}$& $\langle ER \rangle$&$E_{halo}$&$ E_{th}$&$Remarks$\\
 $Refer.$& $Chamber$& & &[$TeV$]& &[$GeV\cdot m$]&
[$TeV$]&[$TeV$] &  \\
&&&&&&&&&\\
\hline\hline
&&&&&&&&&
\\
B154S&Chacal.&$\gamma$&132&818& &140&& &penetr.\\
B133I&&h&24&528& &323&&&cascades and\\
\cite{Ch18B154SB133I}&2-storey&tot&&1346&0.39&&&&mini-cl.\\
&&&&&&&&&\\
\hline
&&&&&&&&&\\
P3C1&USSR-&$\gamma$&34&383& & & &4&penetr.\\
G48H57&Japan&h&5&347& & & &4&clusters\\
\cite{Has_Tam,Chirons_Calgary}&&tot& &730&0.48&3141$^{1}$&&4 &and\\
&standard    &     & &   &    &          & &&cascades\\
&carbon    &&&&&&&&\\
&&&&&&&&&\\
\hline
&&&&&&&&&
\\
113S084I&Chacal.&$\gamma$&67&513&&&&2&penetr.\\
\cite{Has_Tam}&&h&25&839&&&&2&\\
              &2-storey   &tot&&1352&0.62&1362$^{1}$
&&&\\
&&&&&&&&&\\
\hline
&&&&&&&&&
\\
P3C4&USSR-&$\gamma$& 107&1337&&&&4&penetr.\\
G369H370&Japan&h&26&898&&&&4&multi-\\
\cite{Has_Tam}&&tot&&2235&0.4&701$^{1}$&&4&cluster\\
&standard&&&&&&&&\\
&carbon&&&&&&&&\\
\hline
\end{tabular}
\end{scriptsize}
\end{center}
\hspace*{0.6cm}
{\scriptsize $^{1}$ measured by showers of $E(\gamma) \geq$ 20 TeV}
\vspace*{-2mm}
\end{table}

  The examples  of Chiron-type families, which have been found more
recently, 
 are presented
 in  Table~8.  
 These are: 

\vspace*{0.2cm}
1. {\em Ch18B154S-B133I \cite{Ch18B154SB133I}}.\\
     The event was recorded in Chacaltaya chamber No. 18.
     Despite of the large total visible energy (1346 TeV) the family
     has no halo. It has large lateral spread and it is much more rich in
     hadron component than usual events. Hadrons  carry about 40\% of
the total
     visible energy. The most striking feature of the family is
     the existence of two exotic hadrons within an extremely
     collimated cluster of showers located in its central part.
     One gives rise to a Pb-jet in lower chamber. It was found in the
downstream of the 
     shower observed in the upper chamber and
consists of at least four cores, remaining a mini--clusters.
Their average visible transverse momentum is about 2 GeV/c (from
     measurements of the geometrical convergence between cores at
different
      depths in the chamber).
     The
     other gives rise to a pizero-less C-jet (no pair of showers
      which couples into a $\pi^{0}$ produced in the target layer).
Both jets are hardly
explained as well as being of ``usual''  hadron  or  single
``$\gamma$-ray''
origin.

 The event contains also
several
anomalously collimated bundles of showers which
 penetrate through the whole chamber and 
     their observed darknesses in the lower chamber are appreciably
     larger than those expected for pure electromagnetic shower (see
figures in ref. \cite{Ch18B154SB133I}).

\vspace*{0.2cm}
2. {\em P3C1G48H57 \cite{Has_Tam,Chirons_Calgary}.}\\
The event found in the Pamir-joint chamber C1 series P3 is the example of
the
quasi-clean family. The analysis of both longitudinal ($f =
E/E_{vis}$)  and lateral ($E(\gamma) \cdot R$)  spectra
shows that the majority of high energy secondary particles arrive at the
chamber
directly. Provided that the interaction height is $\sim $1 km ($\simeq$
one collision mean free path) or less, the present event is the example of
 particle production with large transverse momenta, $\langle
p_{T}(\gamma)\rangle$
of the order of 2-3 GeV/c or more. Multiplicity of the secondaries
at the main interaction is $\sim$ 10-15. Among high energy showers there
are the strongly penetrating ones which after starting
development in the gamma block enter further into the lower chamber.
 Two penetrating clusters with a small lateral spread were also detected.

\vspace*{0.2cm}
3. {\em  C22-113S-84I \cite{Has_Tam}.}\\
 This is another example of the quasi-clean Chiron-type family
 found in the Chacaltaya
chamber no. 22.  In Figure~\ref{Chiron_I}  transition 
curves of high energy showers are shown.
 They are
 far beyond the fluctuation of
simple electromagnetic cascades, even though they start shower development
in the upper chamber.

\vspace*{0.2cm}
4. {\em P3C4G369H370 \cite{Has_Tam}.}\\
     It is the Chiron type family registered in the Pamir joint chamber C4 
series P3, marked as P3'-C4-368 in \cite{Baradzei}. Several penetrating
mini-clusters
     emitted with abnormally large $p_{T}$ values were found among
secondaries. The energy
     weighted spread of these clusters is small,
     $\langle E(\gamma)R \rangle \simeq$ 10 - 20 GeV$\cdot$m, while 
     the energy spread for the whole family, when it is measured by
cluster
energy, shows very large $p_{T}$ emission, namely
$p_{T}(\gamma)\simeq$ 2 GeV/c or more ($\langle E(\gamma)R
\rangle \simeq$ 2800 GeV$\cdot$m).

\vspace*{3mm}
Since the time of the first finding of the exotic cosmic-ray interaction
named ``Centauro'' it has been even-standing problem whether the produced 
secondaries from such an exotic events are ``ordinary'' hadrons or
something
new.
In fact, {\it  secondary hadrons from the
Chiron-type interaction reveal  very exotic
characteristics}:
\begin{itemize}\vspace*{-2mm}
\item {\it Single or multi-core structure and two very different 
transverse momentum components. }\\
  About one half of hadronic
cascades is single core structure
while the another half  develop a
multi-core structure at  the lateral spread of $\sim$  0.1 -- 1 mm, what
can be
connected with the intrinsic $p_{T}$ of  10 $\sim$ 20 MeV/c. These
narrowly collimated jets of cascades are called {\it mini--clusters}.
 The ratio of $\langle E(\gamma)R \rangle$ of hadronic cascades
in the Chiron families to the $\langle\langle E(\gamma)r\rangle \rangle$
in mini-clusters is $\sim$ 300.
This
is a surprisingly large ratio, telling us that {\it secondaries in the
parent interaction  are produced with extra 
large $p_{T}$ or mini-clusters are  connected with very small transverse
momenta phenomena, or both}.\vspace*{-0.2cm}
\item {\it Strongly penetrating power.}\\
 Majority of  mini-clusters can 
not be of electromagnetic origin
because of their {\it penetrating power}. About one half of
{\it mini-clusters}
is strongly penetrative.

\begin{figure}
\begin{center}
\hbox{
\epsfxsize=400pt
\epsfysize=420pt
\epsfbox[100 200 536 758]{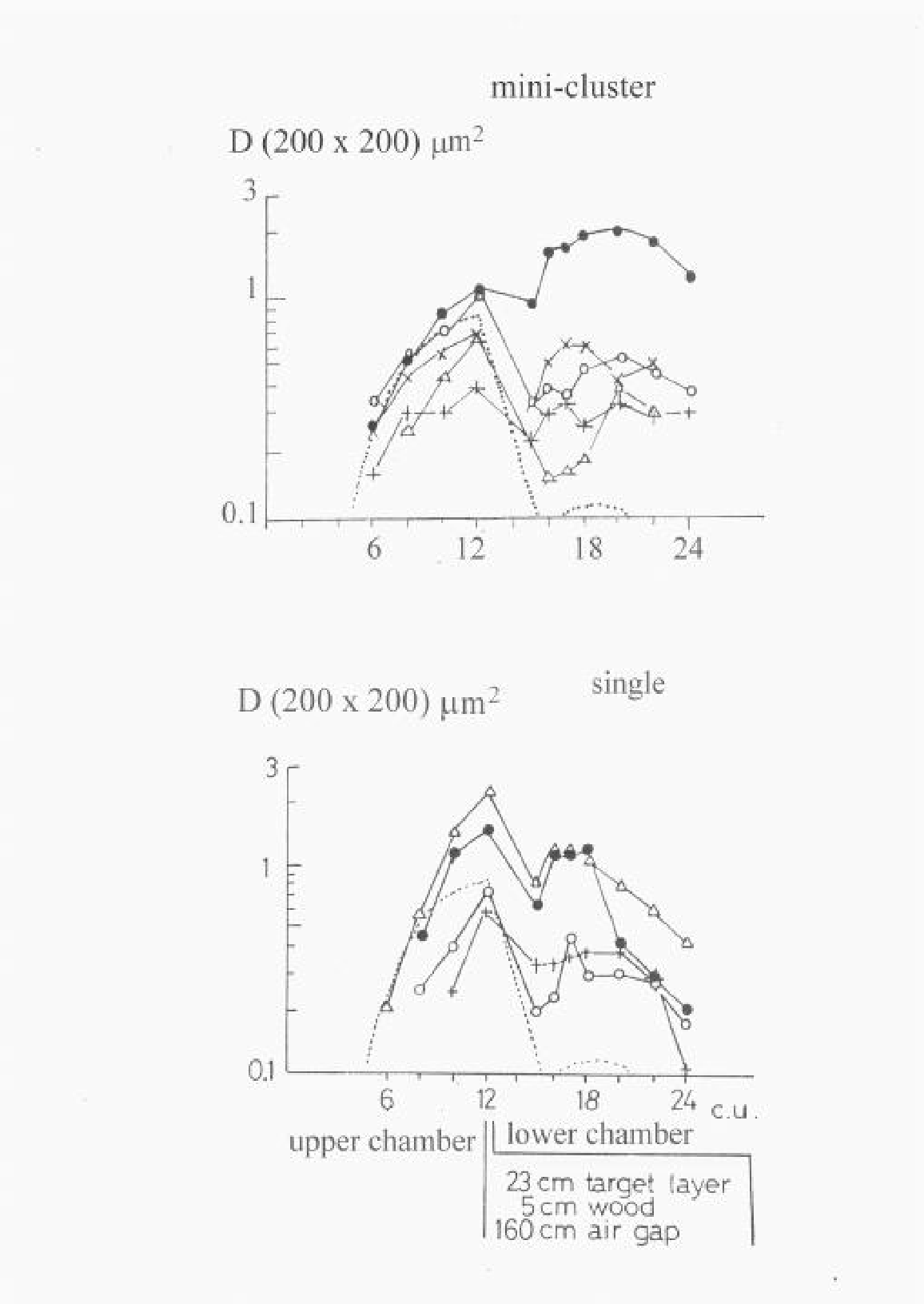}}
\vspace{3.5cm}
\caption[Examples of transition curves from Chiron type families.]{
 Examples of transition curves of cascades
  from Chiron type families \cite{La Paz}.\\Upper figure: mini-clusters:
  ($\bullet$) no. S-1 in family 150-90I,
 ($\circ$)  no. S-1 in family 131S-109I,
 ($\bigtriangleup$) no. S-2 in family 181S-139I,
  (+)                 no. S-2 in family 131S-109I,
 ($\times$)         no. S-10 in family 198S-154I;
  \\Lower figure: single cascades:
 ($\bullet$) no. S-1 in family 155S-136I,
 ($\circ$)  no. S-3-3 in family 131S-109I,
 (+)   no. S-2   in family 123S-90I,
 ($\bigtriangleup$) no. S-1 in family 198S-154I.}
 \label{penetr-single-cl}
\end{center}
\end{figure}

Figures~\ref{Chiron_I} and \ref{penetr-single-cl} show several examples 
of  
 transition curves of  individual cascades from various Chiron  events. 
Figure~\ref{penetr} shows the average transition curves of high energy
showers, with $\Sigma E(\gamma) \geq 10 $ TeV, from Chiron-type families.
For comparison,  the average calculated
transition curve was shown by dotted line. In calculations it was assumed
that 
100
gamma-rays of
energy greater than 10 TeV enter into the chamber from the atmosphere. 
The  power index  of an  integral spectrum,
$\gamma$ = -1.3, the same form
as
observed in the experiment was assumed. One finds that experimental 
showers
demonstrate far stronger
penetrating power than that expected in the case of
electromagnetic particle incidence.
 A possibility of the
showers being gamma-rays from the ordinary type of meson production have
been made implausible as well as from the argument on their penetration,
as from 
isolation
and absence of accompanied air showers. If they would be gamma-rays
we should  expect accompanied air cascades from neutral pions
associated
with them.
\begin{figure}[h]
\begin{center}
\mbox{}
\epsfig{file = 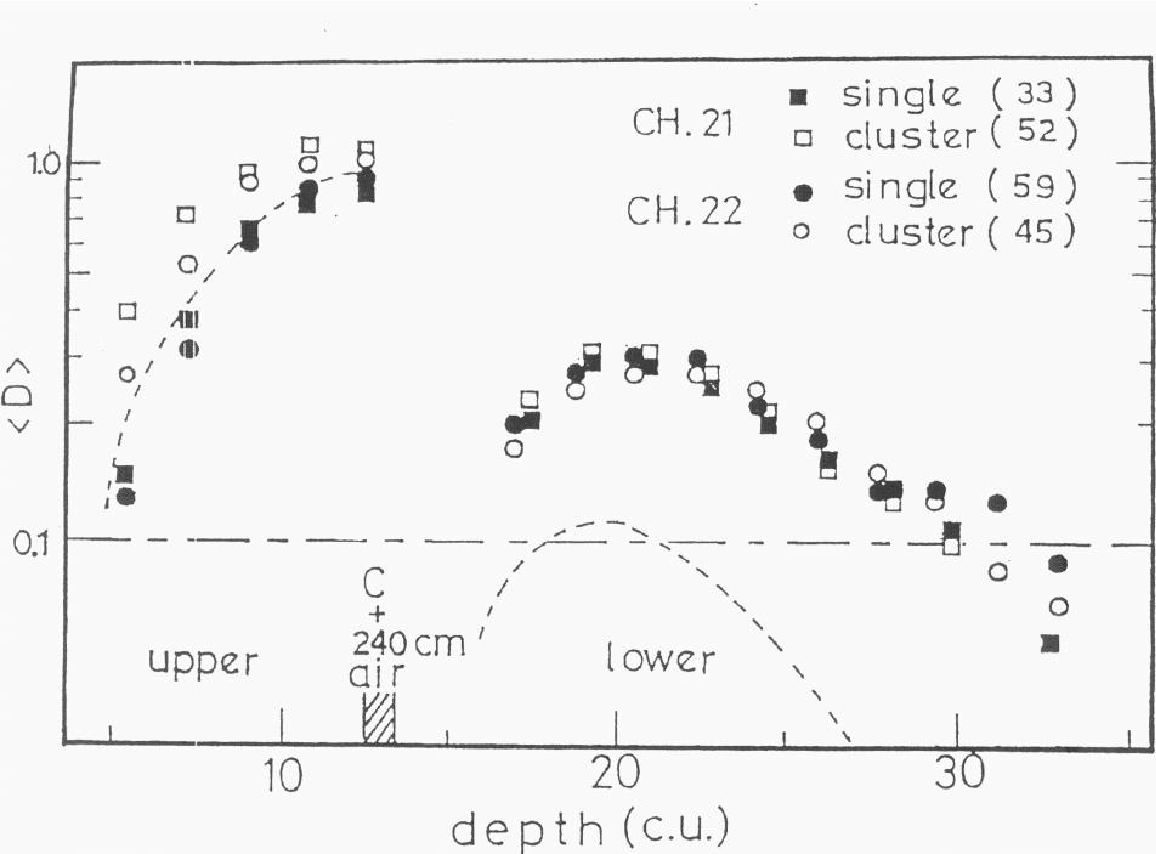,
bbllx=21,bblly=20,bburx=576,bbury=430,width=10cm}
\caption{Average transition curves of high energy
 showers of $\Sigma E_{vis} \geq $ 10 TeV from Chiron type families
  \cite{Has_Tokyo}.}
\label{penetr}
\end{center}
\end{figure}

\item {\it Very wide lateral spread.}\\
 It seems  that the high--energy showers from Chiron--type
events
 { \it are also not usual
hadrons} produced through the ordinary
multiple
meson production, high in the atmosphere. Assumption of  $p_{T}(\pi) \sim$
400 MeV/c for such pions and $k_{\gamma}\sim$ 0.3 for their secondary
interactions gives heights H $\sim 4 \sim 8$ km (i.e.  3-5 nuclear
mean free paths) for typical
$\langle E(\gamma) \cdot R\rangle$ measured in Chiron-type families. Some
hadrons could
survive such distances, by
chance escaping the secondary interactions, but not their majority.
Besides that, hadrons produced at such altitudes should
be accompanied by air cascades
from neutral
pions associated with them, what is not observed.\vspace*{-0.2cm}
\item {\it Rapid attenuation (large interaction cross section).}\\
Another surprising observation is a rapid attenuation of
 both single high
energy ``hadrons''  and mini-clusters.
  Majority of them start the shower development
just after entering  the chamber and their
collision mean free
path turns to be much smaller than the geometrical
value. It is
illustrated in Figure~\ref{Chirons_distr} which
 gives the distribution of shower starting
position in
Chacaltaya two-storey chambers, no. 19, 21 and 22.
222 hadronic showers of $E_{vis} \geq$ 10 TeV were selected  from  82
 Chiron--type families
of $\Sigma E_{vis} \geq $ 100 TeV.  The dotted line in the figure
shows  the attenuation expected by assumption of  geometrical mean free
path in the chamber
materials. The experimental results indicate the collision mean free path
as
small as $\sim $ 1/2-1/3 of the geometrical value.\vspace*{-2mm}
\end{itemize}

\begin{figure} [h]
\begin{center}
\mbox{}
\epsfig{file = 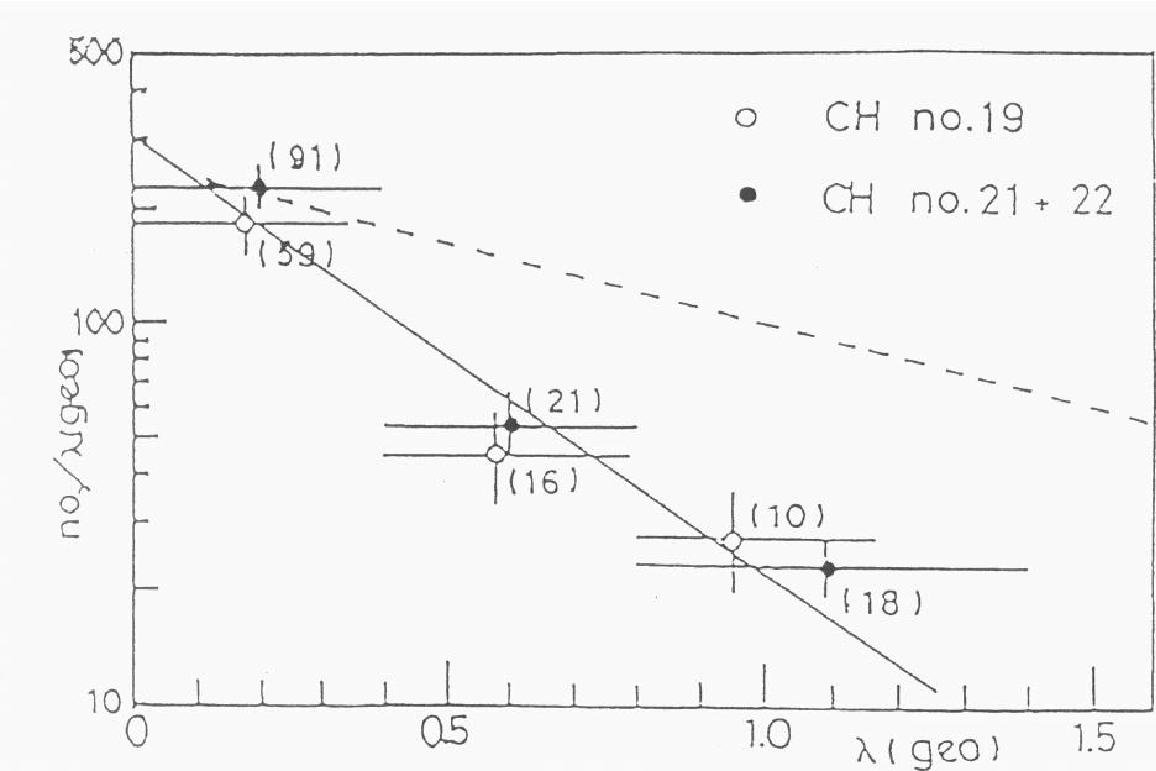,
bbllx=21,bblly=23,bburx=575,bbury=391,width=9cm}
\vspace{0.5cm}
\caption{Distribution of shower starting positions of high energy
 showers, measured by $\lambda_{geo}$
  \cite{Has_Tokyo}.}
\label{Chirons_distr}
\end{center}
\end{figure}

It is worth to note that a similar surprising behaviour
  have been also found
  in the study
of shower development of the high energy cascades  from 17
super-families of visible energy greater than 700 TeV detected in 
homogenous type lead chambers \cite{Arisawa} which
have fair advantage for the study of overall transition behaviour
of shower development.
The high energy hadrons in super
families
showed shorter attenuation length
 than ordinary single arrived cosmic--ray hadrons 
\cite{Has_Tokyo,Arisawa}.
It was reported  \cite{Arisawa} $\lambda_{att} = 170^{+47}_{-26}$
g/cm$^{2}$ for 143
hadrons
of  $E_{\gamma}\geq
10 $ TeV, and
$\lambda_{att} = 137^{+57}_{-26}$ g/cm$^{2}$ for 68 hadrons of
$E_{\gamma} \geq $ 20 TeV for hadrons in super--families in comparison
with $\lambda_{att}
 = 252 \pm 30$ g/cm$^{2}$ 
  obtained for single-arrived ordinary
cosmic--ray
hadrons detected in thick lead chambers.
This result seems to be especially surprising,
if one supposes  that
 high energy hadrons in superfamilies
are dominantly pions. Then their  attenuation length should be larger than
 that for
 ordinary cosmic-ray hadrons which are mainly protons \cite{Arisawa}.
 The
experiment gives the opposite  result  and one can  estimate
 that $\lambda_{coll}$ of high energy hadrons in
super-families is about one half of that the one of ordinary cosmic-ray
hadrons.

It should be noted, however, that
  up to now the unexpectedly short mean free path
of hadrons was observed only in very high energy superfamilies.
 Preliminary results \cite{Kopenkin} of the analysis of
showers from 58 families 
with the visible energy greater than 100 TeV, detected in the Pamir thick
lead chamber, gave the value of attenuation mean free path equal
 to 233 $ \pm$ 40 g/cm$^{2}$, for
hadrons
with energy greater than 10 TeV.

  Results presented here  seem to indicate 
 the existence of ``the new
state of hadrons'' which are emitted in  the high energy
Centauro--type interactions.
 Authors of ref. \cite{Chirons_19}
suggested, basing on thermodynamical arguments from the fire-ball
model, that the secondary particles observed in these exotic events
 could be some heavy and
long--lived particles (with masses
$\sim$ 10 GeV and
  lifetimes  $\tau_{0} \geq 10^{-9}$~s).

\subsection{Penetrating clusters and Halo}

Penetrating component, accompanying  the    exotic events,
has been observed in the form of {\it strongly penetrating cascades,
clusters or ``halo''.}

The term  {\it mini-cluster} \cite{clusters}  is used for stregthening
the cluster
characteristics which is different from ordinary atmospheric
electromagnetic cascade showers, even though the spread of
the clusters have a similar dimensions as air cascades. As it
has been already said in the previous section, mini-clusters
 are
very narrow collimated shower clusters (of a lateral spread $\sim$ 1 mm
or less, $\langle E(\gamma) \cdot R \rangle \sim $ a few TeV$\cdot$mm, and
$p_{T}
\sim $ 10--20 MeV/c). 
At a first glance they look like pure
electromagnetic cascade from the atmosphere. The characteristic which
distinguishes them from pure electromagnetic cascade is their 
strongly penetrative power.
  Mini--clusters are observed in the most forward angular
region
and, assuming the production height of the order of  $\sim$ 1 km, their
emission angles
 turn out to be of the order of  $\sim 10^{-6}$ radians.
They  show a strong  concentration of energy in this very forward 
region.
The fraction of cluster energy relative to  the total visible energy of
the
family is substantially large and the essential amount  of energy flow
is concentrated within a circle of a radius of a few to several
millimeters from the cluster axis.
 This phenomenon has been discovered in Chiron-type families but
it appears also in other Centauro-species (e.g. in Mini-Centauro
\cite{C-jets}).

 Besides the ordinary mini-clusters (called sometimes also 
``uni-clusters''), being the isolated single clusters and
characterized by
relatively small multiplicity, the {\it  giant-mini-clusters (called
``multi-clusters'')} have been 
also observed (e.g. giant mini-cluster in Chiron-type
family no. 174S-134I \cite{clusters}, found in Chacaltaya chamber no. 19).
These are unusual
shower core bundles with exceptionally
large multiplicities and they are suggested to be the ensembles
of ordinary mini-clusters \cite
{clusters}.
The analysis \cite{clusters_Hasegawa} of the huge shower cluster spectra
suggests its low original multiplicity. A shower cluster starts from a
small number
($\sim$ 4-5) of high energy particles with small primordial transverse
momenta.
Subsequent enhancement of shower core multiplicity, accompanied by
softening of its energy spectrum is the consequence of the passage
through
the atmosphere. The nature of parent  particles is unknown.
The study of their penetrating power indicates that
they are not of pure electromagnetic origin, even if they show such small
spread as expected from electromagnetic processes. On the other hand, it
seems that
they are also not  ordinary hadrons. Majority of these showers start
developing as soon as they enter into the lead in the upper chamber.
 
\begin{figure}
\begin{center}
\mbox{}
\epsfig{file = 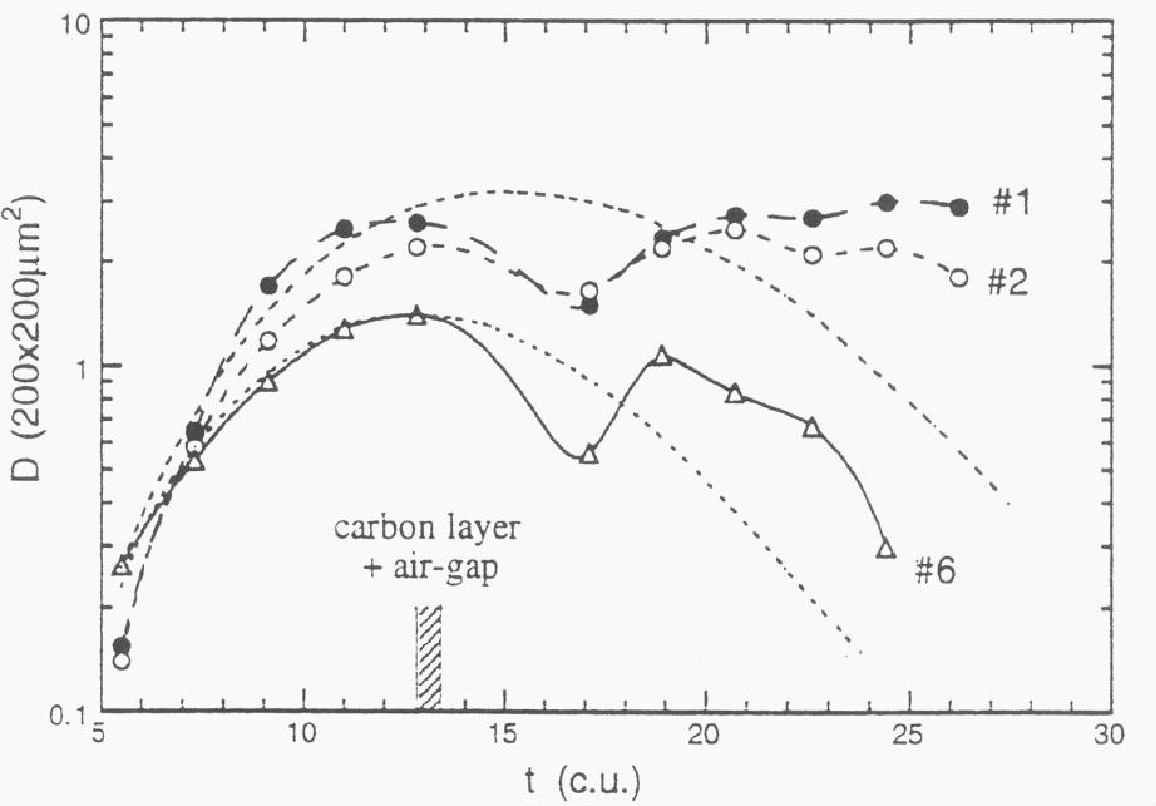,
bbllx=19,bblly=19,bburx=575,bbury=406,width=9cm}
\vspace{0.5cm}
\caption{Transition curves of the three highest energy showers from
the event C22-178S139I \cite{Has_Tam}. Dotted curves are simulated
electromagnetic cascades.}
\label{C22178S139I}
\end{center}
\end{figure}

 Some  examples of hadron--rich giant
mini-clusters are shown in Table  
~\ref{Mini-clusters}.
 Among them  are:
\vspace*{0.3cm}

 1. {\it C22-178S-139I \cite{Has_Tam}}.\\
It was found in the two-storey Chacaltaya chamber no. 22.
The shower cores
at the cluster area were scanned in nuclear emulsion plates under the
microscope. 
It is a typical example of the family confined in small dimensions of
a spread of the order of several millimeters. It is composed
of
only three very high energy showers, starting in the upper chamber, about
30 low energy showers
near the detection threshold energy and one C-jet located approximately
at the position corresponding to the one of high energy upper-chamber
shower. The transition curves of these three high energy cascades
are shown in the Fig.~\ref{C22178S139I}.

In this chamber there is 30 cm of plastic target and 230 cm of air--gap
between the upper and lower chambers. Usually, such air--gap strongly
disturbs the transition curve, i.e. significantly reduces the numbers of
electrons in the lower chamber, due to the electron scattering through the
air--gap.
The observed transition curves demonstrate unexpectedly  strong
penetrating power, indicating that
  they can not be of electromagnetic origin. It is also 
questionable if they could be caused by interactions of usual
 hadrons. The analysis of longitudinal and lateral spectra allows to
suspect
the existence of a new type particle production, characterized by very
small transverse momentum, of the order of $p_{T}(\gamma)\simeq$ 30 MeV/c,
 assuming that the main interaction occured at about one collision
 mean free path, i.e. $\sim$ 1200 m at Chacaltaya, above the detector.

\vspace*{0.3cm}
2. {\it P3C4G454H454 \cite{Has_Tam}}.\\
      Very interesting event with the high energy hadron-rich
cluster (of visible energy $\sim$ 1000 TeV) in the centre of the family
  was found in the Pamir-joint chamber C4 series P3. One observes more
      than ten high energy shower-cores of visible energies greater than
10
TeV, concentrated within a small area of a radius of $\sim$ 3 mm.
      A striking feature is the  strong penetrating power of the
cluster which was observed  deeply in
      the H-block. The longitudinal profiles of  penetrating showers
      show similar transition behaviour  as that  recognizable in
      Fig. \ref{C22178S139I}.
       The
estimated $p_{T}(\gamma$) of high energy showers
      inside the cluster is as small as $\simeq $ 20 MeV/c, if we
      assume that the interaction height equals to one collision
      mean free path at the Pamir altitude.
\vspace*{2mm}

\begin{table}
\begin{center}
\caption[\scriptsize{Giant-mini-clusters, examples.}]
{Giant-mini-clusters, examples.}
\label{Mini-clusters}
\begin{scriptsize}   
\vskip0.3cm
\begin{tabular}{|cccccccccc|}
\hline
&&&&&&&&&
\\
 $Event$ & $Collab.$ &  & $N$ & $Energy$ &
 $Q_{h}$& $\langle ER \rangle$&$E_{halo}$&$ E_{th}$&$Remarks$\\
 $Refer.$& $Chamber$& & &[$TeV$]& &[$GeV\cdot m$]&
[$TeV$]&[$TeV$] &  \\
&&&&&&&&&
\\
\hline\hline
&&&&&&&&&\\
C22&Brasil-&$\gamma$&43&229& & & &2&single\\
178S139I&Japan&h&6&533& & & &2&penetr.\\
\cite{Has_Tam}&&tot& &762&0.70&40$^{1}$
&&2&cluster\\
 &2-storey&&&&&&&&\\
&&&&&&&&&\\
\hline
&&&&&&&&&\\
P3C4&USSR-&$\gamma$&22&326& & & &4&hadron\\
G454H454&Japan&h&23&633& & & &4&rich\\
\cite{Has_Tam} &&tot& &959&0.66&97$^{1}$
&&4&str.penetr.\\
 &standard   & &     & &   &    &  & & cluster\\ 
&carbon &&&&&&&&\\
&&&&&&&&&\\
\hline
&&&&&&&&&\\
C19&Brasil-&$\gamma$&72&417&&&&2&penetr.\\
11S021I&Japan&h&10&321&&&&2&cluster\\
\cite{Has_Tam}              &   &tot&&737&0.44&180$^{1}$&&2&
($\Sigma E_{vis}$ = 465 TeV)\\
& 2-storey&&&&&&&&\\
&&&&&&&&&\\
\hline
\end{tabular}
\end{scriptsize}
\end{center}
\hspace*{0.5cm}
\scriptsize{$^{1}$ measured by showers of $E(\gamma) \geq$ 20 TeV}\\
\vspace*{-2mm}
\end{table}

The unusual penetrative nature of mini-clusters have been studied by
Chacaltaya
, Pamir 
and also by
Chacaltaya-Pamir Collaborations. The
results
of the study of penetrating component in
 17 families with $\Sigma E(\gamma) \ge $ 100 TeV, observed in the
Chacaltaya chamber no. 19 ( 24 penetrating showers out of 37 ones in
total)
and in the chamber no. 18 (16 penetrating showers out of 30 ones in total)
supported the
 existence of unusual showers with strong penetrating
power \cite{clusters_Hasegawa,clusters_Chacaltaya}.
 Similar conclusions have been obtained from the study of penetrating
showers registered in the  Pamir carbon chamber (``Pamir 79/80'' with one
carbon
block of 60 cm thick) where  187 penetrating showers
from
37 families with $\Sigma E_{vis} \geq 100$ TeV were picked up
\cite{clusters_Pamir}.
The detailed analysis  of high energy shower-clusters of visible energy
beyond 100
TeV observed in Pamir-joint chambers (analysed 173 families of
$\Sigma E_{vis} \geq $ 100 TeV) is described
 in
\cite{Baradzei,clusters_Dublin}.
The general conclusions from this study as well as the
single-core showers 
and  cluster--structure ones, are the same for
 three
experiments. They agree that there exist unusual showers with
unexpectedly strong  penetrative
power. Besides that, two very different components of  transverse
momentum are observed. Shower inducing secondaries are produced with
large
$p_{T}$ ($\langle p_{T}(\gamma)  \sim$ 2-3 GeV/c), far beyond that
expected
from the ordinary type. A  low $p_{T}$
phenomenon,
of the order of the electromagnetic one, seems to be responsible for 
cores generation inside the cluster.

 {\it Mini-clusters} are considered to be the premature stage of other
phenomenon called {\it halo}.
 Halo is the diffusion dark spot with
dimensions $\sim$ 1-1000 mm$^{2}$ observed in the center of families.
 Sometimes it consists of 
 several hadron cores, spaced very closely together. The first
gigantic halo
event named ``Andromeda'' was discovered in the Chacaltaya
chamber no. 14 in 1969.
 Since that time the statistics of halo events
have steadily increased and  events with different halo
configurations have been found by mountain 
experiments at Mts. Chacaltaya, Pamir, Fuji and Kanbala \cite{Yamashita}.
 About 50\% of $\gamma$--hadron  families with energy $\Sigma
E_{\gamma}
\ge 500 $ TeV are halo events.
 The examples of cosmic-ray families in which
shower
spots surrounding the central halo show abundantly hadron rich
composition, the same behaviour as seen in cosmic-ray families of Centauro
species, were also found. The halos of anomalously strong penetrative
nature (e.g. Tatyana \cite{Tatyana}) were observed. 
 The  five Chacaltaya families: Andromeda, Ursa
Maior, M.A.I, M.A.II and M.A.III   \cite{Yamashita} are
 examples of
fully analysed halo events.

Many calculations were performed to understand the mechanism of halo
formation and its surprising features such as for instance the strong
penetration capability.
Generally the data show a contradiction between the longitudinal and
lateral halo development which cannot be resolved assuming the ordinary
cosmic ray composition.

 In principle, the nature of
super-families
show  similar characteristics with Centauro species, i.e. unusual
hadron rich composition for showers which  surround the central
halo.
 There are also strong indications 
 that the relative $p_{T}$ of hadrons within 
mini-clusters and halos are in $\sim$  20-30 MeV/c range, the same as in
 Chirons. Such properties as: strong penetrative nature and two different
components of the transverse momentum invoke the idea
 that
in the extremely high-energy interactions the  strongly
collimated bundles  of particles are frequently produced. They
develop in the ``halo''
after atmospheric degradation.

The other  striking feature of halo events is the existence of
many-center halos and their  alignment along a straight line.
Parton-parton scattering naturally leads to the alignment of the final
state nucleon fragments with two or more parton jets. However, the
observed
alignment is reported to be significantly greater than that
expected from QCD. 
 Fraction
of aligned events in superfamilies was claimed to be  between 26-43 \%
dependently on the
type of the chamber (higher for deep Pb chambers) \cite{alignment}. It
has been revealed not only in the Pamir experiment but  also in
the Tien--Shan large ionization calorimeter and in the 
emulsion chamber exposed at the Concorde board in the stratosphere.
According to  recent
calculations \cite{alignment}, the explanation of the abnormal fraction of
aligned events
needs  some mechanism of complanar production of hadrons as well as
 {\it  the existence of highly penetrating particles}. Penetrating
component could prevent
a  destruction of the coplanarity during the development
of nuclear cascade in the thick atmospheric target above the emulsion
chamber. It  has been checked  in simulations that assumed
mechanism
of the coplanar emission is lost in the ``normal'' process of the
nuclear-electromagnetic atmospheric shower development.

 There are 
many other examples of  strongly penetrating
clusters or halos  not listed  here. Unfortunately, 
serious experimental
difficulties
in measurements and  analysis of halo/cluster  events occur
frequently. They
do not  allow for reliable identification  of electromagnetic and hadronic
parts and estimation of 
 the fraction of  
hadronic
component.

\subsection{Anomalous transition curves}

\begin{figure}
\begin{center}
\mbox{}
\epsfig{file = 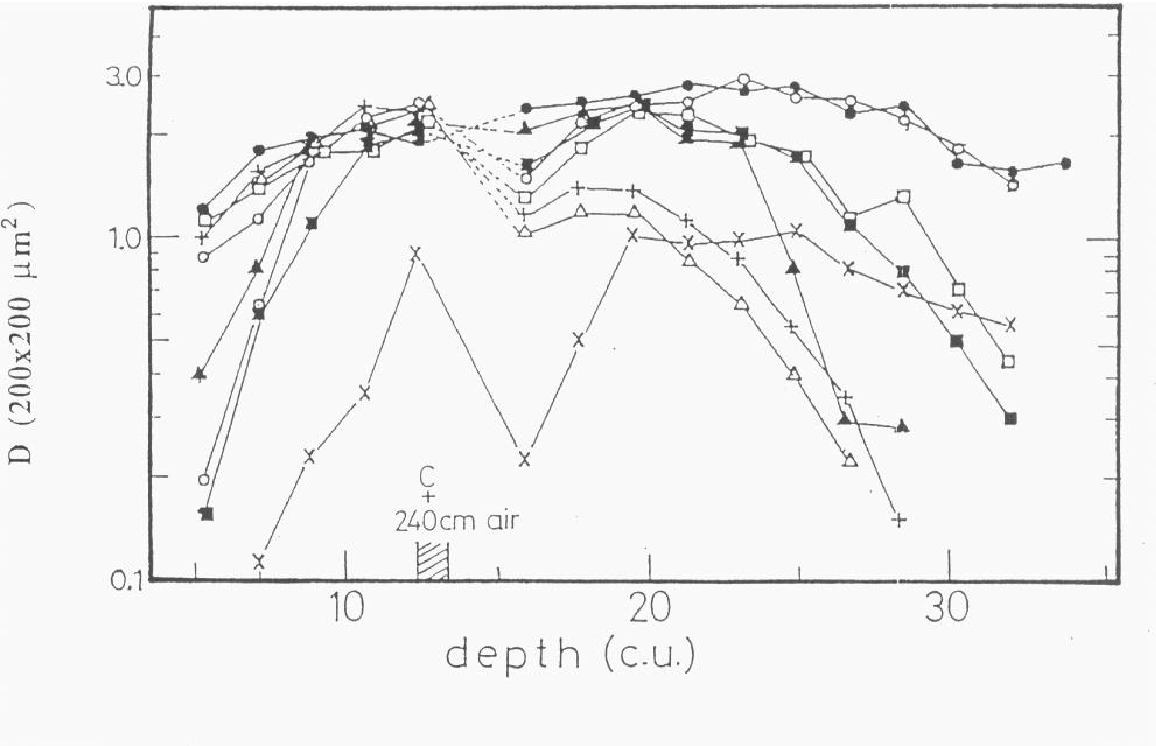,bbllx=21,bblly=36,bburx=574,bbury=375,width=10cm}
\caption{Examples of anomalous transition curves registered
 in Chacaltaya two--storey chambers
  \cite{Has_Tokyo}.}
\label{penetr-Chac}
\end{center}
\end{figure}

Anomalous cascade transition curves have been firstly noticed 
during the study of Chiron-type families (see subsection 2.3).
Later, they have been encountered also in other events
\cite{Has_Tokyo,Baradzei} (see subsections 2.1, 2.4). 
 Typical examples of long-range cascades (clusters) registered in
Chacaltaya two--storey chambers are shown in Figure~\ref{penetr-Chac}.

The common feature  for all such showers  is that 
they
start their development
just after entering the top of the chamber
and penetrate through the whole  apparatus
without significant attenuation.
Unfortunately, due to
  relatively small depth of these chambers and their
inhomogenous structure it is impossible  to conclude unambiguosly
about the exotic nature of  individual cascades. In many  cases the
shape of transition curves  
could 
be simply explained by  consecutive interactions of the same hadron.
Only statistical analysis of groups of showers may indicate that
something unusual happens (see subsection 2.3).
Recently, a penetrating nature of  cascade showers observed in the
 two--storey carbon type chamber  was compared with  simulated 
$\gamma$--ray--induced and hadron--induced cascade showers.
 Using QGSJET and modified UA5 model it was
shown \cite{Tamada_carbon_ch} that about 34\% of penetrating showers
observed in the two--storey chamber no. 19 are neither
$\gamma$-ray-induced  nor hadron-induced showers. A possible explanation
is proposed in connection with ``mini--clusters''.
 
A many--maxima structure of cascades have been also observed  in
  carbon  chambers of the Pamir experiment. 
 Although thick multi-block carbon chambers 
 constitute only a small
fraction of all exposed apparatus some interesting events have been also
 detected there \cite{Cedzyna,N830}.
 The
 example is the
family N830 \cite{N830}
 detected in the Pamir 76/77
chamber,
consisting of a standard gamma--block and four  identical sections of 
hadron
blocks (each consisting of  25 cm of  rubber  and 5 cm of  lead
with X-ray films as sensitive layers). Among hadron cascades there were
observed 
 three high energy ones  revealing  a multihump structure
and traversing tens cascade units through the chamber material. 
  
\begin{figure}
\hbox{
\epsfxsize=200pt
\epsfbox[18 37 576 797]{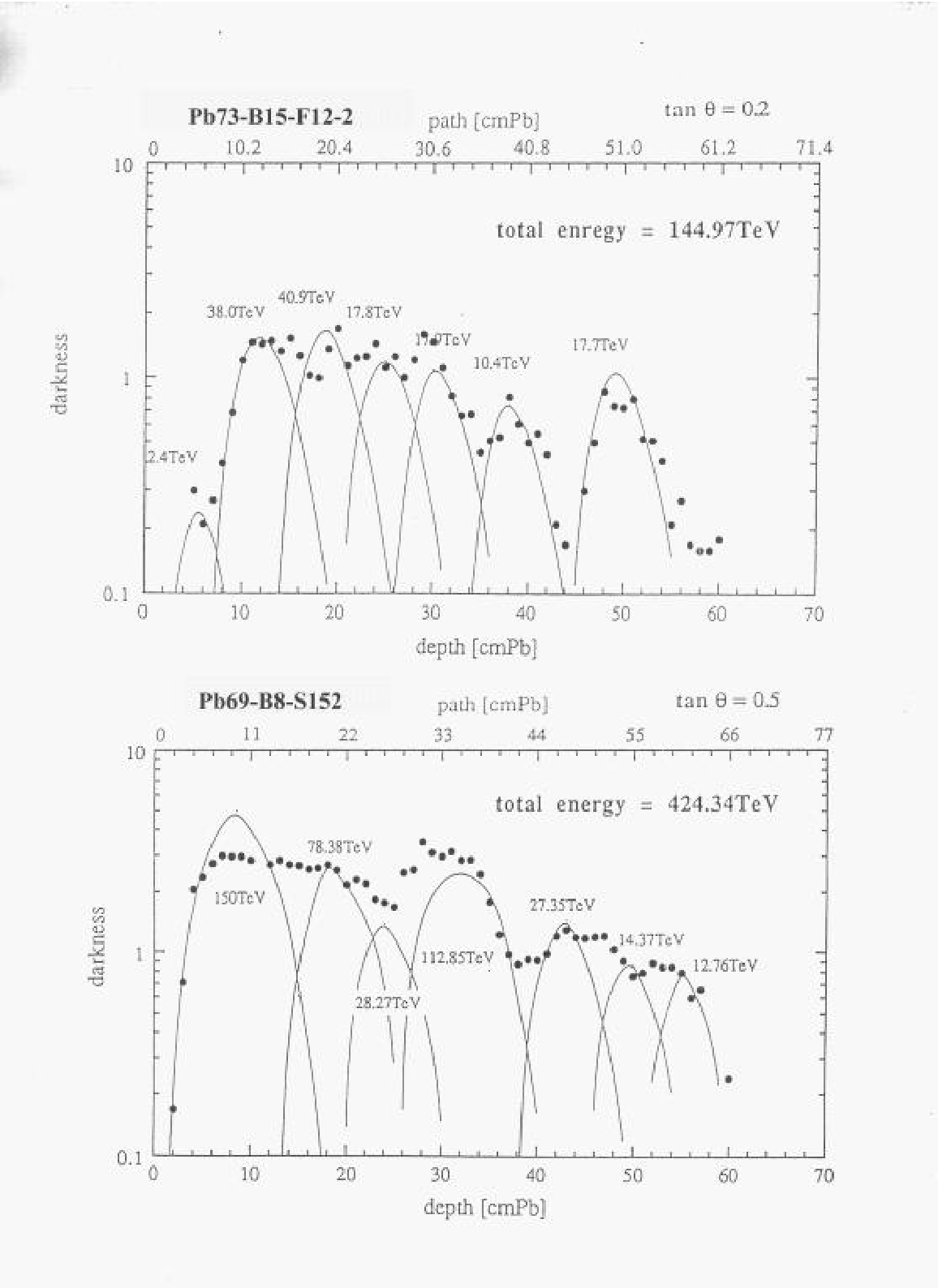}
\epsfxsize=200pt
\epsfbox[18 37 576 797]{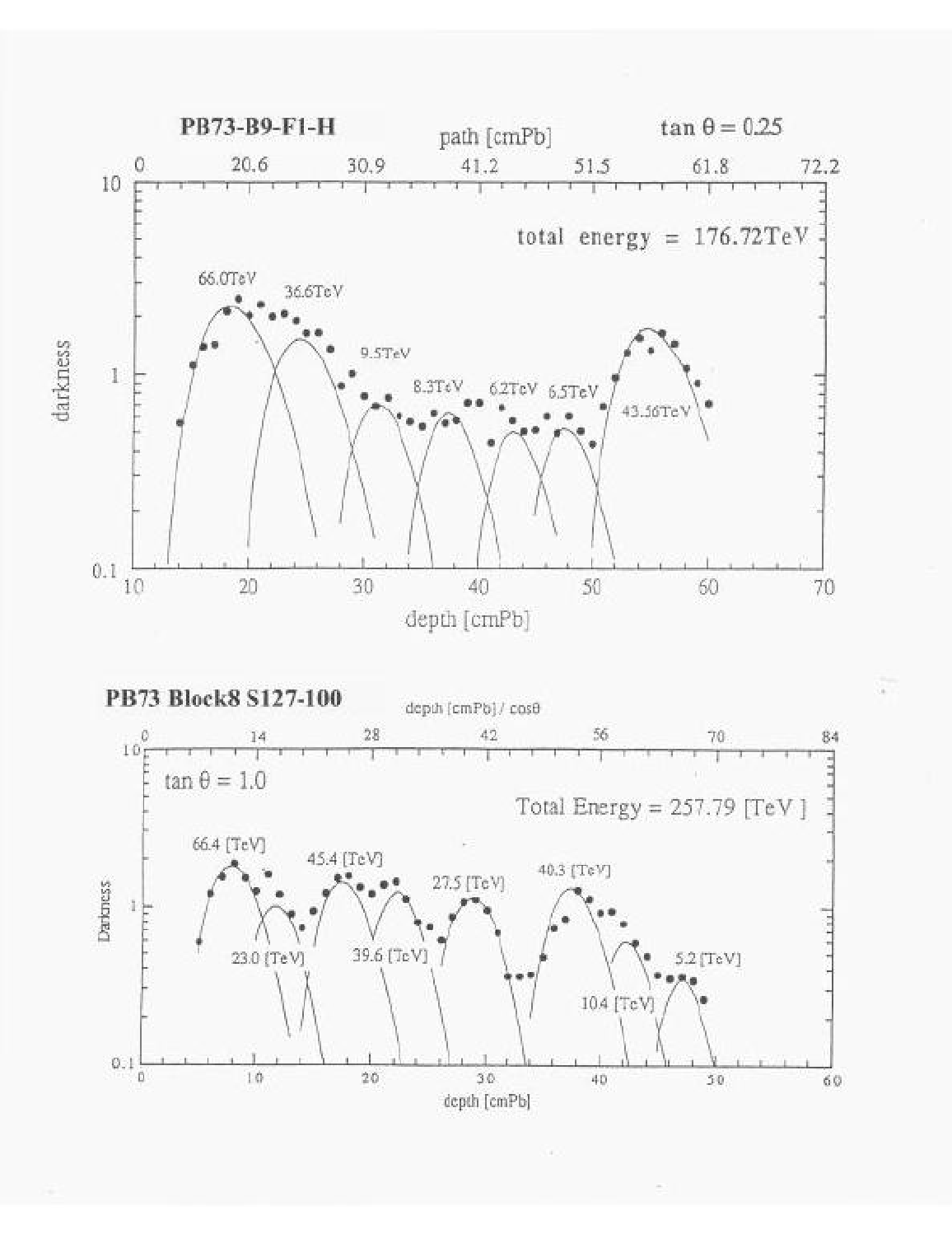}}
\caption{Examples of anomalous transition curves registered
 in thick lead  chambers
 \cite{Has_Tokyo}.}
\label{penetr-lead}
\end{figure}

Homogenous type thick lead chambers are the most appropriate apparatus for
the study
of  penetrability  and for  looking at anomalies in cascade
development.
The spectacular examples are exotic cascades detected in the
Centauro--like event C--K (see subsection 2.1).
Unfortunately, up to now, rather  small area of
these chambers have been exposed and analysed. Most of experimental
material comes from several thick lead chambers installed at the  Pamir in
the years 1988-1991 by 
the MSU (Moscow State University) Group.
 Some extremely
interesting events  have been found there and reported
in \cite{Has_Tokyo,Arisawa}. The examples  are shown in
Figure~\ref{penetr-lead}.
The presented cascades exhibit  surprising features, such as many maxima
structure
and very slow attenuation. Some of them penetrate through the
whole apparatus without noticeable attenuation, sometimes even indicating
 a growing tendency.

\begin{figure}
\begin{center}
\mbox{}
\epsfig{file = 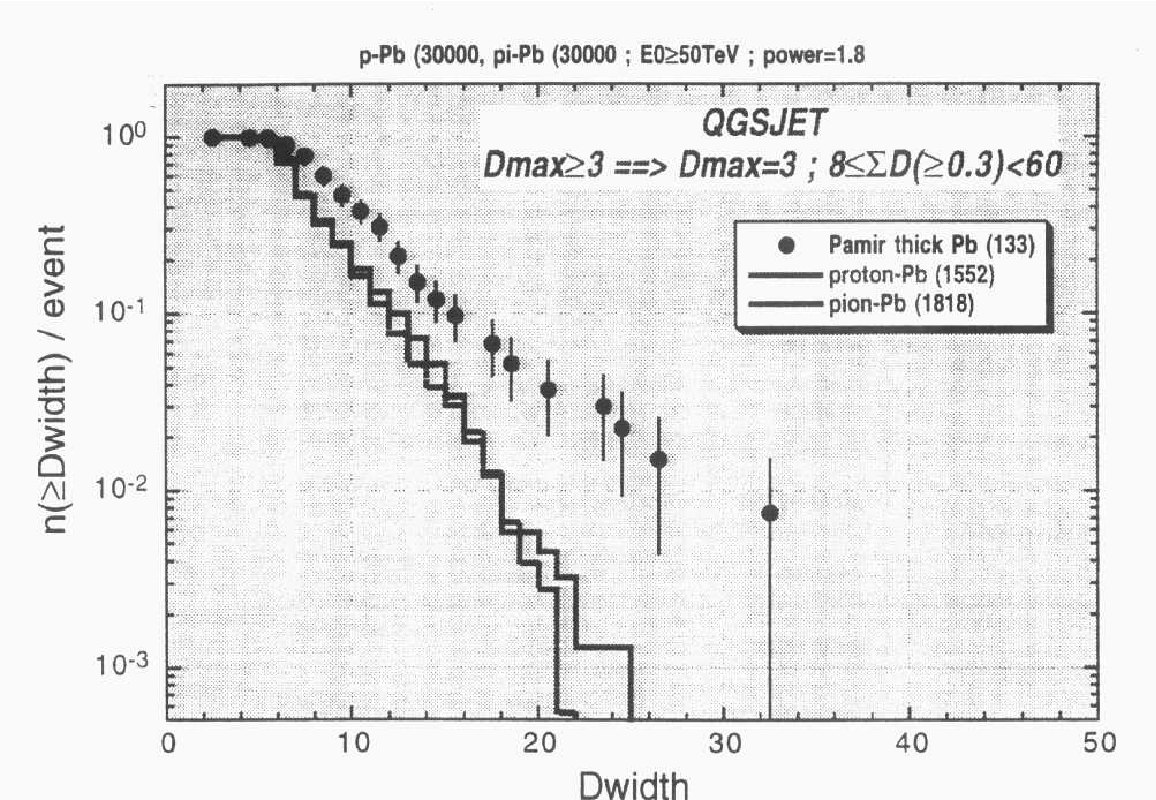,
bbllx=21,bblly=19,bburx=572,bbury=403,width=9cm}
\caption {Distribution of  widths of simulated and observed cascades
  \cite{Tamada_Lodz}.}
\label{cascades_width}
\end{center}
\end{figure}

 Simulation calculations of transition curves in homogenous thick lead
chambers have been performed \cite{Tamada_Lodz,Tamada,Tamada_new}
with the purpose to compare the transition behaviour of ordinary
hadrons with the experimental one.
Some artificial cascades revealed wave--shaped transition
curves, showing the successive maxima separated by the depth
corresponding to about one collision mean free path or so ($\sim$ 15 cm)
in  lead. It is, however,  much longer distance than experimentally
observed one.
Generally, the simulated pictures are  at the first glance qualitatively
different from
the experimentally
observed curves. More recent simulations,  assuming four  models 
of hadron-nucleus interactions (VENUS 4.12, QGSJET, HDPM and modified UA5,
all widely accepted as standard models)
  confirmed the unusual character
of long-penetrating cascades \cite{Tamada_Lodz, Tamada_new}. In
particular,
the widths of experimentally observed cascades and the distribution of the
ratio of energy
released in the first peak to the total energy of the cascade disagree
with those obtained in simulations. As an example,
the distribution of the widths of 133 cascades detected in the Pamir thick
lead chamber in comparison with the simulated one, assuming the QGSJET
model,  is shown
 in Figure \ref{cascades_width}.
It is easily seen that both proton and pion induced simulated cascades
are much narrower than the experimental ones.

 Investigation of the mechanism of  transfering the energy
 into
$\gamma$--rays, during the passage of the hadron through the chamber, 
indicate  that in
simulated events most of energy
is released in the first interaction. It again  apparently disagrees
with
experimental
observations. Authors of \cite{Tamada_Lodz,Tamada, Tamada_new}
conclude
that  extensive simulation studies support the existence of exotic
long-penetrating cascades. They suggest,  as a possible explanation,
the extremely collimated hadron bundles, saying, however, nothing about
their origin.

It is plausible that the observed  strongly penetrating cascades are the
same phenomenon 
as the so--called  long-flying component. This effect
was firstly noticed \cite{Tien-Shan,Yakovlev} in the Tien-Shan lead
calorimeter.
It was observed that the attenuation length (extension) of the cascade  is
almost constant up to
calorimeter cascade energies of $\sim$ 10 TeV, increasing twice in the
energy interval 10-300 TeV. This observation means a  nontrivial hadron
energy deposit at
larger depths in the lead calorimeter. 
The effect has been  later confirmed by the investigation of 
attenuation of hadrons in
deep
lead Pamir chambers \cite{long_fl_MSU}. At depths corresponding to 3-6
hadron attenuation paths  an  excess of  cascades,
which cannot be explained in the framework of a present knowledge
about development of hadron-induced cascades in lead, have been
observed. An abundance of cascades detected at large depths  ($\sim
78-192$ c.u.)  constitutes
$\sim $ 33\% of the total intensity.
 Distributions of cascade origin points cannot be described
as $dN/dt \sim exp(-t/\lambda)$ with the same slope for all depths:
$\lambda_{meas}=212\pm 19$ g/cm$^{2}$ at depths 22-78 c.u.,
 and $\lambda_{meas}=310\pm 36$ g/cm$^{2}$ at depths 78-192
c.u. \cite{long_fl_MSU}. There were considered two  explanations
of the observed phenomenon:
\begin{enumerate}\vspace*{-2mm}
\item {\it Hypothesis of copious production of leading unstable
particles, having relatively long lifetimes.}
\\
Generally the phenomenon could be explained by adding to the normal
hadron component 
 some particles decaying inside  the chamber, at depths  somewhere between
several
tens centimeters and two meters, with energies E $\geq$ 20 TeV. They
should 
 carry
their energy deeply into  lead absorber, practically without spending
it in nuclear interactions. Particles with heavy quarks ($c,b,t$) 
satisfy
these criteria. Among them charmed particles were considered  the
best candidates:
their masses are around 2 GeV/c$^{2}$, lifetimes  $\sim
10^{-12}-10^{-13}$ s and
the inelasticity coefficient in interactions of charmed particles with
nuclei
is small.
However, for obtaining a satisfactory description of experimental data
 very  large charm cross section
production must be assumed.
 This, so called, ''leading charm''  hypothesis needs $\sim $ 10
times larger cross section for
charm hadroproduction, than resulting from extrapolation of accelerator
data, and additionally a small inelasticity coeficient. \vspace*{-2mm}
\item {\it Hypothesis of some heavy (m $\gt$ 10 GeV/c$^{2}$) and 
long-lived
      $(\tau_{0} \sim 10^{-8}-10^{-6} s)$   particles, weakly absorbed in
the 
      atmosphere.}\\
       Such particles, if  consisting of heavy and light quarks
      should  interact with a cross section approximately similar to
that for ordinary 
      hadrons but with a very small inelasticity
coefficient.\vspace*{-2mm}
\end{enumerate}

\section{Centauro species statistics}

\subsection{Mts. Chacaltaya and Pamir experiments}

The basic question is what is the intensity of Centauro species.
The well known numbers, cited in many papers
(see for example
 \cite{Shaulov}), are  intensities  measured by
Pamir-Chacaltaya
Collaboration and claimed to be of the order of $\sim 10^{-2}-10^{-3}$
m$^{-2}$year$^{-1}$ for Centauros and 
$\sim 10^{-1}$ m$^{-2}$
year$^{-1}$ for Chirons, at the Chacaltaya altitude.

\begin{figure}[h]
\begin{center}
\mbox{}
\epsfig{file = 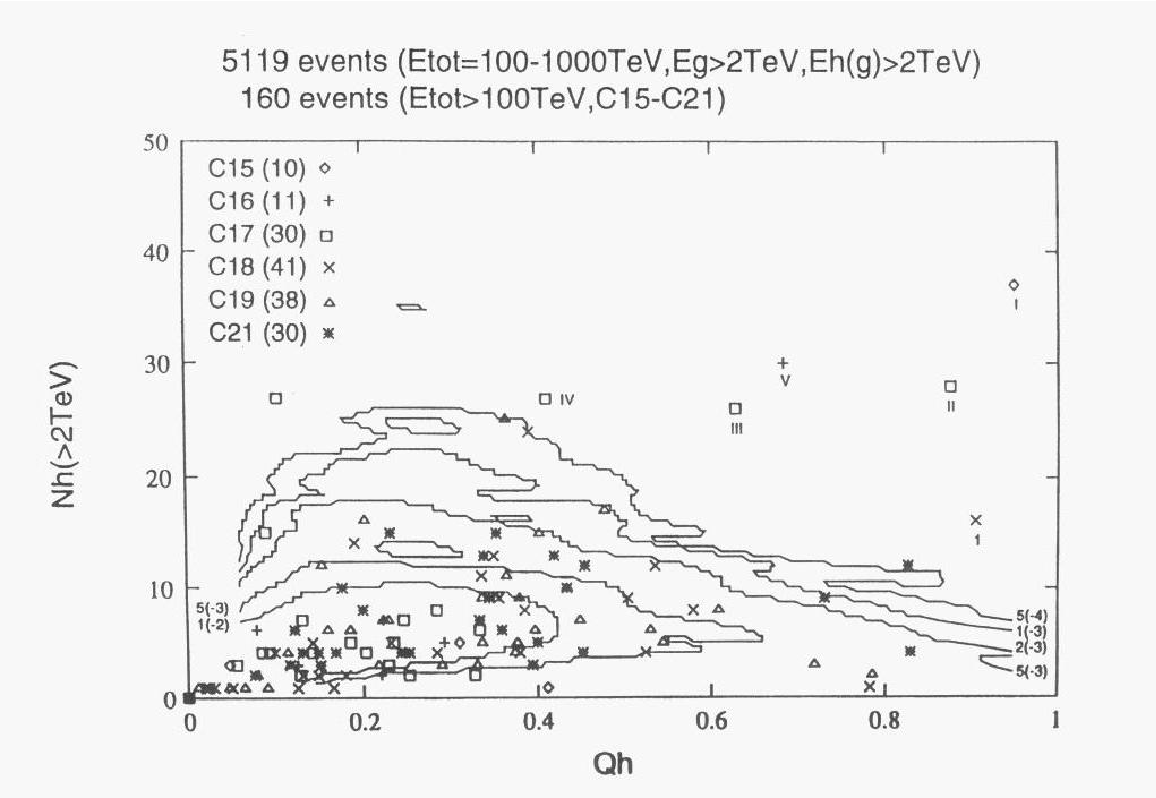,
bbllx=20,bblly=21,bburx=572,bbury=400,width=10cm}
\caption{$N_{h}-Q_{h}$ diagram of families ($\Sigma E_{\gamma} \ge$ 100
TeV)
 observed at Mt. Chacaltaya (Ch.15 - Ch.21).
Centauros are shown by  marks (I-V and 1).  Contours give the normalized
density of the simulated events
  \cite{Ohsawa}.}
\label{Cent_densities}
\end{center}
\end{figure}

However, it should be mentioned that before giving the statistics of
Centauro-type events, the precise definition of such objects should be
formulated. If we use very sharp criteria defining such events
their numbers will be not very large. Up to now there are reported only
two super clean Centauro events, without observed presence of any
$\gamma$' s,
found
in the two-storey Chacaltaya chambers, of the total exposure
ST = 3.49$\times 10^{2}$ m$^{2}$yr (for the sensitive solid angle of the
emulsion chamber $\Omega$ = 0.7).
 A question of appearance of  super clean Centauros have been
 considered  in \cite
{Ohsawa}.
 The intensity of cosmic-ray events at   energies corresponding to
 the observed  Centauro events was calculated  from the
formula:
\begin{equation}
I(\gt \Sigma E_{vis}) = 0.9 ( \Sigma E_{vis}/100  TeV)^{-1.25 \pm
0.10}
/(m^{2}\cdot yr\cdot sr).
\end{equation}
Comparing  the  calculated expected number  of cosmic ray events (with
$\Sigma
E_{vis} = 100 \sim
3000$ TeV) \cite{Baradzei} 
  with 
the total number of observed clean Centauros (i.e. Centauro I and
Centauro-New) the probability of appearance
of such super pure  species have been  estimated to be $\gt 10^{-3}$
\cite{Ohsawa}.
 Simulations (based on 13714 events)  gave probabilities $1.0
\times 10^{-5}$ and $2.0\times
10^{-6}$ for Centauro I and New Centauro respectively.
The difference between the experimental and simulated probability of 
observation of such events indicates that  Centauro events cannot be
produced by a fluctuation in the multiple particle production and/or in
collision mean free path. Figure \ref{Cent_densities} is the diagram
of the $N_{h}$ vs. $Q_{h}$ for the Chacaltaya
families (with $\Sigma E_{vis} \ge $ 100 TeV) with the marked contours
giving the normalized densities of
simulated events.

Similar events, i.e. characterized by the appearance of a family  of 
showers
at the certain apparatus layer, deeply inside the chamber, without
accompanied
showers 
in the upper part of the apparatus, have been observed
\cite{Gladysz-Thesis} also
in the deep lead chamber ``Pamir 74/75''. Unfortunately, in this case 
it was impossible to exclude the trivial explanation that the observed
phenomenon is simply a  ``usual'' family, reaching the chamber
during the time of its assembly. 

The sharp definition of Centauro phenomenon can be a little released 
to  include
  the events in which the observed electromagnetic component is very
small.
 In this case
the
following numbers as the lower limits of their frequency can be quoted: 7
Centauros, 21
Chirons and
15 Mini-Centauros found in 305 families with $\Sigma E_{\gamma}\geq$ 100
TeV (from Chacaltaya and Pamir joint chambers) \cite{Tarbes}.

\begin{figure}[h]
\vspace*{-1cm}
\begin{minipage}{9cm}
\mbox{
\epsfxsize=250pt
\epsfbox[25 90 600 850]{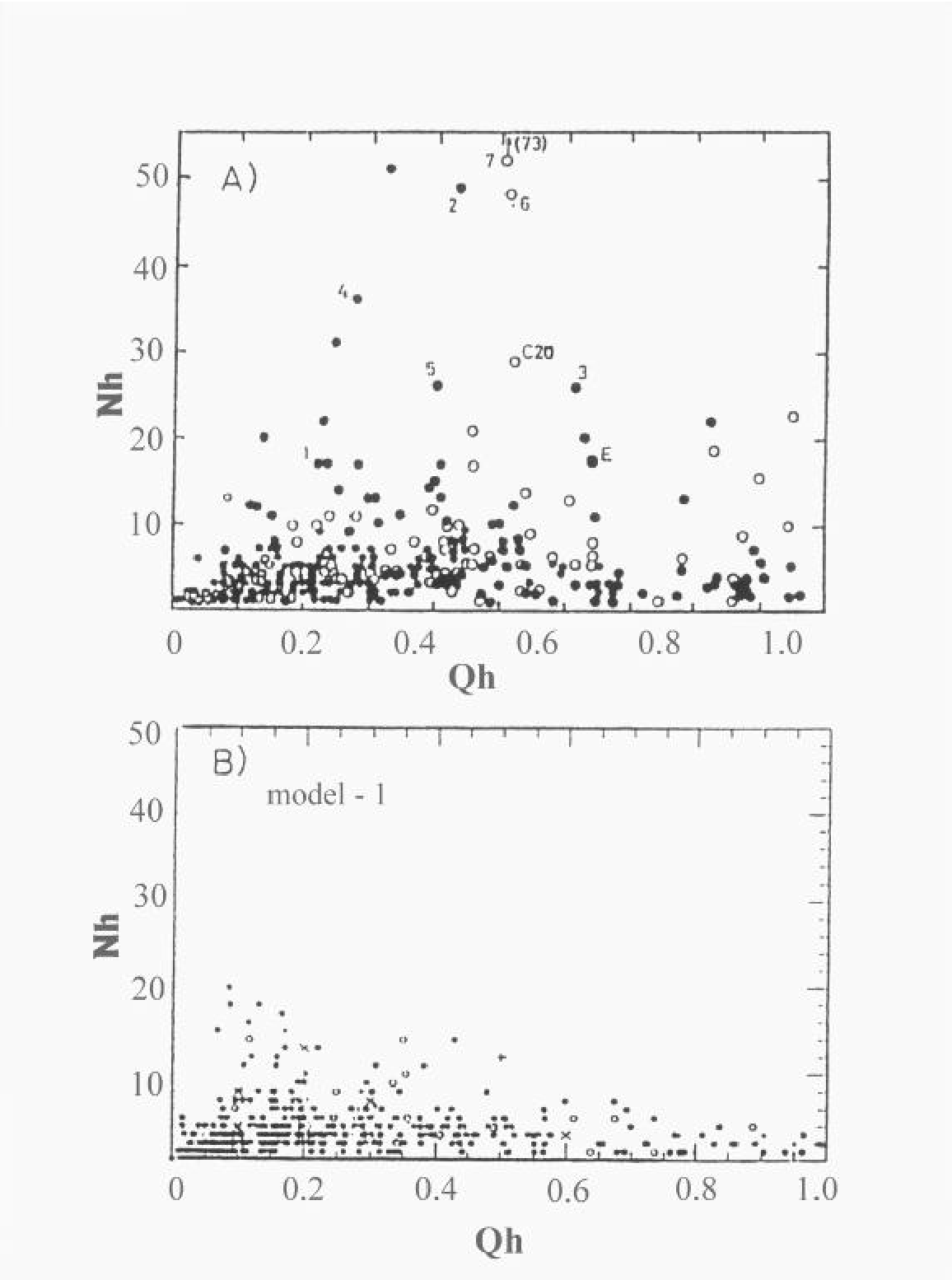}}
\end{minipage}
\begin{minipage}{5.5cm}
\vspace{3cm}
\caption[ (A) $N_{h}-Q_{h}$ diagram of families 
  detected in Pamir,
   Chacaltaya and Pamir-Joint chambers.] 
{ (A) $N_{h}-Q_{h}$ diagram of families 
  detected in Pamir,
   Chacaltaya and Pamir-Joint chambers,
  (B) The same for the simulated families. Different marks
 signify the different primary cosmic-ray nuclei: ($\bullet$) proton,
 ($\circ$)  $\alpha$, ($\diamond$)  CNO, ($\times$)  heavy, (+)  Fe,
  \cite{Baradzei}.}
\label{Centauro_statistics}
\end{minipage}
\vspace*{3mm}
\end{figure}

Further extension of the definition to the objects with anomalously high
fraction
of hadronic component causes  that  $\sim$ 20\% of all families
with the total visible energy $E_{vis}\geq 100$ TeV should be
recognized as
Centauro--species.  This question
has been studied carefully in \cite{Has_Tokyo,Baradzei} and it is
illustrated
in
Figure~\ref{Centauro_statistics}.
 The analysis was based on the unbiased sample of 429 families
from Chacaltaya (open circles), 173 from the Pamir-Joint chambers
and 135 from a part of the Pamir chambers of 500 m$^{2}$yr (closed
circles) with  total visible energy greater than 100 TeV. There is
shown
a scatter diagram of $N_{h}$ vs. $Q_{h}$ where $N_{h}$ denotes the number
of hadrons in a family with visible energy greater than 4 TeV.
 Figure~\ref{Centauro_statistics} b shows  523
simulated families with the total visible energies greater than 100 TeV.
 In simulations  the structure of the chambers was taken into
account and normal chemical composition of  primary cosmic
rays has been assumed.
 Different marks on the plot refer to families due to different primary
cosmic--ray nuclei. The basic
mechanism of particle interaction was the production
of hadronic clusters and their decay with parameters chosen to reproduce
  of the CERN UA-5 experiment results.
One sees that in the experimental data there exist abundant families with
anomalously rich
hadron
content, sometimes in both  the number and  the
energy fraction. They are  beyond expected fluctuations in the
distribution
from UA-5 type hadronic interactions. None of the simulated families
is found to have $Q_{h} \ge$ 0.75 and $N_{h} \ge$ 5, though the
experiment shows the existence of families of much richer hadron
composition.
Among the families there are events with a very large number of hadrons,
expressed by a circle and a number in Figure~\ref{Centauro_statistics} a.
All those are superfamilies with $\Sigma E_{vis} \ge$ 1000 TeV.
Among them the family marked as (7) is called Centauro VII, the family
marked by C20 is called Centauro VI, both found in the Chacaltaya
chambers. 
Family ``Elena'' found
in the Pamir thick-lead chamber is marked by (E). Other events marked
(1)-(6) 
 are superfamilies: P3'-C1-90,
 P3-C5-505,
 P2-C96-125,
 P3-C2-201,
 P3`-C4-369,
 found in Pamir-joint chambers, and  M.A.III  coming from
Chacaltaya chamber.
Some of them are connected with
a strongly penetrative huge halo, what suggests its  strong hadronic
nature. However, because of the technical problems with
measurements
of the halo, 
presented results  include
only the analysis of the off-halo part. The characteristics of both the
off-halo part
and estimations of the halo region are published in
\cite{Has_Tam,Baradzei}. 
The off-halo part of superfamilies shows tendency of hadron-dominant
nature in either $N_{h}$ and $Q_{h}$ or both, even though these
superfamilies are estimated to have been produced at high altitude.
From the study of the off-halo part the large value of $p_{T}$ can be also
concluded.
 
It is important to note that there are no significant differences among
three
experiments.
 The figures (a) and (b) tell us that such anomalously rich hadron content
is neither caused by the incidence of heavy nuclei in the primary
cosmic rays, nor by the superposed fluctuations of ordinary-type
hadronic interactions. The anomalously hadron-rich families constitute a
substantial part of the unbiased observed samples,  at least
$\sim$ 20\%. The effect of the anomalous hadron dominance in families is
much
enhanced if the analysis is restricted to the events generated
at the distances not far from the top of the chamber. It was shown in
\cite{Baradzei}, where
 the events  with rather small lateral
spread,
i.e. with $\langle E^{*}R^{*} \rangle \le $ 300 GeV$\cdot$m, 
after applying the ``decascading'' procedure for $\gamma$-rays, have been
selected.

These conclusions have been supported by the more recent, extensive
simulation studies \cite{Tamada_Lodz} based on four different  
models of 
hadron-nucleus interactions (VENUS, QGSJET, HDPM and modified UA-5) and
by using the CORSIKA code for simulation of nuclear--electromagnetic
cascade
development in the atmosphere. The enhancement of hadron--rich families
can not be explained by widely accepted models of ``normal''
interactions.
The question of abnormal dominance of hadron component has
been studied
separately also for ultra high energy events. In \cite{Has_Tam} has been
presented the systematic analysis 
 of 75 families with  $\Sigma E_{vis} \geq
$ 500 TeV detected in Chacaltaya two-storey chambers (300 m$^{2}$yr),
Pamir-joint chambers ($\sim 530$ m$^{2}$yr) and in the part of the Pamir
exposition ($\sim 500$ m$^{2}$yr). Among presented families, four
Centauro--type, two Chiron-type, and $\sim$ 10 other
 hadron-rich events could be
noticed.
The families coming from homogenous type lead chambers (110, 60 and 40
cm of Pb
thick from a  total exposure $\sim 450$ m$^{2}$yr) have
been also separately analysed.
 The list and characteristics of 17 superfamilies with
$\Sigma E_{vis} \geq 700$ TeV and 40 families with $100 \leq \Sigma
E_{vis} \leq 500 $ TeV are shown in
  \cite{Arisawa}. Among them at least three events with $Q_{h} \geq 0.5$
are present.

\subsection{Mts. Kanbala and Fuji experiments.}

A Centauro like event named ``Titan'' has been reported in 1977 by 
Mt. Fuji experiment \cite{Lattes,Titan}. The authors emphasized its
large
$p_{T}$
and hadron--rich  character. It was suggested that
all  secondaries are hadrons.
A systematic search for Centauro events has been done later \cite{Ren}  in
the
thick-type
lead chambers \footnote{Mt. Kanbala and Mt. Fuji groups  mostly used
 Pb
and
Fe
chambers
with the total thickness of 10-70 c.u.. The flat type chambers which
constitute the most part of the exposed  apparatus are not suitable
for the Centauro problem study.} by the
China-Japan Collaboration.  
The total exposure was 
$\sim 130$ m$^{2}$yr at Mt. Kanbala (5500 m a.s.l.) and 
$\sim 380$ m$^{2}$yr  at Mt. Fuji (3750 m a.s.l.). 
Among 30 hadron families (with the total visible energy
greater than 100 TeV)  coming from  thick lead chambers exposed at
Mt. Fuji and  100 families coming from Mt. Kanbala chambers no candidates
of Centauro events have been found. The upper limit of the fraction
of such events was  estimated to be 3\% (95\% c.l.) in the hadron families 
with energy greater than 100 TeV.
The puzzle of the nonobservation of Centauro events by these experiments
still remains a mystery.
Several reasons, such as differences in experimental conditions,
in emulsion chamber designs  and data
analysis procedure used by different Collaborations can be suspected.
In particular, the following reasons  should be mentioned: 
\begin{enumerate}\vspace*{-2mm}
\item {\it Some differences in hadron identification procedure.}\\
  Mt. Kanbala and Mt. Fuji groups classified
 hadrons and $\gamma$--rays only statistically by referring to the
starting
depth of showers. In the Chacaltaya and Pamir experiments  
additional
criteria have been used, based on differences in development (lateral and
longitudinal) 
of
hadronic  and electromagnetic cascades.
Not only 
 showers observed in the lower chambers were regarded as
      hadrons.
Showers from upper chambers have been studied  under the microscope and
some of them 
       consisting  of well--resolved cores or revealing many--maxima
structure in the longitudinal 
development were also
 identified
      as  hadrons.\vspace*{-2mm}
\item {\it Difference in  exposure altitude.}\\ Mt. Fuji laboratory,
       giving the main part of the experimental data to the Centauro
       problem study,
       is located at the much lower altitude than Mt. Chacaltaya
       or Pamir. If Centauro species were born in nucleus-nucleus
collisions  or if they 
         are the ``strongly penetrating
       objects'' produced at the top of the atmosphere or
       somewhere in the extra-galactic region then 
       the decrease of their flux with the
       atmospheric depth is quite plausible.
\end{enumerate} 

\subsection{JACEE Experiment.}
The Japanese-American Cooperative Emulsion Chamber Experiment,
JACEE, has flown emulsion chambers with baloons near the top of the
atmosphere. Despite of a small area and short time of exposure, as
compared to
Chacaltaya/Pamir Experiment,
a few 
 events of anomalous $\gamma$/charged ratio have been
observed by JACEE Collaboration. However, these events  differ in some
essential points
from classical Centauros. The anomalies were noticed at incident
energies lower than that estimated for ``classical'' Centauros
 and unusual $\gamma$/charged ratios  were   observed only in the
      limited ($\eta-\phi$)  phase space  region. Besides that,
  an excess of photons (anti--Centauro),
       in contrary to the hadron excess observed in Centauros, was
claimed.
The examples are:
\begin{enumerate}\vspace*{-2mm}
\item  4L-II-27 event \cite{Jacee_4LII27} of incident energy of 80
       TeV, yielded 149 charged particles and 120 $\gamma$'s. Almost all
        $\gamma$--quanta  were produced in a narrow jet in the extreme
forward
       direction. The $\gamma$/charged ratio is 2.6 $\pm$ 1.1 in the
region
       of pseudorapidity 5.5 $\leq\eta\leq 7.5$, what is a significant
deviation
from the expected ratio of  $\sim$ 1. The analysis presented in
\cite{Jacee_4LII27} showed an anomaly
       at the 5-10\% level among 41 studied events with $E_{0} \geq 40$
TeV.\vspace*{-2mm}
\item The event, with $\Sigma E_{\gamma}$ = 15.4 TeV, described in
\cite{Bjorken,Jones}
 was initiated by a singly charged
primary. The
collision occurred  within the detector. Almost all  leading particles
were
$\gamma$--quanta. Photons appear to cluster into two groups. The leading
cluster
 consisted of about 32 $\gamma$'s
 with $\langle p_{T}
\rangle \simeq$ 200 MeV and only one accompanying charged particle. A
possibly distinct cluster had three times as many photons as charged
hadrons (about 54 photons versus 17 charged).  This
event is one out of
a sample of about 70.\vspace*{-2mm}
\item The event presented in \cite{Wilczynski} is a peripheral
collision of
      Fe nucleus (E $\simeq$ 9 TeV/nucleon) in emulsion. There were found  
27 $\gamma$--quanta
      with $\eta \leq$ 6. As they came  from pair conversions at
only 0.8
      radiation lengths, one  can  expect  that the total number of
      photons was about  50. At the same time, 
only 6 charged particles (out of 21 charged tracks detected in the
whole angular region) falled in
the same
kinematical range. 
\end{enumerate}
In all these events there was observed a tendency to a group emission
of $\pi^{0}$ mesons. Such $\pi^{0}$ groups, having similar directions and
momenta, could be signs of a  formation  and a subsequent decay of
the chiral condensates.
It should be mentioned, however, that these  events were  found in 
emulsion by scanning for the leading
photon showers, so there was a ``trigger bias'' in favour of a large
neutral fraction. 
It would be interesting to hear something about anti-Centauros from the
mountain-top emulsion chambers. Here, there is, however, even much more
stronger ``trigger bias''  in favour of gamma families, and thus the
interpretation of data, from this point of view, is a complicated
exercise.
 It is rather  difficult to identify anti--Centauros unambigously, with
exception of  unusual and rare
events in which the interaction
vertex is close to the top 
and clearly resolved in the chamber.
 
\subsection{Summary of the Centauro species statistics}

Summarizing  results of cosmic ray emulsion chamber experiments,
 it can be stated that Centauro--type anomalies have been observed by
several
different  Experiments and  Collaborations,
working under different
 experimental conditions and using various types of chambers. Examples
of 
detected unusual events were given in the previous sections. However,
the systematic search for  Centauros, by using the same criteria,  have
been
made only for the part of the exposure from the three experiments :
Chacaltaya Collaboration, Pamir-Collaboration and Pamir-Joint Chambers.
China-Japan (Mt. Kanbala) and Mt. Fuji Collaborations, in their
systematic
Centauro search,  
used different type of chambers and different 
criteria of data analysis.
\begin{table}
\label{Cent_stat}
\caption[\scriptsize{Centauro statistics}]
{Centauro statistics}
\begin{scriptsize}
\begin{center} 
\vspace{0.3cm}
\begin{tabular}{|clccccc|}
\hline
& & & & & &   \\
 $Laboratory$& $Altitude$ & $Chambers$ & $Exposure$ &
$No.$ $of$ $families$ &
$No.$ $of$& $Ref.$\\
 & $m$ $(g/cm^{2})$& & $m^{2}yr$&$E_{vis}\geq 100 TeV$ &$Centauros$ &\\
&&&&&&\\ 
\hline
&&&&&&\\
Mt. Chacaltaya&5200 (540)&two-storey& 300 &121&$\sim 8^{*}$ 
&\cite{Baradzei,Has_Tam}\\
(Brasil-Japan) &  &carbon    &       &      &    &\\
&&&&&&\\
Pamir         &4300 (600)&carbon type&500 &135&$\sim 3^{*}$ 
&\cite{Baradzei,Has_Tam}\\
(USSR-Poland)  &or 4900&          &     &       &   &\\
&&&&&&\\
Pamir         &4300    &carbon type &530 &173&$\sim 2^{*}$
&\cite{Baradzei,Has_Tam}\\
(Russia-Japan)&         & or thick Pb & 
    &       &   &\\
&&&&&&\\
 Mts. Kanbala  & 5500 (520)  &thick-Pb &130 &30&- & \cite{Ren}\\
(China -Japan)&&&&&&\\
&&&&&&\\
Mt. Fuji & 3750 (650) & thick-Pb& 380& 100 &-& \cite{Ren}\\
(Fuji Coll.)&&&&&&\\
&&&&&&\\
\hline
\end{tabular}
\end{center}
\hspace*{1cm}
$^{*} \sim$ 20\% hadron rich families in the sample
\end{scriptsize}
\vspace*{-2mm}
\end{table}
 Table~10
 shows the
statistics of ``unambigious'' Centauros  found only in that part of
experimental material where systematic and uniform searches were done.

 It should be emphasized, however, that extension of  the ``Centauro''
definition to  all hadron--rich species causes that
 about 20\% of events 
 among the families with
the total visible energy $\Sigma E_{vis} \geq$ 100 TeV can be regarded
as Centauro--like anomalies.

\section{Centauro explanations}

The possibility that  fluctuated air showers mimic Centauro type events
 seem to be the most natural suspicion. This question has been
studied
by many authors (see for example
 \cite{Cent_sim,Acharaya}). Recently, very carefull analysis  of
this problem, by
using
the most modern simulation tools, as CORSIKA code (and four different
models of hadron-nucleus interactions: VENUS, HDPM, QGSJET and modified
UA5), 
has
been done
 by M. Tamada \cite{Tamada_Lodz}.
 All considered models fail
in describing Centauro species.
 In particular none of models was able  to
explain the experimentally observed fluxes of hadron-rich events. They 
fail also
in describing  many characteristics of events, such as 
 anomalous hadron-gamma correlations (e.g. $N_{h}$ vs $Q_{h}$),
mini-cluster and
giant-mini-cluster structures etc.

Especially,
no one has
succeeded in reproducing the Centauro I event. The possibility that a
heavy
nucleus interacting in the lower part of  the atmosphere could
give rise to Centauro
I type event was considered for example by Acharaya and Rao
\cite{Acharaya}. They have
shown that, in principle,  it would be  possible to reproduce the event,
though the total number
of such events expected in the global data sample is $\sim$  2$\times
10^{-5}$ what is in apparent discrepancy with the observed flux of
Centauros.
 Similarly, in \cite{Ohsawa} (see also the previous section)
the probability of finding a ``clean'' Centauro event, like
Centauro I or Centauro New, among simulated events was found to be several
order of magnitude smaller than  
 the experimental one.

The   suspicion that 
Centauro like phenomena would arise from  
 nucleus-nucleus collisions was examined  in many works.
Keeping in mind
that there is a 
negligible probability that nucleus would penetrate so deeply into the
atmosphere  B. Mc Cusker \cite{Mc Cusker} suggested that a small fraction
of iron nuclei from a primary cosmic-rays would survive passage
through several hundred g$\cdot$cm$^{-2}$ of the atmosphere and  produce
Centauro events. But P. B. Price et al.
\cite{Price} showed (taking into account various chains of fragmentation)
that the flux of surviving heavy nuclei is too low by a factor of
$\sim 10^{-10}$  to account for Centauros.

So, it is widely believed that Centauro related phenomena could not be due
to
any kind of statistical fluctuation in the hadronic content of normal
events.

Therefore, new type of interaction or the creation of a new kind of
matter is conjectured to be responsible for these extremely unusual
phenomena. 
 In the last years many interesting models
 have been proposed by different authors. It would be difficult to
describe or even list all of them.
Some of them
are amenable to experimental verification in accelerator conditions, the
other ones seem to be
unamenable. Some of them (e.g. \cite{Bjorken_Cen}) assume that the
exotic objects
of unknown
origin are 
present in the primary cosmic ray spectrum and they are seen as
Centauros during their penetration
through the  atmosphere. The other ones assume that the exotic events  are
produced in
extremely high energy
hadron-hadron (e.g. \cite{Goulianos}) or nucleus-nucleus 
(e.g. \cite{Panagiotou,8}) interactions.
 Many of them incorporate the
strong penetrability as the main feature of the phenomena. For instance,
  the problem of very small flux of heavy nuclei
 at the mountain level was avoid in \cite{8}, by conjecturing that the
initial collision of
the heavy cosmic-ray nucleus does occur very high in the atmosphere.
The fireball created in the central collision of a heavy nucleus with air
nucleus
 is a  very dense object which 
 penetrates several hundred g$\cdot$cm$^{-2}$ of the air before exploding
into
fragments.

 The widespread opinion
that {\em the  likely mechanism for Centauro production is the
formation of a
quark-gluon  plasma}  was incorporated in a lot of 
proposed models. Other exotic attempts, as for example 
 the color-sextet quark model
\cite{White}, based on Pomeron physics in QCD were also developed. It
 needs  adding
to the Standard Model an
additional flavour doublet of color sextet quarks.

 All proposed explanations are based on two different believes. In the
first case it is assumed that mostly  baryons are the products of 
Centauro type events.  Such picture is incorporated for example in 
diffractive-type  fireball models (see subsection 4.2  and refs.
\cite{Navia,Goulianos,Navia_model}) and
 in  scenarios with strange quark matter (see subsection 4.3 and
refs.~\cite{Panagiotou,8,9,Shaulov,Bjorken_Cen,Wlodarczyk_str}).
 In the second case, particles produced by the Centauro mechanism
 are suppose to be mainly mesons and there are numerous attempts to
explain Centauros as different types of isospin fluctuations
\cite{Andreev}-\cite{DCC_Wang}.
 According to refs.~\cite{Lam, Pratt}
 large isospin fluctuations could be due to the
Bose
nature of the emitted pions and a laserlike (so-called PASER) mechanism
is  considered to be responsible for Centauro formation.
The other theoretical speculations predict large isospin fluctuations
arising from formation of localized regions of misalignment vacuum which
become
coherent sources of a classical pion field.
 In particular,
formation of disoriented chiral condensate (DCC) 
 \cite{Bjorken,DCC,DCC_Krzywicki,DCC_ions,DCC_Gavin} is suggested to be
 a possible explanation of Centauro-like phenomena (see subsection 4.3).
 It seems, however, that all these
attempts have at least one common difficulty. It has been shown in
ref.~\cite{Wlodarczyk_cen}  that families produced at 
mountain altitudes are insensitive to any isospin fluctuations.

 In this section some of the models will be shortly described.
The most attention will be given to the strange quark matter fireball
\cite{Panagiotou,8} 
model which, in our opinion,  gives the best chance for simultaneous
explanation of different  Centauro-related phenomena. It has
been
used also as the basis for the designing  and
simulation calculations done
 for  the CASTOR experiment at the LHC. 

\subsection{ Exotic extraterrestrial glob of matter}

In \cite{Bjorken_Cen} it was postulated by Bjorken and Mc Lerran that
Centauro could be a metastable glob of highly compressed quark matter
present in the primary cosmic ray spectrum. These globs could have radii
of several
fermi and contain several hundred quarks.
Since the mean transverse momenta  of the Centauro decay products are
several times
larger than the
normal value typical of a nucleus it can be expected  that the glob
would have
 radii $\sim$ 3 - 5 times smaller than that of an ordinary nucleus and 
a very high density, $\sim 30 - 100$ times that of ordinary nuclear
matter.
 If the binding energies of  the constituents
in   the dense nuclear matter are the same order as the transverse
momenta, then 
the glob  should be characterized by both the binding energy much
higher than that of conventional nuclear matter and  by the reduced
geometrical cross section,  allowing it  to penetrate deeply into the
atmosphere.
To develop this idea authors of ref. \cite{Bjorken_Cen} used the crude
liquid-drop
model. They formulated the stability condition and  described the
evolution of
the glob during its penetration through
the atmosphere.

 The energy of the glob is generally a function of  
its baryonic number $N$  and it was taken to be  sum of 
the three terms :\vspace*{-0.2cm}
\begin{itemize}
\item {\em the volume term}, $\alpha N$, where $\alpha$ is an extensive
parameter, arising from repulsive quark interactions and  being for the
bag model with
no interaction $\sim$ 900 MeV,\vspace*{-0.2cm}
 \item {\em the surface term}, $\beta N^{2/3}$, which reduces the
strength of 
the repulsive quark interactions owing to the presence of a surface. It
 destabilizes globs of arbitrary large $N$  and
induces
 a condensation down to globs of finite baryon number.\vspace*{-0.2cm}
 \item 
{\em $\Delta(N)$ term}, which summarizes all the finite-size corrections
to
$E(N)$, not already included in the previous terms.\vspace*{-0.2cm}
 \end{itemize}
\begin{equation}
E(N) = \alpha N - \beta N^{2/3} + \Delta(N)
\end{equation}
If $E(N) \gt Nm_{p}$, the globs are not absolutely stable with respect to
decay into nuclear matter. Thus the stability condition 
 with
respect to single-nucleon emission is:
\begin{equation}
\frac{dE}{dN} \lt m_{p}
\end{equation}
and hence 
\begin{equation}
\frac{2}{3}\beta N^{-1/3} - \frac{d\Delta}{dN} \gt \alpha -m_{p}
\end{equation}
 The globs are always unstable for baryon number $N$ greater than some
 limiting value $N_{0}$. For $ N \lt N_{0}$ they are stable with respect
to nucleon emission unless, for some baryon number $N_{c}$ the term
$d\Delta /dN \gt \frac{2}{3} \beta N^{-1/3} + m_{p} -\alpha$. In this case
the system would be metastable for $N_{c} \lt N$ and $N \lt N_{0}$
and would spontaneously decay for $N \lt N_{c}$ or $N \gt N_{0}$.

 Upon
entering the atmosphere, the glob collides with air nuclei with the
collision length $\sim 30$ g/cm$^{2}$ (for $R_{glob} \sim R_{air}$).
Such  process heats the glob, and it cools either by radiation of mesons
or evaporation of baryons. The boiling of baryons decreases the baryon
number, what can remove  the glob from the metastability region into
stable on unstable region. In the former case  the glob will explode
  and
in the latter one it will collide with air nuclei until it is evaporated.
Thus in some cases the glob could be
characterized by {\it a large
degree
of penetration and short interaction length}, leading to a considerable
deposition of energy in the atmosphere.
To estimate both the energy transferred to the glob per collision and the
evaporation, a simple model of a glob
consisting of an ideal degenerate Fermi gas of quarks was used.
 Taking the quark-proton cross-section to be $\sigma_{qp}
\sim$ 12 mb, $k_{F} \sim$ 1 GeV and assuming the collision of the glob
of the baryon number  $N \sim 100$ with the air nucleus ($N_{air} \sim
14$) authors of ref. \cite{Bjorken_Cen}
obtained that the fractional energy loss of a glob  $\Delta E/E$
is $\sim$ 3 \%.
For the
 gamma factor $\sim
10^{3}-10^{4}$, the glob loses somewhere between $10^{3}-10^{5}$
GeV/collision
and (omitting the possibility of explosion and assuming a collision
length  $\sim 30$  g/cm$^{2}$ as appropriate for $R_{glob} \simeq 
R_{air}$) it
could
easily penetrate a distance $X \sim 6000-8000$ g/cm$^{2}$ in the
atmosphere and reach the sea level.


 In ref. \cite{Bjorken_Cen} were also  discussed  two other variants: the
glob
containing an unconfined massive quark or being a fractionally charged
hadron with large baryon number. In the former  case,
  the
glob would have the quantum numbers
of a quark. As  the quark in such a model is expected to be a complex 
object of a relatively large mass and size, its long-range color field
should attract nucleons via  an induced color dipole moment and help
to compress the quark matter to the required high density.
Below a
certain baryon number, the binding energy
per nucleon will monotonically increase, causing that the ``stripped
glob'' consisting of an unconfined  quark and a few
tightly bound nucleons would penetrate to sea level. There are, in
fact, rather serious experimental constraints on this hypothesis as
it leads as well  to unacceptable flux of quarks at sea level 
as to
an unacceptable rate of horizontal air showers (with zenith angle 
$\gt 70^{0}$).

The picture proposed in \cite{Bjorken_Cen} has the following difficulties:
\vspace*{-0.2cm}
\begin{itemize}
\item Problem of the glob origin;\vspace*{-0.2cm}
\item Unacceptably large rates of fractionally charged quarks and/or
horizontal air showers at sea level; \vspace*{-0.2cm}
\item Adequate  explanation of the absence of
pions.\vspace*{-0.2cm}
\end{itemize}

\subsection{Diffractive fireballs}

Numerous phenomenological models
(e.g. \cite{Panagiotou,Goulianos}) assume that
a fireball
plays a role of an intermediate
state for Centauro production. They were inspired by some  
experimental
characteristics of exotic  events which can be satisfactorily explained
by means of 
 the fireball scenario. All of them  postulate the production of 
``exotic'' clusters, the differences concern the type  
 of
 a projectile 
and the mechanism of evolution and decay of the fireball.

 An example is the scenario  \cite{Navia,Goulianos}
 in which  Centauros and
Mini-Centauros are proposed to be  a result of isotropic decay of
``exotic''
fireballs coherently produced in a diffractive
dissociation process.
 In this phenomenological model a nucleon-nucleon collision at
$\sqrt{s}\sim $1.8 TeV can create a diffractive superheated fireball with
 a mass $\sim$ 180 GeV in the forward region centered around
pseudorapidity $|\eta_{cms}|$ =
2.2 and   with a spread $\Delta \eta = \pm$ 0.7. The high
 $p_{T}$
reflects the temperature of the phase transition. Within this model the
unusual
hadron-electromagnetic asymmetry is assumed to be caused by a phase
transition, similar to the DCC hypothesis, due to the superheated fireball
conditions (with, however, the identification of the hadrons as
nucleons). In principle, there is no explanation why pion emission should
be suppressed in the quark-gluon plasma. The cross-section for a
diffractive event with this mass was predicted to be $\sim$ 0.33 mb.
Full Monte Carlo simulation of Centauro like events was carried out using
this model \cite{Navia,Navia_model}.
 The code consists of two parts : {\it Code Nucleus + Exotic
Algorithm}. {\it The Code Nucleus} describes the nucleus-nucleus
collisions,
using the superposition model. The nucleon-nucleon collision was
described by the UA5 algorithm, based on
 codes for non difractive and single diffractive
interactions.
 {\it The
Exotic
Algorithm} was a code that included the exotic channel, i.e. a diffractive
production of a baryonic fireball and its isotropic decay into baryons. 
The total inelastic cross section
was assumed to be:
\begin{equation}
\sigma_{in} = \sigma_{normal} + \sigma_{exotic}
\end{equation}
with
$\sigma_{normal} = \sigma_{ND} +\sigma_{DD} +\sigma_{SD} \simeq
\sigma_{ND}+\sigma_{SD}$
where $\sigma_{ND}, \sigma_{SD}, \sigma_{DD}$ and $\sigma_{exotic}$ are
the cross
sections for non-diffractive, single diffractive,   double diffractive
 and exotic processes respectively.\\
The branching ratio of ``exotic diffractive channel'' (Centauro and
Mini-Centauro) was chosen to increase with incident baryon energy as
\begin{equation}
P_{exotic} = 0.333 log(E_{N}/100 TeV)
\end{equation} 
This model successfully reproduces the kinematics of cosmic ray exotic
events. Morever, it also
 permits to reproduce the experimental data for
electromagnetic family flux. In opposite to ``normal'' diffractive
events where the majority of secondary particles are pions, in exotic
diffractive events the secondary charged particles were
assumed to be only nucleons and antinucleons.  Multiplicity
 of nucleons was generated using a Poisson distribution with $\langle N
\rangle $ = 100
for Centauro and $\langle N \rangle $ =15 for Mini-Centauro. The
 inclusion of exotic process in nuclear interaction of cosmic ray
particles with atmospheric nuclei, reduces the expected electromagnetic
family flux, because the production of $\pi^{0}$'s in Centauro like events
is suppressed by some unknown mechanism.
 It
has been shown that in the extremely high
energy region ($E_{0} \ge 10^{5}$ TeV) a  dominant exotic channel is
consistent with experimental data from  emulsion chamber
experiment. The model  allows  to understand the negative search
for Centauro events at $Sp\overline{p}S$ CERN collider.
 It introduces the
 energy threshold  for Centauro production connected with
the point of the phase transition $(\sqrt{s}_{th} \sim 2000 \pm
500 $ GeV), which is much higher than energy of $Sp\overline{p}S$
experiments
($\sqrt{s}$ = 540 - 900 GeV) and comparable to that of the Fermilab
collider ($\sqrt{s}$ = 1.8 TeV). The negative results for Centauro search
in some Tevatron experiments operating in the central rapidity region 
(e.g. CDF experiment) were also expected
from this model \cite{Navia_model}, as 
the
Centauros are assumed to be produced close to  diffractive
dissociation
region.

\subsection{DCC}
\subsubsection{Centauros as DCC manifestation}

The QCD phase transition from normal hadronic matter to the Quark-Gluon-
Plasma, 
manifests itself in two forms: deconfinement transition and chiral
symmetry restoration.
 One of the interesting consequences of the chiral
transition is the formation of a chiral condensate in an extended domain,
such that the direction of the condensate is misaligned from the true
vacuum direction. The formation of these so called Disoriented Chiral
Condensate (DCC) domains in high energy collisions of both hadrons and
heavy
ions, has been proposed by many authors 
\cite{Bjorken,DCC,DCC_Krzywicki,DCC_ions,DCC_Gavin}, however,
 this subject is still a quite speculative one. There is no compelling
argument
that DCC must exist, but there is no compelling argument that it does not
exist.

The imporant question is, what are the basic  signatures of DCC?
The main features of that disoriented piece of quark matter result
directly from its definition.
In both descriptions, the linear and non-linear sigma models, DCC can be
defined
as a cluster of pions,
coherently produced,
with anomalously large fluctuations in the neutral fraction.
Pions from a DCC domain will be emitted at low $p_{T}$ and will have a
distinct distribution pattern compared to the normal pion production
mechanism without DCC. An important feature of this radiation 
 is coherence, which means that the
multiplicity distribution of produced particles should be 
 of Poisson--type,
and there will be no Bose-Einstein enhancement. A  DCC almost by
definition
will consist of a cluster of pions with almost identical momenta and in
its
own rest frame will have approximate spherical symmetry. Assuming
some transverse boost for DCC, and the internal relative velocities of the
pions  within a  DCC cluster, smaller than the transverse boost velocity, 
 it should look like in the laboratory frame as a colorless  minijet.

 A  convenient
quantity to characterize a DCC is the fraction of neutral pions per
event\\
\begin{center}
$ f = \frac{N_{\pi ^{0}}}{N_{\pi^{0}} + N_{\pi^{+}} + N_{\pi^{-}}}$.
\end{center}

In a DCC model the probability of finding a given neutral fraction is\\
\begin{equation}
P(f) \simeq \frac{1}{2\sqrt{f}}
\end{equation}
in the limit of large number of pions. 

In generic particle production, that is, production by mechanisms other
than DCC, producing a pion of any given charge is equally likely to
isospin symmetry, so that $f$ is binomially distributed with mean $\langle
 f \rangle =\frac{1}{3}$.
Hence, a basic signature of DCC production is the presence of very
large
event-by-event fluctuations in the fraction of produced neutral pions.
 A significant low $p_{T}$ enhancement should be also observed.

 Other DCC
signatures
include various effects on pion pair correlation.
 Searching for DCC  is connected  with the correlation
question, although it is not understood completely what  correlation
structure of the distribution chiral order parameter is likely to be.
Z. Huang and X. Wang \cite{DCC_Wang} 
found (basing on a linear $\sigma$ model) that
 a rapidity interval in which the pion field is correlated
in isospin direction could be as large as  $\sim$ 2 - 3 rapidity units and
the cluster
structure of pions radiated from the coherent pion field may occur
anywhere in the whole
rapidity region. According to \cite{Bjorken} the natural correlation
length for    
DCC is of the order of 1-2 units of rapidity. However, this is yet an
unsolved
theoretical issue. One  can imagine, for example, a piece of a DCC
centered
at rapidity of +1 ,
 and another one centered at -1 with a different chiral order
parameter and having some small overlap at rapidity 0. The question
is how do the two pieces interact? Will there be a tendency to create 
a common alignment, or will they form independent domains? As it was
suggested in \cite{Bjorken} there is even a remote possibility that the
nonlinearities of the linear sigma model are strong enough to promote
long-range correlations in rapidity.
 The formation time and corresponding
correlation length  have been also estimated
in \cite{DCC_Krzywicki} by non-linear sigma model and quenching
approximation.
The  correlation length 
was estimated as $\sim$ 8 fm in nuclear collisions at 200 GeV/nucleon
(taking WA80 data).
This large value reflects the coherence of the source. The ratio of
``anomalous'' correlation length, related to a DCC  mechanism, to
the ``ordinary'' correlation length was estimated as $\sim$ 5.6
\cite{DCC_Krzywicki,DCC_Chac_1}.
It should be mentioned that
 some signs of the existence of the long range correlations  
have  been already reported    in 1991, in
NA35
data
(sulphur-sulphur 
central collisions at 200 GeV/n), where the so--called coarse grain
fluctuations have been studied by using different kinds of multiplicity
moments \cite{NA35_Gladysz} and fractal analysis \cite{entropy}.
 The results  may
indicate the
appearance of weak, long range correlations,
with the length $\xi \gg $ 3 rapidity units
and  suggest the appearance of
a  coherent domain
 connected  with fragmentation region.

To answer  the basic question, if a DCC scenario could be able to
explain all
Centauro related phenomena
 needs, however, very carefull studies. Some doubts
can be mentioned  at once. At first, it seems
that using a DCC scenario it would be rather difficult to
explain simultaneously both the large transverse momenta observed in the
Centauro species
and the existence of a strongly penetrating component.
The  anomalously high $\langle p_{T} \rangle$ of the particles
produced in Centauro type events
seems to be the main problem. Assuming even that the value of the
gamma inelasticity factor, $k_{\gamma}$ is closer to 0.4 than 0.2, as
it was taken in cosmic-ray papers, the obtained $\langle p_{T} \rangle
\simeq  0.875 \pm 0.375$  GeV. This  value seems to be still too high to
be
connected 
with a DCC mechanism. On the other
hand, one can  not  rule out   that the narrow and strongly
penetrating clusters,
observed in Chiron type events, are produced by  a 
transversally boosted DCC. 
It cannot be excluded that the 
coherently emitted groups of DCC pions may have a net transverse drift
velocity,
 forming the ``colorless''
jet. J. Bjorken \cite{Bjorken_Chiron} speculates that the relative $p_{T}$
of the pions
in the ``colorless'' jet may be under 100 MeV, what is very close of that
measured for cosmic-ray mini-clusters observed in Chiron type events.
 
The
other question concerns the existence of charged DCC, i.e.
the events in which the orientation of the vacuum is orthogonal to the
$\pi^{0}$ direction. Bjorken
\cite{Bjorken} considered this idea, as more speculative than ordinary DCC,
however, he did not exclude it.  He proposed the scenario
where for example the positive charge DCC is produced at negative $\eta$ ,
and the negative charge DCC is produced at positive $\eta$, with a ``domain
wall'' in between. He found that the width of the dipole layer could be
about 1-2 units of rapidity.
  Assuming that charged DCC could be formed, both 
 Centauros and anti-Centauros should be observed in the same experiment.
As it has been already mentioned, the experimental situation is
surprising. Both extremes are claimed
to be seen in cosmic ray data but by two various experiments and
in different energy range. There
were found some
anti-Centauro like events by JACEE Collaboration and no Centauros.
There are Centauros and no reported anti-Centauro events in Chacaltaya and
Pamir
experiments. 
Fig.~\ref{Centauro_statistics} 
do not show any excess of $\gamma$--rich  events when  the observed 
families are compared with the simulated ones.  
It allows to
 doubt if Centauro and DCC are the same phenomenon. On the other hand, it
should be
mentioned  that  the analysis of $\gamma$-hadron families detected at
mountain
laboratories from the point of view of disoriented chiral condensates 
 done in ref.~\cite{DCC_Chac_1,DCC_Chac_2,DCC_Pamir} 
does not
exclude 
 a
small contamination of DCC clusters among normally produced particles
(see the next subsubsection). However,
this result is inconsistent with  the conclusion contained
in
\cite{Wlodarczyk_cen} where authors demonstrate that artificial families
produced at 
mountain altitudes are  insensitive to any isospin fluctuations.
Their simulations showed that all scenarios of isospin fluctuations lead
to essentially the same 
characteristics of families registered at mountain altitudes because
the cascading process in the atmosphere ''kills`` the
characteristic features of
the main interaction. It suggests  that Centauros
 should  originate
from strongly penetrating projectiles.
\subsubsection{DCC search in $\gamma$-hadron families.}
In 1996, a Brasil emulsion group reported \cite{DCC_Chac_1} an
analysis
of several tens cosmic ray events of visible energy greater than 100 TeV
from
Mt. Chacaltaya. They plotted the number of observed hadrons against the
energy fraction of hadronic component (N$_{h}$ vs. $Q_{h}$) and compared
the experimental diagrams with the simulated ones. Artificial families
were constructed for two different models.
 The first scenario  which does not include an anomalous channel was
called
     ''ordinary interaction'' model. It was based on two separate 
algorithms 
  for non--diffractive and single
diffractive
     interactions used by the UA5 group for the
    description of nucleon-nucleon collision.
The second model  included an anomalous production (DCC) in the
diffractive
channel, i.e.
 a coherent emission of pions from a large domain of disoriented
chiral condensates in the {\em leading particle region}.

It was shown that $N_{h}$ vs. Q$_{h}$ distribution
does not match Monte Carlo simulation based on UA-5 data. Experimental
distributions were better simulated by a Monte Carlo which included
24\% DCC production rate in the
leading particle region.
On the other hand,  a  significant discrepancy
between the experimental family flux and  results
 of simulations should be mentioned. Even the model including
the anomalous
channel (DCC)
fails in description of such global characteristics as the family
flux observed at mountain altitudes. The question of unusual 
transverse momentum observed in hadron-rich families was not investigated
by  authors of \cite{DCC_Chac_1}.
 
More recently,  a more advanced analysis,  using the robust observables
have been done \cite{DCC_Chac_2,DCC_Pamir}.
The robust observables $r_{i,j}$ are constructed through the ratios of
normalized bivariate factorial
moments $F_{i,j}$.
\begin{equation}
r_{i,j}= F_{i,j}/F_{i+j,0}
\end{equation}
where
\begin{equation}
F_{i,j}=\frac{\langle
N_{ch}(N_{ch}-1)...(N_{ch}-i+1)N_{\gamma}(N_{\gamma}-1)...(N_{\gamma}-j+1) 
\rangle }{\langle N_{ch} \rangle ^{i} \langle N_{\gamma} \rangle ^{j}}
\end{equation}
 As was shown in \cite{DCC_robust}, the robust observables  are
sensitive to
DCC admixtures in multiple production.
The basic difference between generic production and DCC production is that
in the former case the generating function depends only upon one variable,
while for DCC it depends nontrivially upon two. Thus the factorial
cumulants are  very convenient tools for
 distinction between the distributions.
 If a distribution of ratio
of neutral to all pions, $f$,  is governed
by a binomial distribution, as it is observed in standard pion production,
called  generic pion production, then
\begin{equation}
F_{i,j}= F_{(i+j),0}.
\end{equation}
Therefore many ratios of the $F_{i,j}$ are expected to be unity:
\begin{equation}
r_{i,j}(generic) = \frac{F_{i,j}}{F_{(i+j),0}} = 1, 
\end{equation}
and particularly
 for $i \ge 1$ and j=1.
\begin{equation}
r_{i,1}(generic) =1.
\end{equation}
If on the other hand a distribution of $f$ is of the form  $p(f) =
1/(2 \sqrt{f})$
,
as predicted by
a DCC mechanism, then  in the semiclassical limit
\begin{equation}
r_{i,1}(DCC)=1/(i+1).
\end{equation}
Thus the values of $r_{i,j}$ below 1 could be the  indicators of
 a DCC
formation.
 The results presented in \cite{DCC_Chac_2} were  based on analysis of  
 59 $\gamma$-hadron
families, with $\Sigma E_{vis} \gt 100 $ TeV,
detected in thick lead chambers, of 57 m$^{2}$year exposure.
 The robust observables have been calculated
for  real and 
simulated families and compared one to the other. It was concluded 
 that there are 
peculiar clusters of pions with large asymmetries in the neutral pion
fraction distribution, absent in the artificial families.

A similar conclusion has been obtained in \cite{DCC_Pamir} where 139
$\gamma$-hadron families with energies 100 TeV $\lt \Sigma E_{vis} \lt
$700
TeV, from Pamir experimental data, have been analysed. Here
 the behaviour of robust observables as a function of the visible
 energy of a leading jet, obtained after decascading procedure, was
studied.
 The authors of refs.
\cite{DCC_Chac_2,DCC_Pamir} explain the observed  deviation from the 
simulated characteristics  as a 
copious 
production of pions from a DCC domain, formed in the far-forward angular
region. It should be noted, however, that the production
rate of DCC, estimated in \cite{DCC_Chac_1} as 21-24\%
 for the
leading particle region,  is not sufficient
 to explain the global characteristics, such as the
 energy spectra
 of families.
For the whole inelastic channel this value was estimated to be less than
5\%.

 The analysis
\cite{DCC_Chac_1,DCC_Chac_2,DCC_Pamir} 
has been done in the whole region of the experimentally accessible phase
space.
More information could be achieved from the study of robust
observables in pseudorapidity and/or  azimuthal angle  bins of different
sizes.
Some hints can be taken from the  earlier studies of
multiplicity fluctuations in pseudorapidity bins
\cite{Intermittency_Gladysz}. The analysis has been done for several
events, coming from the similar type thick lead  chamber of the Pamir
experiments. Strong fluctuations have
been revealed in the high energy  ($\Sigma E_{vis} \gt 100$ TeV)
photon and hadron
families. Here, the multiplicity fluctuations have been studied, by means
of scaled factorial moments, 
separately for photonic and hadronic component. Such result, however, 
may also
indicate on  fluctuations of the ratio of photon  to hadron
multiplicity in analysed bins (DCC domains). 
Some difficulties in interpretation of experimental data are connected
with 
  discrepancies between
 various  simulation studies.
Authors of \cite{DCC_Chac_1, DCC_Chac_2} conclude about the existence
of non-statistical fluctuations in the neutral fraction. This
conclusion is based on  disaccordance of experimental and
 simulated characteristics, based on the ``normal'' model
production and Monte-Carlo calculations of the cascade processes 
through  the whole atmosphere. 
 On the other hand, authors of 
\cite{Intermittency_Wlodarczyk} claimed that 
strong fluctuations observed in atmospheric families are primarily
determined by fluctuations in the atmospheric cascades themselves, and
there is no need for changing  the act of elementary production.
 The problem needs further 
investigation.
\subsection{Strange Quark Matter }
Colour single states observed so far consist of three quarks (baryons),
three antiquarks (antibaryons) or quark--antiquark pairs (mesons).
 Such objects are described by the Standard Model which does not forbid
the 
existence of colour single states in a bag containing an integer multiple
of three quarks.
In such quark matter bags all the quarks are free within the hadron's
boundary, so such states are inherently different from nuclear ones that
are
composed of a conglomerate of A = 1 baryons. Quark matter states composed
of only $u$ and $d$ quarks are known to be less stable than normal nuclei
of the same baryon number A and charge Z since they do not decay into
quark
matter.
The strange quark matter (SQM) is a matter with strangeness per baryon of
 the order of
unity, thus containing a comparable fraction of up, down and strange
quarks. If it ever exists, may be a stable ground state of a QCD
instead of the nuclear matter.
 The fact that SQM is absolutely stable does
not contradict the ordinary experience that matter is, for the most part,
made of ordinary nuclear matter. The reason is that changing the ordinary
matter to SQM one, a large number of $u$ and $d$ quarks need to convert
to $s$ quarks, which would require a very high order weak interaction
and therefore it is  highly improbable \cite{Witten}. On
the other hand, the existence of ordinary nuclear matter shows that quark
matter consisting of only $u$ and $d$ quarks is unstable. Adding  
 a third flavour introduces another Fermi well and thus reduces the energy
relative to a two--flavour system what can make the system stable. The
most stable configurations would have roughly equal numbers of up, down
and
strange quarks with charges of +2/3e, -1/3e and -1/3e, respectively,
therefore minimizing the surface and Coulomb energies. A major
destabilizing factor is the large mass of the strange quark.
 The anticipated mass range for this kind of
matter may lie between the masses of light nuclei and that of neutron
stars of $A \simeq 10^{57}$.
Some astrophysical mechanisms can convert very large stars into strange
stars.
Strange matter being the part of the cosmic radiation is called sometimes
 strange quark {\em nuggets} or {\em nuclearities}.
 Smaller amounts of strange quark matter are
usually called {\em droplets} of strange quark matter, or simply
{\em strangelets}. 
 The existence of stable SQM was postulated by Witten
\cite{Witten}. He also suggested the possibility of  production of 
small lumps of SQM, strangelets with $10^{2} \lt $ A $ \lt 10^{6}$, in
today's universe by quark(neutron) stars, which could convert to  more
stable SQM stars. 
They
 could permeate the Galaxy and reach the
Earth.
 SQM could have
important cosmological consequences for today's universe which arise
from the possibility that the early universe has undergone a first-order
phase transition from SQM to  nuclear matter.

Many authors investigated conditions for SQM stability (see for example
 refs. \cite{Fahri, Berger,Chin}. The practical measure of  stability of
a strangelet is provided by the so called separation energy $dE/dA$, i.e.
the energy which is required to remove a single baryon from a strangelet.
 If $dE/dA \gt m_{N}$ then strangelet can evaporate neutrons
from its surface.
Contrary to normal
nuclei, SQM stability increases with A and the threshold of its stability
is
close to $A_{crit} \sim 300$. Some calculations, based on QCD and the
phenomenological bag model \cite{Chin,Schaffner} (up to the baryon number
A = 40) suggest that strange quark matter may be metastable or  even
completely stable for a wide range of bag model parameters values
($B^{1/4} \sim$ 150-170 MeV). Generally, for higher bag parameter values
there are less long--lived strangelets and they are shifted towards higher
values of baryon number A and  strangeness factor $f_{s}$.
 There are also predictions
that quite  small strangelets might gain stability due to shell effects
\cite{Gilson,Madsen}. They are called ``magic strangelets''.
However, due to the lack of theoretical constraints on bag model
parameters and difficulties in calculating colour magnetic interactions
 and finite size effects, experiments are necessary to help answer the
question of the stability of strangelets. 
The properties of some forms of hypothetical strange matter, as small
lumps of strange quark matter (strangelets) or hyperon matter (metastable
multihypernuclear objects MEMO's) have been discussed by many authors
(see for example \cite{Berger,Chin,Schaffner,Greiner}) with special
emphasis
on their relevance to the present and future heavy ion experiments.
Different aspects of strange quark matter physics are described in the
  reviews \cite{neutron_star,Madsen_rev,Greiner_rev}.

Searches for SQM have been made on terrestrial matter \cite{str_terr},
cosmic rays and
astrophysical objects \cite{str_astr}. The searches resulted in low limits
for strangelets in terrestrial matter, but on the other hand
 some
 data  could be understood by assuming the
presence of strangelets in cosmic rays \cite{8,9,Shaulov,
Wlodarczyk_str, Zhdanov, muon_bundle} (see the
next section). It is postulated that the millisecond pulsar SAX
J1808.4-3658 is a
good candidate for a strange star \cite{strange_star}.
 Several strangelet candidates have been reported
in cosmic ray experiments \cite{Saito,Price_str,Ichimura}.
According to the picture proposed
in \cite{8,9}  strangelets could be remnants of the
decay of slightly-strange quark matter fireball, produced in
 central collisions of primary cosmic ray nuclei.

 There are many
experiments looking for strange quark matter in heavy ion collisions.
They are mostly based on
such discerning property of strangelets as an unusual charge to mass
ratio ($Z/A \ll 1$).
 The
strange counterparts of
ordinary nuclei are searched for in high-energy collisions at Brookhaven
\cite{BNL_str} and at CERN \cite{CERN_str}.
 To date no experiment has published results indicating
a
clear positive signal for strangelets.
Although   experiments were able to set  upper limits on the
existence of strangelets in the range of sensitivity of the experiments,
they  were not able to answer the question concerning their existence.   
There are several definite reasons for this. The main ones are  that
the
experiments are sensitive only to metastable strangelets with proper 
lifetimes greater than $\sim 5\cdot 10^{-8}$ s and they look for
strangelets
 in the midrapidity region. Most of experiments (except the E814) used
focusing
spectrometers which, for a given magnetic field setting, have a good
acceptance only for a fixed momentum and charge of the produced particle.
Therefore, the production limits obtained in these experiments are
strongly dependent upon the production model assumed for lumps of strange
quark matter.
The present status of strange quark matter searches, both in cosmic ray
and in accelerator experiments is presented by R. Klingenberg in his 
  review \cite{Klingenberg}.

\subsubsection{Extraterrestrial  strangelets}
Strangelets have been discussed, in the context of Centauros, 
  by several authors
\cite{8,Shaulov,Wlodarczyk_str,Wlodarczyk_cen}. However, in contrary  
 to the picture
 \cite{8}, where they are postulated to be produced in nucleus--nucleus
collisions, some
authors
have   considered the possibility that  
{\it strangelets are a
component of primary cosmic rays}. After penetration
 deeply into the
atmosphere they  could decay, evaporating nucleons and producing
Centauro-type
events. In principle, the ideas presented in
\cite{Shaulov,Wlodarczyk_str,Wlodarczyk_cen} 
are
very close to the old picture of Bjorken and
McLerran  \cite{Bjorken_Cen}. The main difference is
the proposed Centauro-fireball (strangelet) cross section.

 S. B. Shaulov
in
\cite{Shaulov} has presented 
a rather qualitative discussion of the presence of strangelets in
the primary cosmic rays. 
Similarly as in \cite{Bjorken_Cen} the evaporation of neutrons was assumed
to be the main channel of strangelet energy loss and the energy fraction
evaporated in a single act of collision
was estimated in similar way.
 It was obtained that in one  act of a collision
several neutrons can be evaporated and a strangelet loses energy of
several
GeV
(in its own rest frame).
In this work it was  postulated that
 the new penetrating
component, carrying a large part of energy through the atmosphere,
could qualitatively explain a chain of anomalous phenomena.
 Extensive Air Shower (EAS) spectrum ``knee'' naturally originates if the
strangelet spectrum has the maximum at the energy higher than 3 PeV and
the strangelet intensity in the maximum is similar to the nuclear one. 
Large variety of anomalies observed in individual events, such as: 
 Centauro, Chiron,
      Geminion, halo, penetrating component etc.
 could be
 natural
manifestation  of the strangelet evaporation during its passage through
the
atmosphere. Also mini-clusters frequently observed in Chirons could be  
 a consequence of small $p_{T}$ evaporation processes. 
 Within this scenario, however, it is hard 
 to understand the appearance of
high transverse momenta secondaries in such events.
For consistency of this picture the beginning of strangelet spectrum
 near
$E_{0} \sim $ 10 PeV with the power index $\ge$ 3 should be assumed.

G. Wilk and Z. W\l{}odarczyk 
presented in \cite{Wlodarczyk_str,Wlodarczyk_cen} 
the  picture of propagation of
lumps of Strange Quark Matter
through the atmosphere, similar to the one described in
\cite{Shaulov,Bjorken_Cen}. In opposite to \cite{8} and \cite{Bjorken_Cen} 
 they argue, however, that strangelets
are of almost  normal
nuclear sizes. This statement results from
 the value of strangelet
quark chemical potential assumed by authors of
\cite{Wlodarczyk_str} to be $\mu_{q} \sim$ 300 MeV. Such value disagrees
with that
 estimated in \cite{Panagiotou}, from
Centauro experimental characteristics, and being $\mu_{q} \sim$ 600 MeV.
 According to refs. \cite{Wlodarczyk_str,Wlodarczyk_cen} the strong
penetrability of
strangelets is 
not caused by their small geometrical radii. Strangelets penetrating
deeply into
the atmosphere are formed in many consecutive interactions with air
nuclei, when the mass number of the incident, very heavy lump of
SQM is successively diminished. 
 Big primary strangelets can penetrate
very deeply into the atmosphere until their baryon number $A_{str}$
exceeds
some critical value  $A_{crit} \sim 300-400$. Below this value they simply
disintegrate into nucleons.
 Such decay could imitate Centauros.
According to \cite{Wlodarczyk_str} Centauro events observed at the 
mountain altitudes would require a primary strangelet of $A_{0} \simeq
1000-2000$. For the mass spectrum of the form $N(A_{0}^{str}) \sim
exp(-A_{0}^{str}/130)$
the model explains the smaller number of Centauros detected at Pamir
altitude ($\sim$ 600 g/cm$^{2}$) than at Chacaltaya (540 g/cm$^{2}$).
Simultaneously, as it was already mentioned, the authors of ref.
\cite{Wlodarczyk_cen} claimed that
their simulations indicate the insensitivity of families detected at
mountain
altitudes to any isospin fluctuations. It could mean that the
explanation of
Centauros
needs the presence of some very strongly penetrative projectiles 
in primary cosmic rays.  
The model passes over in silence the question of high $p_{T}$ of
Centauro-like phenomena.

It is important to note that the proposed pictures of  the strangelet
penetration through  the matter  and its successive destruction in the
consecutive collision acts  
 should be connected with the observation of the large cloud of the
low
energy nucleons from the destroyed target nuclei. Interesting, the 
 extremely long delay ($\gt $ 0.5 msec) neutrons have been recently
observed
\cite{Zhdanov}
in  large Extensive Air Showers ($N_{e} \gt 10^{6}$) by the neutron
monitor working in
conjunction with EAS instalation ``Hadron''. This phenomenon appears at
primary energies higher than $3\times 10^{15}$ eV and it is observed 
 close to the EAS axis. As the tentative explanation of this
phenomenon
one can propose the arrival of a new type of primary cosmic ray particles,
like strangelets, with gradual dispersion of their energy along the whole
atmosphere.

Also muon bundles of extremely high multiplicity observed by ALEPH
detector (in the dedicated cosmic--ray run) could originate from
strangelets collisions with the atmosphere \cite{muon_bundle}.
 
The old experimental results are also worth to recalling. Anomalous
massive (A=75...1000) and relatively low charged objects (Z=14...46),
which could be interpreted as strangelets, have
been observed. These are:\vspace*{-0.2cm}
\begin{itemize}
\item Two anomalous events,  with  charge $Z\simeq 14$
and  mass number A$\simeq$ 350 and $\simeq$ 450 (what can be 
consistent with  theoretical estimate for  Z/A ratio for  SQM),
observed in primary cosmic ray by counter
experiment with the balloon (at the depth $\sim$ 9 g/cm$^{2}$)
\cite{Saito};\vspace*{-0.2cm}
\item The so--called Price's event \cite{Price_str} (detected at the
depth
$\sim$ 3-5 g/cm$^{2}$) 
with Z $\simeq$ 46 and A
$\gt$ 1000,
regarded previously as a possible candidate for a magnetic monopole. It
turned
out to be also consistent with the small Z/A ratio in SQM;\vspace*{-0.2cm}
\item The so--called Exotic Track event with Z $\simeq$ 20 and
A $\simeq$ 460
      \cite{Ichimura}. It was observed in the emulsion chamber, exposed
      to  cosmic rays with the  balloon at the atmospheric depth of only
11.7
g/cm$^{2}$ at zenith angle of $87.4^{0}$. It means that the
projectile
causing that event traversed $\sim$ 200 g/cm$^{2}$ of the atmosphere. It
is  
in
contrast with the previous events where the corresponding depths were of
the
order of 5-15 g/cm$^{2}$ of the atmosphere only.\vspace*{-0.2cm}
 \end{itemize}
 If the Cenaturo-like events are really caused by strangelets then 
 the Exotic Track event could be an argument  supporting 
their
  strongly penetrative power. It is rather impossible for  normal
  nuclei of A $\sim$ 460 to traverse
   $\sim$ 200 g/cm$^{2}$ of matter without fragmentation.
 The Exotic Track event motivated the balloon JACEE \cite{JACEE_str} 
   and the Concorde aircraft \cite
 {Capdevielle}  experiments (at the depth
110 g/cm$^{2}$) intended to search for
strangelets with long mean free
 paths.

\subsubsection{Strangelets formed in the Centauro fireball}

At present, several types of  models are used to describe the
strangelet
production in heavy ion collisions
\cite{Greiner,Baltz,Braun-Munzinger}.
They can be classified into two categories:  strangelet production
by  coalescense of hyperons or by production following  a quark-gluon
plasma creation.
In a  very popular
coalescense model, an ensemble of quarks, which are products
of nucleus-nucleus collisions, form a composite state which fuses
to form a strangelet \cite{Baltz}.  
 The
formation of a quark-gluon plasma
is not needed in this scenario, as hyperons coalesce during the late stage
of the collision forming a doorway state for strangelet production.
 Such scenario  favours the  production of low mass strangelets
 in
the
midrapidity region, formation of
strangelets
with $A \geq 10$ is rather unlikely.
Thermal models further assume  that chemical
and
thermal equilibrium are achieved prior to final particle production
\cite{Braun-Munzinger}.
Coalescense and thermal models usually predict lower strangelet cross   
section than models postulating a formation of  a QGP
state.
 Greiner et al. \cite{Greiner} suggest that once a QGP droplet is
formed, almost every QGP state evolves into a strangelet by means of the
strangeness distillation mechanism.  Distillation process
provides a possibility for producing more stable large
strangelets since the QGP would lose energy by meson emission possibly
resulting in a strangelet of approximately the same A as the QGP droplet.

 The model of Centauro production
in nucleus-nucleus collisions has been proposed in \cite{Panagiotou} and 
 it has been developed 
in \cite{8}.
 The formation of  incident quark-matter fireball, in 
collisions of the primary cosmic ray nuclei with air nuclei, at the
top of the atmosphere, and
its
subsequent
transformation into relatively long-lived strange quark-matter droplet,
was postulated.
In order to survive passage from the upper atmosphere downward to the
Chacaltaya (Pamir) altitude, the Centauro
fireball should have  a large mean free path ($\lambda \ge 150$
g/cm$^{2}$)
and/or lifetime  $\tau_{0} \sim 10^{-9}$ sec. Then the hadronization
of the Centauro fireball will occur close to the detector.
  The picture \cite{8,9} is based on models involving an intermediate
 quark-gluon plasma state and the distillation process as separation
mechanism of strangeness
and anti-strangeness, as described for heavy ion collisions in
\cite{Greiner}. The  baryon-rich environment
 is considered.
The model avoids the problem of the very small flux of cosmic ray
nuclei at the mountain top level and additionally contains very exciting
suggestion of the simultaneous interpretation of both Centauros and
long-flying component.
 The strongly penetrating component could be
the sign of the passage of strangelets, formed at the last stage of the
Centauro fireball evolution, through the matter. 
 This hypothesis has been studied 
in \cite{9}. 
The model explains most features of Centauro events. 
 It  could likely
explain also the observed  low family flux. 
Such
features of the exotic fireball, as  its strong penetrating power
and  its  decay predominantly into baryons could cause  a decrease of
the electromagnetic
family flux.

\paragraph{Characteristics and time evolution of Centauro fireball}

\begin{figure}
\begin{minipage}{9cm}
\begin{center}
\mbox{}
\epsfig{file = 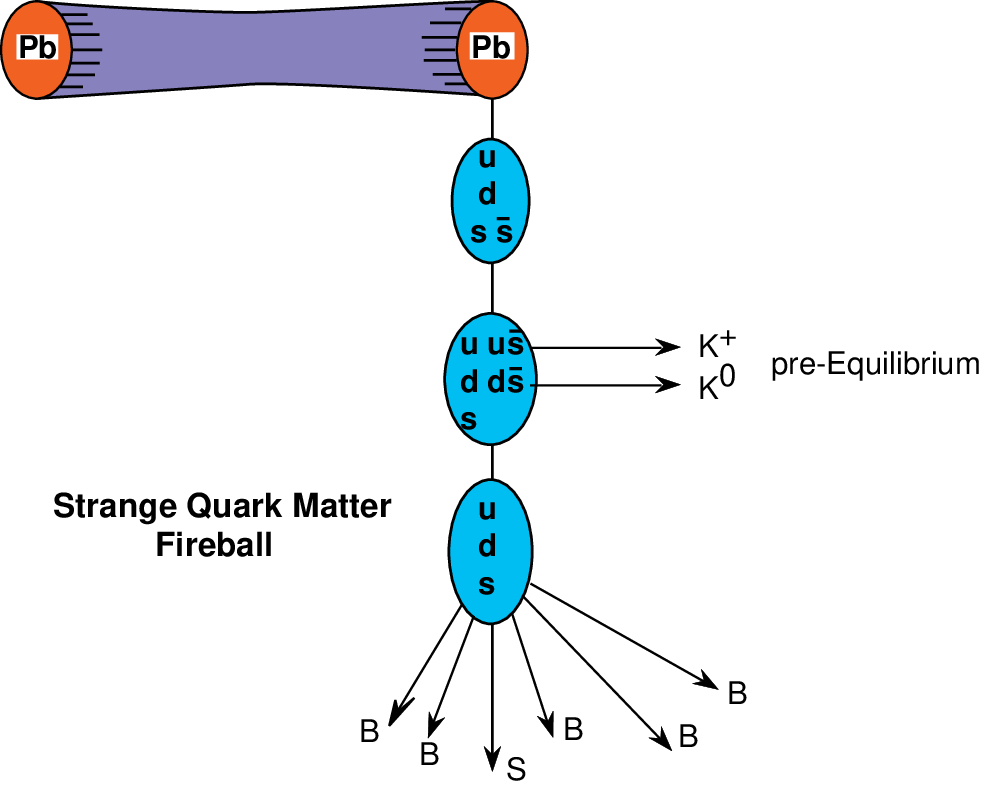,
bbllx=50,bblly=441,bburx=520,bbury=819,width=5cm}
\end{center}
\end{minipage}
\begin{minipage}{5.5cm}
\vspace*{1cm}
\caption{Centauro fireball evolution scheme \cite{CASTOR}.}
\label{Cen_evolution}
\end{minipage}
\end{figure}

The main stages of the development in time of the Centauro fireball
 (see Fig.~\ref{Cen_evolution}) are the following;

\begin{enumerate}\vspace*{-2mm}
\item
{\em Formation of a quark--matter fireball}. \\
The fireball is created in central collisions of ultrarelativistic
($ E_{Lab} \sim 1740 $ TeV)  medium--heavy cosmic--ray nuclei
with air nuclei.  At the 
moment of creation it consists of $u$, $d$ quarks and gluons only. It has
very
high matter density, hence high binding energy. As the fireball is formed 
in the baryon--rich, projectile fragmentation region, the
 high baryochemical potential prohibits the creation of $ u \overline{u}$
and $d \overline{d} $
quarks,
resulting in gluon fragmentation $ g \rightarrow s \overline{s} $ mainly.
\vspace*{-2mm}
\item
{\em Chemical equilibrium} \\
The state of (partial) chemical equlibrium may be
 approached in a time interval
$ \Delta \tau > 10^{-23} $ s, the relaxation time for
$ g \rightarrow s \overline{s} $ . The larger amount of
up and down quarks with respect to the anti-quarks results in a higher
chance for a $\overline{s}$ quark to find $u$ or $d$ quark and  form a
$K^{+}$
or $K^{0}$  than for $s$-quark to form the anti-particle counterparts.
Emitted kaons 
 can carry away  all strange antiquarks and positive charge,
  lowering somewhat the
initial temperature and entropy. 
\vspace*{-2mm}
\item
{\em Strange quark matter state}\\
After emitting kaons 
  {\em the initial
quark--matter
fireball} is transformed into {\em a lightly strange quark matter state}
(with the value
of
strangeness to baryon ratio $\rho_{s}/\rho_{b} \sim 0.06$). 
The fireball is a mixture of $u$, $d$ and $s$ quarks and it is still
characterized by  very large density, low temperature and
the value of charge to
mass ratio $ (Z/A \sim 0.4) $
less than that of the original quark--matter fireball.
 The fireball has a finite
 excess of s--quarks and due to their stabilizing
 effects \cite{Witten,Greiner} it could become
{\em a long--lived state} capable  to reach
the
mountain top level before  decaying.\vspace*{-2mm}
\item
{\em Hadronization}\\
The mechanism of strangeness distillation 
\cite{Greiner}, can cause  the strange quark content of the Centauro
fireball to form low--mass strangelet(s). The total energy per baryon
can be lowered assembling the nonstrange quarks into pure nucleonic 
degrees of freedom and  leaving the strange quarks in a strange matter
droplet, its strangeness fraction being enriched to $f_{s}\simeq 1$.
 Such state would
be a mixture of two phases: pure nucleonic matter and  strange
quark--matter cluster.
The Centauro fireball finally could decay into non--strange baryons
 ($<N_{h}>\sim 75$)
and strangelet(s) having very high strangeness--to--baryon ratio
($f_{s} = N_{s}/A\sim 1$),
very low charge--to--baryon ratio ($Z/A \sim 0$), and a small mass number
($A_{str} \sim 10 - 15$).
\end{enumerate}
 
This scenario is based on
 the experimental Centauro characteristics derived from five
``classical Chacaltaya
Centauros'' (see section 2.1, Tables~\ref{classical_cen} 
and \ref{Cen_kin}).
Using these characteristics the {\em thermodynamic
quantities} have been estimated in \cite{Panagiotou},
 assuming an
ideal two--flavour QGP.
\begin{table}[h]
\label{Cen_characteristics}
\caption[{\scriptsize
 Summary of estimated 
 quantities
 characteristic of the cosmic--ray Centauro events \cite{Panagiotou,8}.}]
{
 Summary of estimated 
 quantities
 characteristic of the cosmic--ray Centauro events \cite{Panagiotou,8}.}
\begin{center}
\begin{scriptsize}
\begin{tabular}[l]{|ll|}\hline
&\\
\multicolumn{2}{|c|}{"Centauro" fireball}\\&\\\hline
&\\
Mass of fireball & $M_{fb}$ = 180 $\pm$ 60 GeV\\
Volume of fireball & $V_{fb} \leq$ 75 - 100 fm$^{3}$ $^{(*)}$\\   
Energy density of fireball & $\varepsilon_{fb}\geq$ 2.4 $\pm$ 1 GeV
 fm$^{-3}$ $^{(*)}$\\
Baryochemical potential of fireball & $\mu_{b}$ = 1.8 $\pm$ 0.3 GeV\\
Temperature of fireball & $T_{fb}$ = 130 $\pm$ 6 MeV\\
Quark density of fireball & $< \rho_{q}> $ = 8 $\pm$ 3 $fm^{-3}$\\
Baryon density of fireball & $< \rho_{b} >$ = 2.7 $\pm$ 1 $fm^{-3}$\\
Strange quark density & $\rho_{s} \sim$ 0.14 $fm^{-3}$\\
Antiquark density & $\rho_{\overline{q}} \sim$ 3.6 x 10$^{-3} fm^{-3}$\\
Gluon density & $\rho_{g} \sim$ 0.6 $fm^{-3}$\\
Entropy density & $S \sim$ 16.4 $fm^{-3}$\\
Entropy/baryon density & $S/\rho_{b} \sim$ 6\\
Strangeness/baryon density & $\rho_{s}/\rho_{b} \sim$ 0.06\\
Final charge/baryon & $(Z/A)_{f} \sim$ 0.4\\
Net strangeness & $N_{s} - N_{\overline{s}} \sim 14$\\
Predicted particle ratio & N(pion)/N(nucleon) $\simeq$ 7x10$^{-6}$\\
\hline
&\\
\multicolumn{2}{|c|}{
"Centauro" strangelet}\\&\\\hline &\\
Mass & A $\sim$ 10 - 15\\
Charge/baryon & Z/A $\sim$ 0\\
Strangeness/baryon~~~~~~~~~~~~~~~~~~~~~~~~~~~~~~~~~~~ & $f_{s} \sim$ 1 \\
\hline
\end{tabular}
\end{scriptsize}
\end{center}
\hspace{1cm}
$^{*}$ {\scriptsize according to \cite{Panagiotou_Note} $V_{fb} = 27 \pm
16$
fm$^{3}$ and
$\varepsilon_{fb} = 6.7 \pm
3.6$
 GeV/fm$^{3}$}\\
\vspace*{-2mm}
\end{table}
 Estimated main features of both the Centauro fireball and the  strangelet
are
summarized in  Table~11.

 A more realistic analysis,  introducing {\em three
flavours
(u,d,s) interacting quarks}, realistic {\em strangeness equlibration
factor}
and using the full equation for
$\varepsilon(T,\mu_{q},\alpha_{s},\gamma_{s})$ was done in \cite{8}.
The energy density have been calculated from the formula
\cite{3flavor_curve,Panagiotou_phase_curve}:
\begin{eqnarray}
\varepsilon_{qgp} &=&(\frac{37}{30}-\frac{11a_{s}}{3\pi})\pi^{2}T^{4}+
(1-\frac{2a_{s}}{\pi})3\mu_{q}^{2}T^{2}+\frac{3}{2\pi^{2}}
                  (1-\frac{2a_{s}}{\pi})\mu_{q}^{4}+
\nonumber\\       &+&\gamma_{s}(\frac{18T^{4}}{\pi^{2}}(\frac{m_{s}}{T})^2K_{2}
                  +6(\frac{m_{s}T}{\pi})^{2}(\frac{m_{s}}{T}K_{1}).
\end{eqnarray}
 The entropy per baryon
\begin{equation}
S_{b}=S/\rho_{b}
\end{equation}
where the baryon density
\begin{equation}
\rho_{b}=\frac{2}{3}(1-\frac{2a_{s}}{\pi})(\mu_{q}T^{2}+
\frac{\mu_{q}^{3}}{\pi^{2}}),
\end{equation}
and the entropy
\begin{eqnarray}
S &=& \frac{1}{3}\frac{\partial \varepsilon}{\partial T}=\nonumber
   (\frac{74}{45}-
\frac{44a_{s}}{9\pi})\pi^{2}T^{3}+2(1-\frac{2a_{s}}{\pi})\mu_{q}^{2}T+
\nonumber\\
  &+& \gamma_{s}(\frac{3m_{s}^{3}}{\pi^{2}}K_{3}+
\frac{12m_{s}^{2}}{\pi^{2}}TK_{2}
+\frac{m_{s}^{4}}{\pi^{2}T}K_{2}+\frac{5m_{s}^{3}}{\pi^{2}}K_{1}+
\frac{m_{s}^{4}}{\pi^{2}T}K_{0}),
\end{eqnarray}
 where $K_{i}(m_{s}/T)$ are i--order modified Bessel functions
 and $\gamma_{s}$ is the strangeness equlibration factor.

\begin{figure}[ht]
\begin{center}
\hbox{
\mbox{}
\epsfig{file = 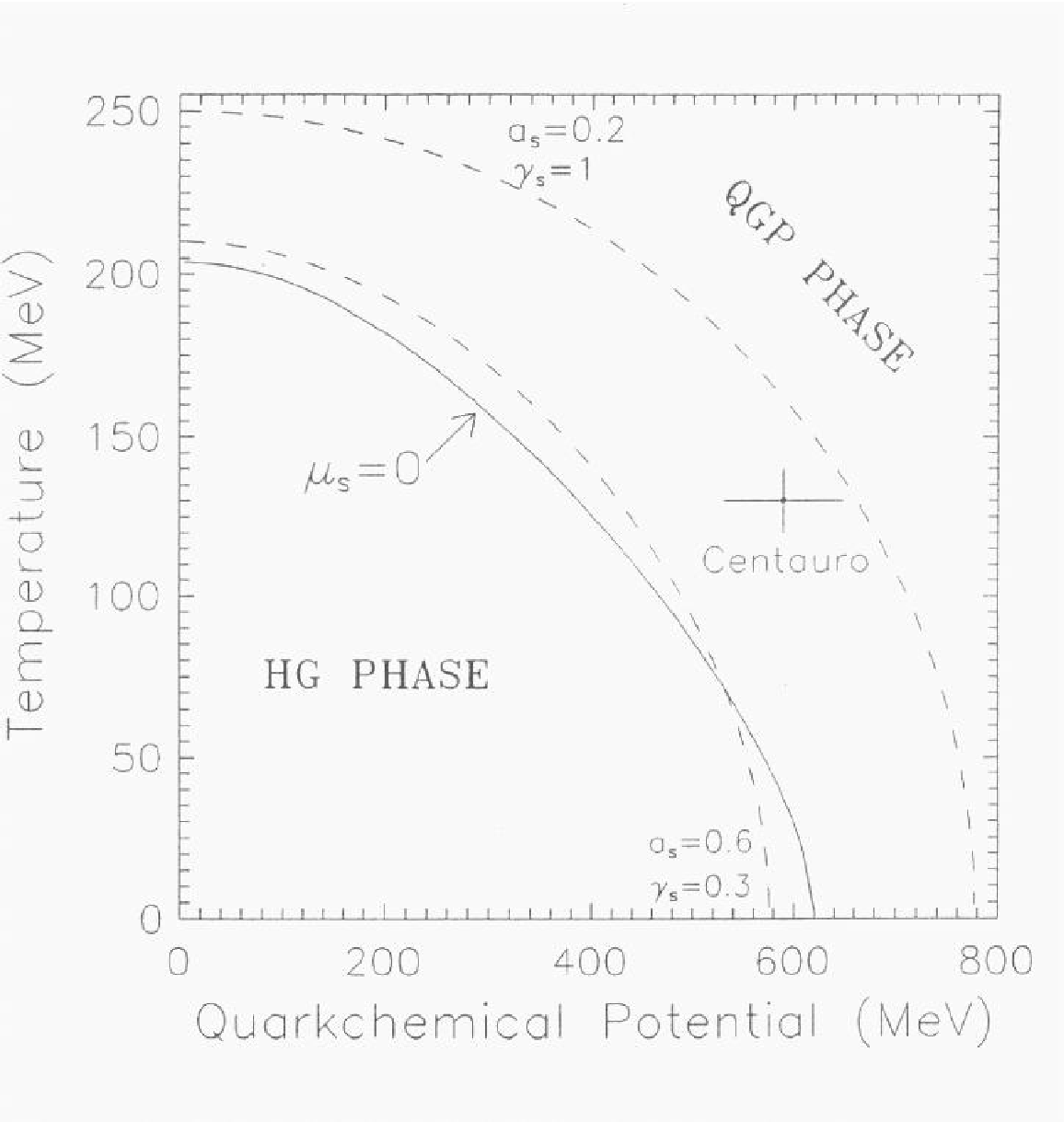,
bbllx=19,bblly=36,bburx=574,bbury=600,width=7cm}
\mbox{}
\vspace*{4mm}
\epsfig{file = 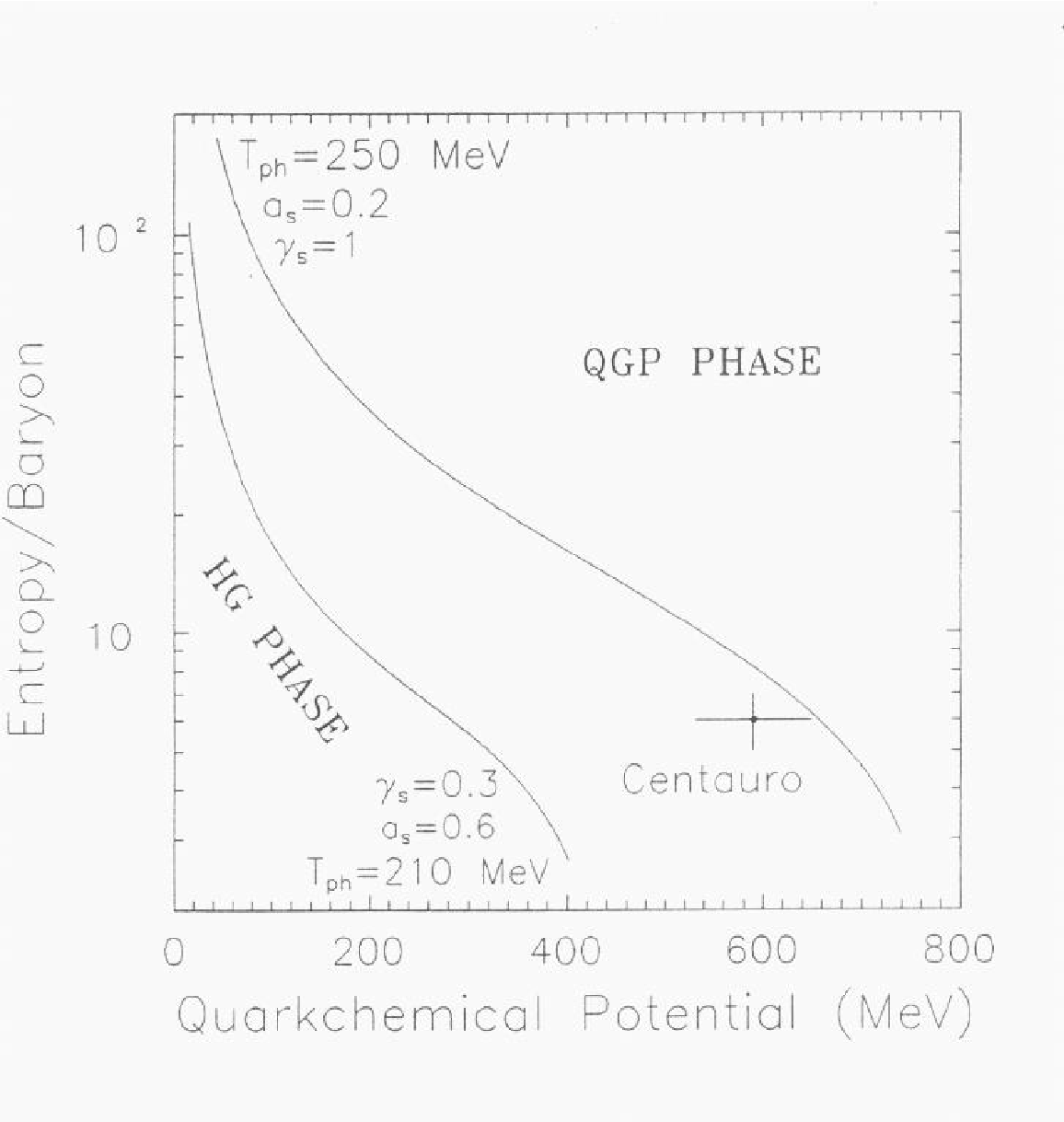,
bbllx=19,bblly=36,bburx=574,bbury=600,width=7.3cm}}
\caption{Location of the Centauro point $(T,\mu_{q})$ on the
phase
 diagram for two temperatures:  $T(\mu_{q}=0) $= 250 and 210 MeV (dashed
curves). Solid curve is for $\mu_{s}=0$ in the hadron gas. The plot at
the right
side  shows
entropy per baryon versus quarkchemical potential, for two
 phase curves
\cite{Panagiotou_phase_curve}.}
 \label{Cen_phase_curve}
\end{center}
\vspace*{-.8cm}
\end{figure}

Using these formulas, it has been shown in \cite{8} that the
Centauro
fireball with values 
of the temperature, quarkchemical potential and energy density 
estimated previously to
be $T$ =130
MeV,
 $\mu_{q}$ = 600 MeV and  $\varepsilon $ =
2.4
GeV/fm$^{3}$, is not the ideal QGP fireball because the strangeness
equilibration factor is estimated to be less than unity and strong
coupling constant is expected to be greater than zero.
 Taking for this quark matter state  the strong
coupling constant $\alpha_{s} \simeq 0.5-0.6$
and $\gamma_{s} \simeq 0.4$ one obtains
 the Centauro point  located at the phase
diagram close to the QGP phase, at the in-between region of the ideal
QGP phase connecting with chiral phase transition and the Hadron Gas
Phase as it is shown in Figure~\ref{Cen_phase_curve}.

{\centerline {\bf Penetrability question}}

According to this model, Centauro events are produced in central
collisions of ultra-relativistic cosmic-ray nuclei with air nuclei.
 A collision of the medium-heavy cosmic-ray nuclei occurs at the
top of the atmosphere. Thus the Centauro fireball  must possess
features allowing it to
reach the mountain-top altitudes ($\sim$ 500 g/cm$^{2}$) without
interaction with air nuclei  or spontaneous decay
during its flightpath of about 15 km. This requirement creates the
following questions:\vspace*{-0.2cm}
\begin{itemize}
\item  what is the magnitude of the interaction length $\lambda$ and
 the lifetime
$\tau$ of
   the Centauro fireball, resulting from the existing experimental data?
\vspace*{-0.2cm}
\item
    do they  not contradict  the requirement of  traversing
   of about 15 km before the fireball decays? \vspace*{-0.2cm}
\item  what are the theoretical reasons which could allow  such
  long flightpath?\vspace*{-0.2cm}
\end{itemize}

These questions have been  considered in \cite{8} and the
stability question has been studied separately 
in \cite{Panagiotou_Note}.
 
\vspace*{2mm}

{\centerline {\em Experimental constraints on $\lambda$ or $\tau$}}
\vspace*{2mm}

An  estimate of the value of the interaction length of the Centauro
fireball can be obtained from the comparison of the observed rate  of
 Centauro events, detected
 at the mountain top level,  after Centauro fireballs decaying,
 with the results of experiments
looking for heavy penetrating particles, assuming that Centauro
fireballs before their decays may be identified with such objects.
 On the one hand,
it seems that the large flux of Centauro events detected in Chacaltaya and
Pamir emulsion chambers 
  suggests a long interaction length of the Cenaturo fireball, allowing
it
to reach the mountain top level. On the other hand, the negative
results
of the experiments looking for heavy penetrating particles
at White Mountains and Mt Norikura by means of
CR--39 films
could be explained, by assuming a short interaction length of Centauro
fireball,
allowing for its decay before reaching the mountain top altitude
  \cite{lit_Kinoshita}.
However, the careful analysis of that seeming
contradiction  \cite{8} shows that the
 interaction length of the Centauro fireball,  $\lambda_{Cnt}$
larger than
150 g/cm$^{2}$, is in accordance with the  results of both kind
of experiments.

The long interaction length of Centauro fireball could be connected
with the very high binding energy
 and the reduced geometrical
cross section (the estimated volume of the fireball \cite{8} is more than
six times smaller than that of the ordinary nucleus with A=75).
If the Centauro fireball is an unstable state spontaneously decaying with
the
probability P(x) after travelling a distance x $\approx$ 15 km, then the
lifetime $\tau_{0}$, in its own frame,
 assuming $ P(x) \sim 10\%-50\%$, 
 is  obtained to be
 of the order of 10$^{-9}$ s.
This means that the fireball could  decay via weak interactions,
since strong decays occur within $\sim 10^{-23}-10^{-24}$~s. Even so, the
lifetime of
$10^{-9}$ s is unrealistically long for a nonstrange, normal
quark matter state.

\vspace*{2mm}

{\centerline {\em Theorethical justifications of the  long flightpath,
i.e. stability
question}}
\vspace*{2mm}

It was estimated in  \cite{8} that the interaction length of the Centauro
fireball in
the atmosphere
 $\lambda_{Cnt} \ge 150$ g/cm$^{2}$ is 
in accordance with different experimental observations. This is
an extremely
long
interaction length.  It is about 1.6 times larger
than that of a nucleon and about 35 times larger than that of a normal
nucleus of a comparable mass number.
The long lifetime and small cross section 
  could be attributed to its
very high binding energy and much reduced geometrical volume. The very
high baryochemical potential and the transformation of the intial quark
matter state into a partially-strange quark matter one, with
increased spatial concentration of quarks due to the extra fermion
flavour, result in an increase of the binding energy. Extremely large
density and binding energy and the small volume may result in the
formation of a (meta)stable state.
However,
the
problem is opened and
according to our present knowledge rather far from satisfactory
and quantitative
solving.
The main questions are:
\vspace*{2mm}

1. {\em Radius (and volume) question}.\vspace*{-0.2cm}
\begin{itemize}
\item The long penetration length of the Centauro fireball through
      the atmosphere
 suggests the small
       geometrical cross section of that object, i.e. very small
       radius. Comparing the values of 
 the
      average interaction mean free path
      of nucleons in the air with that for the Centauro fireball one can
expect that 
  $R_{Cnt} \sim $ 0.9 fm and  $r_{0}^{Cnt}
\sim$ 0.22 fm  (for A = 75).
 Thus, 
the
Centauro
fireball radius could be about 5.5 times smaller than that of 
  a normal nucleus of corresponding A.\vspace*{-0.2cm}
\item Another estimate of the Centauro fireball radius can  be done
      by using its volume value, $V_{Cnt} \simeq 75$ fm$^{3}$ as obtained 
     in \cite{Panagiotou}. Assuming the spherical shape of the
fireball one gets $R_{Cnt} = 2.6 $ fm and $r_{0}^{Cnt} = 0.62$ fm.
       These values seem to be overestimated in comparison  with
the previous ones.
\vspace*{-0.2cm}
\item Another estimate taken from  \cite{Panagiotou_Note} gives
      $R_{Cnt}=1.86 \pm 0.36$ fm and $r_{0}^{Cnt}$ = 0.44 $\pm$ 0.08 fm.
\vspace*{-0.2cm}
\end{itemize}
 Different estimates locate
the Centauro fireball radius in the wide  range:
$R_{Cnt} \sim 0.9-2.6$ fm and $r_{0}^{Cnt} \sim 0.22 -0.62 $
fm.\vspace*{2mm}

 2. {\em Stability curves}.

\begin{figure}[ht]
\begin{center}
\mbox{}
\epsfig{file=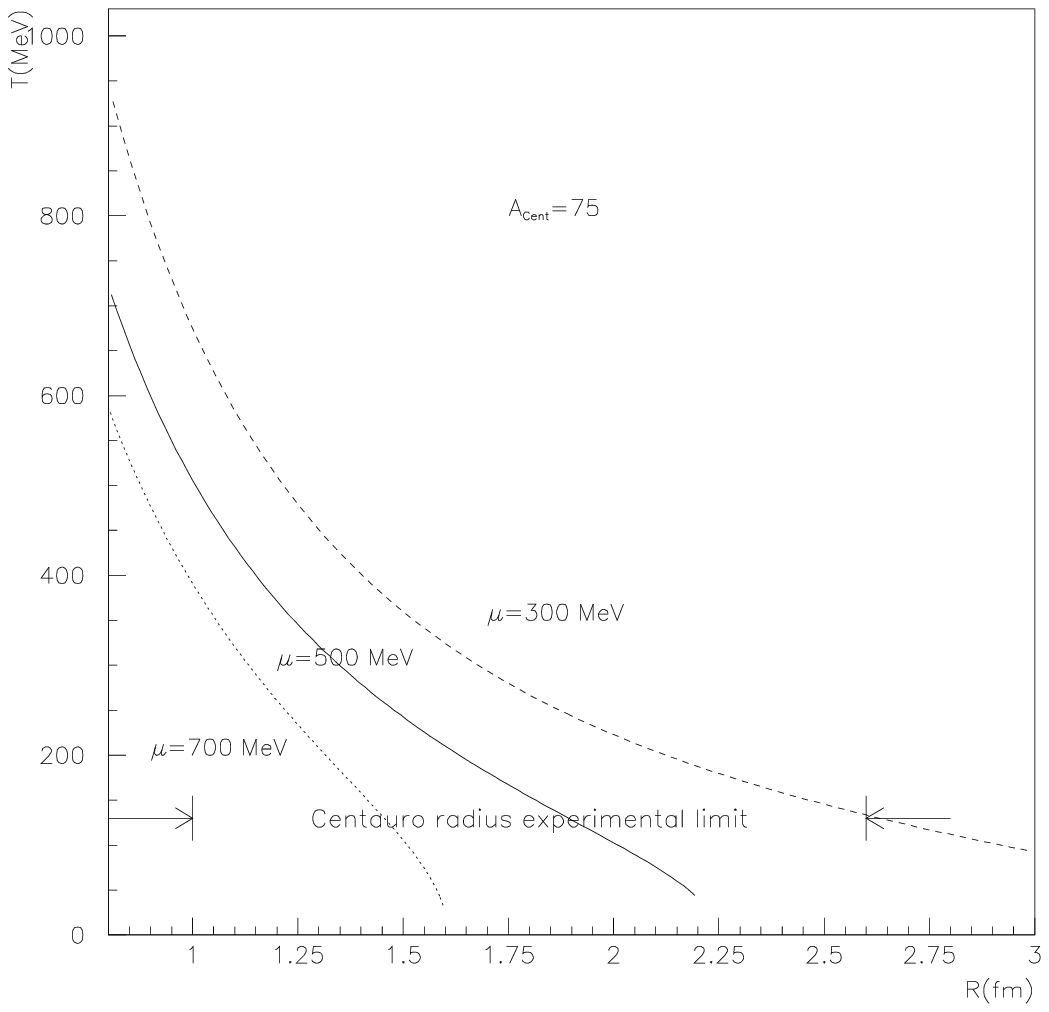,
bbllx=3,bblly=171,bburx=588,bbury=684,width=9cm}
\caption{ Stability curves plotted as T versus R for fixed
$\mu_{q}$ = 300, 500, 700 MeV for a strangelet with A=75.}
\label{r_t}
\end{center}
\end{figure}

The main question is if  extreme conditions,
such as a very high
quarkchemical potential and 
 reduced volume, can cause the (meta)
stability of the object.  A
general condition for DQM (Deconfined Quark
Matter) bag stability comes from the equalization of the internal (quark-
gluon) and external  (bag) pressure. The formulas for
thermodynamical quantities  of strangelets 
  being in
mechanical equilibrium can be found for example in
\cite{Madsen_rev}. They have been derived assuming the  massless
$u,d$ and $s$-quarks of equal
chemical
potentials $\mu$,  and $\alpha_{s}$ = 0.
Strangelets are in mechanical equilibrium at fixed temperature $T$ and
baryon number $A$ when
\begin{equation}
BV=(\frac{19\pi^{2}}{36}T^{4}+\frac{3}{2}\mu^{2}T^{2}+\frac{3}{4\pi^{2}}
\mu^{4})V-(\frac{41}{216}T^{2}+\frac{1}{8\pi^{2}}\mu^{2})C
\end{equation}
where extrinsic curvature $C=\oint(\frac{1}{R_{1}}+\frac{1}{R_{2}})dS$
( $C = 8\pi R$ for a sphere), and V and B  are  the volume and bag model
value
constant respectively.
 \\
In this case
\begin{equation}
A=(\mu T^{2} +\frac{1}{\pi^{2}}\mu^{3})V - \frac{\mu}{4\pi^{2}}C
\end{equation}
The stability curves obtained from this formula 
are
shown in Fig.~\ref{r_t}.

 There is plotted the temperature T as a
function of the radius R for three values of quark chemical potential
($\mu$ = 300, 500, 700 MeV) and baryon number A=75. There is drawn also
the experimentally estimated Centauro fireball radius limit. It is 
 well inside the ``stability region'' for the
Centauro
fireball with parameters close to experimentally obtained 
values, i.e.
T=130 MeV and $\mu$ = 600 MeV.\\
According to the model \cite{Panagiotou,8} the Centauro fireball is
 supposed to be,
however,  only the slightly strange quark matter glob.
 It has been concluded in ref. \cite{Panagiotou_Note} that
 Centauro fireball quantities,
such as T = 130 MeV, $\mu_{q}$ = 600 MeV and 
 $R_{Cnt} = 1.86$ $\pm$ 0.36 fm are within one standard deviation from the
stability conditions, thus it might be possible that the Centauro fireball
is a metastable state.\\
In the context of discussing of  Centauro fireball stability it
 should be stressed that there are  also some predictions concerning the
possibility
of the existence of stable collapsed nuclei \cite{collapse}. A collapsed
nucleus is a highly dense nuclear matter state, the nucleon density being
more than 20 times larger than that corresponding to a normal nucleus,
with binding energy per nucleon of hundred MeV and even up to
the nucleon mass (for T=0). In this condition, a collapsed nucleus is
a cold deconfined quark matter state, with the radius $r_{0}$ estimated
to be 0.4 fm, which  is close to that of  the Centauro
fireball.
It should be mentioned, however, that these attempts need about two times
larger bag pressure than the value required for stability of a strangelet
at T = 0. For instance,
 B$^{1/4} \sim$ 375 MeV is needed for the
stabilization of the quark-gluon plasma bag with T=130 MeV, $\mu_{q}$ =
600
MeV, A=75 and R= 1.62 fm \cite{Panagiotou_Note}.
\vspace*{3mm}

{\centerline {\bf Strangelets from the Centauro-fireball}}
\vspace*{3mm}

 At the last stage of the Centauro
fireball evolution, due to the strangeness distillation mechanism 
 a strangelet can be formed. The characteristic features of such
quark--matter nuggets 
can be estimated  by using the above mentioned quark--gluon
plasma
model describing the experimental Centauro characteristics. On the other
hand they
 can be derived independently from the analysis  of the
strongly penetrating cascades, assuming that they are due to
a strangelet passage through the apparatus.

The excess of  strange quarks in the Centauro fireball is mainly
caused by the pre--equilibrium emission of the $K^{+}$ and $K^{0}$ mesons
which carry $\overline{s}$ quarks away. One can assume that the total
number of
$s\overline{s}$ pairs created in the Centauro fireball volume V
 approximately equals to the number of emitted $K^{+}$, $K^{0}$ mesons
and it can be estimated from the formula:
\begin{equation}
  N_{K^{+}+K^{0}} \simeq N_{s\overline s} = \rho_{\overline s}V
\end{equation}
where $\rho_{\overline s}$ is $s$--antiquark density and depends on the
temperature by the relation \cite{Biro}:
\begin{equation}
\rho_{\overline s} \simeq 0.178(\frac{T}{200  MeV})K_{2}(\frac{150
  MeV}{T})
\end{equation}
For T=130 MeV which is the temperature for cosmic-ray
Centauros,
 $\rho_{\overline{s}}
\simeq 0.14$ fm$^{-3}$.
 Taking a volume of the Centauro fireball V $\sim$ O(100 fm$^{3}$)
the number of created $s\overline{s}$ pairs (and thus emitted kaons)
 will be $\sim$ 14.
After the pre--equilibrium emission of kaons,
 a strange quark matter metastable object with a small Z/A ratio
  is formed. The change of Z/A ratio, because of kaon emission
 can be calculated from the formula
\begin{equation}
  (\frac{Z}{A})_{f}=(\frac{Z}{A})_{i}-\frac{\Delta Q}{A}
\end{equation}
where the net change of the charge
\begin{equation}
 \Delta Q = N_{s\overline s}((Q_{u}+Q_{d})/2+Q_{\overline s}).
\end{equation}
Assuming the quark charges $Q_{u} = 2/3, Q_{d} = Q_{s} = -1/3$,
 $(Z/A)_{i} \simeq 0.5$ and the temperature T
$\sim $ 130 MeV it was evaluated $(Z/A)_{f} \simeq 0.4$.
In the extreme case when all produced strange quarks became constituents
of the strangelet and 
$N_{s}/A_{str} \simeq 1$,  
the final explosion of the Centauro fireball into non-strange baryons and 
accompanied  {\it emission of a small
strangelet
characterized by $A_{str} \sim 10-16$} can be expected. Its very high
strangeness-to-baryon ratio implies also very
low  charge-to-baryon ratio (Z/A $\sim$ 0).
\vspace*{3mm}

{\centerline {\bf  Strongly penetrating cascades as signs of
strangelets}}
\vspace*{3mm}

The  hypothesis, that the strongly penetrating cascades can be produced
by a strangelets,
  announced in \cite{8},
was developed in \cite{9} where
 the  possible connection between the very
penetrating component,
frequently accompanying the cosmic ray exotic phenomena, and the
hypothesis of
the formation of strangelets in the process of strangeness distillation
was studied.
Several possible decay \cite{Witten,Berger,Chin,Greiner,Madsen_Kreta} and
interaction \cite{Wlodarczyk_str}
modes of a small strangelet were considered and  its
travelling through the  
homogenous--type thick lead chamber was simulated.

\begin{large}
{\centerline {\em Unstable strangelets}}
\end{large}

 Objects  decaying  via strong interactions
in the timescales typical
for strong processes ($ \tau_{0} \stackrel{<}{\sim} 10^{-20} s$) were
called unstable strangelets. Although 
a
variety of such processes is expected \cite{Witten,Berger,Chin},
the most important one is a {\it strong neutron emission} $S \rightarrow
S' + n$
in which a strangelet $S$ emits a neutron, yielding a daughter strangelet
$S'$ with
parameters changed by $\Delta A = -1, \Delta Y = -1, \Delta Z = 0$.

\begin{figure}[ht]
\begin{minipage}{7cm}
\hbox{
\epsfxsize=180pt
\epsfbox[20 19 575 364]{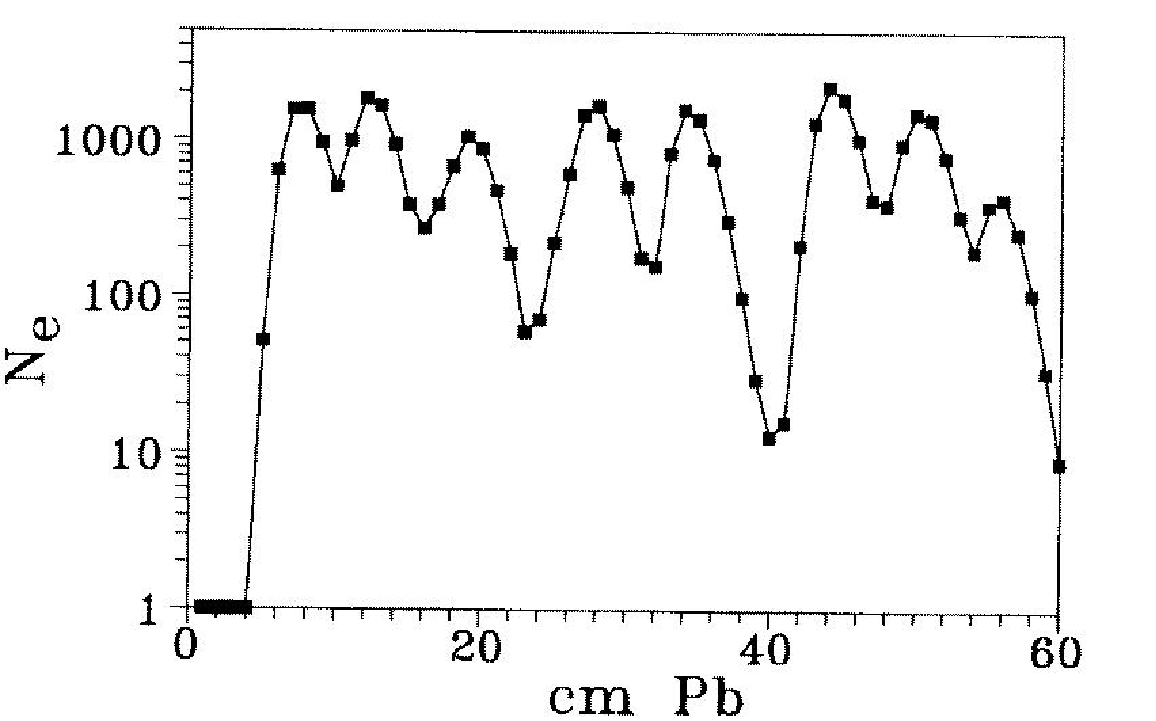}}
 \begin{center} {\scriptsize Unstable
 strangelet decaying into a bundle
 \vspace*{-1mm} of 7 neutrons
 ($E_{n} \simeq  E_{str}/A_{str} \simeq
$
 200 TeV)}.
\end{center}
\end{minipage}
\begin{minipage}{7cm}
\vspace*{-4mm}
\hbox{
\epsfxsize=180pt
\epsfbox[20 21 573 381]{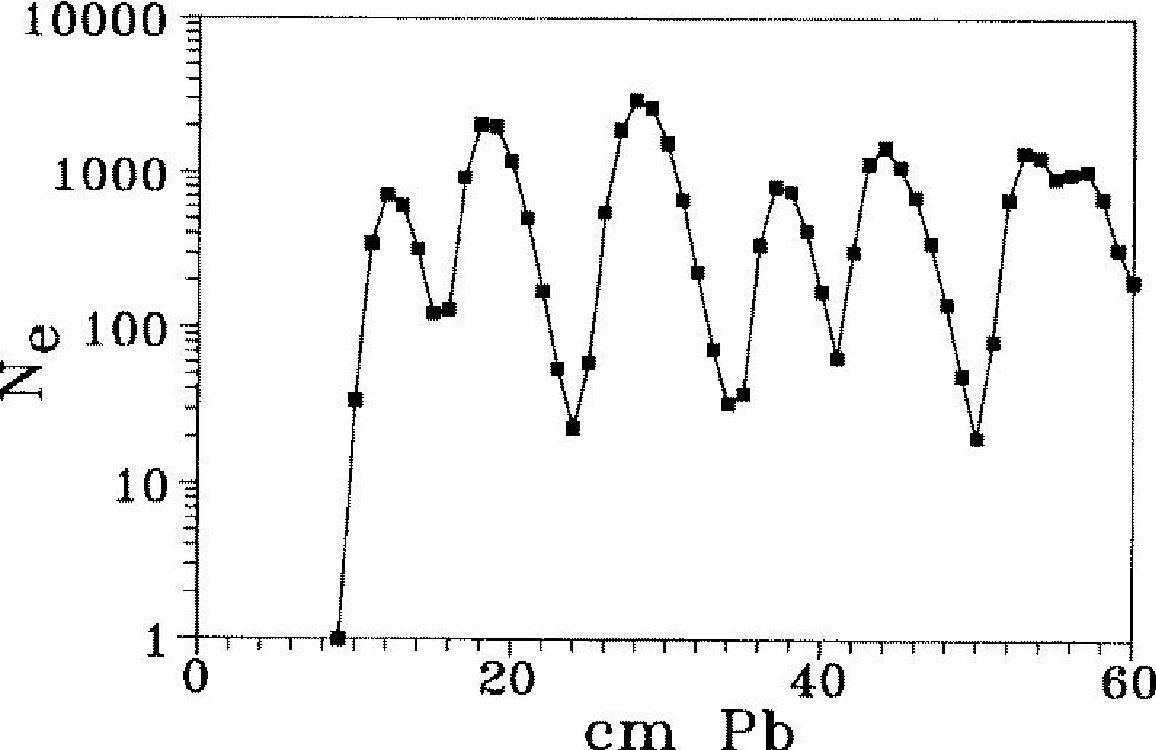}}
\vspace*{-1mm}
\begin{center} {\scriptsize Metastable strangelet\\\vspace*{-1mm}
  $(A_{str}$=15, $E_{str}=200$
 A TeV, $\tau \sim 10^{-15}$ s).}
\end{center}
\end{minipage}
\begin{minipage}{7cm}
\vspace*{3mm}
\hbox{
\vspace*{4mm}
\epsfxsize=180pt
\epsfbox[20 21 576 376]{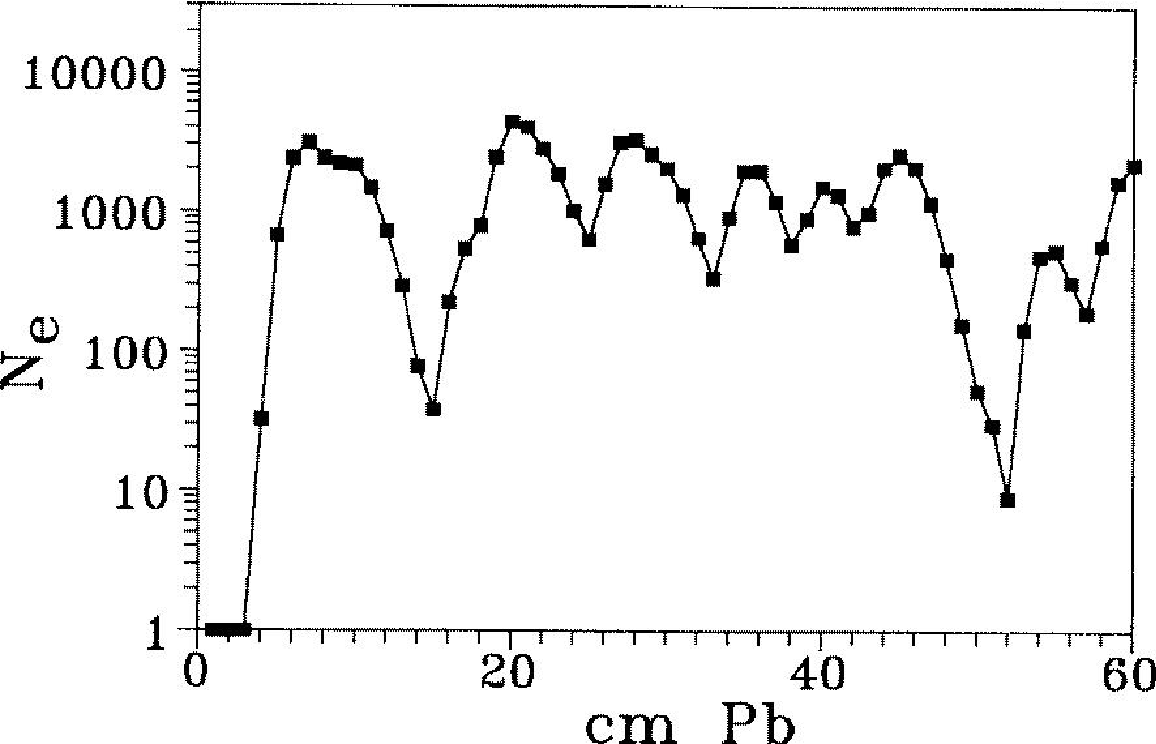}}
\begin{center}
\vspace*{-2mm}
\scriptsize{
 Long-lived strangelet\\
 ( $ A_{str} = 15$, $\mu_{q} =$ 600 MeV).}
\end{center}
\end{minipage}
\hspace*{1cm}
\begin{minipage}{6cm}
\caption[{ Examples of simulated transition curves recorded
in the lead chamber and produced by various strangelets.}] 
{Examples of simulated transition curves recorded
in the lead chamber and produced by various strangelets. Numbers of
electrons $N_{e}$ are counted within the
radius of
 50 $\mu$m.}
\label{stable_str}
\end{minipage}
\end{figure}

Unstable strangelets decay very fast, practically at the point  of their
 formation.
 For strangelets with $\tau_{0} \sim10^{-20}s$ 
and a Lorentz factor $\gamma \sim 10^{4} - 10^{5}$ the  estimated length
of the 
decay
path will be  of the order of $\sim 10^{-5} - 10^{-6}$ cm, what is very
small in comparison with the estimated heights of the Centauro fireball
explosion (H $\sim$ 100 - 1000 m above the chamber). Thus the considered
picture resolves into a simple case of a bundle of nucleons entering the
chamber. If all nucleons are evaporated with  
momenta close one to the other, a bundle of strongly collimated nucleons
is obtained.
If the relative transverse momenta of neutrons 
 are of the same order as those observed
in mini-clusters, i.e. $p_{T}(\gamma) \sim 10$ MeV/c, then
 the average lateral distances between
neutrons, at the chamber level, are estimated to be of the order $\sim$ 100 - 
1000 $\mu m$.
Thus, depending on the formation height and the energy, a decaying
strangelet  will produce in the chamber  the 
picture resembling  a mini-cluster or one single long-range
cascade (if 
the relative distances between the decay products 
are smaller then the lateral resolution of the used 
detectors).
 The development of nuclear-electromagnetic cascades, caused
by such neutron bundles, in the typical homogenous lead chamber with detection 
layers placed every 1 cm of the lead absorber was simulated.
It has been found that the bundle of several ($\sim$7) neutrons
generates the 
many--maxima long-range cascades, very similar to the  long-flying
component
observed in cosmic-ray experiments.
A typical simulated event is shown in Fig.~\ref{stable_str}.

The impression of the strong similarity between the simulated and observed
transition curves has been confirmed by investigation of the relative distances
and the distribution of  energy contained in the successive humps
\cite{9}.

\vspace{3mm}
\begin{large}
{\centerline {\em Metastable strangelets}}
\end{large}

Metastable strangelets are commonly assumed to decay with a lifetime 
$\tau_{0} \stackrel{<}{\sim} 10^{-4}$s. 
They decay via weak interaction processes, of which the most important is
{\it the weak neutron decay}. As it involves a weak interaction
flavour-changing
process, $s + u \leftrightarrow u + d$, it should be much slower than a strong
neutron decay. Generally, however,  lifetimes of small metastable strangelets
are not predicted precisely  at present and they are still a matter of
debate \cite{Greiner,Madsen_Kreta}. Thus, in principle, 
a wide range of possible lifetimes of metastable strangelets should be
taken into account.
 In the extreme situation, i.e. sufficiently long(short) strangelet
lifetime the problem simply reduces to the stable(unstable)
 strangelet scenario.
 In the
 intermediate case a strangelet penetrating through the chamber can 
successively evaporate neutrons. Neutrons interact with nuclei of the lead
absorber producing the many-maxima cascades in which   the  distances
between the successive humps are strongly correlated with the time interval
 between
the successive evaporation acts. Assumption of a strangelet lifetime
$\tau_{o} \sim 10^{-15}$s  leads to the formation of the long-range
many-maxima cascades similar to these observed in the experiment. Example
of the transition curve, obtained under the assumption of the
evaporation of neutrons from a metastable strangelet with $A_{str}$ = 15,
 passing through the
lead chamber,  is shown in
Fig.~\ref{stable_str}.

This hypothesis involves, however, some hardly acceptable points:
\begin{itemize}\vspace*{-2mm}
\item Required lifetime $\tau_{0} \sim 10^{-15}s$  seems to be too short 
for the weak neutron decay process (much shorter than for example the 
lifetimes of both weak semileptonic and weak nonleptonic decays, estimated 
in \cite{Crawford} on the basis of the Berger - Jaffe mass formula
\cite{Berger} 
for strangelets of $A_{str} = 15$).\vspace*{-0.2cm}
\item The picture assumes that a strangelet decays successively inside the 
chamber. Thus, it concerns  objects produced at a very small distance from 
the chamber or just inside it. In the case of strangelets produced in 
typical cosmic-ray families (at  heights $H \sim 100 - 1000$ m above the chamber) it
 seems unclear why they do not decay in the air layer, above the
apparatus.\vspace*{-0.2cm}
\end{itemize}

\begin{large}
{\centerline {\em Stable strangelets}}
\end{large}

The weak radiative decays $(u + d \longleftrightarrow s + u + \gamma)$ and weak
leptonic decays $(d \longleftrightarrow u + e^{-} + \overline{\nu}_{e}$,
$s \longleftrightarrow u + e^{-} + \overline{\nu}_{e}$) are expected to be 
slower than a weak neutron emission. The rate of the radiative decays is
inhibited
by the electromagnetic coupling constant and all $\beta$-decays by the
 three-body
phase space. Additionally, the strangeness changing $\beta$ process is
suppressed by the Cabibbo factor.
 Such long-lived objects,
called ``stable'' strangelets, if
produced in the 
high energy cosmic-ray families should pass without decay through the apparatus.
 The simplified picture of their possible interaction in the
chamber matter was assumed.

 The strangelet is considered as an object with the radius
\[R = r_{0} A^{1/3}_{str}\]
where the rescaled radius
\begin{equation}
r_{0} = ({\frac{3 \pi}{2(1 - \frac{2 \alpha_{c}}{\pi})[\mu^{3} + 
(\mu^{2} - m^{2})^{3/2}]}})^{1/3}
\end{equation}
$\mu$ and $m$ are the chemical potential and the mass of the strange quark 
respectively and $\alpha_{c}$ is the QCD coupling constant.

 The mean 
interaction path of strangelets in the lead absorber
\begin{equation}
\lambda_{s-Pb} = \frac{A_{Pb} \cdot m_{N}}{\pi(1.12 A^{1/3}_{Pb} + 
 r_{0}A^{1/3}_{str})^{2}}
\end{equation}
Penetrating through the chamber a strangelet collides with lead nuclei. In
each act of  collision, the "spectator" part of a strangelet
survives 
 continuing the passage through the chamber and the "wounded"
part is destroyed.
 Particles,
generated at the 
consecutive collision points, interact  with lead nuclei in usual way,
resulting in the electromagnetic-nuclear cascade, developing in the 
chamber matter.

 The penetration of a strangelet with mass number 
$A_{str} = 15$ through the chamber, assuming two different values of the 
chemical potential: $\mu$ = 300 MeV and 600 MeV 
 was simulated.
In both cases, many maxima cascades can be produced.  
Example of the 
transition curve obtained in the process of successive interactions of
a strangelet in the chamber matter is shown in Fig.~\ref{stable_str}.
\vspace*{2mm}

 The calculations done in \cite{9} showed that the long-range cascades
observed in
the thick
homogenous lead/emulsion chambers could be the result of a
 strangelet
penetration  through the apparatus. Their strong penetrating power can be
connected both with the small interaction cross section (in comparison
with nucleus of comparable A) of the strangelet
and with the big concentration of its energy in a narrow region of phase
space. This energy could be liberated into conventional particle
production in many consecutive evaporation or interaction acts.
In this context,
 the Centauro-like event with abnormally long cascades found
among its
secondaries,
described in \cite{Cen_Krakow,Cedzyna}, could be regarded as an event, in
which
strangelets were generated as 
remnants of the strange quark matter fireball.
It seems that also numerous hadron-rich families accompanied by highly
penetrating cascades, clusters or halo  could be explained by
assuming the
same mechanism of the formation of a strange quark matter fireball and its
successive decay into predominantly baryons and strangelet(s).
In principle, three
possible pictures of a strangelet passage through the chamber,
 which give the
long-range many-maxima cascades similar to these observed in the cosmic-ray
experiments, exist. 
 The most plausible are scenarios assuming a penetration of 
 a  short-lived (unstable or metastable
ones) or 
 a long-lived (``stable'')  strangelet.
 However, to distinguish between these two scenarios 
  a more 
precise detector and  higher event statistics are necessary.  
\section{Search for Centauros in accelerator experiments}
Centauros and/or related phenomena have attracted a lot of attention
since the early 80's. Several accelerator experiments have been
already performed to search for such unusual events.
 Also
some 
current and future experiments  are going to look for them.
\subsection{Past experiments at CERN SPS collider}
The first accelerator  searches
for the Centauro phenomenon have been
performed many years ago (in 1981-1986) at the SPS CERN
$p\overline{p}$ collider 
by the {\bf UA1} experiment \cite{UA1} (at energy $\sqrt{s}$ =630 GeV) and
by the {\bf UA5} Collaboration \cite{UA5}
 (at energy
$\sqrt{s}$ = 546 and 900 GeV).  Both experiments looked for Centauros
in the central rapidity region (UA1 at $|\eta|\leq 3$
region and UA5 in
 wider pseudorapidity interval). The UA5 detector consisted of two
streamer
chambers,
 placed on opposite sides of the beam pipe. Charged particles left there
tracks
which were photographed for analysis. Photons were detected through
conversions in a lead-glass plate, placed between the beam pipe and the
upper streamer chamber.
 In the last search the UA5 detector 
was significantly improved. The geometrical acceptance of the large
streamer chambers was about 95\% for $|\eta| \leq 3$ , falling to zero at
$|\eta|$ = 5.
 The lead-iron-scintillator calorimeter was
situated at $90^{0}$ and covered the interval $|\eta| \leq$ 0.9 and
 azimuthal angle $\Delta \phi \simeq 30^{0}$. In the latest analysis upper
limits of
0.1-0.5\% at 95\% c.l. have been placed on Centauro production. No
indication of Centauro production
was observed.

 The negative results of UA1 and UA5 experiments have  several possible
alternative 
explanations:\vspace*{-0.2cm}
\begin{itemize}
\item The  energy threshold for the  production of such objects is
higher
      than $\sqrt{s}$ = 900 GeV, what corresponds to  the incident energy
      in the laboratory frame $E_{lab}$ higher than 431 TeV,   when
assuming a nucleon projectile. This explanation seems to be  especially 
      convincing, as 
       the average incident 
      energy of cosmic--ray Centauros has been estimated to be $\sim 1740
$ TeV.
      The separate question  concerns Mini-Centauro species which, in
 fact, have
been
observed
      in cosmic ray experiments in a lower energy region. In this case the
reason of their nonobservation by UA1 and UA5 experiments may be
      the existence of
      ``genetic relations'' between different exotic phenomena
      (see e.g. \cite{C-jets}).
\vspace*{-0.2cm}
\item Centauros are produced in the projectile fragmentation 
region, thus
      the above mentioned experiments, limited to the central region, were
unable to find them.  
\vspace*{-0.2cm}
\item  Centauros can not be born in  nucleon-nucleon
collisions. This would support  the hypothesis of the formation of
Centauros in
nucleus-
 nucleus collision \cite{Panagiotou} or its exotic (extra-galactic) origin
\cite{Bjorken_Cen}.
\vspace*{-0.2cm}
\end{itemize}
\subsection{Current experiments at Tevatron and CERN SPS}

\begin{large}
\begin{center}
{\em  Experiments at Fermilab ($p\overline{p}$ Tevatron
      at $\sqrt{s}$ = 1.8 TeV)}
\end{center}
\end{large}

 At Tevatron energies the situation is
delicate 
      because the expected energy threshold for Centauro production
is roughly
      consistent with $p-\overline{p}$ collider  energy,
 when assuming the Centauro formation in nucleon--nucleon collisions 
 ($\sqrt{s}=1.8$
TeV is equivalent to about
 1.7$\times 10^{14}$ eV in laboratory frame for  nucleon-nucleon 
collisions). However, if
Centauros were produced in
       nucleus-nucleus collisions the situation will be  different, as
 the
total interaction energy
       of the average cosmic-ray Centauro in ``60 + 14''
c.m. system is $\sqrt{s}\simeq$
 6.8 TeV.

The other question, mentioned already in the case of CERN SPS
experiments, is
the different kinematical range for cosmic ray experiments and
accelerator
ones. 
Cosmic ray experiments are primarily sensitive to energy flow, and
generally detect particles from the fragmentation rapidity region
whereas the accelerator studies are mainly  focused on the central region
of
phase space in the c.m. system \cite{Jones}.

      In some Fermilab experiments the Centauro phenomenon is
(or is going
to be) searched
       in nucleon-nucleon  collisions. These
are:

 {\bf Mini-Max (T-864)}

 This 
 is a small experiment situated at the C0
interaction region of the Tevatron, with the primary goal
      to search for DCC and possibly related exotic phenomena such as
Centauro, in the forward region. It has been designed to
measure the ratio of
      charged to neutral pions produced at $\eta\simeq$ 4.1.
      The detector was designed as a telescope of multi-wire
      proportional chambers (MWPC's) together with scintillation
 counters  along the
beam pipe, and 
      the lead converter 
 inside the
telescope. Charged particles can 
      be observed in the chambers before and after the converter, and
photon conversions in the chamber behind the converter. The
electromagnetic calorimeter is placed behind the MWPC telescope.
      Mini-Max is able to observe both charged particles and photons
      in the region $3.4 \lt \eta \lt 4.2$ and it is sensitive to
low-p$_{T}$
      particles. The distribution of photons to charged particles ratio is
expected to be
different for the generic binomial and DCC particle production models
(see subsection 4.3).
      The main problem of the experiment is the background from the
accelerator beam pipe and the small acceptance
      for both $\gamma$'s from $\pi^{0}$ (acceptance covers a circle of
radius $\sim 0.75$ units in 
       $(\eta,\phi)$  at $\eta = 4.1$).  The
combined ratios of the
factorial
moments, which result in experimentally robust variables ( independent
of acceptance and various efficiency factors), were used in analysis.
      A comparison of ratios of measured robust variables
      and those expected from various models showed that data
      are consistent with the generic production
mechanism\cite{DCC_robust,Mini-Max_results}. 
 Limits on DCC production in various models
      are  claimed to be at the $\sim$ 5-20\%
      level.

The C0 group has also designed the Zero Degree proposal for studying far
forward physics. Unfortunately this proposal was not approved.

 {\bf CDF}

Also this experiment \cite{CDF} reports preliminarily a negative search
for
Centauros 
      in the central rapidity region. According to
ref. \cite{Jones_Delphi}
 the CDF Collaboration has put an upper limit of 10 microbarns on the
 production of Centauros in $1.3 \lt \eta \lt 4.1$.
 Within each event particles were
      detected and identified as either hadronic requiring $E_{T} \gt$ 0.4
      GeV, or electromagnetic requiring $E_{T} \le$ 0.2 GeV, using
calorimeter towers out to $|\eta| \lt$  4.2.
The search for Centauro-like
      events was based primarily on their unique particle kinematics:
      particle multiplicities $N \sim$ 75, $\langle p_{T} \rangle \sim$
1.7 GeV/c
      and  $\langle \eta \rangle $ of hadrons centered around 2.2 with
$\sigma_{\eta}
      \sim 0.7$ accordingly to phenomenological interpretation of
Centauros as diffractive fireballs \cite{Navia,Goulianos}. In
addition,
looking
for Centauro candidates with
      unusual hadronic to electromagnetic asymmetry, as predicted by
      DCC hypothesis, was
done. This  analysis, however,  does not consider the possibility that
Centauros
      decay to protons and neutrons what can be the additional
reason of the negative result.

{\bf D0 Detector}

The main purpose of D0 detector \cite{D0}  is the study of high mass
states,
large
$p_{T}$ phenomena and  the
rapidity gap fraction. It basically consists of three elements: the
central detector, the liquid argon-uranium calorimeter and the outer muon
detector. It has charged particle tracking capability up to 3.2 units of 
pseudorapidity.
The uranium-liquid argon calorimeters  have full coverage
for the psudorapidity range of $|\eta|$ \lt 4.1. The calorimeters are
azimuthally symmetric and have electromagnetic and hadronic resolution of
15\%/$\sqrt{E}$ and 50\%/$\sqrt{E}$ respectively. The transverse
segmentation of the  calorimeter towers is typically $\Delta\eta
\times\Delta\phi=0.1\times0.1$.
Thus the detector is well suited to search for Centauro type  events.
The strategy of the search for Centauro events in the D0 detector depends
on the interpretation of the event. Assuming the 
 isotropic decay of a fireball with
mass
$\sim $ 180 GeV/c into baryons, a large deposit in the hadron calorimeter
with no energy in the EM calorimeter is expected. Monte Carlo study shows
that the detector is suitable for Centauro phenomenon investigation.
\begin{large}
\begin{center}
 {\em Fixed target experiments at  CERN SPS} 
\end{center}
\end{large}

Some present CERN SPS experiments searched for Centauro--like phenomena
in
heavy ion
collisions at energy 158 A GeV. Among them:

 {\bf  WA98}

The WA98 \cite{WA98} experiment emphasizes on high precision, simultaneous
measurement
of both hadrons and photons. The experimental setup consists of large
acceptance  hadron and photon spectrometers,
detectors for charged particle
and photon multiplicity measurements, and calorimeters for transverse
and forward energy measurements. Among these detectors there are two
which are
well suited for the DCC search. One can measure charged particle
multiplicities with a Silicon Pad Multiplicity Detector 
 ($2.35 \lt \eta \lt 3.75)$ 
and photons
with a Photon Multiplicity Detector 
 covering $3.0 \lt \eta \lt 4.2$.
WA98  uses various methods of DCC analysis, such  as:\vspace*{-0.2cm}
\begin{itemize}
\item Global event characteristics, i.e. the total number of photons
      and charged particles over the entire phase space covered by the
photon and charged particle detectors;\vspace*{-0.2cm}
\item Methods for DCC domains, when the available
phase space is divided into several $\eta - \phi$ bins. Among them
the wavelet analysis (multiresolution scanning the entire phase space, no
averaging over events or $\eta-\phi$ space),
various moments and their combinations, i.e. ``robust observables''
were  calculated from the 
distribution of photons and charged particles in each bin 
 \cite{WA98}.
\vspace*{-0.2cm} \end{itemize}
The measured results were compared to those
from simulations and to those from various types of
mixed events to isolate the source of non-statistical
fluctuations \cite{WA98_new}. The comparison indicates the presence
of non-statistical fluctuations in both charged particle and photon
multiplicities in limited azimuthal regions, however, no correlated
charged-neutral fluctuations are observed \cite{WA98_new}.
To date they have observed no events with a large charge to neutral
fluctuation from among 200K events, and reported no
significant DCC
signal in Pb+Pb collisions. Within the context of simple DCC model,
upper limits on the presence of localized non-statistical DCC-like
fluctuations of the order $\sim 10^{-2}-10^{-3}$ were extracted.

{\bf NA49}

The experiment NA49 \cite{NA49_app} is a large acceptance hadron detector
at the
CERN SPS. Four large volume time projection chambers (TPC) record
the trajectories of particles. Two of them
(VTPC1,2) are placed inside two superconducting magnets and two are placed
further downstream and symmetrically on both sides of the beam line.
The VTPC2 and MTPC acceptances complement each other around mid--rapidity
and the combination of both detectors covers a major fraction of the
available phase--space.
The set-up also includes four time-of-flight (TOF) walls and a set
calorimeters for triggering and $E_{T}$ measurements. Among many topics
which are investigated in the experiment, also a statistically significant
determination of momentum space distributions and particle ratios
can be performed for single events, allowing for a study of event-by-event
fluctuations.   The first results of analysis of fluctuations in the
average transverse momentum of individually measured charged particles
from event to event, done in a region of $0.005 \lt p_{T} \lt 1.5$ GeV/c
and rapidity $4 \lt y_{\pi} \lt 5.5$ have been published in
\cite{NA49_DCC}. DCC models suggest that pions emitted from DCC domains
will be preferentially produced at low transverse momenta. This provides
for a translation of the number--fluctuations predicted by the DCC models
into $p_{T}$ fluctuations. The isospin fluctuations of pions
production from DCCs lead to multiplicity fluctuations of charged
pions at low transverse momenta and therefore to non-statistical
fluctuations in the distribution of the mean transverse momentum. Using
the model in which the DCC production
is characterized by the probability to form a single DCC domain in the
event and the fraction $\xi$ of pions coming from the DCC it was concluded 
 that assuming DCC's occuring in every event the
fluctuations observed in the data rule out DCC sizes of $\xi \gt 3.5\%$.
  
{\bf EMU16}

It is a Magnetic-Interferometric Emulsion Chamber where it is possible to
study
isospin fluctuations (DCC) at small y $\le$ 2.

\vspace*{5mm}

 Negative results of the Centauro search at collider energies (CERN
and Fermilab)  allows one  to suspect  that
 either the investigated energy region is below the threshold for
      such events formation, 
 either the methods of looking for Centauro events are not fully
      adequate because of some
       misunderstanding the 
experimental observation (different rapidity regions should be
explored, different types of produced particles should be looked for,
 different projectiles should be used, etc.).
       If  Centauro fireballs
      are created in nucleus--nucleus
       collisions, as it is suggested in \cite{Panagiotou,8}  
 then in the  future collider experiments, at RHIC and LHC,  the 
       appropriate conditions for formation of  these objects are 
expected.       
\subsection{Current and future experiments at RHIC and LHC}
Relativistic Heavy Ion Collider (RHIC) at Brookhaven has  started in June 
2000. During Run I the maximum energy was 130 GeV per nucleon pair.
 In the nearest
future the collisions 
 of relativistic heavy ions (i.e. Au-Au) with a center of mass
energy of 200 A GeV are expected to be investigated.
  Few years later
(in 2006) it is planned to start the Large Hadron Collider (LHC) at
CERN, 
 with the Pb+Pb beams, carrying the  center of mass energy of about  5.5
 A TeV.
Majority of the  experiments
 has a   wide program with the main purpose  to find the
quark--gluon plasma by looking at many different observables. Among
them
some exotic signatures related to Centauro phenomena are planned to be
used in the following  experiments  at RHIC:  

{\bf PHOBOS}

 PHOBOS  detector \cite{STAR} 
 consists of two parts. The first one is  
a silicon multiplicity detector covering
almost the entire pseudorapidity range of the produced particles, and   
measuring  total charged multiplicity only, $dN/d\eta$. The second one is 
a two-arm spectrometer at mid-rapidity.
The PHOBOS configuration is that it will give fairly subtle measurements
in the midrapidity (0 $ \lt \eta \lt $ 2) and low $p_{T}$ region. It
will be able to
look for DCC, produced in the central rapidity region, by signatures  
such as the anomalies (unusual fluctuations) in the pseudorapidity
distribution of
charged particles. (No direct comparison between the electromagnetic and
hadronic component will be possible.)

{\bf STAR}

 STAR \cite{STAR} (Solenoidal Tracker at RHIC)
 consists of high resolution tracking detectors, trigger detectors, and
partial coverage of electromagnetic calorimetry inside a
 0.5 T solenoid.
The measurements will be carried out at midrapidity, over a large
pseudorapidity range $(|\eta| \lt 4)$ with full azimuthal coverage
($|\Delta
(\phi)| = 2\pi$). The tracking detectors are a silicon vertex tracker
covering $|\eta| \lt 1.7$, and a forward radial-drift TPC
 covering
$ 2.5 \lt |\eta| \lt 4$. In addition to the tracking detectors, the
electromagnetic calorimeter
 will measure the transverse energy of
events, and trigger on and measure high transverse momentum photons.
 The STAR detector
system will simultaneously measure many experimental observables to study
signatures of the QGP phase transition as well as the space-time evolution
of the collision process.

 A highly granular photon multiplicity detector
is being planned for
the STAR, which in combination with charged particle detectors and
forward TPC,
will be quite adequate for DCC search, in event-by-event mode.
Here one would be able to select on
the low $p_{T}$ particles, characteristic of DCC pions, and to study
$\pi^{+-}$ spectra at low $p_{T}$.

{\bf PHENIX}

 The physics goals of PHENIX \cite{STAR,PHENIX} (Pioneering High Energy
Interaction
eXperiment) are to measure as many potential signatures of the QGP as
possible. PHENIX will measure lepton pairs, photons and hadrons, being
sensitive to very small cross section processes, as the production of the 
$J/\psi , \psi^{'}$ and high $p_{T}$ spectra. The PHENIX detector consists
of three spectrometers: two muon spectrometers covering the full azimuth
for $1.1 \lt |\eta|\lt 2.4$ and a central spectrometer consisting of two
arms
 each subtending $90^{0}$ in azimuth and with $|\eta| \lt 0.35$.
 DCC search in PHENIX will be possible by correlating signals from charged
particle detectors and photons from electromagnetic calorimeter. Global
event characterization is achieved via a  silicon
multiplicity-vertex
detector
covering $|\eta| < 2.5$ around  midrapidity.

 {\bf BRAHMS}

Small experiment \cite{STAR} with a forward and midrapidity hadron
spectrometer. The detectors used include TPC's, wire chambers, Cerenkov
counters, and a time--of--flight system. Their experimental emphasis is
particle identification over a broad rapidity range, covering nearly
12 units of rapidity, well beyond the other experiments capabilities.
This may allow to explore the more baryon dense matter at forward and
backward
rapidities.
 BRAHMS experiment is considering to add a photon arm to look for DCC.

\vspace*{3mm}

Among the experiments planned for LHC there was {\bf FELIX} 
  (Forward ELastic and Inelastic eXperiment) \cite{FELIX}. 
 It was a proposal for a full-acceptance detector and
experimental program for the LHC dedicated to study  QCD in all its
aspects, hard and soft, perturbative and non-perturbative, particularly to
investigate also exotic phenomena. Unfortunately it
was rejected.

The other one is  {\bf CASTOR} \cite{CASTOR_Delphi,CASTOR}.\\
It is a  dedicated detector for Centauro and strangelet search in the very
forward rapidity region, in nucleus-nucleus collisions. It will be
operated as the part of the ALICE experiment \cite{ALICE}. For more
details see
the next section.

Generally,  different strategies of looking for Centauro
related phenomena are proposed by different experiments, dependently on
the 
preferred model. The present and future experiments prefer mostly
the DCC mechanism, where the large imbalance in the production of charged
to neutral pions could be the result of the approximate restoration of the
chiral symmetry. Sophisticated analysis methods are being developed to 
disentagle DCC events out of the large background of events with
conventionally produced particles.

\section{CASTOR: a detector for Centauro And STrange Object Research}

 The main part of the ALICE \cite{ALICE}
heavy ion experiment at CERN LHC
 will be fully instrumented for hadron and photon
identification only in limited angular region around midrapidity,
covering the pseudorapidity interval $\mid\eta\mid \leq 1$.
It constitutes only a small part of the available phase space which, at
the  beam energy of 2.75 A TeV for Pb ions at LHC, extends to
$\mid\eta\mid$ = 8.7. Therefore, in
addition to the main detector system
some smaller detectors covering more forward rapidity region are also
foreseen. These are:\vspace*{-0.2cm}
\begin{itemize}
\item Muon detector \cite{muon_spec}  which will be installed on one side
and will cover the
pseudorapidity interval $2.5 \leq \eta \leq 4.0$;\vspace*{-0.2cm}
\item  A pre--shower photon
multiplicity detector \cite{PMD} which will also be installed on one side
and will
cover
the pseudorapidity interval $2.3 \leq \eta \leq 3.3$;\vspace*{-0.2cm}
\item  Small--aperture zero degree calorimeters \cite{ZDC}, located far
downstream 
in the machine tunnel (neutron calorimeter at a distance of
about
116 m and proton calorimeter at about 115.5 m from interaction point)
which
 will provide fast information about the centrality of the
collisions;\vspace*{-0.2cm}
 \item A specialized detector system CASTOR \cite{CASTOR_Delphi, CASTOR}.
\vspace*{-0.2cm}
\end{itemize}

\begin{figure}[ht] 
\begin{minipage}{7cm}
\hbox{
\epsfxsize=180pt
\epsfbox[-80 160 660 633]{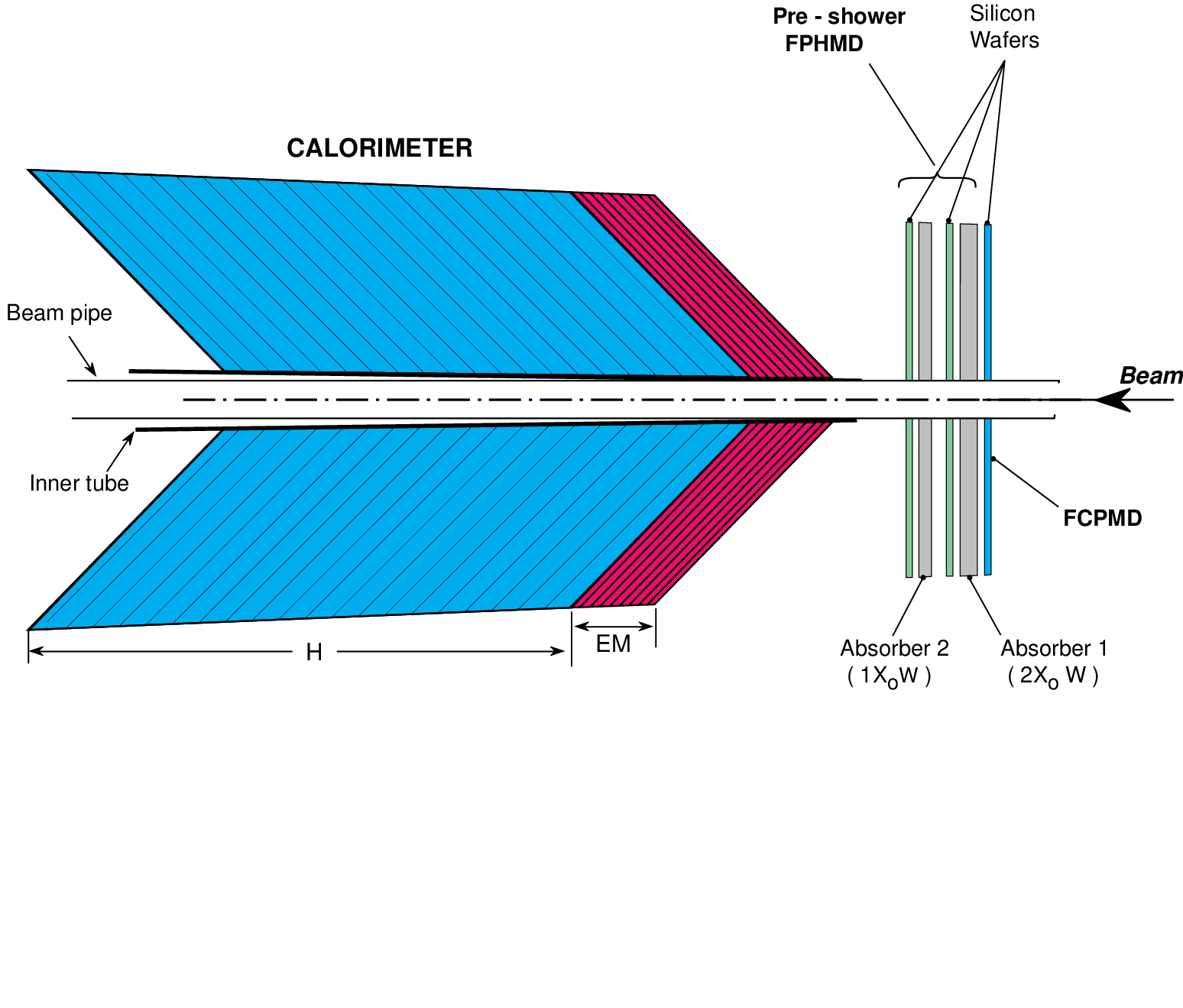}}
\caption{Scheme of the CASTOR detector.}
\label{Castor_scheme}
\end{minipage}
\hspace*{0.3cm}
\begin{minipage}{7cm}
\vspace*{0.2cm}
\hbox{
\begin{turn}{-90}
\epsfxsize=120pt
\epsfbox[20 72 575 738]{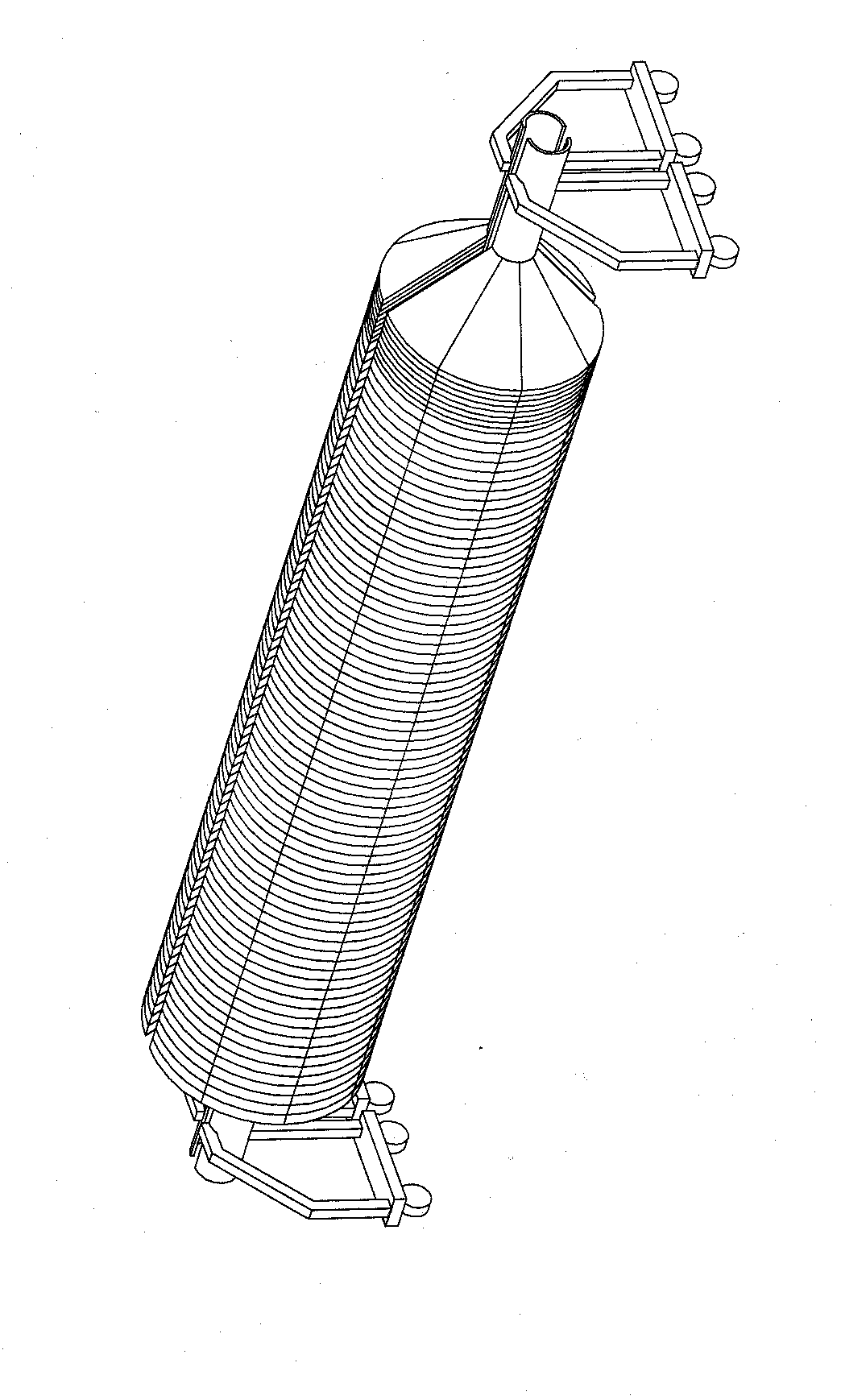}
\end{turn}}
\vspace*{-0.5cm}
\caption{View of the CASTOR calorimeter.}
\label{Castor_cal}
\end{minipage}
\end{figure}

Already at the early stages of the preparation of the ALICE proposal it
was
pointed out \cite{midrap_physics} that the interesting physics beyond
midrapidity  should be the additional subject of
ALICE investigations. From these considerations evolved the
idea of the  CASTOR detector. CASTOR  is dedicated to study
the novel phenomena expected to appear in the high baryochemical potential
environment produced in Pb+Pb collisions at LHC energies, in particular
the formation of Deconfined Quark Matter (DQM), which could exist e.g.
in the core of neutron stars, with characteristics different from those
expected in the much higher temperature baryon--free region around
midrapidity. Its signatures could be
Centauro species and strongly penetrating objects.  
 CASTOR will cover
the very forward rapidity region ($\sim 5.6 \lt \eta \lt 7.2$) and
 it was originally 
proposed
to consist of a silicon charged particle multiplicity detector, a silicon
photon multiplicity detector, and a quartz fibre tungsten calorimeter with
electromagnetic and hadronic sections.
 The scheme of the
apparatus is shown in Figure~\ref{Castor_scheme}.

The LHC, with an energy equivalent to $\sim 10^{17}$ eV for a moving
proton
impinging on one at rest, will be the first accelerator to effectively
probe the highest cosmic energy domain. From Figure~\ref{multiplicity}
it is seen that CASTOR is located at the region of maximum baryon
number, high energy flow
and low multiplicity, similarly as majority of cosmic--ray experiments.

The basic experimental aim of CASTOR is to identify events with
characteristics
similar to those of the exotic cosmic--ray events, in an event-by-event
mode. It will  look for:
\begin{itemize}\vspace*{-2mm}
\item extreme imbalance between the hadronic and photonic components,
      both in terms of the particle multiplicity and the energy content of
the event,\vspace*{-0.2cm}
\item non-uniform (in azimuthal angle $\phi$) deposition of a large amount
      of energy,\vspace*{-0.2cm}
\item highly penetrating objects, far beyond the range of normal hadrons,\vspace*{-0.2cm}
\item abnormal shapes and structures seen in the transition cascade curves
in the
calorimeter.\vspace*{-0.2cm}
\end{itemize}

The project is motivated by the experimental evidence of exotic
cosmic--ray events on the one side  and by theoretical expectations
on the other hand \cite{8,Castor_idea}.
 The model
 of ref. \cite{Panagiotou,8} 
was used
to
simulate the detector system performance. 

\subsection{The design of the CASTOR calorimeter}

At the first stage of project realization the building of only
the calorimeter is planned. Here will be considered the basic version of
the calorimeter.  Some further refinements of this design are
being considered, but these are not expected to alter its
response in an essential way.
The general view of the CASTOR calorimeter  is shown
in Figure~\ref{Castor_cal}.

The calorimeter will be made of layers of active medium sandwiched between
tungsten absorber plates. The active medium consists of planes of silica
fibres and the signal will be  the Cherenkov light produced as they are
traversed by  charged particles from the shower. The fibres are
inclined
at 45 degrees relative to the incoming particles to maximize light output.
The calorimeter is azimuthally divided into 8 octants. Each octant is
longitudinally segmented into 80 layers, the first 8 ($\simeq$ 14.7
X$_{0}$) comprising the electromagnetic section and the remaining 72
($\simeq 9.47 \lambda_{I}$) the hadronic section. The light output from
groups of 4 consecutive active layers is proposed to be coupled into the
same light
guide, giving a total of 20 readout channels along each octant.
Such a calorimeter works as a ``shower core'' detector, sampling
essentially only the part of the shower which lies within a narrow cone
around the direction of the particle. Thus the energy of a hadronic shower
can be reliably measured also very close to the calorimeter's edge. The
transverse width of the hadron shower is about  7 mm (at 200 GeV).
The calorimeter will be   located at a distance of $\sim$ 1740
cm from
the beam interaction point. 
\begin{table}
\label{cal_spec}
\caption[\scriptsize{CASTOR calorimeter specifications}]
{CASTOR calorimeter specifications}
\begin{center}
\vspace*{0.3cm}
\begin{scriptsize} 
\begin{tabular}{|ccc|}
\hline
 & & \\
 &$Electromagnetic$&$Hadronic$\\
 & &                \\
\hline
& &\\
$Material$& Tungsten + Quartz Fibre& Tungsten + Quartz Fibre\\
$Dimensions$& $\langle R_{in} \rangle =$ 26 mm,& $\langle R_{in} \rangle
=$ 27 mm,\\
 & $\langle R_{out} \rangle$ = 129 mm& $\langle R_{out} \rangle$ =134 mm\\
$Absorber$ $Plates$& Thickness = 5 mm& Thickness = 10 mm\\
$at$ $45^{0}$& Eff. thickness = 7.07 mm& Eff. thickness = 14.1 mm\\
$No.$ $layers$& 8& 72\\
$Eff.$ $length$& 56.6 mm $\simeq 14.7$ X$_{0} \simeq 0.53
\lambda_{I}$&1018.1 mm $\simeq 9.47 \lambda_{I}$\\
$Quartz$ $Fibre$&$\sim$ 0.45 mm&$\sim$ 0.45 mm\\
$No.$ $of$ $QF$ $planes$& 2 per sampling& 4 per sampling\\
$Sampling$ &$\simeq 1.84 $X$_{0}$& $\simeq 0.13 \lambda_{I}$\\
$Reading$&Coupling of 4 samplings& Coupling of 4 samplings\\
$No.$ $Readings$& 2 & 18\\
$No.$ $Channels$&$2\times 8 =16$& $18\times 8 = 144$\\
$QF/W$ $vol.$& 10\%& 10\%\\
& &\\
\hline
\end{tabular}
\end{scriptsize}
\end{center}
\vspace*{-0.2cm}
\end{table}
The detailed
specifications of the basic version of the calorimeter are given in
  Table~12.

\subsection{Probes and observables of LHC Centauros and Strangelets}

According to this model, Centauros are
formed
in central collisions of ultrarelativistic nuclei. Thus the future
collider experiments with heavy--ion beams seem to be appropriate to look
for them. Some predictions concerning Centauros which could be produced
at RHIC have been given in \cite{Panagiotou}. The kinematical
conditions accessible at RHIC will be very close to these at which
cosmic ray Centauros are produced. Almost the same value of energy per 
nucleon ($\sqrt{s}^{RHIC}$ = 200 A GeV in comparison with
$\sqrt{s}^{CENT}_{cos. ray}$ = 233 A GeV) allows one in easy way to
predict
the
typical features of Centauros expected to be produced at RHIC.
In particular, the
assumption that  transverse momentum of the fireball decay products 
will be roughly the same for cosmic ray and RHIC Centauros seems to be
reasonable.
The question is, however, the existence of the baryon rich environment.
The first measurements of the ratio of anti--protons to protons
\cite{PHENIX} indicate that the central region, which is mostly explored
by all four RHIC experiments, is meson--dominated. 
 At the LHC the situation will be different,
although
 the crucial conditions for Centauro events
production seem to be well
 fulfilled.
 The energy accessible in Pb+Pb central collisions at the LHC
($\sqrt{s}^{LHC} \simeq$ 5.5 A TeV, i.e. $\sqrt{s_{TOT}}^{LHC} \simeq$
1150 TeV)
will be much higher than the threshold energy for Centauro species
production.
 At the LHC  a region of vanishing baryon density is expected at
midrapidity and the pronounced baryon--rich region with maximum baryon
density at rapidity between 5--7 (see Figure~\ref{multiplicity}).
 Thus 
 the {\it 
existence of
baryon rich environment}, which is
the essential requirement  for Centauro
species production,
 is also expected. The baryon rich environment  is expected
to be
placed at LHC in the forward rapidity region, what is consistent with
the observation  of cosmic ray Centauros in the projectile fragmentation
 range
(see Table~\ref{Cen_kin}).  

\subsubsection{Rate of Centauros in  CASTOR}

The important point is how many Centauros can  be produced
and detected in CASTOR. The  evaluation of the possible  rate of
Centauros
 $N_{Cent}$ can be obtained  from the formula:
\begin{equation}
N_{Cent} \simeq  N_{coll}^{centr}\times P_{Cent}^{prod} \times
P_{Cent}^{decay}
\end{equation}
where $N_{coll}^{centr}$ is the number of central Pb+Pb collisions
during one year of ALICE run,
$P_{Cent}^{prod}$ is a probability of producing Centauro event
in such collision,
and $P_{Cent}^{decay}$ is a probability of Centauro fireball decay
before the CASTOR detection system.
 To observe the distinct Centauro characteristics
in the calorimeter,  
 a decay of the  fireball  at a distance shorter
than
1 m from the interaction point is required. Then
\begin{equation}
P_{Cent}^{decay} = 1- exp(-1m/(c \tau_{0} \gamma))
\end{equation}
Assuming the Centauro fireball gamma factor $\gamma$
$ \simeq 300$ and its lifetime $\tau_{0} = 10^{-9}$ s
 one  obtains  that a probability of Centauro  decay at the path shorter
than
 1 m is greater than 1\%.  
 In opposite case
it could be possible to observe a very big amount of energy released
somewhere in the calorimeter and concentrated within  very narrow
angular cone.
Assuming further $N_{coll}^{centr} \simeq 10^{7}$ per ALICE running year,
and very conservatively that Centauros are produced only in $10^{-2}$
of all central collisions it can be  expected that  at least $\sim$ 1000
Centauro events will be produced
and
detected in the CASTOR during one year.

\subsubsection{Average characteristics estimates}

The main features of Centauros and strangelets produced at the LHC
can be evaluated by   extrapolating the  characteristics of
these objects registered in cosmic ray energy range or by doing 
analytical calculations based on the model. More detailed characteristics
and the distributions of the characteristic quantities can be predicted
by using  Monte Carlo methods (see
the next section).

In particular, keeping in mind that  shapes of the angular and
energy fraction spectra of Centauro secondaries are consistent 
with  isotropic emission of particles
 from a fireball, we can expect  a Gaussian--type
pseudorapidity distribution for  decay products of ``LHC'' 
Centauros.     
The center of the distribution, being the function of both  the
nuclear stopping power  $\Delta y_{stop}$ and a rapidity
shift caused by  transverse momentum of
emitted particles $\Delta y_{PT}$, 
is generally
expected to be placed close to the maximum of the baryon number
distribution,
i.e.
\begin{equation}
\langle y \rangle ^{LHC}_{Cent} \simeq y_{beam}-\Delta y_{stop} -
\Delta y_{PT} \simeq 8.7-2.0-0.9 \simeq 5.8
\end{equation} 
The first rough evaluation of the optimal
position
and  size of the CASTOR detector and  the estimation of its 
geometrical acceptance \cite{CASTOR,Castor_acc} were done in this way.

 Evaluation of other characteristic quantities, as for example 
the types and multiplicities of produced particles needs, however,
some additional assumptions.
 The expected multiplicities of kaons
$N_{K^{+}+K^{0}}$,
non-strange
baryons $N_{h}$, the total number of nuclear active particles $N_{n}$, and
the baryon 
number of a strangelet $A_{str}$ emitted from
a decay of Centauro fireball produced in central Pb+Pb collisions
at LHC energies were estimated in \cite{Castor_acc} for two
different, extreme scenarios:

\vspace*{1mm}

 1. {\it Lower limit of energy density (and temperature).}

   In 
the most conservative case it can be  assumed that the
transverse momentum
        of  decay products of  LHC  Centauro fireball is the same as
for cosmic--ray Centauros, i.e. $\langle p_{T} \rangle \sim$ 1.7 GeV/c.
        Then the average Centauro fireball mass can be estimated from 
 the relation:
\begin{equation}
\langle M_{fb} \rangle = \langle N_{n} \rangle \langle E_{n} \rangle
\end{equation}
where
$\langle N_{n} \rangle$ is the mean number of nuclear active particles
emitted from the Centauro fireball and $\langle E_{n} \rangle$ is the
mean particle energy in the fireball rest system. One can  assume  that
$\langle N_{n} \rangle$ equals to the number
 of participating projectile nucleons
$\langle N_{p} \rangle$, i.e.
it is in the
range of 150-207 for central Pb+Pb collisions and 
\begin{equation}
\langle E_{n} \rangle  = \sqrt{((4/\pi)p_{T})^{2} + M_{N}^{2}} \simeq
2.4  GeV
\end{equation}
where $M_{N}$ is the nucleon mass.

The volume $V_{fb}$ of the fireball and the number of participating
nucleons $N_{p}$ can be evaluated from simple geometrical considerations.
For the volume $V_{fb} \sim 117$ fm$^{3}$,
corresponding to a central collision with the number of
participating nucleons $N_{p} \sim 150$ ( b
$\leq$ 5 fm)
\cite{Panagiotou}, these   assumptions lead to  $M_{fb}
\simeq$ 350 GeV and to  energy density
of  LHC quark matter fireball 
\begin{equation}
\varepsilon_{fb} = M_{fb}/V_{fb} \simeq 3   GeV/fm^{3}
\end{equation}
which is close to that reached in cosmic ray
Centauro events.
Using the phase curve given by eq. (14) with the
reasonable strangeness equilibration factor ($\gamma_{s} \sim 0.4$)
one can find that such energy density corresponds to the temperature
T $\sim$ 130 MeV and to the quarkchemical potential $\mu_{q} \sim$ 590
MeV, and the average characteristics of LHC Centauros
could  be close to these observed in cosmic ray experiments.
In particular, 
 the maximal number of emitted $K^{+}$ and $K^{0}$ 
 for T = 130 MeV and the volume
corresponding to the numbers of participating nucleons in the range
of $N_{p} \sim$ 150 - 207, is  estimated to be  $N_{K^{+}+K^{0}} \sim $
16--20
 (from equations  (20) and (21)).
The maximal baryon  number of a strangelet will be reached when in the
process of strangeness distillation all
$s$--quarks will be absorbed in the strangelet, and  then $A_{str} \simeq$
16 - 20.
Consequently the number of emitted non-strange baryons
\begin{equation}
N_{h} \simeq N_{n} - N_{K^{+}+K^{0}} \simeq 134-187.
\end{equation}  

 In general, assumption of constancy of $\langle p_{T} \rangle$
 allows one to expect that the values of
characteristic quantities  ($N_{h}$, $M_{fb}$, $\varepsilon$, $T$, $\mu$)
for Centauros possibly produced  at RHIC (Au+Au) and  at LHC
(Pb+Pb) should be similar one to the other. These values should not 
also be very different from these found for cosmic ray Centauros.

\vspace*{1mm}

  2. {\it Upper limit of energy density (and temperature).}

 Assumption of  the same $\langle p_{T} \rangle $ of decay products of
cosmic ray and
RHIC Centauros seems to be reasonable because $\sqrt{s}^{RHIC}_{NN}
\simeq \sqrt{s_{NN}}^{cos.ray}$. However, it could be no longer valid for
LHC
because $\sqrt{s}^{LHC}_{NN} \gg \sqrt{s}^{cos.ray}_{NN}$. We can expect
$\varepsilon^{LHC}_{Cent} \gg 
\varepsilon^{cos.ray}_{Cent}$.
J. Schukraft quoted in \cite{Schukraft} 
 the values of $\varepsilon_{LHC} \sim$ 8--27 GeV/fm$^{3}$ (averaged
 over different
papers) as expected at Pb+Pb
central collisions.
K. Geiger \cite{Geiger} claimed even 31(17) GeV/fm$^{3}$ as upper limit of
energy density, averaged over the whole central volume, at
Au+Au (S+S) central collisions. In such a case the assumption of 
constancy of the mean transverse momentum can be unjustified.
In connection with it, much higher values of  temperature and
  quarkchemical
potential  can be  reached. It would probably allow to reach the
region of
the {\it ideal QGP}, however, it is not easy to predict the quantitative
features of Centauros at these conditions. 
 Estimates of  the ranges of characteristic quantities and indications of 
the most
typical signatures of ``LHC'' Centauros was presented in 
\cite{CASTOR_Delphi,CASTOR,Castor_acc}. Generally,
much
higher numbers of 
produced strange quarks ($\rho_{\overline{s}} \gg 0.14$ fm$^{-3}$, e.g.
$\rho_{\overline{s}} \simeq$ 0.5 fm$^{-3}$ for T = 190 MeV),
and hence higher multiplicities  of produced
kaons and  bigger strangelets can be expected.
At a temperature T $\sim$ 190 MeV the number of emitted kaons
could be  $N_{K^{+}+K^{0}} \sim$ 55-68, the number of non-strange
baryons $N_{h} \sim$ 95-139, and  strangelets with baryon numbers as
large as 
$A_{str} \sim$ 55-68 could be formed.

\subsubsection{Expected signatures}
 From the above  considerations it is clear that
 `` LHC'' Centauros  should
be  characterized by signatures dramatically different from  other 
``normal'' events. The most typical signatures will be the following:
\begin{enumerate}\vspace*{-2mm}
\item 
 {\em  Abnormal  photonic to hadronic ratio}.\\   
It is  expected that
  both:\vspace*{-2mm}
\begin{itemize}\vspace*{-2mm}
\item hadronic to photonic energy ratio $E_{h}/E_{\gamma}$,\vspace*{-0.2cm}
\item hadron to photon multiplicity ratio $N_{h}/N_{\gamma}$\vspace*{-0.2cm}
\end{itemize}\vspace*{-2mm}
 should be much
bigger than in " normal " hadronic interactions.
\item  {\em  Very low
total multiplicity} .\\
   The total  multiplicity of ``LHC'' Centauro decay products 
is expected to be  extremely small
in comparison with enormous total multiplicity of particles 
 predicted to be produced 
in ``usual'' events in Pb+Pb central collisions at
LHC energies \cite{Schukraft}.
\item {\em Very specific picture of time development connected with
  abnormal particle composition.}

 The  following products of  time
 evolution of the ``LHC'' Centauro  (at projectile fragmentation
rapidity) could  in principle be observed:\vspace*{-0.2cm}
\begin{itemize}
\item   {\bf Particles  K}$^{+}$ and {\bf K}$^{o}$;
 
$K^{+}$ and $K^{0}$ are expected to be the firstly emitted particles
from the long--lived Centauro fireball. Regardless of the long lifetime
$(\tau_{0}\simeq 10^{-9}$ s) of the fireball these pre--equilibrium
particles will be emitted within very short time interval $(\Delta \tau
\simeq 10^{-22}$ s) and  could be observed in detectors much 
earlier
than other decay products of Centauro
\footnote{They are emitted practically at the point of interaction;
$l=c\gamma \Delta \tau=c\cdot coshy\cdot \Delta \tau \simeq
(0.6-0.9)10^{-11}cm$,
 when assuming $\gamma=coshy=200-300$.
}.
The total number of emitted $K^{+}$ and $K^{0}$ can be maximally
equal to the number of $s{\overline s}$ pairs created in the volume
$V_{fb}$
of the Centauro fireball
 and several
tens kaons could be produced.

\item  {\bf Strange quark matter metastable object with small Z/A ratio};

 This state is formed after the pre--equilibrium emission of kaons.
 Then the fireball is a mixture of $u$, $d$ and $s$ quarks and
experimentally  
 it can be identified  by its Z/A ratio less than that of the ordinary
 nuclear matter. The change of Z/A ratio, because of kaon emission
 can be calculated from formulas (22) and (23).
Assuming $(Z/A)_{i} \simeq 0.5$ and the temperature in the range
$\sim $ 130--190 MeV one evaluates $(Z/A)_{f} \simeq 0.45-0.3$.
This slightly strange quark--matter fireball  decays after a relatively
long time $\tau_{0} \sim 10^{-9}$ s.
Taking into account consequences of the collider kinematics, which
decreases
the lab--frame decay lengths, and assuming $\gamma \sim 200-300$, the
object will decay after traveling the path of the order
$\sim$ 60--90 m. 
\item {\bf  Non--strange baryons
 and  highly penetrating strangelet(s)}.
 
They should be observed at the last stage of the "Centauro" evolution as
the
result of
 the decay of strange quark matter state after
the process of strangeness separation.
 Assuming that all strange  
quarks are absorbed in the strangelet and $\frac{N_{s}}{A_{str}} \sim 1$,
 the formation of small strangelets characterized by
the baryon numbers of several tens should be expected (see the previous
section).
In principle, the formation of one bigger or several smaller strange
droplets could be possible.
 The number of emitted non--strange baryons
 depends on the size of the strangelet   
 and  was  estimated to be in the range
 $\sim$ 70--190.
\end{itemize}
\end{enumerate}
\begin{table}[h]
\label{Cen_LHC}
\caption[\scriptsize{Average characteristics of Centauro events and
Strangelets
 produced in cosmic rays and at the LHC \cite{CASTOR_Delphi}.}]
{Average characteristics of Centauro events and Strangelets
 produced in cosmic rays and at the LHC \cite{CASTOR_Delphi}.}
\begin{scriptsize}
\begin{center}
\vspace{0.3cm}
\begin{tabular}{|ccc|}
\hline
 & & \\
 $Centauro$&$Cosmic$ $Rays$&$LHC$\\
 & &                \\
\hline
& &\\
$Interaction$& ``$Fe + N$''& $Pb + Pb$\\
$\sqrt{s}$&$\geq$ 6.76 TeV& 5.5 A TeV\\
$Fireball$ $mass$& $\geq$ 180 GeV& $\sim$ 500 GeV\\
$y_{proj}$& $\geq$ 11&8.67\\
$\gamma$&$\geq 10^{4}$&$\simeq$ 300\\
$\eta_{cent}$& 9.9&$\sim$ 5-7\\
$\langle p_{T} \rangle$& 1.75 GeV& 1.75 GeV (*)\\
$Life-time$ & $10^{-9}$ s & $10^{-9}$ s (*)\\
$Decay$ $prob.$& 10\% ( x $\geq$ 10 km)& 1\% (x $\leq$ 1 m)\\
$Strangeness$& 14 & 60-80\\
$f_{s}(S/A)$ & $\simeq$ 0.2&$\sim$ 0.1-0.4\\
$Z/A$& $\simeq$ 0.4& $\sim$ 0.3-0.4\\
$Event$ $rate$& $\geq 1$ \%& $\simeq$ 1000/ALICE-year\\
& &\\
\hline
& &\\
``$Strangelet$''& $Cosmic$ $Rays$& $LHC$\\
& &\\
\hline
& & \\
$Mass$& $\simeq$ 7-15 GeV& 10-80 GeV\\
$Z$& $\simeq$ 0& $\simeq$ 0\\
$f_{s}$& $\simeq$ 1& $\simeq$ 1\\
$\eta_{str}$& $\eta_{Cent}$ + 1.2& $\eta_{Cent}$ + 1.2\\
& &\\
\hline
\end{tabular}
\end{center}
\hspace{2 cm} (*) assumed\\
\vspace*{-0.5cm}
\end{scriptsize}
\end{table} 
 Some  average
characteristic
quantities of Centauros and Strangelets produced in cosmic rays and
expected at the LHC are compared in   Table~13.

The events of such type  may be easily observed in  detectors with
particle identification, such as silicon detectors, TPC, etc. (the most
important
is to distinguish between kaons, baryons and pions).
  What regards strangelets, if the hypothesis that cosmic--ray
long--flying
component is a result of passing of strangelets through the
apparatus is true, we could look for them  using
the deep forward calorimeters.
 Penetrating through calorimeters
they could  produce (via their weak decays
or by interactions with absorber nuclei) succesive maxima seen in their
transition curves. Observation of  anomalous transition curves
needs sufficiently deep calorimeters (above 120 cascade
units) with granularity good enough for watching individual cascade
development, and detection layers placed every 2--3 cascade units.
One  could  consider also the idea of using  simple and unexpensive  
CR--39 plastic detectors which could look for both the  high Z
quark matter fireballs
and  for accompanying strangelets.

The important questions, concerning the sensitivity of the CASTOR
detector for these signatures and the influence of the background
on  the distinction of the signals, have been answered by means of the
Centauro and Strangelets generators described in the next section.

\section{Centauro and/or Strangelet  Simulations}
\subsection{Generator for Centauros}
The Monte Carlo generator of Centauro events is based on the
phenomenological model of refs.~\cite{Panagiotou,8} and it was described
 in detail in \cite{Castor_gen}. The model is formulated in terms of
impact parameter $b$ of the ion collisions, two thermodynamical
parameters, baryochemical potential $\mu_{b}$ and temperature $T$, and the
nuclear stopping power $\Delta y_{stop}$. The generator calculates the
Centauro fireball
parameters and produces a full event configuration according to the
fireball evolution scenario.

According to the model the Centauro events occur in the projectile
fragmentation region when the projectile nucleus penetrating through the
target nucleus transforms its kinetic energy into heat and forms hot
quark matter (Centauro fireball) with high baryochemical potential.
 The produced glob of deconfined quark matter initially contains 
$u$, $d$ quarks and gluons only and it  is
characterized by
temperature $T$ and baryochemical potential $\mu_{b}$. Its energy
density $\varepsilon$ and other thermodynamical quantities, such as
baryon/quark 
 number density $n_{q}$ are  calculated from  equations (14) and (16)
 respectively. The number of quarks $N_{q}$ in the 
Centauro fireball is defined from collision geometry as $N_{q} = 3N_{b}$.
 Baryon number of the fireball, $N_{b}$,  is calculated from
the relation:
\begin{equation}
N_{b} = 0.9 V_{ovr}(A_{1}/V_{1})
\end{equation}
The factor 0.9 stands for the central part of the overlapping region
$V_{ovr}$ of colliding nuclei with the atomic numbers $A_{1}$ and $A_{2}$.
Hence one can obtain the 
volume $V_{fb}$ as well as the mass $M_{fb}$ of
the fireball
\begin{equation}
V_{fb}=N_{q}/n_{q}, \, M_{fb}=\varepsilon V_{fb}
\end{equation}

At the second stage of the Centauro fireball evolution a partial chemical 
equilibrium is achieved by coupling $\overline{s}$-quarks with $u$ and  
$d$ quarks and emission of $K^{+}$ and $K^{0}$ mesons, what decreases the 
temperature and entropy. The number of $s\overline{s}$ pairs inside the
fireball and hence the number of emitted kaons  is calculated from
equations (20) and (21).
\begin{equation}
N_{s} = N_{K^{+}+K^{0}} = n_{s}V_{fb}
\end{equation}
After emission of $2N_{\overline{s}}$ quarks with kaons, the mass of
remaining strange quark matter (SQM) fireball is defined by the average
quark
energy and the number of quarks in the Centauro
and SQM fireball:
\begin{equation}
M_{SQM} = M_{fb}(1-2N_{\overline{s}}/n_{q})
\end{equation}
This anti--strangeness emission is described as an isotropic decay of the
Centauro fireball into $N_{\overline{s}}$ kaons and the SQM fireball with
mass $M_{SQM}$.
At the final stage of the evolution the SQM fireball decays into baryons
and strangelets.

 \subsection{Simulations of
Centauro events}

As it was shown in \cite{Castor_gen} the generator reproduces the main
features of Centauro cosmic ray events.
Simulations of Centauro events formed in Pb+Pb central collisions  at
$\sqrt{s}$ = 5.5 A TeV have been firstly performed in
\cite{Castor_gen}.
New simulations, in an extended range of parameters were
done more recently 
\cite{Sowa,Sowa_note}. Different sets, each
consisting  of
10000 events, were generated.
The very central collisions with $0 \lt b \lt 1$ fm were studied.
Simulations were done for the temperature $T$ = 130, 200, 250 and 300 MeV
and baryochemical potential $\mu_{b}$ = 1, 1.8 and 3 GeV.
The nuclear stopping power parameter values were put $\Delta y_{n.s.}$ =
1.5, 1.0, 0.5. Keeping in mind the high values of the temperature,
and in  consequence very high transverse momenta of particles
emitted from the fireball, such values of $\Delta y_{stop}$ lead to
the effective stopping bigger than the assumed $\Delta y_{stop}$ value, at
least
about one rapidity unit. It
gives the position of the maximum of the pseudorapidity distribution
 of baryons
 in accordance with that predicted by HIJING/VENUS generators. 
The strong coupling constant  $\alpha_{s}$ was taken as  0.3.

The characteristics  of simulated Centauro events are apparently different
from
those  obtained from ``usual'' (e.g. HIJING) generators.

In particular, Centauro events are characterized by almost total absence
of a photonic component among secondary particles. The majority of
secondary particles are baryons. Some kaons  emitted from the
primary
fireball  decay   into neutral pions which in turn give
photons, but the neutral pion production is suppressed here strongly.

\begin{figure}[ht]
\vspace*{-1cm}
\hbox{
\epsfxsize=400pt
\epsfbox[43 150 536 658]{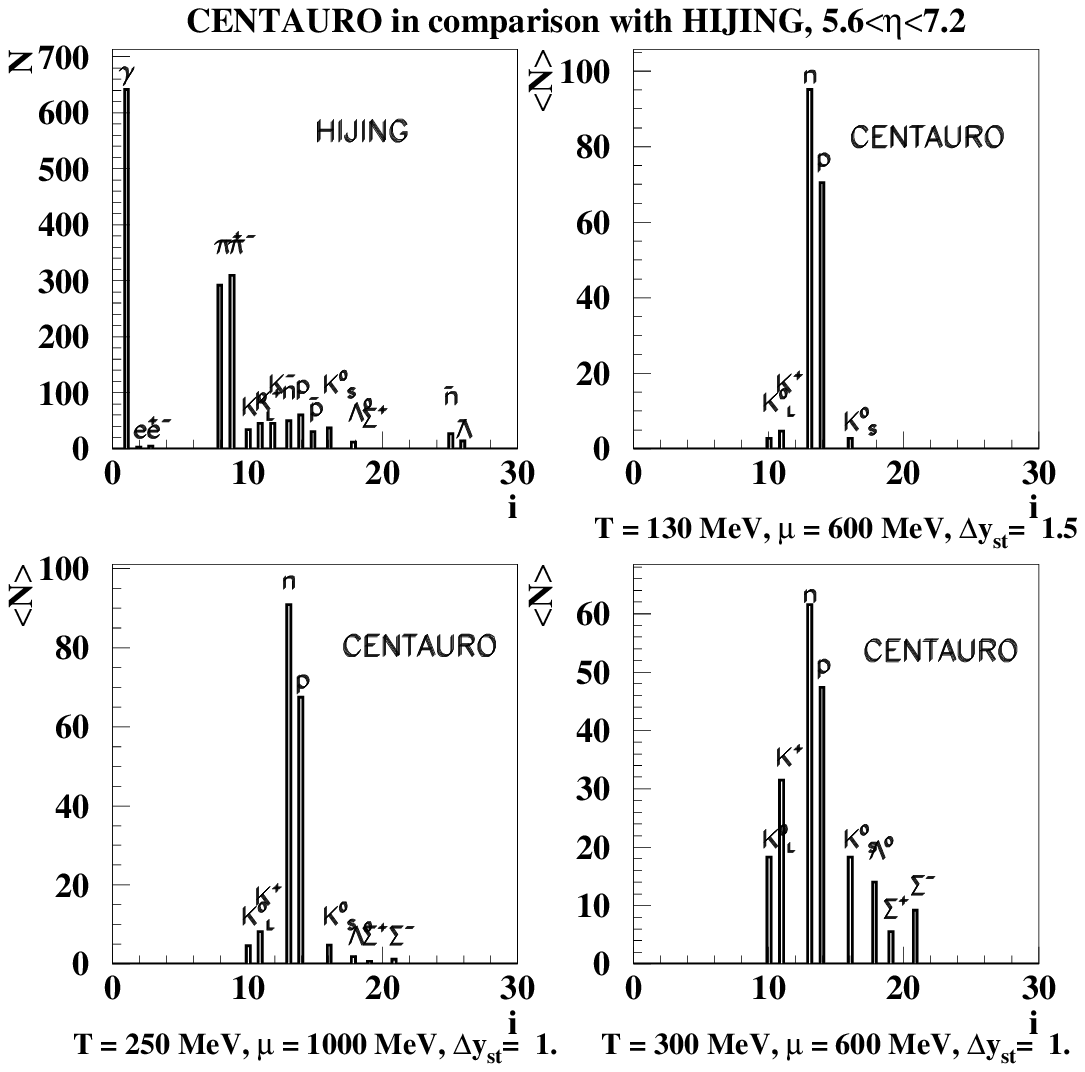}
}
\vspace{5mm}
\caption[Average multiplicities of particles produced in
``usual''
 (HIJING) event and by Centauro mechanism.] 
 {Average
multiplicities of particles produced in
``usual''
 (HIJING) event and by Centauro mechanism. Only particles
 within CASTOR acceptance are shown.}
\label{cen_hij}
\end{figure}

The next surprising feature is a multiplicity of Centauro events which is
much smaller than the one predicted by ``conventional'' generators for
nucleus-nucleus collisions at that energy.
In ref.~\cite{Castor_acc}  the expected average multiplicities of kaons,
non-strange
baryons and  the total number of nuclear active particles
 emitted from
a decay of Centauro fireball, produced in central Pb+Pb collisions
at LHC energies, were estimated in analytical way (see subsubsection
6.2.2).
The results of the  simulations are
 illustrated in
Fig.~\ref{cen_hij} which  shows the average multiplicities of
different kinds
of particles produced  by Centauro mechanism, contained within  the
 geometrical
acceptance of the CASTOR, in comparison with
the HIJING predictions. The average
multiplicities  for three sets of
events,
characterized by  T =  130, 250, 300  MeV and $\mu_{b}$ =  1.8, 3.0, 1.8 GeV
respectively, are shown as the examples.

Secondary particles in the Centauro events have larger mean transverse
momenta in comparison with ordinary hadronic interactions. In usual 
events the average transverse momentum of produced particles $\langle
p_{T} \rangle$
= 0.44 GeV/c, as predicted by HIJING, which is several times smaller than
that of Centauro events \cite{Castor_gen}. 

\begin{figure}[ht]  
\hbox{
\epsfxsize=133pt
\epsfysize=140pt
\epsfbox[43 150 536 658]{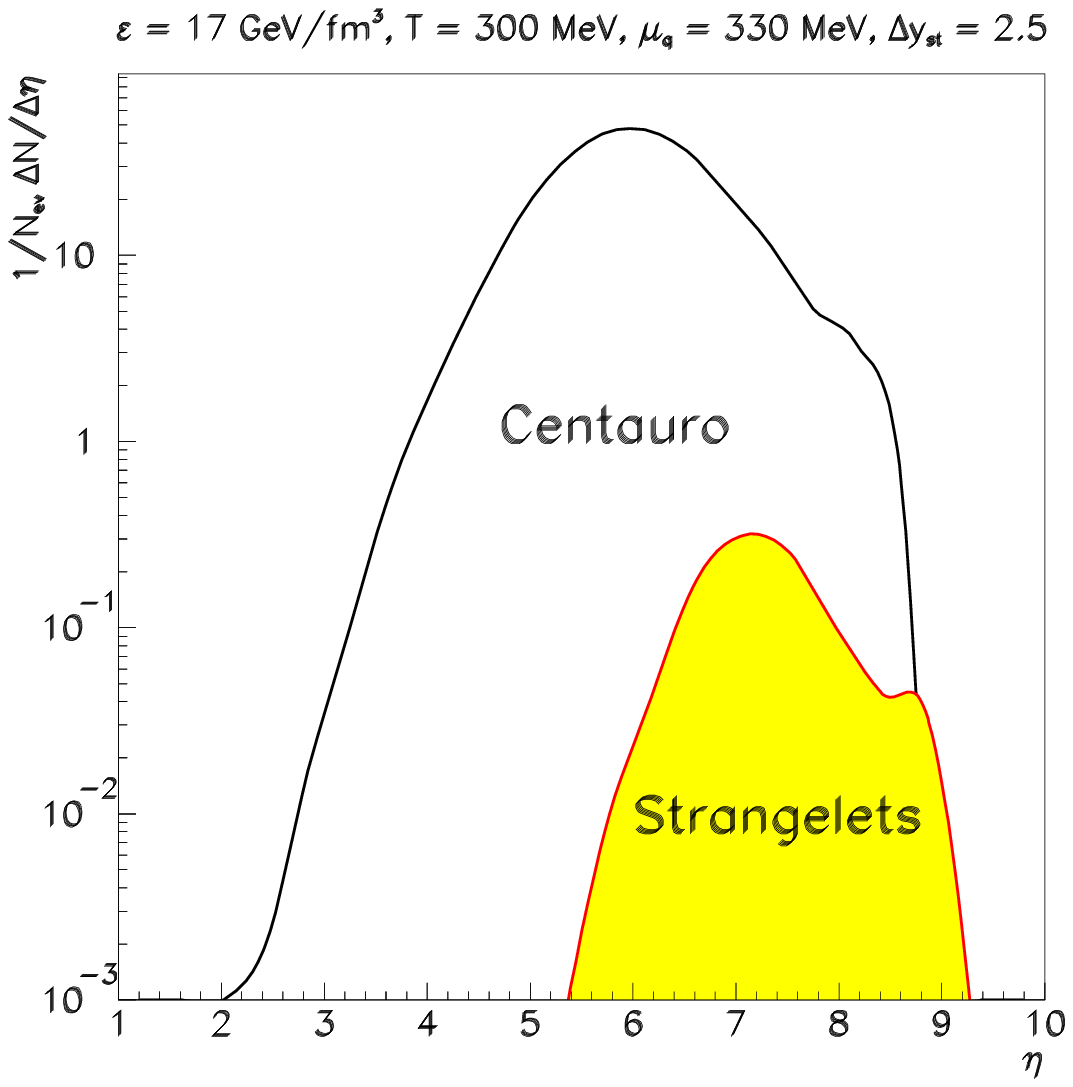}
\epsfxsize=133pt
\epsfysize=140pt
\epsfbox[43 150 536 658]{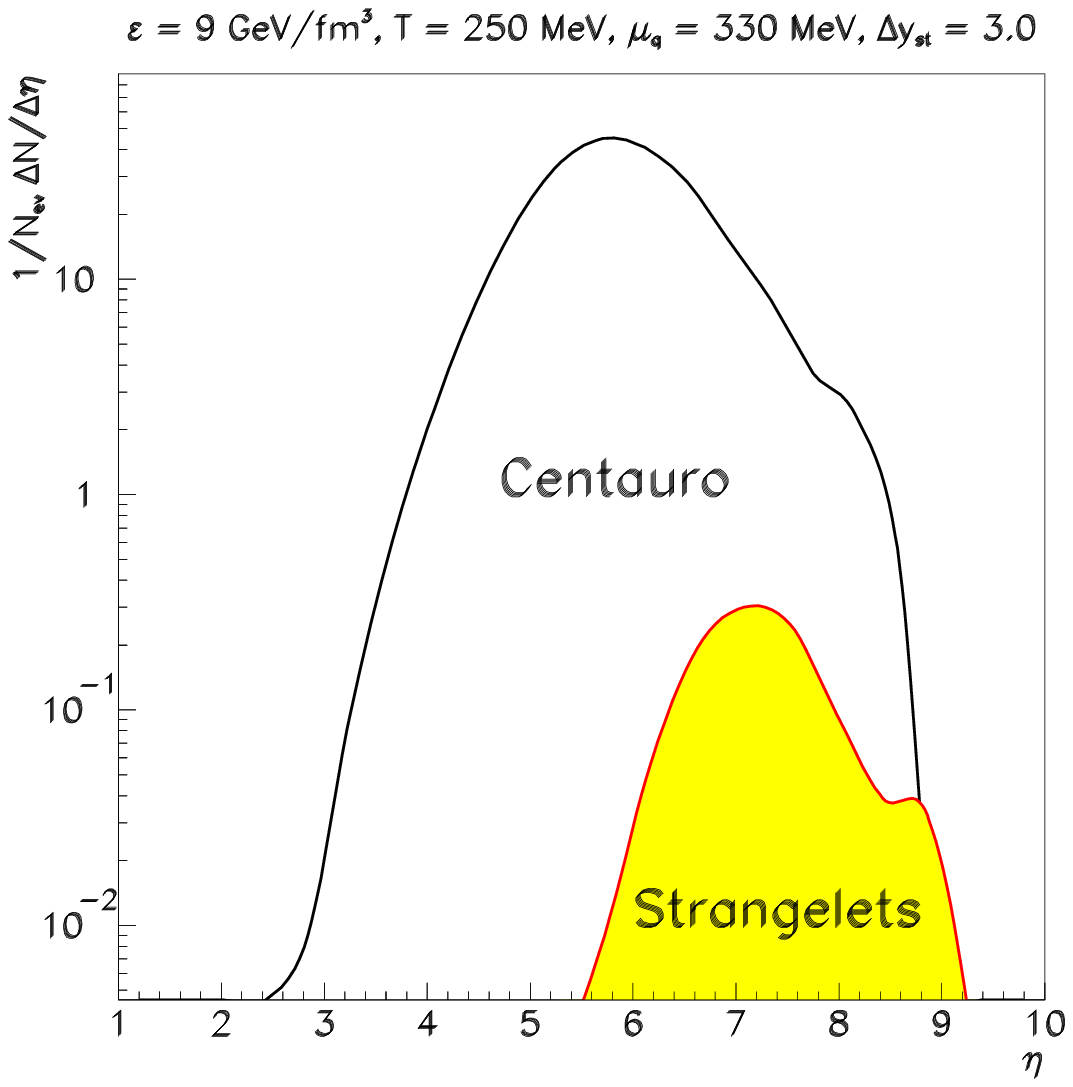}
\epsfxsize=133pt
\epsfysize=140pt
\epsfbox[43 150 536 658]{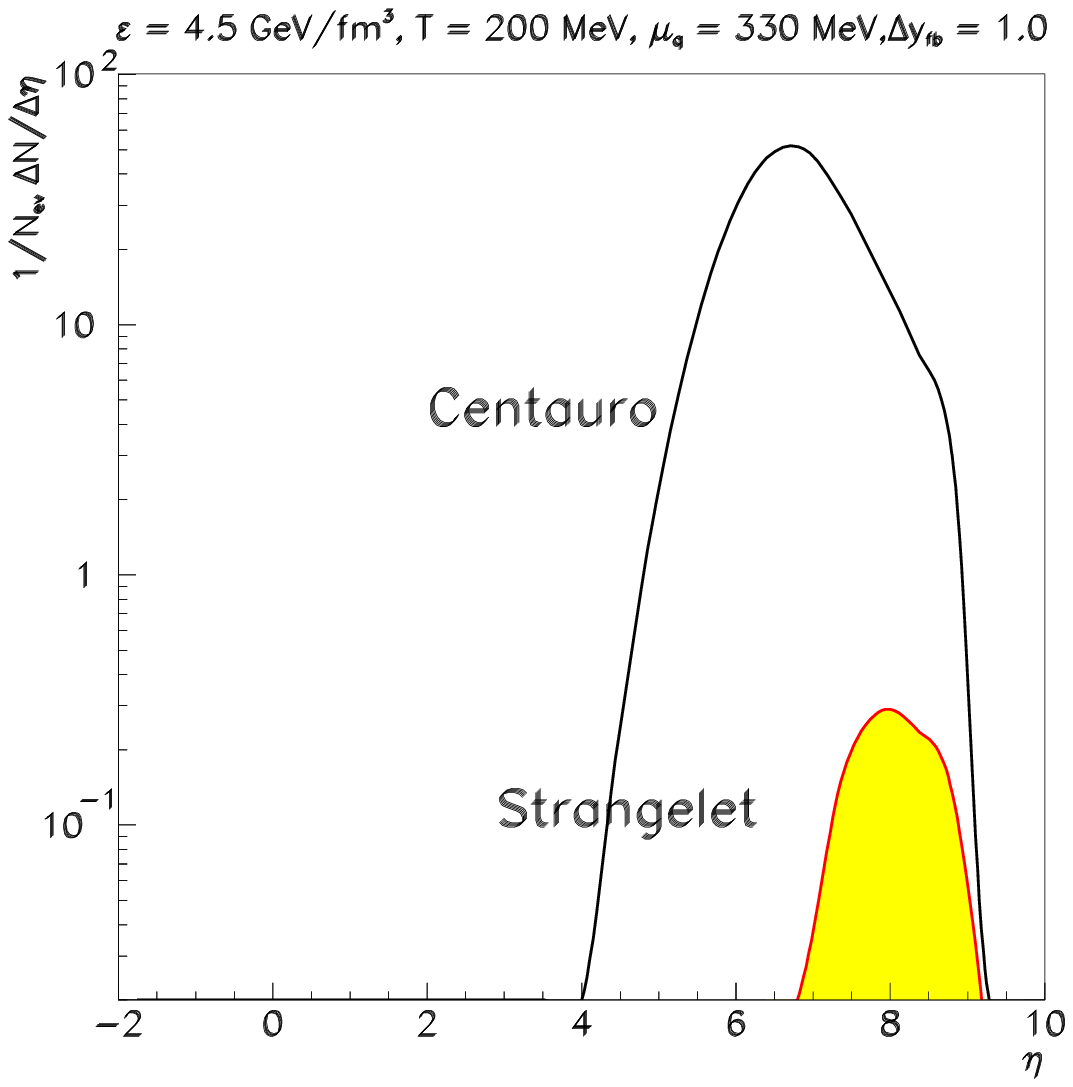}}
\caption [Probability of a strangelet  and other
particles
  production from the decay of Centauro fireball as a
 function
 of pseudorapidity.] 
{Probability of a strangelet (yellow colour) and other
particles
  production from the decay of Centauro fireball as a
 function
 of pseudorapidity.}
\label{Cen_eta}
\end{figure}

The position and shape of rapidity(pseudorapidity) distributions of decay
products of the 
Centauro fireball  depend on thermodynamical variables ($\mu_{b}$
and $T$) and mainly on the nuclear stopping power
$\Delta
y_{n.s}$. Thus
the detection probability of Centauro events in the CASTOR detector
is also the funcion of these variables.
 Fig.~\ref{Cen_eta} shows  the probability
 of the production 
of a strangelet (yellow colour) and other particles from the Centauro 
fireball decay, as a function of pseudorapidity
$\eta$.
Three sets
of events with different values of parameters   were chosen as examples.
It is seen
that a strangelet,
 as it should be expected from cosmic ray observations, always flies
 in more forward direction that other particles. The difference is of the
order of  one pseudorapidity unit. As a consequence, the
simultaneous
detection of a strangelet and other Centauro decay products will
be sometimes excluded. The big chance for  detection of both
species simultaneously is when temperature, quarkchemical potential
and nuclear stopping power values are close to T $\sim$ 250 MeV, $\mu_{b}
 \sim$ 3 GeV and $\Delta y_{stop} \sim 1.5$.

\begin{figure}
\hbox{
\epsfxsize=180pt
\epsfysize=180pt
\epsfbox[1 114 610 714]{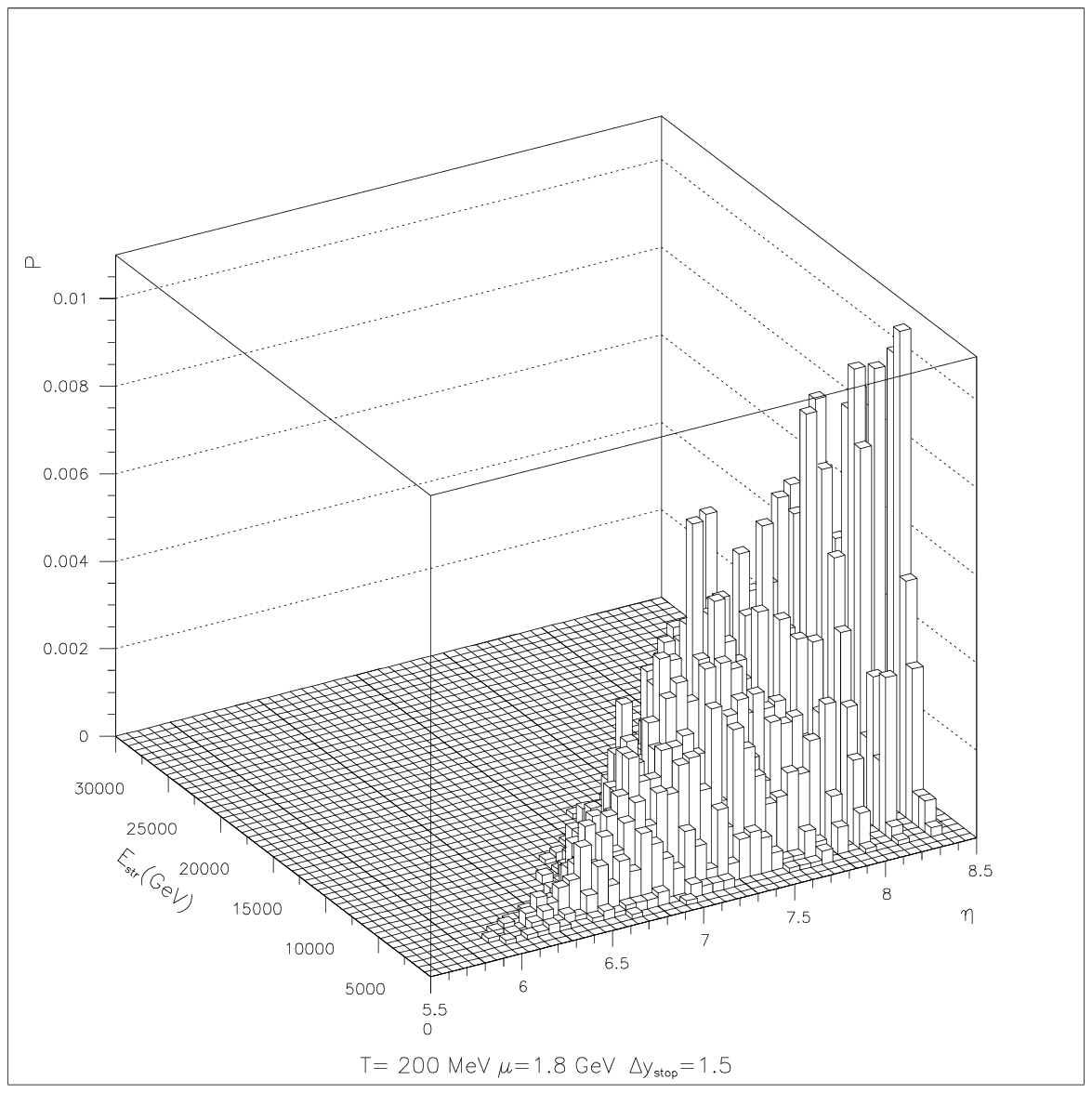}
\hspace*{0.5cm}
\epsfxsize=180pt
\epsfysize=180pt
\epsfbox[1 114 610 714]{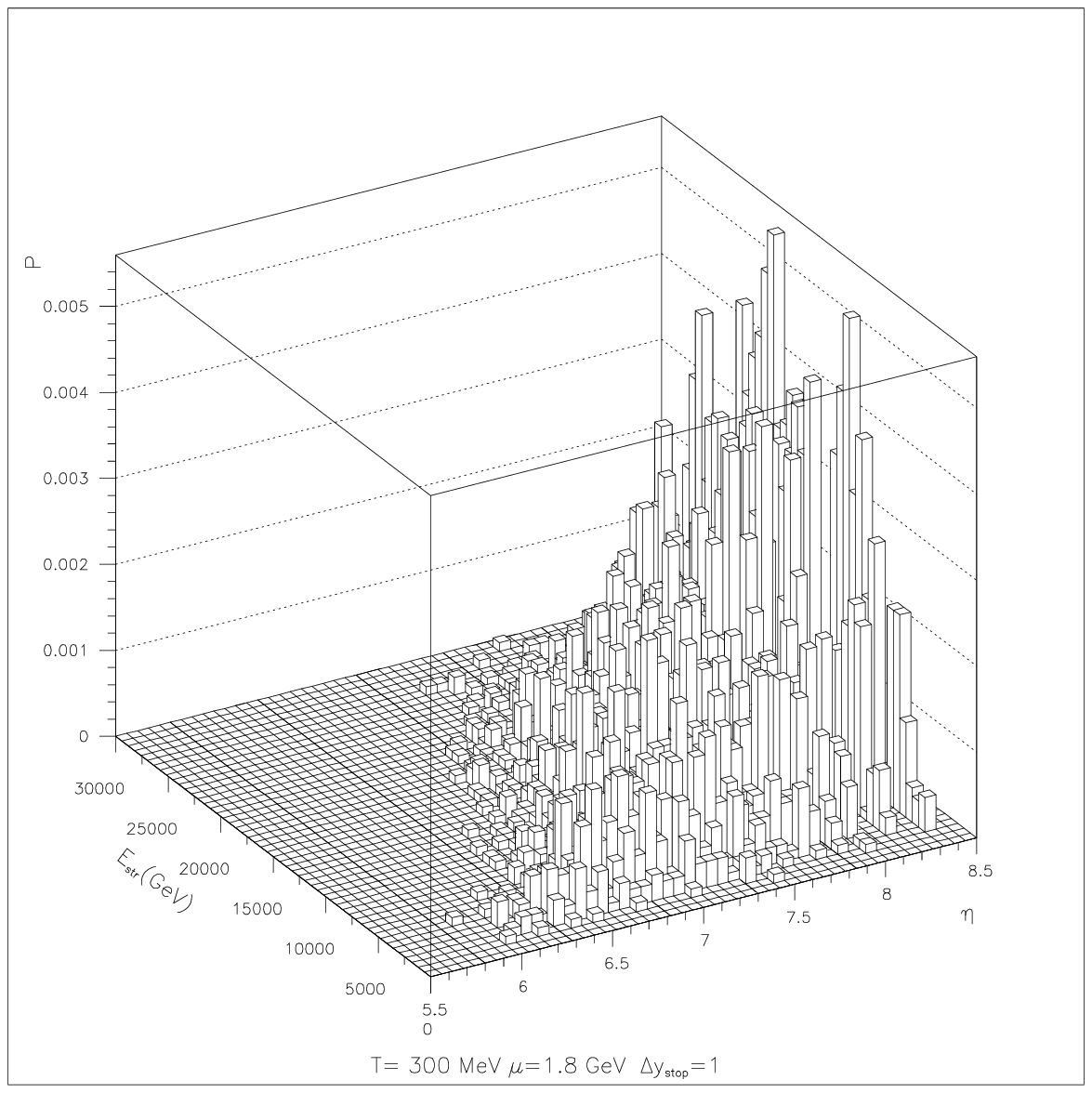}}
\caption{Probability of  
   a strangelet production as a function of its energy and
pseudorapidity.}
\label{estr_eta}
\end{figure}

Fig.~\ref{estr_eta} shows two--dimensional lego histograms which
illustrate the probability of a strangelet production, as a function of
their pseudorapidity and energy.

The  question of acceptance has been investigated in
analytical way in \cite{Castor_acc} and later,  by
simulations,
in \cite{Castor_gen,Sowa,Sowa_note}. For the  
reasonable values of parameters 
more than 50\% of energy (or  the  number of produced particles)  of an
event
 falls into the detector. The maximal degree of containment of Centauro
decay
products is
close to
$\sim$ 0.74.
  
\subsection{Exotic objects in  deep  calorimeters}

 Basing on the model of  strangelet formation
 in the
quark-matter (Centauro) fireball, one can 
expect that at the LHC kinematical conditions the
production of a variety of strangelets, characterized
by a wide spectrum of the  baryon numbers ($A_{str} \sim$ several tens  
for
 temperature $T \sim$ 130-190 MeV and  quark chemical potential
$\mu_{q} \sim$ 600-1000 MeV) should be possible (see section 6.2 and
Table~13).
The important question is what  signals will be produced by such exotic
objects
during their passage through the deep calorimeters,
 and if these signals can be distinguished
from those produced by conventional events. It should be mentioned,
however, that
properties of  CASTOR--type  calorimeters differ
from those of the deep emulsion chambers used in cosmic ray
experiments. The
latter
have very fine lateral resolution ( $\sim 100 \mu m$), allowing
for the observation of the development of  individual cascades through the
whole calorimeter depth. In
contrast, in the CASTOR-type  calorimeter, the signal produced by a
strangelet 
will be detected simultaneously with those generated by other particles
entering the
same
calorimeter octant. Thus, the additional question is the intensity and
 the shape of a possible background and an ability to extract  an
``exotic'' signal from it.

 Two possible ``exotic'' scenarios were investigated. In the first one it
was
assumed that  strangelets were born anywise among other
conventionally produced
particles \cite{9,Castor_str} (see subsubsection 7.3.1). In the second
case, 
 strangelets  produced as 
remnants of the Centauro
fireball explosion, according to the mechanism proposed in \cite{8}
 (see subsubsection 7.3.2) were considered.
In this case the signal will be the sum of a strangelet transition curve
and  the one
 produced by nucleons coming from the isotropic
decay of the Centauro fireball. This case has been preliminarily
investigated in
\cite{9} and recently  more detailed   simulations have
been done \cite{Sowa_note} by
GEANT
3.21.
In both cases  the
 background was estimated by means of the HIJING generator, assuming that 
 a part of energy
going into conventional particle production equals a difference
between the total energy available in the  phase space
region (and being  $\sim$ 150 TeV within the CASTOR
 acceptance)
and energy taken by
a Centauro fireball and/or a strangelet.

\subsubsection{Strangelets in the deep calorimeter}
 The
scenario, in which strangelets are born together with  other
conventionally
produced particles was  investigated
  in
\cite{9,Castor_str}
 for  both short-lived and stable
 strangelets.
\begin{center}
\begin{large}
{\em Short-lived strangelets}
\end{large}
\end{center}
   Unstable objects
which can decay via
strong interactions
$(\tau_{0} \leq 10^{-20} $~s) or the metastable ones decaying via weak
nucleonic decays (see subsubsection 4.4.2) were named  short-lived strangelets.
 Generally, however,
the lifetimes of small metastable strangelets are not predicted precisely
at present \cite{Greiner,Madsen_Kreta}. If their lifetimes
are shorter than $\sim
10^{-10} $~s they could decay before
reaching
the CASTOR
calorimeter and give the  same picture  as  unstable
strangelets. In the opposite case the situation is analogous to the case
of
``stable'' strangelets, considered later.
The  complete decay of a strangelet via strong
processes or its fission into a daughter strangelet and an arbitrary
number of
hadrons is possible. A daughter strangelet will be shifted to a higher
strangeness factor $f_{s}$.
After surviving a strong and possibly also weak nucleonic decay
it can reach the region of a very high strangeness factor
($f_{s} \geq$ 2.2) where it is expected to become a long-lived (stable)
object \cite{Schaffner}. The algoritm of the calculations was the same
as used previously for cosmic ray events (see subsubsection 4.4.2).
 It was assumed, for simplicity, that a strangelet
decays only via neutron emission.
Unstable strangelets decay very fast, practically at the point of their
formation, thus the considered picture resolves into the simple case of a
bundle   
of neutrons entering the calorimeter.

As it has been shown in \cite{9}
 the scenario in which an unstable or metastable strangelet via strong
or weak decays produces a strongly collimated bundle of neutrons,
successfully
describes the long-range many--maxima cascades observed in the cosmic ray
experiments. The successive maxima, seen in the structure of a
transition
curve, could be the result of interactions of such neutrons in the
apparatus.

\begin{figure}
\begin{center}
\vbox{
\epsfxsize=300pt
\epsfysize=120pt
\epsfbox[2 247 566 566]{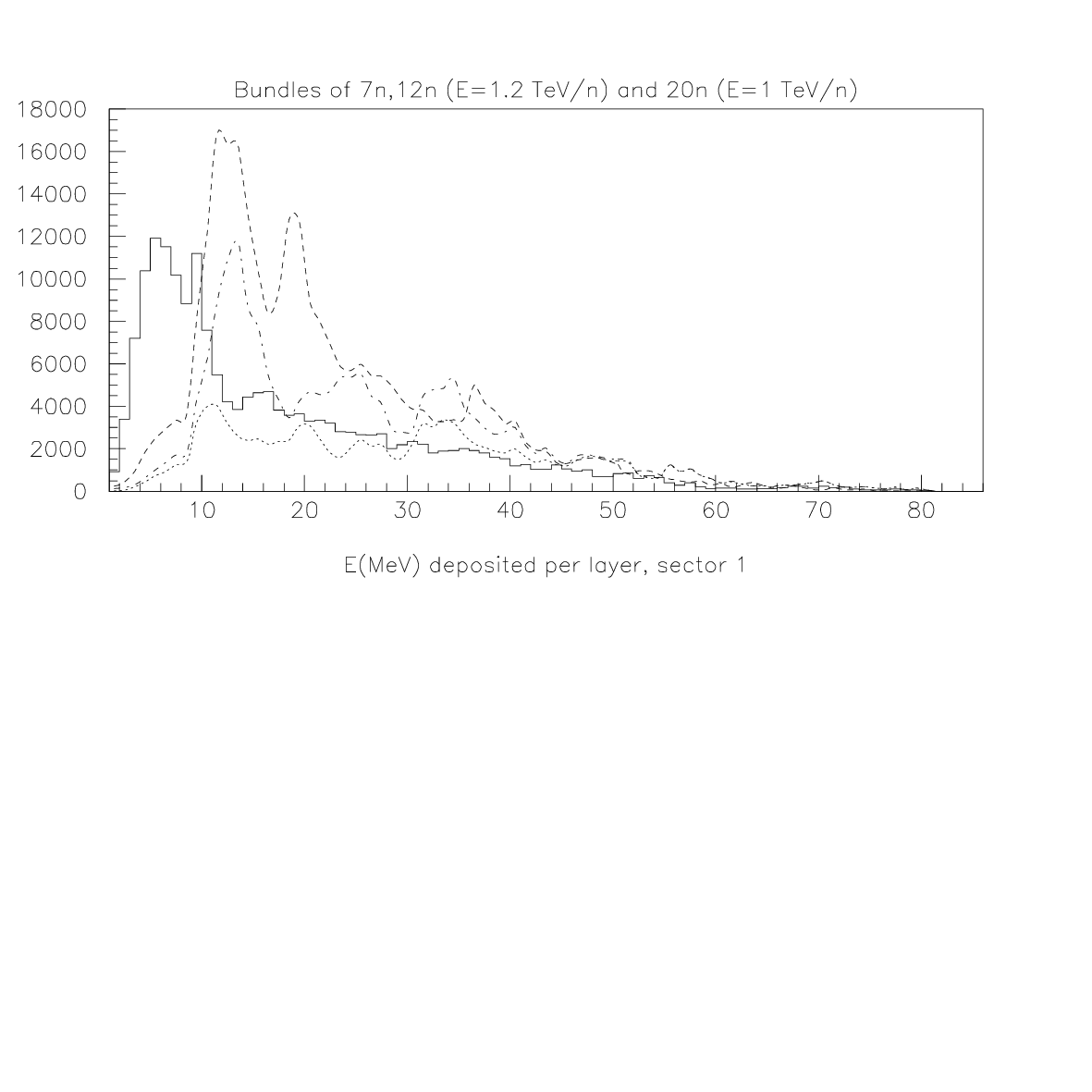}
\epsfxsize=300pt
\epsfysize=200pt
\vspace*{-0.5cm}
\caption[Transition curves produced by bundles of 
 neutrons] {Transition curves produced by bundles of 7
(dotted-dashed
curve)
 and 10 (dotted curve) neutrons of energy $E_{n}\approx$ 1.2 TeV
 and 20  (dashed curve) neutrons of energy $E_{n}\approx$ 1 TeV
evaporated by
  a short-lived strangelet of $A_{str}$ = 40. Full line histogram shows
the HIJING
 estimated background.
 Energy deposit (MeV) in the calorimeter layers, in the octant
containing a strangelet, is shown.}
\label{castor_unst}
\epsfbox[1 0 565 567]{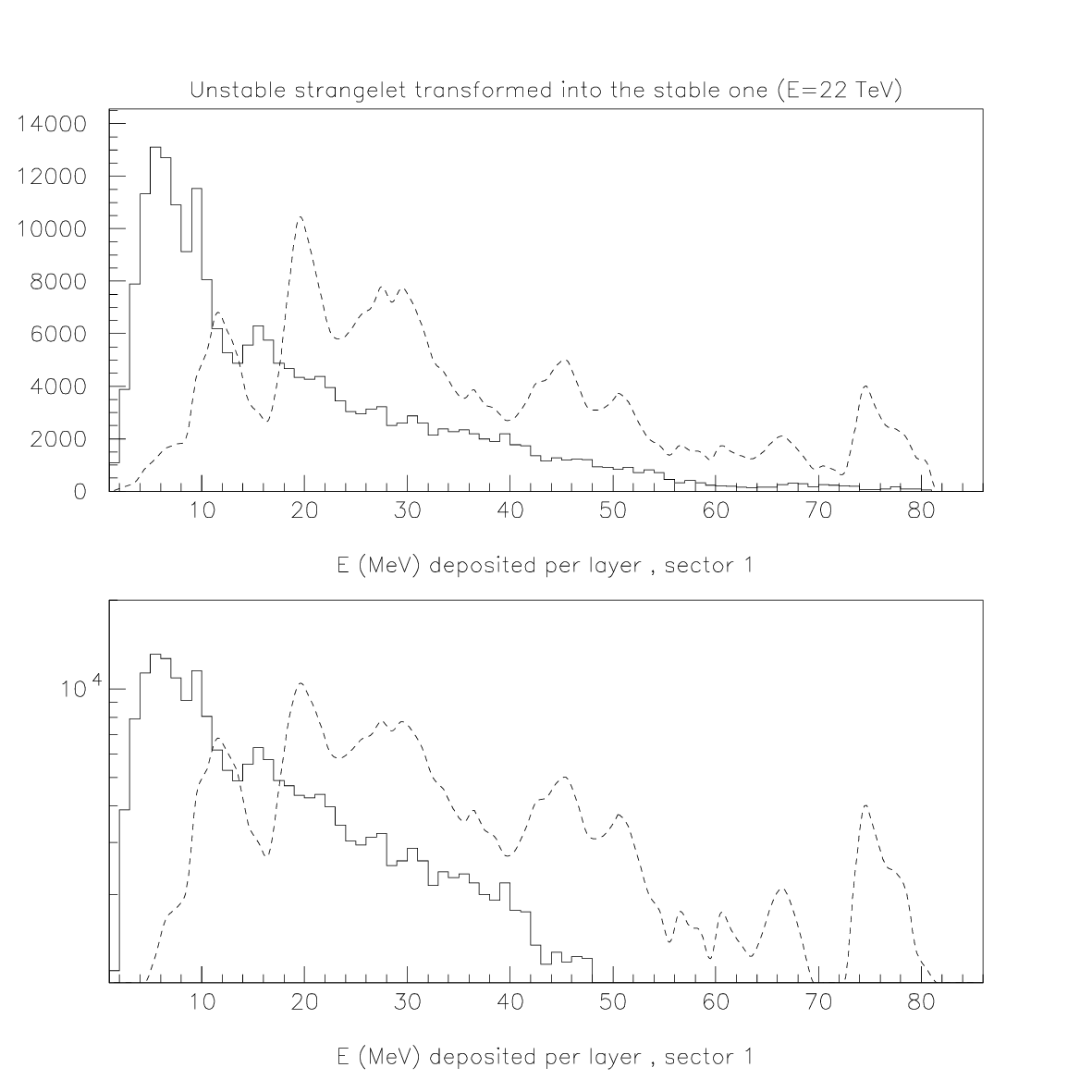}}
\caption [Unstable strangelet
transformed
 into a  stable one.]{
 Unstable strangelet ($A_{str}$ = 22, $E_{str}\approx$ 22 TeV)
transformed
 into a  stable one after evaporation of 7 neutrons.
 Energy deposit (MeV) in the calorimeter layers, in the octant
containing a strangelet, is shown.}
\label{st_un}
\end{center}
\vspace*{-2mm}
\end{figure}

The general conclusion concerning  the signals 
 produced in the
CASTOR calorimeter by
short-lived
 strangelets, formed in Pb+Pb interactions at the LHC
 is the same
as in the study  of the cosmic-ray strangelets. Bundles of
collimated  
neutrons can give in the CASTOR calorimeter the unconventional many-maxima
signal. Its longitudinal structure and extent depend, of course, on the
strangelet
energy and on the number of evaporated neutrons $N_{n}$.
Fig.~\ref{castor_unst}  shows three
examples of transition curves produced by bundles of 7 and 12 neutrons
of energy $E_{n} \approx$ 1.2 TeV and 20 neutrons of energy $E_{n}
\approx$
1 TeV, evaporated by a short-lived strangelet with  baryon number
$A_{str}$ = 40. They are compared with the possible background (full line
histogram)
estimated by  HIJING,  assuming that the rest of the available energy
(i.e. not carried  by the  strangelet) is going into  conventional
particle
production.
 For the comparison  particles produced in one central Pb+Pb collision
were taken.

 It was found  that a bundle of several neutrons ($N_{n} > 7$) possessing
sufficiently
high energies ($E_{n} >$ 1 TeV) produces in the  calorimeter a  signal
which can be distinguished from the conventional event signal.
The strangelet signal is higher, has longer  longitudinal extent and
reveals
a many-maxima structure in contrast to the rather smooth background.
\begin{center}
\begin{large}
{\em Long-lived strangelets}
\end{large}
\end{center}
  ``Stable'' objects,
capable to reach
and
pass through the apparatus without decay, i.e.
 having a lifetime $\tau_{0} \geq  10^{-8}$~s to traverse
 the CASTOR calorimeter, were named  long-lived strangelets.
Similarly, as in the case of cosmic ray strangelets,
 the simplified picture \cite{9} of the  interaction of a stable
strangelet in the
calorimeter obsorber was assumed.
Penetrating through the calorimeter a strangelet collides with tungsten
nuclei.
The mean interaction path of strangelets in the apparatus  absorber can be 
calculated from  equation (25).
In each act of collision the spectator part of a strangelet survives
 and continues
its
passage through the calorimeter and the wounded part is destroyed.
Particles
generated at the consecutive collision points  interact with tungsten
nuclei
in usual  way, resulting in the electromagnetic - nuclear
cascade
which
develops in the calorimeter.

 Penetration of stable strangelets through the
calorimeter, assuming $\alpha_{s}$ = 0.3 and several different sets of the
initial strangelet parameters was simulated
 ( $\mu_{q}$ = 300, 600, 1000 MeV;
   $A_{str}$ = 15, 20, 40;
  $E_{str} \approx$ 8 - 40 TeV (or 400 - 1000
 A GeV ).
Strangelets characterized by such values of the parameters could be
produced
at LHC energies \cite{CASTOR,Castor_gen,Sowa_note} according to the
picture proposed in \cite{8}.

\begin{figure}[ht]
\begin{center}
\mbox{
\epsfxsize=280pt
\epsfbox[1 0 511 566]{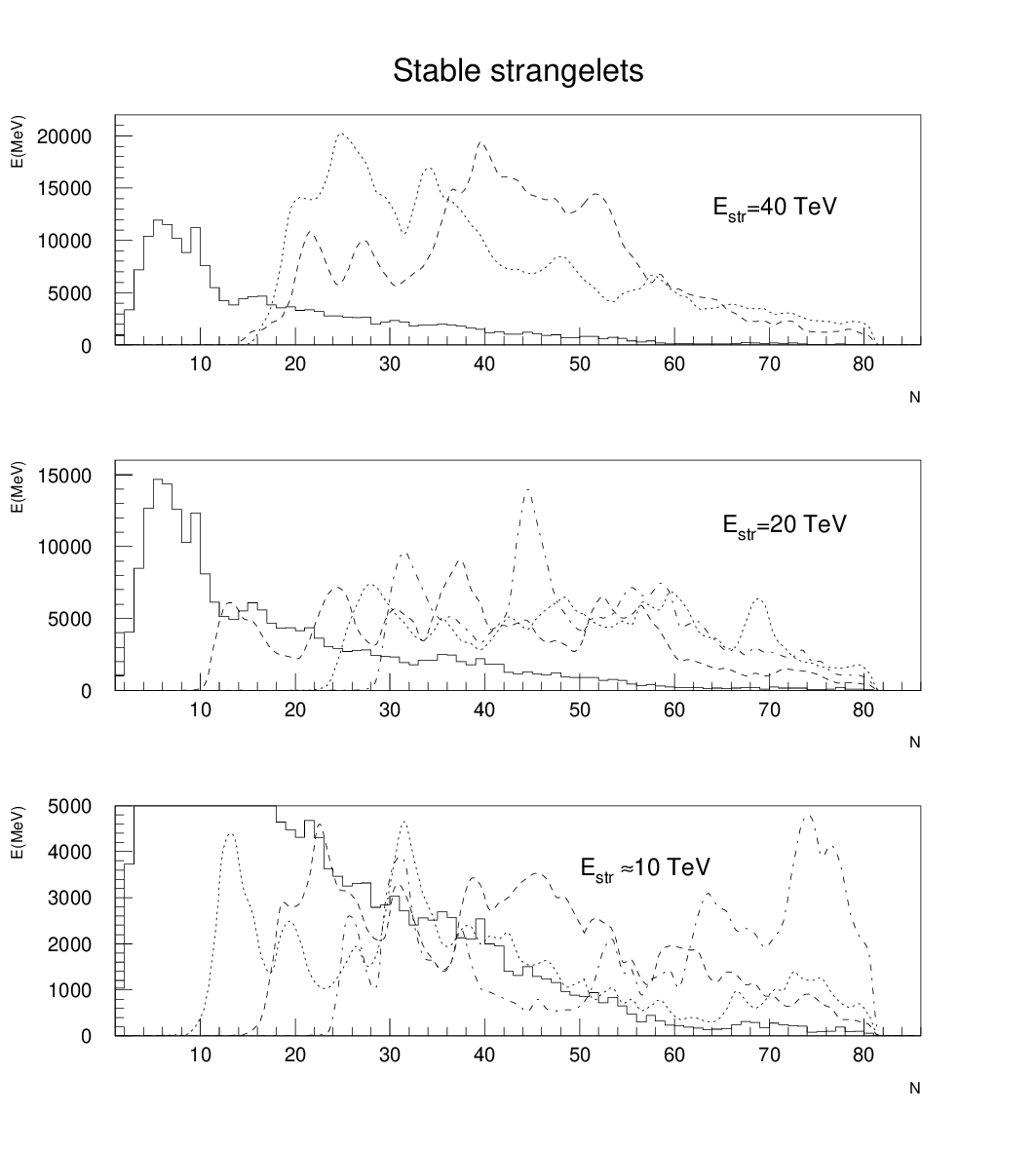}}
\end{center}
\vspace*{-1cm}
\caption [   
 Transition curves of stable strangelets.]
{
 Transition curves of stable strangelets with energy
 $E_{str}$ = 10-40 TeV,
 baryon number  $A_{str}$ = 15-40, quark chemical potential $\mu_{q}$ =
600,
  1000 MeV.
 Energy deposit (MeV) in the calorimeter layers, in the octant
containing a strangelet, is shown.
 Full line
 histograms  show the HIJING estimated background.}
\label{stable_3}
\end{figure}

Examples of transition curves produced in the CASTOR calorimeter by
various
stable strangelets are presented in
Fig.~\ref{stable_3}.
Here are mainly shown strangelets with $\mu_{q}$ = 600 MeV, as such value
of quark chemical potential has been estimated from cosmic ray Centauros.
The
curves
are limited to one  calorimeter octant containing a strangelet. The
strangelet cascade profiles are compared with those of the conventional
background,
produced by particles generated by HIJING (full line histogram), after 
subtraction of  energy carried by the strangelet.

 It has been concluded that:
\begin{enumerate}\vspace*{-2mm}
\item Stable strangelets can produce in the calorimeter long range
many-maxima cascades.\vspace*{-2mm}
\item The strangelet signal is manifestly different from that produced by
background from a conventional event, i.e.:\vspace*{-0.2cm}
\begin{itemize}
\item higher (in the hadronic part of the calorimeter),\vspace*{-0.2cm}
\item less attenuated,\vspace*{-0.2cm}
\item farther extended longitudinally  (some strangelets give a
strong
signal
even at the very end of the calorimeter, i.e. after 
penetration of more than 80 cm of  tungsten absorber (60
calorimeter layers), where the signal from a conventional event is
practically
negligible),\vspace*{-0.2cm}
\item has a different shape (reveals many-maxima structure in contrast to
the
smooth background).\vspace*{-0.2cm}
\end{itemize}  
\item The penetrating power of the signal increases with the value of
  the quark chemical potential $\mu_{q}$
 (strangelet cross sections decrease with increasing $\mu_{q}$), and
  the strangelet baryon number $A_{str}$.\vspace*{-2mm}
\item The longitudinal structure of transition curves depends mainly on
the strangelet baryon
number
$A_{str}$ (a many-maxima structure is more  pronounced for smaller
$A_{str}$).\vspace*{-2mm}
\end{enumerate}

The appearance of the many-maxima structure is the consequence of 
successive collisions of the strangelet with  nuclei of the calorimeter
material.
At each
act of  collision some part of the strangelet energy is transformed
into  energy of the secondary particles which in the process of
 usual  interactions initiate  nuclear
cascades in the calorimeter. Thus the  distance between
 consecutive humps depends both on the value of the mean interaction
path of the strangelet in the tungsten absorber $\lambda_{s-W}$ 
(hence on $\mu_{q}$ and $A_{str}$ ) and on the values of the mean
interaction paths of usual particles.

\vspace*{3mm}

From these simulations it is seen that
 the deep  
calorimeter
 is an appropriate tool for  strangelet detection, independently of the
strangelet
lifetime. Both stable and unstable strangelets  can produce
in the calorimeter signals apparently different from the conventional
background.
The calorimeter will be also  able to detect the possible
{\em strangelet evolution} when the strong nucleon emission process
transforms
a short-lived strangelet into a stable object.
Fig.~\ref{st_un} shows the transition curve of a short-lived
strangelet
($A_{str}$ =
23,
$E_{str}$ = 23 TeV) which after evaporation of 7 neutrons, in the strong
decay
process, becomes a long-lived object.

The probability of  strangelet detection in a calorimeter depends
both on the strangelet properties  and the calorimeter parameters.
Generally, such factors as  large depth, small longitudinal  sampling
length
and
fine granularity (division  into  radial as well as   azimuthal
sectors)
improve the sensitivity to strangelet detection.
The influence of some factors on the sensitivity to strangelets detection
has been studied \cite{Castor_str}. It was shown that
 a deep  calorimeter can be sensitive to the
detection of
strangelets for a  wide spectrum of their parameters. For 
 the considered design,
  stable strangelets with  total energy $E_{str} >$ 10 TeV, or
energy per baryon number $E_{str} \geq$ 500 A GeV  and  baryon number
$A_{str} \geq$ 15,
can be easily identified. Sometimes even less energetic strangelets,
because
of
favourable fluctuations, can produce very deep in the calorimeter
a characteristic
 signal allowing  their identification.

\begin{figure}[ht]
\vspace*{-1cm}
\begin{center}
\mbox{
\epsfysize=280pt
\epsfbox[0 0 547 555]{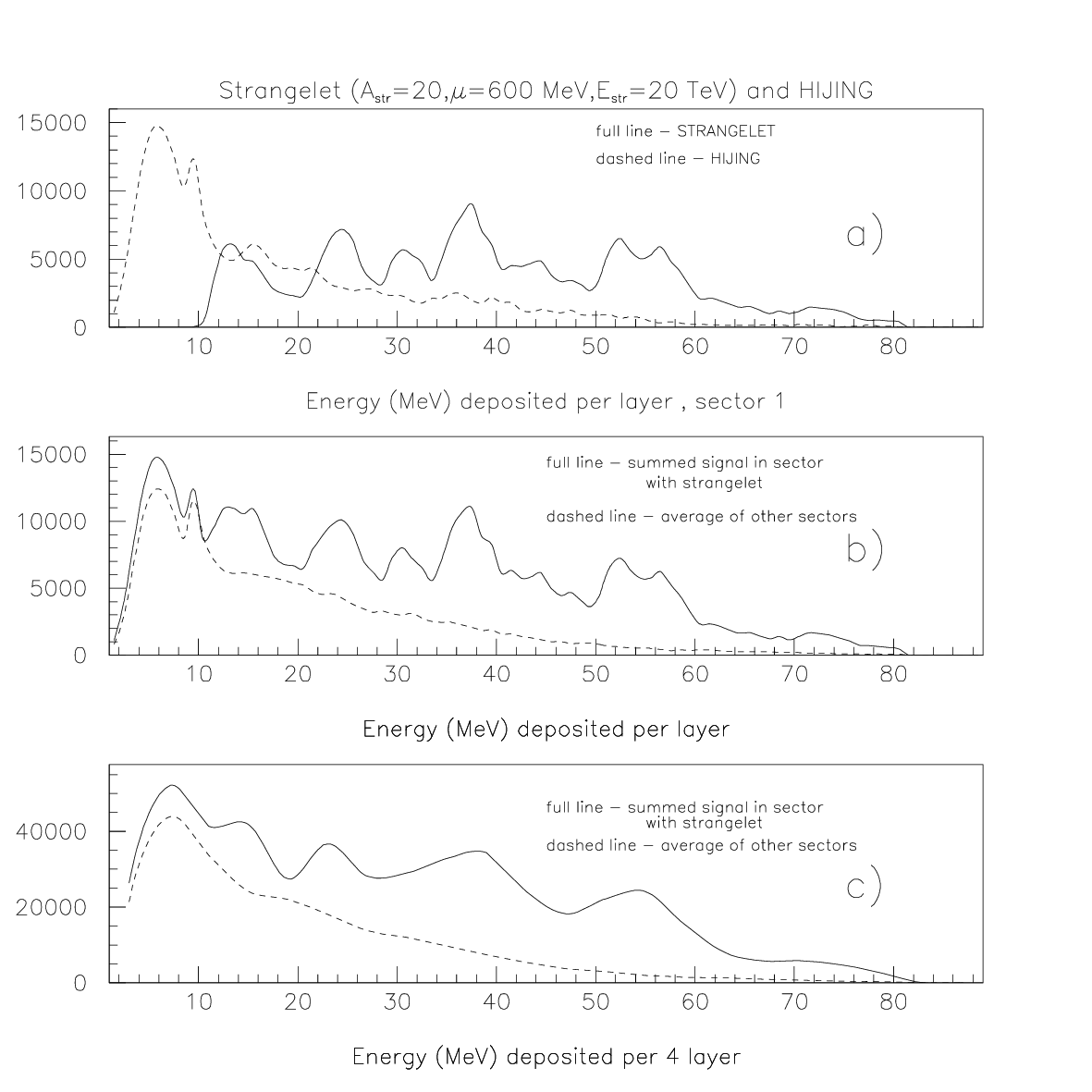}}
\caption[  
  Transition curves produced by a stable strangelet ($A_{str}$= 20,
 $E_{str}$ = 20 TeV, $\mu_{q}$ = 600 MeV) and by the HIJING in the CASTOR
 calorimeter.]
{ 
 a) Transition curves produced by a stable strangelet ($A_{str}$= 20,
 $E_{str}$ = 20 TeV, $\mu_{q}$ = 600 MeV) and by the HIJING
 background separately;
 b) Summed signal in the ``strangelet'' sector  in comparison with the
 average of other sectors assuming readings of
 every layer, i.e. every 5
 mm
 in electromagnetic and 10 mm in hadronic sectors;
 c) same as (b),
 but readings of the summed signal from consecutive groups  of 4
 layers.}
\label{ev}
\end{center}
\vspace*{-3mm}
\end{figure}

In order to identify any unusually penetrating component, 
the
development, intensity (energy content) and propagation of hadronic
cascades
as a function of  calorimeter depth, should be observed. To meet this
requirement, the 
calorimeter
must be sampled along its length, with appropriate sampling steps. The
simulations
presented here have been done for the sampling and reading planes
placed every 5 mm ($\sim 1.94 X_{0}$ of effective thickness) in the
electromagnetic
part and every 10 mm ($\sim 3.88 X_{0}$ of effective thickness) in the
hadronic
part of the calorimeter. Such sampling is similar to that in the cosmic 
ray  
emulsion chambers where the many-maxima long range cascades have been
observed and it 
seems to be also suitable for observation of the  many-maxima character
of
the
transition curves produced by  strangelets formed at the LHC. It has been
also checked that making absorber plates thinner leads only to a small
increase of the light output. An  important question is   how
a
many-maxima structure changes with the increase of the reading unit
thickness.
It is illustrated in
 Fig.~\ref{ev}  where  transition curves produced by a typical stable
strangelet ($A_{str}$ = 20, $E_{str}$ = 20 TeV, $\mu_{q}$ = 600 MeV)
are shown.
Fig.~\ref{ev} a
 shows the signals
produced by the strangelet and the HIJING
background, separately. Figs.~\ref{ev} b and \ref{ev} c
show the summed signal in the ``strangelet'' sector in comparison with the
average output from  the other sectors. Fig.~\ref{ev} a and \ref{ev} b
illustrate the standard
sampling and reading step as described above.
 Fig.~\ref{ev} c is the same
transition curve  obtained for readings of the summed signal from
consecutive groups of 4 layers (i.e. every
$\sim 16 X_{0}$). As it  can be expected, with decreasing the number of
reading
steps, the strangelet signal becomes smoother.
The  many-maxima structure becomes
rather a wave-like structure but
it is still visibly different from the still smoother background.

\subsubsection{``Mixed'' events  in the deep calorimeter}

 Further questions concern the shape of  transition
curves produced in the
calorimeter  by:
\begin{itemize}\vspace*{-0.2cm}
\item   strangelets coming from   the Centauro fireball explosion
and  registered in the apparatus simultaneously  with other
Centauro
decay products,\vspace*{-0.2cm}
\item  Centauro fireball decay products  without accompanying
strangelets emission 
      (or such case when a strangelet escapes
       the  detection in the calorimeter).\vspace*{-0.2cm}
\end{itemize}

To investigate this topic the exotic events generated 
   by means of the Centauro code
 were  passed  through the CASTOR
calorimeter,
by using GEANT 3.21.
 For each event  
the following transition curves were simulated separately:\vspace*{-0.2cm}
\begin{itemize}
\item a curve produced by
the Centauro fireball decay products,\vspace*{-0.2cm}
\item by the accompanying strangelet,\vspace*{-0.2cm}
\item and by background of conventionally produced
particles, possibly accompanying the Centauro event
      (estimated by HIJING).\vspace*{-0.2cm}
\end{itemize}

All these three contributions separately and also their sum,
 constituting the so--called ``mixed'' event,
were compared with the ``usual'' transition curve, produced by HIJING.
 Analysed Centauro events were characterized by  different
values of 
parameters: temperature ($T$ = 250, 300 MeV), quarkchemical 
  potential ($\mu_{q}$ =
 600, 1000 MeV) and nuclear stopping power ($\Delta y_{stop}$ =
 0.5, 1.0, 1.5 what corresponds to the effective stopping in the range
 of about $\sim$ 1.5 - 3.0 pseudorapidity units).
 In some events  strangelets
with baryonic numbers $A_{str} \simeq$  20-40 and energies
$E_{str} \simeq $ 8-20 TeV were formed, what corresponds to the energy per
baryon
 $E/A_{str} \simeq$ 0.3-1.0 TeV.

\begin{figure}
\begin{minipage}{7cm}
\hbox{
\epsfxsize=180pt
\epsfbox[43 150 536 658]{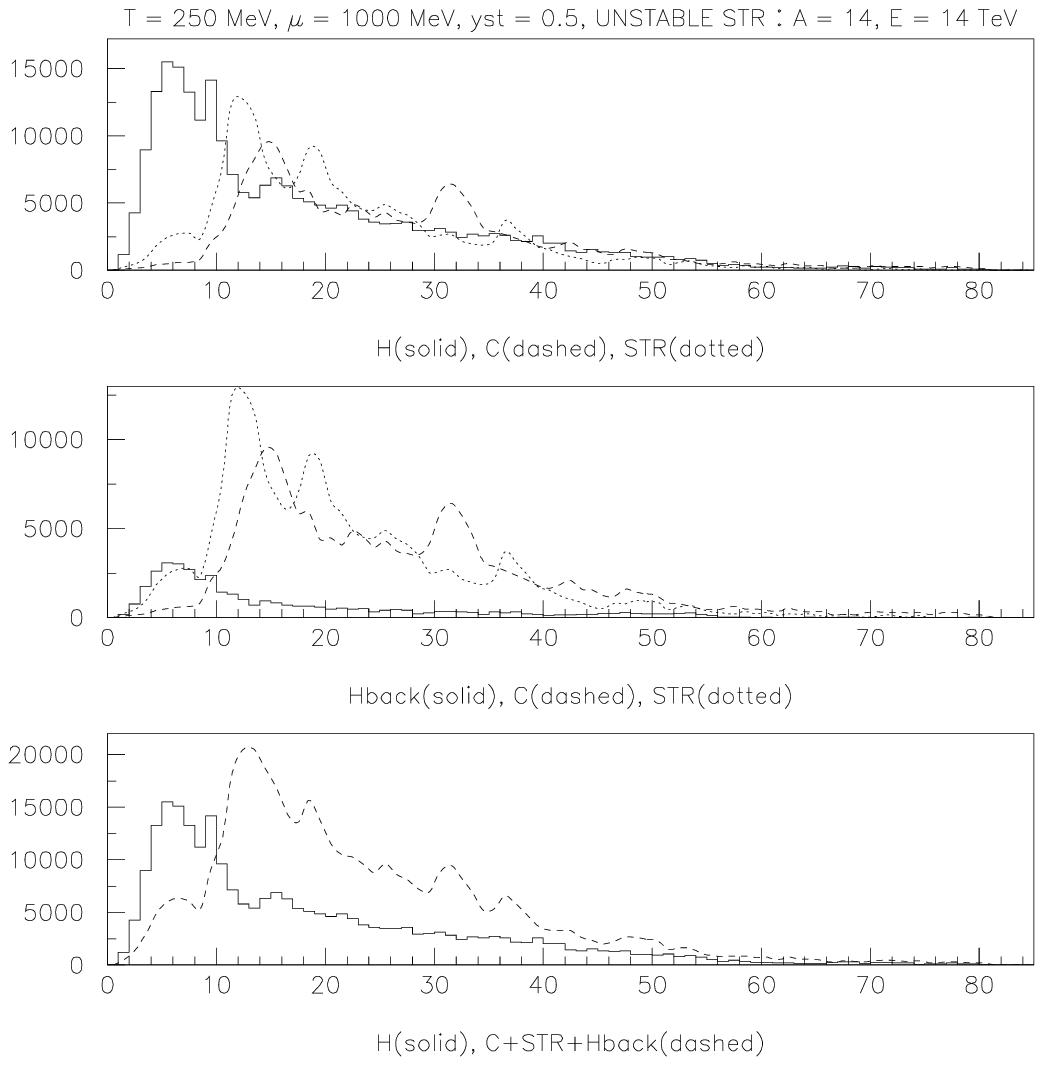}}
\end{minipage}
\begin{minipage}{7cm}
\hbox{
\epsfxsize=180pt
\epsfbox[43 150 536 658]{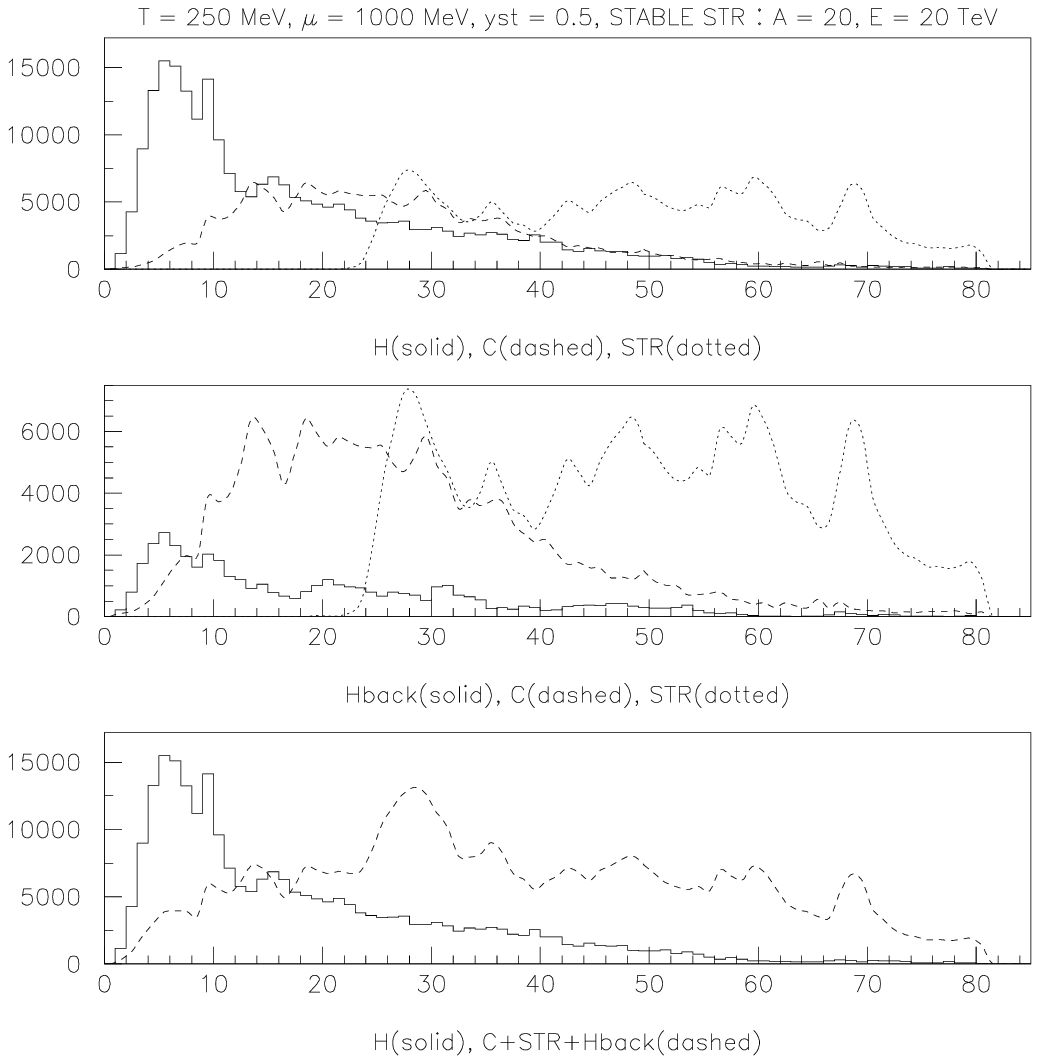}}
\end{minipage}
\caption[
 Transition curves produced by Centauro event 
   with unstable or stable strangelets, in comparison with
HIJING.]
{
 Transition curves produced by Centauro event ``C''
   with unstable or stable strangelets ``STR'', in comparison with
HIJING ``H''. 
 Energy deposit (MeV) in the calorimeter layers, in the octant
containing a strangelet, is shown.}
\label{st_20_20_bis}
\end{figure}

 Examples of resulting transition curves are presented  in
 Figs.~\ref{st_20_20_bis}, and 35.
 Energy deposit in the consecutive calorimeter layers, in the sector
containing a strangelet,  is shown.

\begin{figure}[ht]
\vspace*{-2cm}
\hbox{   
\epsfxsize=400pt
\epsfbox[43 150 536 658]{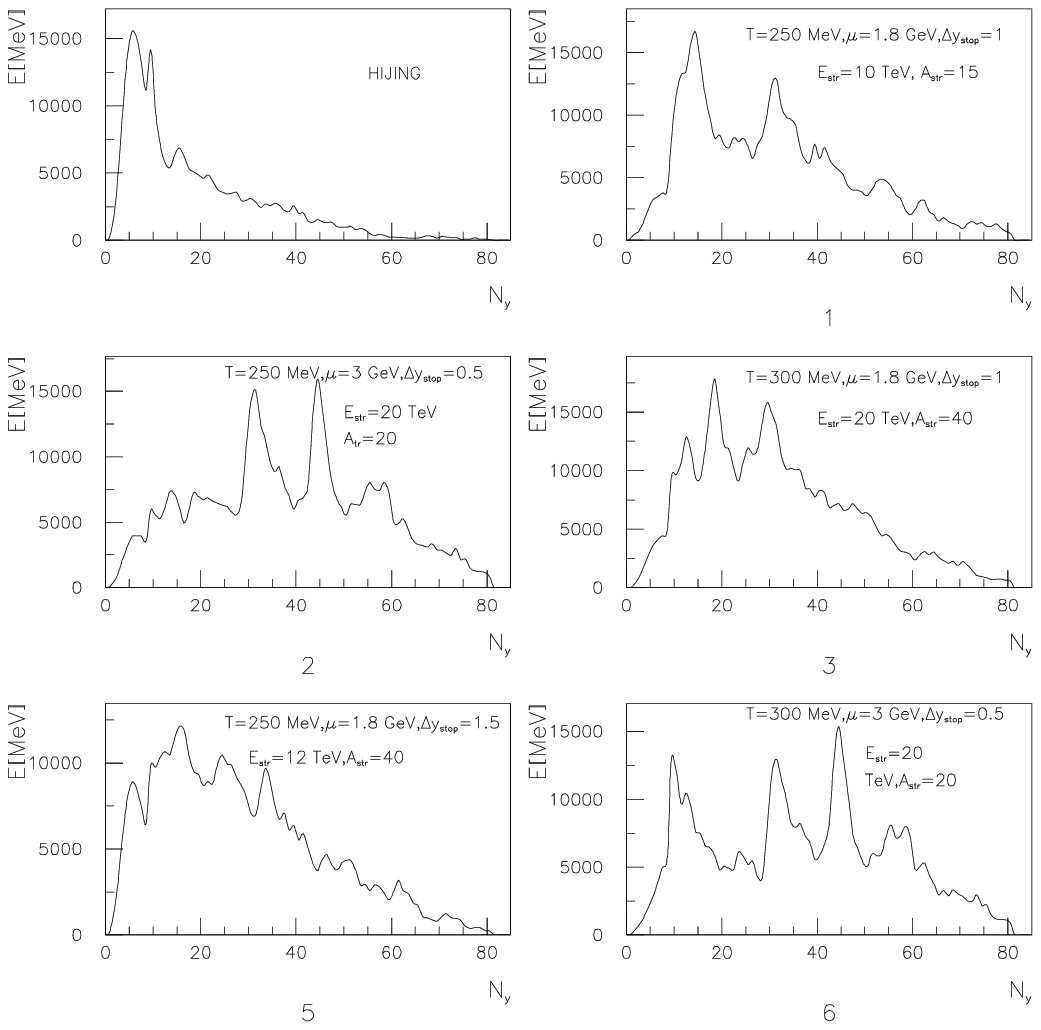}}
\caption[
Examples of resulting transition curves being the sum of
Centauro, strangelet and background contributions, and in comparison 
the  HIJING  event.]
{Examples of resulting transition curves being the sum of
Centauro, strangelet and background contributions, and in comparison 
the  HIJING  event.
 Energy deposit (MeV) in the calorimeter layers, in the octant
containing a strangelet, is shown.} 
\label{st_5ev}
\end{figure}

 Separate
contributions from: Centauro fireball decay products, a strangelet,
 the HIJING central event and a background from  conventionally (HIJING)
produced
particles are plotted in Fig.~\ref{st_20_20_bis} showing two
  Centauro events with unstable and stable  strangelets produced among its
secondaries.
Figs.~\ref{st_5ev}
illustrate the  signals, expected  in the calorimeter, assuming
that stable strangelets are formed during the Centauro fireball decay.

 This analysis indicates that  Centauro events (as well  accompanied
and not accompanied by a strangelet) can be easily distinguishable from  
``usual'' events. Centauro transition curves in the calorimeter
 are expected to have apparently different shape and longer extent
from those produced by  ``normal'' events.
  A Centauro produced signal
has a maximum at about 14th calorimeter layer with
the
average $\langle N_{Cent} \rangle \simeq$ 25. HIJING event produces the
maximum of the signal at  $\sim$ 8th calorimeter layer, with the average
at $\langle N_{HIJ} \rangle \simeq $ 19 layer.
 Generally, Centauro produced
signal is stronger in the deeper (hadronic)  part of the calorimeter,
in opposite to the HIJING generated one, which is peaked in the
electromagnetic section of the calorimeter.

Strangelet transition curves have been analysed in details in the previous
subsubsection, with the general conclusion that they are apparently
different
from those produced by ''normal`` events and that strangelets
produced among other conventional  particles should give easily 
distinguished signal. Here are shown
 signals, being the sum of Centauro, strangelet and background
contributions. These transition curves are again very different from
conventionally produced ones. They have much longer extent (they are
strongly pronounced in the deep hadronic part of the calorimeter) and
many-maxima structure. Their shape depends, of course, on the strangelet
and Centauro parameters. The important   quantity is the Centauro
and strangelet energy falling into the  calorimeter acceptance.
Generally, higher energy of the exotic species bears  smaller
conventional background   and in  consequence gives  
more pronounced signal.
 The Centauro fireball energy depends mainly on
the value of nuclear stopping power. The plot no 2 of
Fig.~\ref{st_5ev} is the example 
of the event in which the Centauro fireball energy, covered by the
calorimeter equals to 158 TeV, exceeding the HIJING predicted value
( $\sim$ 150 TeV). It results in  zero conventional background
and hence  the summed transition curve is  very different from the HIJING
predicted one. On the contrary, the  event illustrated in the plot no
5 of Fig.~\ref{st_5ev}
 in which the Centauro fireball energy equals
to  79.5 TeV, carries only about one half 
of the energy allowed by HIJING in that kinematical region. The remaining
energy could  go into the conventional particle production.
This fact in connection  with the small strangelet energy causes 
  the expected signals to be weaker than  others presented 
in Figs.~\ref{st_5ev}. But also such events have apparently different
characteristics than the ``normal'' ones and should  be easily 
picked up in the process of analysis  of the shape
and extent of transition curves
in the calorimeter. 

\vspace*{0.5cm}

Summarizing the results of simulations of exotic objects, it should be
stressed that the energy deposit in the deep calorimeters seems to be
the new unconventional signature of the quark--gluon plasma. Such objects
as Centauro--like events or strongly penetrating/long--lived particles
 possibly created in the phase transition from the QGP to the hadronic
 matter, should produce the
characteristic long extended signals,  during their passage through the
calorimeter.
 This signature should be
taken into account in the preparation of the new experiments.

 \section {Summary}
  In this work the experimental data from super-high energy cosmic ray
mountain experiments
 have been reviewed. In spite of many experimental uncertainties
 and some doubts, it can be concluded that they give a compelling
evidence  of unusual events which are hardly explained
 by means of ``conventional'' models. Among many theoretical
 attempts to understand these anomalies the more attractive ones
 are the scenarios with the QGP. One of them is the picture 
 assuming the production of a strange quark matter fireball
 in nucleus--nucleus collisions.
 At the last stage of the fireball evolution the Centauro--type events
 and strangelets can be formed. This model  allows to understand
 many different anomalies, such as the existence of hadron--rich events,
 a strongly penetrating component, mini--clusters etc. In
addition, such
picture is consistent with up to now negative results of Centauro search
in accelerator experiments and allows to expect their appearance
in  future, in heavy ion interactions at RHIC and LHC colliders. The
CASTOR, a subsystem
of the ALICE detector is proposed to study  the baryon--rich
 environment formed in  Pb+Pb collisions at LHC energies, and to search
for
exotic
events, such as 
Centauros and
 strangelets.
\vspace*{2mm}

\vspace*{0.5cm} \begin{Large}
{\bf Acknowledgements}
\end{Large}

\vspace*{1mm}
 I would like to thank all of my collegues with whom whenever I worked
  or discussed the topics  concerning  the 
 cosmic ray exotic
 phenomena.
 In  particular, many thanks  prof. Z.
W\l{}odarczyk for our common study of a strangelet question, prof. 
A.D. Panagiotou  and dr A.
Angelis for common investigations of Centauro problem and  
designing the CASTOR detector.
I am  grateful prof. J. Bartke for helpfull discussions  and 
very carefull  reading
of the raw version of the 
manuscript,  plenty important remarks and corrections.
This work was partly supported by Polish State Committee for Scientific
Research grant No.
 2P03B 011 18 and SPUB-M/CERN/P03/DZ1/99.

\end{document}